\pdfoutput=1
\documentclass[11pt]{book}
\usepackage{amsmath, amsthm, amssymb,cancel,color,array,colortbl}
\usepackage{epsfig,sidecap}
\usepackage{hyperref}
\usepackage{wasysym,dsfont}
\usepackage{slashed,braket,stmaryrd}
\usepackage[toc,page]{appendix}
\usepackage{wrapfig}
\usepackage{fancyhdr}
\usepackage{datenumber}
\newcommand{\nn}{\nonumber}
\newcommand{\ii}{\mathrm{i}}
\newcommand{\pr}{\prime}
\newcommand{\op}{{\cal Q}}
\newcommand{\dd}{\mathrm{d}}
\newcommand{\gs}[1]{\slashed{#1}}
\newcommand{\MSbar}{$\overline{\mathrm{MS}}$ }

\newcommand{\be}{\begin{eqnarray}}
\newcommand{\ee}{\end{eqnarray}}

\newcommand{\epsUV}{\epsilon_{\mathrm{UV}}}
\newcommand{\epsIR}{\epsilon_{\mathrm{IR}}}

\newcommand{\uc}{\hat{y}}
\newcommand{\xc}{\hat{x}}
\newcommand{\Brfrac}{\mathrm{Br}^{\mathrm{NLO}}/\mathrm{Br}^{\mathrm{LO}}}

\newcommand{\sy}{\hat{s}}
\newcommand{\uy}{\hat{u}}

\newcommand{\Leff}{{\cal L}_{\mathrm{eff}}}

\newcommand{\bs}[2]{\begin{array}{r}#1\\#2\end{array}}
\newcommand{\LINE}[5]{#1 & #2 & #3 & #4 & #5}

\definecolor{orange}{rgb}{1.0,0.38,0.0}
\definecolor{smurfblue}{rgb}{0.223,0.47,0.85}
\definecolor{nicered}{rgb}{0.7,0.1,0.1}
\definecolor{nicegreen}{rgb}{0.1,0.5,0.1}
\hypersetup{colorlinks=true,linkcolor=smurfblue, citecolor=orange, linktoc=all}

\begin{document}

\pagestyle{empty}

\begin{center}
\LARGE   University of Ljubljana 
\end{center}
\begin{center}
\LARGE  Faculty of Mathematics and Physics
\end{center}

\vspace{3cm}
\begin{center}
\Large Jure Drobnak
\end{center}
\begin{center}
\bf \LARGE Constraints on new physics from top quark decays at high precision
\end{center}

\vspace{0.5cm}
\begin{center}
\large Ph.D. Thesis
\end{center}

\vspace{4cm}
\begin{center}
\Large Advisor: Prof. Dr. Svjetlana Fajfer
\end{center}

\begin{center}
\Large Co-Advisor: Dr. Jernej F. Kamenik
\end{center}

\vspace{2cm}
\begin{center}
Ljubljana, 2012
\end{center}
\thispagestyle{empty}
\newpage
{\color{white}a}
\newpage
%
\pagestyle{empty}

\begin{center}
\LARGE Univerza v Ljubljani
\end{center}
\begin{center}
\LARGE Fakulteta za Matematiko in Fiziko
\end{center}

\vspace{3cm}
\begin{center}
\Large Jure Drobnak
\end{center}
\begin{center}
\bf \LARGE Natan\v{c}no omejevanje prispevkov nove fizike preko razpadov kvarka top
\end{center}

\vspace{0.5cm}
\begin{center}
\large Disertacija
\end{center}

\vspace{4cm}
\begin{center}
\Large Mentorica: Prof. Dr. Svjetlana Fajfer
\end{center}

\begin{center}
\Large So-mentor: Dr. Jernej F. Kamenik
\end{center}

\vspace{2cm}
\begin{center}
Ljubljana, 2012
\end{center}
\thispagestyle{empty}
\newpage



\chapter*{Povzetek}\thispagestyle{empty}
\vspace{-1.cm}
V tem delu raziskujemo kako lahko razli\v{c}na teoreti\v{c}na odstopanja od okvira {\sl Standardnega modela} (SM) vplivajo na lastnosti razpadov kvarka top. Manifestacijo {\sl nove fizike} (NF) onkraj SM, katere energijska skala znatno presega skalo elektro-\v{s}ibkih pojavov, parametriziramo v obliki vrste vi\v{s}je dimenzionalnih operatorjev in z uporabo metod efektivne teorije polja.

V prvem delu obravnavamo NF, ki bi lahko vplivala na nevtralne tokove kvarkov top, ki spreminjajo okus. Ti tokovi so v okviru SM zelo redki in v primeru, da bi jih v bodo\v{c}e eksperimentalno odkrili, bi to pomenilo odkritje NF. Bolj konkretno se zanimamo za dvo-del\v{c}ne razpade $t\to q Z,\gamma$ in tro-del\v{c}ne razpade $t\to q \ell^+ \ell^-$. V analizo dvo-del\v{c}nih razpadov vklju\v{c}imo popravke prvega reda kvantne kromodinamike. Preu\v{c}ujemo tako me\v{s}anje operatorjev pod renormalizacijo, kot tudi kon\v{c}ne popravke matri\v{c}nih elementov, vklju\v{c}no s tako imenovanimi ``bremsstrahlung'' procesi. Izka\v{z}e se, da so popravki kvantne kromodinamke, zlasti v razpadih $t\to q \gamma$, lahko pomembni in vplivajo na interpretacijo eksperimentalnih meritev. Glavni motiv za obravnavo tro-del\v{c}nih razpadov je pove\v{c}anje faznega prostora kon\v{c}nega stanja, kar pomeni, da imamo na voljo ve\v{c}je \v{s}tevilo kinemati\v{c}nih opazljivk, ki nam lahko pomagajo razlikovati med razli\v{c}nimi strukturami NF v nevtralnih tokovih kvarka top, ki spreminjajo okus.

V drugem delu pa se zanimamo za NF, ki bi lahko vplivala na glavni razpadni kanal kvarka top $t\to W b$. Ta je tesno povezan z nabitimi tokovi kvarkov. Z vnosom novih operatorjev, ki spremenijo lastnosti $tWb$ interakcije, v teorijo, ne vplivamo le na glavni razpadni kanal kvarka top. Indirektne posledice, ki se pojavijo v teoreti\v{c}nih napovedih opazljivk v fiziki mezonov so lahko znatne, saj virtualni kvarki top v redkih procesih igrajo vodilno vlogo. V tem delu zato podrobno preu\v{c}imo indirektne omejitve na NF v nabitih tokovih kvarka top, ki izvirajo iz fizike mezonov $B$. \v{S}ele, ko odkrijemo intervale za parametre NF, ki so skladni z meritvami mezonske fizike, se osredoto\v{c}imo na posledice v vi\v{s}je energijskih procesih, razpadih kvarkov top na masni lupini. Osrednje opazljivke na\v{s}e analize so {\sl su\v{c}nostni dele\v{z}i} (ang. {\sl helicity fractions}) bozona $W$, ki nastane pri razpadu kvarka top. Ker se kvantno kromodinamski popravki pri obravnavi teh opazljivk izka\v{z}ejo za pomembne v analizi SM, take popravke obravnavamo tudi ob vklju\v{c}itvi NF. Iz eksperimentalnih meritev, ki za enkrat ka\v{z}ejo ujemanje s SM, lahko postavimo direktne omejitve na NF, ki jih primerjamo z indirektnimi. V tem se nam razkrije zanimiva povezava med fiziko kvaka top in fiziko mezonov $B$.

\vspace{0.3cm}
\small
{\noindent \sf Klju\v{c}ne besede}: Razpadi kvarka top, nevtralni tokovi, kvark top in nova fizika, efektivne teorije in nova fizika, popravki kvantne kromodinamike, su\v{c}nostni dele\v{z}i bozona $W$, indirektne omejitve nove fizike


{\noindent \sf PACS}: 12.15.Mm, 12.38.Bx, 12.60.Cn, 12.60.Fr, 13.88.+e, 14.65.Ha
\normalsize

\chapter*{Abstract}\thispagestyle{empty}
\vspace{-1.0cm}
We study possible theoretical deviations from {\sl Standard Model} (SM) in top quark physics which alter the decay properties of the top quark. Using effective filed theory techniques we parametrize the effects of potential {\sl new physic} (NP) of scales well above the electroweak scale in terms of effective operators. 

On one side we investigate NP manifestation in the form of {\sl flavor changing neutral current} (FCNC) decays of the top quark, which are highly suppressed in SM and potential observation of which would undoubtedly signal the presence of NP. We examine the two-body $t\to q Z,\gamma$ and three-body $t \to q \ell^+ \ell^-$ decays. Our analysis of the two-body FCNC decays is performed at {\sl next-to-leading order} (NLO) in {\sl quantum chromodynamics} (QCD). We examine the effects of operator mixing under the QCD renormalization as well as the finite matrix element corrections along with appropriate bremsstrahlung processes. We find that the effects of FCNC operators mixing under renormalization can be substantial, especially in the case of $t\to q \gamma$ decays and QCD corrections affect the way the experimental measurements are to be interpreted. In the three-body decays we aim to exploit the increased phase space of the final state by defining different types of observables which could help to discriminate between structures of the vertices governing the FCNC transition of the top quark.

On the other side we examine possible deviations from SM predictions in top quark's main decay channel, which is governed by the charged quark current interactions. Introduction of higher dimensional operators that modify $tWb$ interactions, however has additional consequences that have yet not been thoroughly analyzed in the literature. In particular, low energy observables of rare processes in $B$ physics, where virtual top quarks and their charged current interactions play a dominant role, are expected to be affected. We therefore perform a detailed study of indirect constraints on the NP operator basis and only then turn to the study of the effects in the decays of on-shell produced top quarks. The observables of interest are the helicity fractions of the $W$ boson produced in the main decay channel. Since for SM predictions of helicity fractions higher order QCD corrections prove to be crucial, we conduct the analysis of NP effects at NLO in QCD. We confront our predictions with the experimental measurements to obtain the direct constraints on NP further comparing them with the indirect constraints from low energy processes revealing an interesting interplay of top and bottom physics. 

\vspace{0.3cm}
\small
{\noindent \sf Key Words}: Top quark decays, flavor changing neutral currents, new physics in top quark physics, effective theory approach to new physics, QCD corrections, helicity fractions of $W$ boson, indirect constraints of new physics

{\noindent \sf PACS}: 12.15.Mm, 12.38.Bx, 12.60.Cn, 12.60.Fr, 13.88.+e, 14.65.Ha
\normalsize

\begingroup
\hypersetup{linkcolor=black}
\tableofcontents \thispagestyle{empty}
\endgroup

\addtocontents{toc}{\protect\thispagestyle{empty}}

\newpage
\thispagestyle{empty}

\pagestyle{fancy}
\renewcommand{\sectionmark}[1]{ \markright{\thesection.\ #1}}
\renewcommand{\chaptermark}[1]{\markboth{\MakeUppercase{ #1}}{}}
\chapter{Introduction}
\setcounter{page}{1}
\section{The Standard Model}
The goal of theoretical high energy physics is to mathematically describe phenomena occurring at lowest experimentally accessible length scales using as the main tool quantum field theory. Theoretical predictions are put to the test by some of the most sophisticated experiments in the world ranging from different kind of particle colliders, to satellites orbiting our planet. Continuous advances in theoretical insight and experimental techniques in the past century have led to the formulation of the {\sl Standard Model} (SM), theoretical framework describing the content of elementary particles and their interactions with its formulation dating back to the 1960's \cite{Weinberg:1967tq}.

SM is remarkable both in its simplicity and great predictive power which has been put under enemas scrutiny over the last century, most recently by the {\sl Large Hadron Collider} (LHC). 

In this introductory section we aim to give a very brief overview of the main theoretical features of the SM. Our discussion is kept on a purely informative level providing however appropriate references to be consulted for many underlaying details. A bit more time is spent on the discussion of flavor aspect of SM, since the concepts encountered there prove to be crucial for our top quark studies. At the end of the section we try to motivate the need for theoretical explorations beyond the SM, since this is the frontier that we shall be crossing in our work.
\subsection{Particle content}
The corner stone of SM as a quantum field theory is the gauge group under which the theory is to be invariant
\begin{eqnarray}
SU(3)_c \times SU(2)_L \times U(1)_Y\,. \label{eq:sm_gauge_group}
\end{eqnarray}
Here $SU(3)_c$ is the gauge group of {\sl quantum chromodynamics} (QCD), $SU(2)_L$ is the weak isospin group and finally, the Abelian $U(1)_Y$ is called the weak hypercharge. The corresponding coupling constants of the three groups are denoted by $g_s$, $g$ and $g^{\prime}$ respectively. Specifying the gauge group entirely fixes the content of the gauge boson sector. On the other hand, there is more freedom in specifying the scalar and fermionic sector of the theory. In particular we have to assign what representations of the gauge group the fields are to be in.

The only scalar field in SM is the Higgs boson. Its representations under the gauge groups can be written as $\phi(1,2)_{+1/2}$, meaning that it is a singlet under $SU(3)_c$, a doublet under $SU(2)_L$ and it carries a hypercharge of $+1/2$. At present it  remains the only quantum of the SM that has not been experimentally confirmed, however the recently  discovered new particle at LHC \cite{ATLAS:higgs, CMS:higgs} poses to be a very strong candidate. Higgs boson plays a very specific role in the SM, since by acquiring a {\sl vacuum expectation value} (VEV) it spontaneously breaks the SM gauge group
\begin{eqnarray}
SU(3)_c \times SU(2)_L \times U(1)_{Y} \xrightarrow{\langle \phi \rangle} SU(3)_c \times U(1)_Q\,, \hspace{0.5cm} Q = Y + T_3\,,
\end{eqnarray}
where $U(1)_Q$ is the gauge group of {\sl quantum electrodynamics} (QED), with $Q$ the electromagnetic charge and $T_3$ the eigenvalue of the diagonal $SU(2)_L$ generator. QED gauge coupling is $e = g \sin \theta_W = g^{\prime} \cos \theta_W$, where $\theta_W$ is the Weinberg mixing angle.  The described pattern of symmetry breaking gives rise to mass terms for the three weak gauge bosons through the covariant derivative $(D_{\mu}\phi)^{\dagger}(D_{\mu}\phi)$ part of the Higgs Lagrangian and to mass terms for the fermions through the so-called Yukawa interactions, which we shall return to shortly.

Turning to the fermionic sector of the SM, we first note the subscript $L$ of the weak gauge group, which describes an important postulate of the SM, that only left-handed fermions carry weak-hypercharge.

Further we classify fermions depending on what representation of $SU(3)_c$ they are in. Singlets are called leptons and quarks are postulated to be in its fundamental representation.
\begin{table}[h]
\begin{center}
\begin{tabular}{c|c|c}\hline \hline
 &Quarks & Leptons\\ \hline
left handed& 
$Q_L(3,2)_{+1/6} =\Big(\hspace{-0.15cm} \begin{array}{c} u_L\\ d_L\end{array}\hspace{-0.15cm} \Big)$&
$L_L(1,2)_{-1/2} =\Big(\hspace{-0.15cm}\begin{array}{c}\nu_{L} \\ \ell_L\end{array}\hspace{-0.15cm} \Big)$ \\
right handed &$u_R(3,1)_{+2/3}$, $d_R(3,1)_{-1/3}$ &$\ell_R(3,1)_{-1}$  \\ \hline \hline
\end{tabular}
\caption{SM fermions and their representations under the SM gauge group (\ref{eq:sm_gauge_group}). Subscripts $L$ and $R$ denote the chirality of the fields.}
\label{tab:sm_fermions}
\end{center}
\end{table}

The fermionic sector of the SM is further enriched by making three repetitions (generations) of gauge representations described above and given in Tab.~\ref{tab:sm_fermions}. Each of the repetitions is assigned a flavor allowing us to distinguish among them. We define $\{u, c, t\}$ up-type quarks, $\{d,s,b\}$ down-type quarks, $\{e, \mu, \tau\}$ charged leptons and the accompanying neutrinos $\{\nu_e, \nu_\mu, \nu_\tau\}$. Since it is not crucial for our studies, we shall not deal with the leptonic part of flavor physics and concentrate only on the quarks.
\subsection{Flavor}
\label{sec:intro_fcnc}
The only part of the SM Lagrangian that is not flavor universal, meaning that it distinguishes between different flavors, is the Yukawa interaction term 
\begin{eqnarray}
{\cal L}_Y = - \bar{Q}_L^i [Y_d]_{ij} \phi d_R^j-\bar{Q}_L^i [Y_u]_{ij} \tilde{\phi} u_R^j + \mathrm{h.c.}\,. 
\label{eq:intro_yuk}
\end{eqnarray}
Indices $i$ and $j$ denote the flavor and we have introduced the ``up" and ``down'' $3\times 3$ complex Yukawa matrices and $\tilde{\phi} = \ii \sigma^2 \phi$. While there are two Yukawa matrices for the quarks, there is only one for the leptons, because as evident from Tab.~\ref{tab:sm_fermions}, there are no right-handed neutrinos in the SM. 

In the SM the Yukawa sector is the only source of flavor physics. This statement can be put in a more formal group theoretical form by saying that Yukawa interactions break the big global symmetry of flavor
\begin{eqnarray}
\mathcal G^{\mathrm{SM}} = U(3)_Q \times U(3)_u \times U(3)_d\,, 
\label{eq:SM_G_flav}
\end{eqnarray}
obtained when three generations of fermions are introduced to the theory. $U(3)_{Q,u,d}$ are groups of rotations in flavor space $V_{Q,u,d}$ that can be applied to $Q$, $u$ and $d$ quark fields respectively
\begin{eqnarray}
Q_L \xrightarrow{U(3)_Q} Q_L^\prime= V_Q Q_L\,,\hspace{0.5cm}
u_R \xrightarrow{U(3)_u} u_R^{\prime}=V_u u_R\,,\hspace{0.5cm}
d_R \xrightarrow{U(3)_d} d_R^{\prime}=V_d d_R\,,
\end{eqnarray}
where we have suppressed the flavor indices. Omitting the scalar field, a general $\mathcal G^{\mathrm{SM}}$ rotation effects the Yukawa terms in the following way
\begin{eqnarray}
\bar{Q}_LY_d d_R = \bar{Q}_L^\prime\,V_Q Y_d V_d^{\dagger}\,d_R^{\prime}\,,\hspace{0.5cm}
\bar{Q}_LY_u u_R = \bar{Q}_L^\prime\,V_Q Y_d V_u^{\dagger}\,u_R^{\prime}\,,
\end{eqnarray}
which can be seen as a change of basis in the flavor space. We can use the rotations of the broken symmetries to rotate away all the unphysical parameters of the Yukawa sector, since we know that out of all the parameters, there are as many unphysical as there are generators of broken symmetries
$$
N_{\mathrm{phys.}}= N_{\mathrm{all}} - N_{\text{broken gen.}}\,.
$$
We note that there is a remaining $U(1)_B$ global flavor symmetry even after inclusion of Yukawa terms, associated with the baryon number conservation. This means that we start of with  $36$ ($18$ real and $18$ imaginary) free parameters of Yukawa matrices and break $26$ out of $27$ generators of the global symmetry. $17$ of which are rotations containing phases, and $9$ are rotations with no phases\footnote{$U(3)$ can be written as $U(1)\times SU(3)$. The $U(1)$ transformation is rephasing $\mathrm{e}^{i\beta}$. $SU(3)$ transformation can be written as $\mathrm{e}^{\ii \alpha_a T^a}$, where $T^{2,4,7}$ are imaginary and contribute real rotations, while $T^{1,3,4,8}$ are real and contribute rotations with rephasing.}, this leaves us with $10$ physical flavor parameters, $9$ of which are real and $1$ is a complex phase. 

Using $V_{Q,u,d}$ we can rotate quark fields to the basis where one of the Yukawa matrices is diagonal
\begin{eqnarray}
\bar{Q}_L^{\prime}\underbrace{V_Q Y_d V_d^\dagger}_{Y_d^{\mathrm{diag.}}} d_R^\prime + \bar{Q}_L^{\prime} \underbrace{V_Q V_x^\dagger}_{V^\dagger} \underbrace{V_x Y_u V_u^\dagger}_{Y_u^{\mathrm{diag.}}} u_R^\prime\,. \label{eq:down_basis}
\end{eqnarray}
In the second term we had to insert a unitary matrix $V_x$ to obtain an expression with a diagonal $Y_u$, which is however multiplied with a unitary matrix $V$, called the {\sl Cabbibo-Kobayashi-Maskawa} (CKM) matrix, which was formulated in \cite{Cabibbo:1963yz,Kobayashi:1973fv} and will be subject of more discussion later. This basis is usually referred to as the ``down basis'', while we could have performed an analog diagonalization of the $Y_u$ matrix, ending up in ``up basis''.

When the $SU(2)_L\times U(1)_Y$ gauge symmetry is broken by the VEV of the Higgs field $\langle \phi \rangle =(0,v/\sqrt{2})$, $\mathcal L_Y$ can be written as
\begin{eqnarray}
\mathcal L_Y = - \frac{v}{\sqrt{2}}\bar{d}^\prime_L Y_d^{\mathrm{diag.}} d_R^\prime - \frac{v}{\sqrt{2}} \bar{u}_L^\prime V Y_u^{\mathrm{diag.}}u_R^\prime + \cdots \,. \label{eq:breaking}
\end{eqnarray}
Note that we have obtained a mass term for every down-type quark. Since $SU(2)_L$ has been broken, we can rotate the $u_L$ fields with a separate $V_{u_L}$ rotation. By choosing the rotational matrix to be the CKM matrix $\bar{u}_L^{\prime\prime} = \bar{u}_L^\prime V$, we obtain the mass terms for the up quarks as well. This rotation has important consequences in the charged current sector, where it generates flavor changing currents. Removing all the primes from the quark fields and reintroducing the flavor indices the charged current Lagrangian is
\begin{eqnarray}
\mathcal L_{\mathrm{cc}} =- \frac{g}{\sqrt{2}}
\big[\bar{u}_{iL}\gamma^{\mu}d_{jL}\big]V_{ij}W_{\mu}^+ - \frac{g}{\sqrt{2}} 
\big[\bar{d}_{jL}\gamma^{\mu}u_{iL}\big]V^*_{ij}W_{\mu}^-\,.
\label{eq:SMcc}
\end{eqnarray}
On the other hand, due to the unitarity of CKM matrix $V$, there are no tree-level {\sl flavor changing neutral currents} (FCNCs) in SM\footnote{There are no dimension 4 operators in SM that would generate FCNC transitions.}. The elements of the CKM matrix are usually denoted as
\begin{eqnarray}
V = \left(\begin{array}{ccc}
V_{ud}& V_{us}& V_{ub}\\
V_{cd}& V_{cs}& V_{cb}\\
V_{td}& V_{ts}& V_{tb}\\
\end{array}\right)\,,
\label{eq:CKMmat}
\end{eqnarray}
where not all of the parameters are independent. We have accounted for $6$ of the flavor parameters to be quark masses, meaning that the CKM matrix must contain the remaining $4$ of which $3$ are real and $1$ is a complex phase. 

The CKM mechanism has proven to be very successful in describing flavor physics which has been tested with various high precision experiments. The so called CKM unitary triangle has been over-constrained by numerous measurements and is showing remarkable consistency. Furthermore a highly diagonal dominated form of the matrix has been experimentally established and perhaps most importantly, the complex phase has been measured to be non-zero, proving that the discreet symmetry of simultaneous {\sl charge conjugation and parity} (CP) is indeed violated in nature. The CKM phase is the only source of this violation within SM.
For in-depth coverage of the subjects on CKM mechanism and CP violation we refer the reader to the following references \cite{Lavura:CP, Bigi:CP,Charles:2004jd}. 
\subsection{Need to go beyond SM}
\label{seq:strategy}
Despite the unprecedented success of SM it is evident that it cannot be the final theory of elementary particles and their interactions. The discovery of dark matter (see for example \cite{Olive:2003iq,Trimble:1987ee}) and neutrino oscillations \cite{Fukuda:1998mi} show that the particle content of SM is not adequate since there is no dark matter candidate and neutrinos are massless in SM. While the CP violation at low energies is very well described by the CKM mechanism, there are strong indications, that far more violation is needed to explain the high dominance of matter over anti-matter that we observe in the universe today \cite{Shaposhnikov:1987tw,Canetti:2012zc,Kajantie:1996mn}.

What is more, SM does not try to describe gravitational interactions which become relevant at very high energies, of the Planck scale $\Lambda_{\mathrm{P}}=\sqrt{\hbar c/G_{\mathrm{N}}} \sim 10^{19} $ GeV. Assuming that SM could be a valid theory all the way to the Planck scale gives rise to a puzzling situation referred to as the ``hierarchy problem'' \cite{Martin:1997ns,Wells:2009kq}. The word hierarchy is related to the large separation between the electroweak and the Planck scale. Due to the fact that the dimensionality of the Higgs mass operator is $2$, the radiative corrections to the mass $\delta m_H$ turn out to be quadratically proportional to the cutoff scale $\Lambda$
\begin{eqnarray}
m_H^2 = m_H^{(0)2} - \frac{\lambda^2}{(4\pi)^2} \Lambda^2 + \cdots \,.
\end{eqnarray}
Under the assumption that $\Lambda \sim \Lambda_{\mathrm{P}}$, a great deal of fine-tuning ($\sim$ 30 orders) between the parameters entering the $m_H^2$ expression is necessary in order to obtain the appropriately low Higgs mass (of the electroweak order).

The different views about its significance as a problem of the theory has earned the hierarchy problem to be labeled as somewhat controversial. Nevertheless it is at least an aesthetic issue and one that {\sl beyond Standard Model} (BSM) theories (often in the literature referred to as UV completions) try to address and do so in many different manners.

Following the aesthetic drive, the idea to unify interactions by embedding the SM product gauge group~(\ref{eq:sm_gauge_group}) into a single larger group comes naturally. The so called grand unified theories aim to do just that~\cite{Mohapatra:1999vv}.

Because the flavor parameters, masses as well as mixing parameters described by the CKM elements, exhibit strong hierarchy one can not help but wonder if there is an underlaying symmetry manifested at energies above the electroweak scale that could explain it. SM provides no answer to this question which has become known as the flavor problem of SM. 

There is clearly more than enough reasons for theorists to explore the possibilities of {\sl new physics} (NP) BSM and further to look for observables and processes which could help to discover NP and to discriminate between different NP scenarios. In many of them, the top quark provides a preferred search window due to its large coupling to the physics responsible for the electroweak symmetry breaking. A fascinating possibility, that we shall be exploring in our work, is that the top quark properties exhibit deviations from their predictions within the SM.

In the next section we argue that among other places, also the top quark decays are good ``hunting grounds" for NP. In particular decay modes and observables which are within the SM predicted to be highly suppressed are promising for detection of NP effects, since typically a non-zero measured signal of such quantities would, right-away, present a signal of physics beyond SM. 
Important thing to keep in mind is that in addition to directly probing top quark physics at LHC and Tevatron, top quark properties can also be explored in lower energy phenomena of meson physics, where top quark appears as a virtual particle often having the leading role in rare processes. 

In our exploration of BSM top quark physics, we shall not be committing to particular NP models or frameworks. Rather we will take the effective field theory approach, adding to the SM effective Lagrangians with which will be able parametrize our ignorance about the BSM theory. We will then study different observables which might be sensitive to our additions. In principle our results may be applicable to a variety of BSM models.
\section{Top quark decays}
Top quark is the heaviest experimentally confirmed elementary particle. It was discovered at the Tevatron in 1995 \cite{Abe:1995hr,Abachi:1995iq}. The two main features of the top quark are its large mass, experimentally measured to be  \cite{Lancaster:2011wr}
\begin{eqnarray}
m_t = 173.2 \pm 0.9 \,\,\mathrm{GeV}\,, \label{eq:mt}
\end{eqnarray}
and its decay width. In the SM top quark is predicted to decay almost exclusively through the charged weak current (\ref{eq:SMcc}). The $t\to W b$ channel, which we shall refer to as the main decay channel, is highly dominant due to the extreme hierarchy between the CKM matrix (\ref{eq:CKMmat}) elements of the third row $|V_{td}|,|V_{ts}| \ll |V_{tb}|$. Branching ratios of top quark decays are always normalized to the main decay channel. The tree-level decay width computed at {\sl leading order} (LO) in QCD can be written as
\begin{eqnarray}
\Gamma (t\to W b ) =|V_{tb}|^2\frac{m_t}{16 \pi}\frac{g^2}{2}\frac{(1-x^2)^2(1+2x^2)}{2x^2} \sim 1.5 \,\,\mathrm{GeV}\,,\label{eq:SM_MDC}
\end{eqnarray}
where $x= m_W/m_t$, $m_W = 80.1$ GeV, the mass of $b$ quark has been neglected and $|V_{tb}|=1$. Numerical value of $g$ is related to the Fermi constant through $G_F/\sqrt{2} = g^2/(8m_W^2) = 1.167 \,\,\mathrm{GeV}^{-2}$ \cite{PDG}.
Due to its large mass, the average life time of top quark is an order of magnitude below the typical hadronization time scale, causing it to decay before forming bound states \cite{Beneke:2000hk} making theoretical treatment free of non-perturbative QCD effects.

While the production of top quarks is a very interesting area of research as well, especially in the case of $t\bar{t}$ production, where we are witnessing a persisting anomaly in the forward-backward asymmetry from Tevatron~\cite{Aaltonen:2011kc,Abazov:2011rq} that has stimulated various theoretical attempts to reconcile it~\cite{Kamenik:2011wt,Drobnak:2012cz}, we shall be concentrating in this work on two aspects of top quark decays that we describe below within the framework of SM showing that they might be interesting for NP observations and constraints. On one hand we will study the main decay channel exploring the helicities of $W$ bosons produced through the top quark decay. The information on what fraction of $W$s produced in the decays have certain helicities allows us to directly probe the structure of the $tWb$ interaction and its potential deviations from the SM form~(\ref{eq:SMcc}). On the other hand we shall consider the possibility of observing FCNC decays of the top quark. The branching ratios for these decays are highly suppressed within the SM and any observation of such a process would signal a presence of NP. Both analysis, which are given in sections~\ref{sec:fcnc_twobody} and~\ref{sec:hel_nlo} will be conducted at {\sl next-to-leading order} (NLO) in QCD.

We should note that in the last five years, the precision of the experimental top quark mass determination has been gradually improving and the central value (\ref{eq:mt}) of the top quark mass has been continuously changing. As a consequence we will, in this work, encounter a few different values for top quark mass being used in the numerical analysis since the work presented here spans over four years. Deviations are however small and variations within these values have no significant effect on the results presented.

\subsection{Helicity fractions in the main decay channel}
\label{sec:hfSM}
Since $W$ boson is a spin $1$ particle, we can split the decay width of the top quark's main decay channel into three parts depending on which of the three helicity states the produced $W$ is in
\begin{eqnarray}
\Gamma(t\to W b) = \Gamma_L + \Gamma_+ + \Gamma_- \,,
\end{eqnarray}
where $L$ stands for longitudinal, while $+$ and $-$ denote positive and negative transverse helicities respectively. We further define the helicity fractions as
\begin{eqnarray}
\mathcal F_{L,+,-} = \frac{\Gamma_{L,+,-}}{\Gamma}\,,
\end{eqnarray}
telling us what fraction of $W$ bosons produced in top quark decays have certain helicity. The main reason why helicity fractions are interesting for NP searches is that they are sensitive to the structure of the $tWb$ vertex governing the decay.

\begin{SCfigure}[3.5][h!]
\includegraphics[scale=0.8]{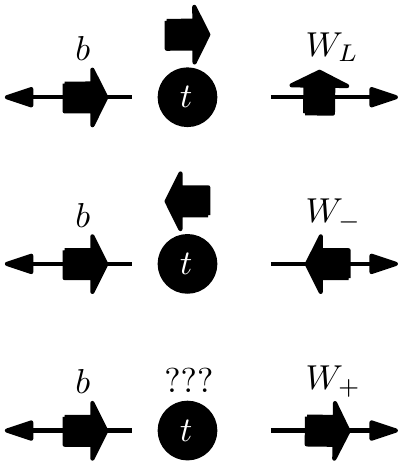}
\vspace{0.2cm}
\caption{An illustration of $t\to W b$ decay in the rest frame of the top quark where the limit $m_b =0$ is taken. Wide arrows represent third component of the spin with respect to the horizontal axes, while narrow arrows represent the direction of momentum. Because helicity and chirality of the massless $b$ quark coincide, its helicity is always negative. Third pictures represents a situation that is not possible since top quark would have to have spin greater than $1/2$ to accommodate the spin conservation, indicating the helicity suppression of $\mathcal F_+$. }
\label{fig:illust}
\end{SCfigure}

On the computational side, there is more than one way to extract a certain helicity of the final state vector boson. We shall be making use of the covariant helicity projectors \cite{Kadeer:2009iw}, which are particularly useful when computing loop diagrams for QCD corrections. To define them we write down the squared matrix element for the $t\to W b$ decay as
\begin{eqnarray}
|\mathcal M|^2 = H_{\mu\nu} \epsilon^{\mu}(q,\lambda)\epsilon^{*\nu}(q,\lambda)\,.
\end{eqnarray}
$\epsilon^{\mu}(q,\lambda)$ are the polarization vectors of the $W$ fields with $\lambda=1,2,3$ labeling their basis and $q$ denotes the momentum of the $W$. We have put everything else into the $H_{\mu\nu}$. When going from the squared matrix element to the decay width we can, even when considering particular helicity final state, perform the summation over the polarizations of the $W$ boson $\sum_{\lambda}\epsilon_{\mu}(q,\lambda)\epsilon^{*}_{\nu}(q,\lambda)$ replacing it with appropriate helicity projector given in Tab.~\ref{tab:projectors}.
\begin{table}[h]
\begin{center}
\begin{tabular}{r|l|c}\hline\hline
Helicity & Projector: $\sum_{\lambda}\epsilon^{\mu}(q,\lambda)\epsilon^{*\nu}(q,\lambda)\to\mathbb{P}^{\mu\nu} $ & SM LO $\mathcal F_i$ with $m_b=0$ \\\hline
Unpolarized: $\Gamma$&$ \mathbb{P}_{\mathrm{U}}^{\mu\nu}=-g^{\mu\nu}+\frac{q^{\mu}q^{\nu}}{m_W^2}$\\
Asymmetric: $\Gamma_F$ &$\mathbb{P}_{\mathrm{F}}^{\mu\nu}=\frac{1}{m_t}\frac{1}{|{\bf q}|}\ii \epsilon^{\mu\nu\alpha\beta}p_{t\alpha}q_{\beta}$ \\
Longitudinal: $\Gamma_L$& $\mathbb{P}_{\mathrm{L}}^{\mu\nu}=\frac{m_W^2}{m_t^2}\frac{1}{|{\bf q}|^2}\big(p_t^{\mu}-\frac{p_t\cdot q}{m_W^2}q^{\mu}\big)\big(p_t^{\nu}-\frac{p_t\cdot q}{m_W^2}q^{\nu}\big)$ & $\frac{1}{1+2 x^2}$\\
Positive transversal: $\Gamma_+$& $\mathbb{P}_{\mathrm{+}}^{\mu\nu}=\frac{1}{2}\big(\mathbb{P}_{\mathrm{U}}^{\mu\nu}-\mathbb{P}_{\mathrm{L}}^{\mu\nu}+\mathbb{P}_{\mathrm{F}}^{\mu\nu}\big)$&$0$\\
Negative transversal: $\Gamma_-$& $\mathbb{P}_{\mathrm{-}}^{\mu\nu}=\frac{1}{2}\big(\mathbb{P}_{\mathrm{U}}^{\mu\nu}-\mathbb{P}_{\mathrm{L}}^{\mu\nu}-\mathbb{P}_{\mathrm{F}}^{\mu\nu}\big)$&$\frac{2x^2}{1+2x^2}$\\ \hline\hline
\end{tabular}
\caption{Covariant projectors extracting different helicities of a final state massive vector boson in a three-body decay. Presented projectors are for $t\to W b$ decay, where $p_t$ is the momentum of the decayed top quark, and $q$ is the momentum of the $W$. The three-vector length $|{\bf q}|$ is assumed in the top quark rest frame and $\epsilon_{0123} = 1$. Appropriate projector $\mathbb{P}_i^{\mu\nu}$ is to replace the sum over polarization vector basis depending on what helicity state we wish to project out. Last column shows the tree-level SM helicity fractions at LO and in the limit of the massless $b$ quark.}
\label{tab:projectors}
\end{center}
\end{table}
In the last column we show the LO helicity fractions in the limit of the massless $b$ quark. We can see that within SM helicity fraction $\mathcal F_+$ is $0$ in the presented approximation. The suppression is not hard to understand and is illustrated in Fig.~\ref{fig:illust}. If we consider $b$ quark to be massless and produced in the weak interaction, which strictly involve left-handed components of the fermionic fields, its helicity has to be negative, since for massless fermions helicity and chirality coincide. From a simple consideration of spin conservation, the situation where $W$ boson would have a positive helicity is not possible. 

This simple picture is altered if we consider the mass of $b$ quark or the process involves more than just three particles, which is the case once higher order quantum corrections to the decay are considered. All of these effects have been analyzed within the SM including $m_b$ and finite width of top quark effects, NLO QCD and electroweak corrections as well as {\sl next-to-next-to leading order} (NNLO) QCD corrections \cite{Fischer:2001gp,Czarnecki:2010gb,Do:2002ky,Fischer:2000kx}. We summarize the theoretical prediction to be 
\begin{eqnarray}
\mathcal F_L^{\rm SM} =  0.687(5)\,,\hspace{0.5cm} \mathcal F_+^{\rm SM} = 0.0017(1) \label{eq:e22b}\,. 
\end{eqnarray}

Even with inclusion of these corrections  $\mathcal F_+$ remains highly suppressed, and a measurement of the positive helicity fraction of the per-cent order would undoubtedly signal the presence of NP. How these predictions get altered by the presence of NP governing the $t\to W b$ decay, is the subject of section \ref{sec:hel_nlo}.
\begin{figure}[h]
\begin{center}
\includegraphics[scale=0.6]{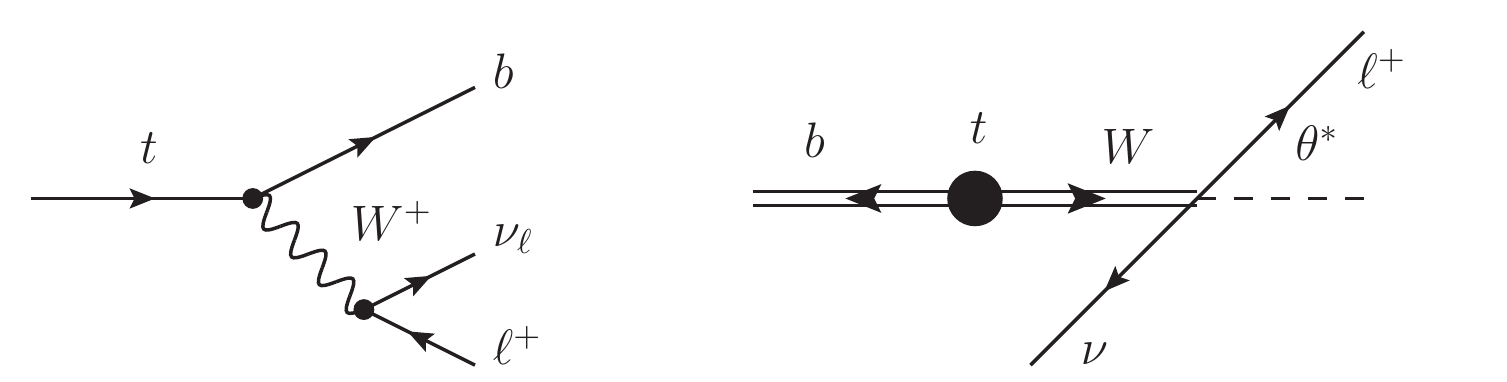}
\caption{Leptonic final state of the top quark's main decay channel used for the extraction of helicity fractions. Angle $\theta^*$ is defined as the angle between the direction of the lepton in the $W$ rest frame and the direction of the $W$ in the rest frame of the decayed top quark. }
\label{fig:hel_exp}
\end{center}
\end{figure}

Helicity fractions are experimentally accessible through the measurement of angular distributions of the leptonic final state to which the $W$ boson decays $t\to W b \to b \ell \nu$ as indicated in Fig.~\ref{fig:hel_exp}. The extraction is based on the distribution of decays over $\cos\theta^*$, where $\theta^*$ is the angle between the momentum of the charged lepton in the $W$ boson rest frame and the momentum of the $W$ boson in the top quark rest frame \cite{Aaltonen:2010ha}
\begin{eqnarray}
\frac{\dd \Gamma(t\to b \ell \nu)}{\dd \cos\theta^*} \sim \frac{3}{4} (1-\cos^2\theta^*)\mathcal F_L + \frac{3}{8}(1+\cos\theta^*)^2\mathcal F_+ + \frac{3}{8}(1-\cos \theta^*)^2 \mathcal F_- \,.
\end{eqnarray}

At the moment the most precise measurements of the helicity fraction still come form the combined D0 and CDF analysis \cite{Aaltonen:2012rz}
\begin{eqnarray}
\mathcal F_L^{\mathrm{CDF+D0}}  & =  0.722 \pm 0.081 \label{e1} \,,\hspace{0.5cm} \mathcal F_+^{\mathrm{CDF+D0}}  = -0.033 \pm 0.046 \,, \label{eq:hel_exp}
\end{eqnarray}
which are for now showing agreement with the SM predictions (\ref{eq:e22b}). Helicity fractions are also being measured at the LHC \cite{ATLAS:hel} and the sensitivity expected to be reached is \cite{AguilarSaavedra:2007rs}
\begin{eqnarray}
\sigma(\mathcal F_+) = \pm 0.002\,,\hspace{0.5cm} \sigma(\mathcal F_L)=\pm 0.02\,,
\end{eqnarray}
where $\sigma(\mathcal F_i)$ denotes the predicted absolute error on the measurement of the helicity fraction with $10 \,\,\mathrm{fb}^{-1}$ of accumulated data.

\subsection{FCNC top quark decays}
\label{sec:top_fcnc}
As mentioned in section \ref{sec:intro_fcnc} there are no FCNC decays possible at tree-level in the SM. They can however occur at one-loop level where through two insertions of charged flavor changing currents we can obtain a neutral flavor change. 
\begin{figure}[h]
\begin{center}
\includegraphics[scale=0.65]{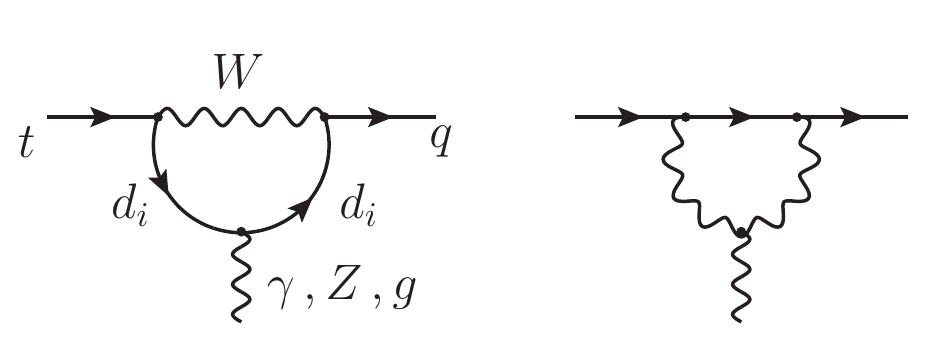}
\caption{Feynman diagrams contributing to FCNC decays $t\to q \gamma,Z,g$ in the unitary gauge for the weak interactions. Quarks running in the loops are down-type and their flavor is denoted by $i$. }
\label{fig:top_pingos}
\end{center}
\end{figure}

Top quark FCNC decays that we will be interested in $t\to q V$, where $V=\gamma,Z,g$ and $q=u,c$, proceed through the so-called penguin diagrams shown in Fig.~\ref{fig:top_pingos}. If we choose to work in the unitary gauge for the electroweak interactions, there are indeed only two diagrams to consider. 
By computing these diagrams we obtain the decay widths of the following form
\begin{eqnarray}
\Gamma (t\to q V)\propto \Big|\sum_{i}\tilde{\lambda}_i^{(q)} f^{(V)}(x_i)\Big|^2\,, \label{eq:fcnc2}
\end{eqnarray}
where $i$ denotes the flavor of the down-type quark with mass $m_i$ running in the loops and $x_i = m_i^2/m_W^2$. The loop functions $f^{(V)}$ are specific for the $V$ gauge boson in the final state. Further we have defined 
\begin{eqnarray}
\tilde{\lambda}_i^{(q)} = V_{ti}^{*}V_{qi}\,. \label{eq:skalprod}
\end{eqnarray}
Note that due to unitarity of the CKM matrix, the following equation holds
\begin{eqnarray}
\tilde{\lambda}_{d}^{(q)} + \tilde{\lambda}_s^{(q)} + \tilde{\lambda}_b^{(q)} = 0 \label{eq:CKMuni}\,,
\end{eqnarray}
since there is an orthogonal relation between different rows of CKM matrix~(\ref{eq:CKMmat}). Consequently, when computing the decay widths (\ref{eq:fcnc2}), the sum over the flavors enables us to drop any term in $f^{(V)}$ that is not dependent on $x_i$. This is a property of loop functions often encountered when dealing with one-loop induced FCNC processes. Neglecting the $m_{d,s}$ and making use of unitarity relation (\ref{eq:CKMuni}), the branching ratios for these decays are proportional to
\begin{eqnarray}
\mathrm{Br}[t\to q V] \propto |V_{qb}|^2|f^{(V)}(x_b)|^2\,.
\end{eqnarray}
The suppression of the branching ratios is two fold.  Firstly $x_b \ll 1$, so the loop functions give small contributions and secondly $|V_{qb}|\ll 1$. The resulting values for $t\to c V$ are \cite{Eilam:1990zc, AguilarSaavedra:2004wm}
\begin{eqnarray}
\mathrm{Br}[t\to c \gamma]\sim 10^{-14}\,,\hspace{0.5cm}
\mathrm{Br}[t\to c Z]\sim 10^{-14}\,,\hspace{0.5cm}
\mathrm{Br}[t\to c g]\sim 10^{-12}\,,
\end{eqnarray}
while the results for $t\to u V$ are an additional order of magnitude smaller. Within many BSM models, like Two Higgs Doublet Models, Minimal Super Symmetric Standard Model, models with up-type quark singlets, etc., the suppression of FCNC top quark decays can be lifted ~\cite{AguilarSaavedra:2004wm,Yang:2008sb,deDivitiis:1997sh,delAguila:1998tp}. It has been pointed out recently, that top quark FCNC phenomenology is crucial in constraining a wide class of NP scenarios where new flavor structures are present but can be aligned with the SM Yukawas in the down sector~\cite{Fox:2007in, Gedalia:2010zs, Gedalia:2010mf, Datta:2009zb}. 

Top quark FCNCs can be directly probed both in production and in decays of the top quark and all three FCNC decays presented here are being experimentally searched for. There has been no observation made so far, thus upper limits at 95\% {\sl confidence level} (C.L.) have been set, most stringent of which are 
\begin{eqnarray}
\mathrm{Br}[t\to \{u,q\} \gamma]&<& \{5.9,32\} \times 10^{-3} \cite{Chekanov:2003yt,Abe:1997fz}\,,\label{eq:brs}\\
\nn\mathrm{Br}[t\to q Z]& < &3.4\times 10^{-3} \hspace{0.2cm}\cite{CMS:tcZ}\,,\\
\nn\mathrm{Br}[t\to \{u,c\} g]&<& \{5.7, 27\} \times 10^{-5}\hspace{0.2cm} \cite{Collaboration:2012gd}\,.
\end{eqnarray}
With the increasing accumulation of data LHC will be able to probe branching ratios of lower orders \cite{Carvalho:2007yi}, in particular in the case of no signal, ATLAS projects to improve the upper bounds (\ref{eq:brs}) to
\begin{eqnarray}
\mathrm{Br}[t\to c\gamma] \lesssim 10^{-5} \,,\hspace{0.5cm} \mathrm{Br}[t\to cZ]\lesssim 10^{-4}\,. \label{eq:ATC1}
\end{eqnarray}


\section{Effective field theories}
\label{sec:effec}
The concept of effective field theories is highly applicable in high energy physics where we often encounter problems involving widely separated scales. 

On one hand effective theories are very useful when the underlying theory is unknown and allows us to parametrize its effects on the physics at lower energies in a systematic fashion. On the other hand, it is also useful when the underlying theory is known, since in general the full theory can be quite complicated and going to an effective theory simplifies matters greatly. In particular, going to an effective theory can manifest approximate symmetries that are not visible in the full theory and increased symmetry means increased predictive power. 

Furthermore, when the full theory contains several disparate scales $m\ll M$, perturbation theory can be poorly behaved as typically, when considering higher order quantum corrections, one generates logarithmic terms of the form $\log(m^2/M^2)$. When these logs are large, they need to be re-summed in a systematic fashion in order to keep perturbation theory under control. Working within an effective theory simplifies the summation of these logs.

In this section we aim to briefly introduce the effective theory techniques that we shall be employing in our analysis. It is worth mentioning that the strength and applicability of effective theories in particle physics goes far beyond what we shall be presenting. For detailed explanations we refer the reader to the following pedagogical works \cite{Collins:Ren,Collins:1995hda,Buras:1998raa,Rothstein:2003mp}. 

\subsection{Operator product expansion}
The {\sl operator product expansion} (OPE) \cite{Wilson:1969zs} translates a time-ordered product of two operators to a series of local operators
\begin{eqnarray}
\hat{T}\mathcal O_1(x)\mathcal O_2(y)\xrightarrow{x\to y} \sum_{i} C_i^{(12)}(x-y) \mathcal O_i(x)\,,
\end{eqnarray}
where the spatial separation $x-y$ is assumed to be small. Wilson coefficients, which are c-numbers, capture all the short distance $x-y$ dependance. We can apply this expansion when we are computing amplitudes for different processes, which is typically done in the momentum space. In particular, when we encounter a Feynman diagram with a virtual heavy particle of mass $M$ and we are interested in the external momenta $p$, where $p\ll M$ we can perform a Taylor expansion of the amplitude $A$ in the parameter $p/M$. We can then ask ourselves what kind of {\sl effective Lagrangian} which does not include the heavy field we would need to write down in order to be able to reproduce the full theory result. This leads us to an expansion of the form \cite{Rothstein:2003mp}
\begin{eqnarray}
\mathcal L_{\mathrm{eff.}} = \sum_{i,d} C_i^{(d)} \frac{1}{M^{d-4}} \mathcal O_i^{(d)}\,,\label{eq:ope2}
\end{eqnarray}
where the sum runs over $d$, representing the dimensionality of the local operators, and over $i$, the basis of operators with a given dimensionality. Typically the basis consists of more than just one operator, and each operator comes accompanied with its own Wilson coefficient, which is in this equation assumed to be dimensionless. OPE reveals a very important point, that contributions of higher dimensional operators come suppressed with higher powers of the high scale $M$, representing the scale of the physics that has been ``integrated out''.

The procedure of integrating out the heavy fields reduces the number of dynamical fields in the Lagrangian. If the underlying theory is known, we are able to match the effective and full theory, thus obtaining the Wilson coefficients. If, on the other hand, the full theory is not known, the matching procedure can not be performed, but we can still rely on the OPE to write down an appropriate basis of operators and analyze how they affect certain observables and most importantly, truncate the series at a certain dimensionality of the operators, knowing that higher dimensional operators are accompanied with higher suppression as indicated in (\ref{eq:ope2}).



\subsection{Running of the Wilson coefficients}\label{sec:running}
As already mentioned, computation of quantum corrections within a theory containing two scales $m,M$ (which we will refer to as the full theory) will typically introduce the following form of logarithmic terms in the amplitude
\begin{eqnarray}
A_{\mathrm{full}} = \cdots + \Big(a + b \log \frac{M^2}{m^2}+\cdots\Big)\langle \mathcal O_i\rangle + \cdots\,,\label{eq:fulll}
\end{eqnarray}
where $a$ and $b$ in general denote a product of different couplings. We assume $a$ to come from LO diagrams, while the $b$ term comes from some NLO corrections. For the NLO part we have written out only the logarithmic term, making our analysis a leading-log approximation. Finally, $\langle \mathcal O_i\rangle$ is the matrix element of a certain operator. If the two scales are widely separated $m\ll M$, the logarithm of their ratio can be big and we might encounter a problem, since even if $b$ is small, comprised of parameters appropriate for perturbative expansion, the logarithm creates a potentially large enhancement, rendering the perturbative expansion at least questionable. 

The best way to resum the large logs is to employ the effective theory approach. We compute the amplitude for the same process and to the same order in perturbation theory in the effective theory $\nolinebreak{\mathcal L_{\mathrm{eff}}= C_i \mathcal O_i}$, from which the heavy degree of freedom (having mass $M$) has been integrated out. The amplitude will be UV divergent and the factor in front of the divergence will exactly match the factor in front of the large log in the full theory
\begin{eqnarray}
A_{\mathrm{eff}}= \cdots + C_i \Big(1+ \frac{b}{a}\big(\frac{2}{\bar{\epsilon}} - \log \frac{m^2}{\mu^2} \big) \Big)\langle \mathcal O_i\rangle + \cdots \,, \hspace{0.5cm} \frac{2}{\bar{\epsilon}} = \frac{2}{\epsilon} - \gamma + \log 4\pi\,.\label{eq:efff}
\end{eqnarray}
Here we have chosen the dimensional regularization \cite{'tHooft:1973mm,'tHooft:1973us,Leibbrandt:1975dj} of the UV divergence, working in $d=4-\epsilon$ dimensions, which necessitated an introduction of an arbitrary scale $\mu$ and $\gamma$ is the Euler constant. Performing the matching procedure between (\ref{eq:fulll}) and (\ref{eq:efff}) order-by-order in perturbative expansion,  thus gives us the Wilson coefficient
$$
C_i= a + b\Big( \frac{2}{\bar{\epsilon}} + \log \frac{M^2}{\mu^2}\Big) + \cdots\,,
$$
which is UV divergent. We can renormalize it using the {\sl modified minimal subtraction} ($\overline{\mathrm{MS}}$) renormalization scheme, very appropriate for the renormalization group applications (for detailed discussion see \cite{Buras:1998raa}). Notice, that had we performed the matching only at LO, the extracted Wilson coefficient would have been the same as the leading-log coefficient, with the matching scale is set to $\mu = M$. 

The renormalization of the effective theory is said to involve operator renormalization, stating that either the operators or the Wilson coefficients in the Lagrangian are bare objects, denoted by $(0)$, and need to be renormalized by the appropriate renormalization matrices denoted by $\hat{Z}$
\begin{eqnarray}
\text{Bare operators: } & C_i \mathcal O_i^{(0)}=C_i Z_{ij} O_j \,,\\
\text{Bare W. coefficients: }&C_i^{(0)} \mathcal O_{i} = C_i Z^{c T}_{ij} \mathcal O_j=C_i Z_{ij}^{cT} Z^{-1}_{j k}\mathcal O_k^{(0)}\,. 
\end{eqnarray}
Comparing the last expression of the second line with the first expression of the first line we find the relation among the two renormalization matrices 
\begin{eqnarray}
\hat{Z}^{cT} = \hat{Z}^{-1}\,.
\end{eqnarray}
The extraction of $\hat{Z}$ in the minimal subtraction scheme is achieved by finding the UV divergent parts of the effective theory amplitude. Operators are composed of fields and other parameters which themselves need to be renormalized. Usually we consider these objects appearing in the operators to be renormalized, so when looking for the operator renormalization matrix, this needs to be taken in consideration (for details see Ref.~\cite{Buras:1998raa}).
To obtain {\sl renormalization group equations} (RGE) for the Wilson coefficients we use the fact, that the bare quantities cannot depend on $\mu$  
\begin{eqnarray}
\mu \frac{\dd}{\dd \mu} \mathcal O_i^{(0)} = 0& \Longrightarrow& 
\mu\frac{\dd }{\dd \mu} \boldsymbol{\mathcal O} = - \Big[\hat{Z}^{-1}\mu\frac{\dd}{\dd \mu}\hat{Z}\Big]\boldsymbol{\mathcal O}\equiv -\hat{\gamma} \boldsymbol{\mathcal O}\,,\\
\mu \frac{\dd}{\dd \mu} C_i^{(0)}=0&\Longrightarrow&
\mu \frac{\dd}{\dd \mu} \boldsymbol{C} = \Big[\hat{Z}^{-1} \frac{\dd}{\dd \mu} \hat{Z}\Big]^T \boldsymbol{C} \equiv \hat{\gamma}^T \boldsymbol{C}\,.\label{eq:runC}
\end{eqnarray}
We have defined the anomalous dimension matrix $\hat{\gamma}$ which governs the running of Wilson coefficients. It is dependent on some perturbative coupling, which we generally denote $g$, and is in the cases that we will be considering always QCD coupling constant $g_s$. Important thing to note is that the coupling constant is itself dependent on $\mu$ and its running is determined by its beta-function
\begin{eqnarray}
\mu \frac{\dd g}{\dd \mu} = \beta(g)\,,
\end{eqnarray}
where QCD beta-function is known to 4 loops  \cite{vanRitbergen:1997va}. Taking that into account, the solution to the differential equation (\ref{eq:runC}) can be expressed in the following way
\begin{eqnarray}
\boldsymbol{C}(\mu_2) = \hat{U}(\mu_2,\mu_1) \boldsymbol{C}(\mu_1)\,,\hspace{0.5cm} U(\mu_2,\mu_1) =\exp\Big[\int_{g(\mu_1)}^{g(\mu_2)}\frac{\dd g\prime}{\beta(g\prime)}\hat{\gamma}^T(g\prime)\Big]\,.
\end{eqnarray}
We can now depending on which scale we want to evaluate the matrix element of the operator at, run the Wilson coefficient to that scale and by doing that re-sum the large logarithms promoting our result to renormalization group improved perturbation theory prediction.


\section{Top quark in meson physics}
\label{sec:SMmix}
Top quark plays an important role also in the physics of energies lower than its mass. In such processes there is not enough energy to produce an on-shell top quark, rather it appears as a virtual particle in loop diagrams. 
Similarly as the existence of the charm quark was predicted by the Glashow-Iliopoulos-Maiani mechanism~\cite{Glashow:1970gm} before its experimental discovery, there was a strong belief in the existence of the top quark well before its discovery at the Tevatron. What is more, due to good theoretical and experimental understanding of rare processes in kaon and $B$ meson physics, its mass was well estimated \cite{Ginsparg:1983zc, Buras:1983ap, Albrecht:1987dr, Buras:1993wr}.

As pointed out in Ref.~\cite{Fox:2007in} when searching for BSM physics in top quark sector, which we have set out to do, one should also consider the meson physics observables and the indirect effects that NP in top quark physics might cause.
 
Employing the OPE and effective theory techniques presented in section \ref{sec:effec}, our study of such indirect effects is reduced to finding the Wilson coefficients of lower energy theory where the top quark and the heavy vector bosons have been integrated out. Our modification of the SM Lagrangian will impact only physics of scales at which QCD is perturbative, and will therefore all be contained in the Wilson coefficients. Once we shall compute the Wilson coefficients, the procedure of obtaining low energy meson observables will be exactly the same as in SM since no modification to physics of low energies is made.

In this section we go to some detail in explaining the matching procedure at LO in QCD for $|\Delta B|=2$ and $|\Delta B|=1$ processes, in which top quark turns out to play an important role. The same computational approaches that we introduce here for the SM case shall be employed when we consider NP manifestation in charged quark currents in chapter \ref{chap:CC}.

\subsection{$|\Delta B| = 2$ transitions}
\label{sec:dB2}
Mixing between $B_q$ and $\bar{B}_q$ mesons, where $q$ stands for either $d$ or $s$ down type quarks, is a $|\Delta B| =2$ process since a $b \leftrightarrow \bar{b}$ transition occurs. It is a FCNC process highly suppressed in the SM and sensitive to NP effects. For pedagogical description of theoretical treatment as well as insight into the experimental aspects of meson mixing we refer the reader to the following references \cite{Lavura:CP, Bigi:CP, Lenz:2010gu}.


The $B_q$ and $\bar B_q$ states are flavor eigenstates and they oscillate between each other. Within the Wigner-Weisskopf approximation, the oscillation is governed by the Schr\"odiner equation
\begin{eqnarray}
\ii \frac{\dd }{\dd t}
\left(\hspace{-0.2cm}\begin{array}{c}
|B_q(t)\rangle\\
|\bar{B}_q(t)\rangle
\end{array}
\hspace{-0.2cm}\right)= [M^q - \frac{\ii}{2}\Gamma^q]
\left(\hspace{-0.2cm}\begin{array}{c}
|B_q(t)\rangle\\
|\bar{B}_q(t)\rangle
\end{array}
\hspace{-0.2cm}\right)\,,\label{eq:osc}
\end{eqnarray}
where $M^q$ and $\Gamma^q$ are Hermitian mass and decay matrices. The physical eigenstates $|B_L\rangle\,, |B_H\rangle$, having $M_{L,H}^q$ masses and  $\Gamma_{L,H}^q$ decay widths, are obtained by diagonalizing the $M^q-\ii\Gamma^q/2$ matrix. The oscillation (\ref{eq:osc}) between the flavor eigenstates involves three physical quantities
\begin{eqnarray}
|M_{12}^q|\,,\hspace{0.5cm} |\Gamma_{12}^q|\,,\hspace{0.5cm} \phi_q = \arg\Big(-\frac{M_{12}^q}{\Gamma_{12}^q}\Big)\,,\label{eq:mixnorm}
\end{eqnarray}
which are the off-diagonal mass and width matrix terms and the CP phase respectively. On the computational side, $M_{12}^q$ is obtained from the dispersive part of the transition amplitude between the meson and anti-meson,
\begin{eqnarray}
M_{12}^q = \frac{1}{2m_{B_q}}\langle B_q|\mathcal H_{\mathrm{eff},q}^{|\Delta B|=2}|\bar B_q\rangle_{\mathrm{disp}}\,,
\end{eqnarray}
where $m_{B_q}$ is the mass of the $B$ meson. On the other hand $\Gamma_{12}^q$ is obtained from the absorptive part of the same matrix element, but we shall not be considering it further. $\mathcal H_{\mathrm{eff},q}^{|\Delta B|=2}= -\mathcal L_{\mathrm{eff},q}^{|\Delta B|=2}$ is the effective Hamiltonian governing the $|\Delta B|=2$ transitions and in the SM we have
\begin{eqnarray}
\mathcal L_q^{|\Delta B|=2}=- \frac{G_F^2 m_W^2}{4\pi^2}(V_{tq}^*V_{tb})^2 C_1(\mu)\mathcal O_1^q \,,\hspace{0.5cm} \mathcal O_1^q = \big[\bar q_L \gamma^{\mu} b_L\big] \big[\bar q_L\gamma_{\mu}b_L\big]\,.
\label{eq:LSMmix}
\end{eqnarray}
In order to find the Wilson coefficient at high scale of the $W$ boson and top quark mass, which we denote as $\mu_W$, we need to perform the matching at LO in QCD.

\begin{figure}[h]
\begin{center}
\includegraphics[scale=0.6]{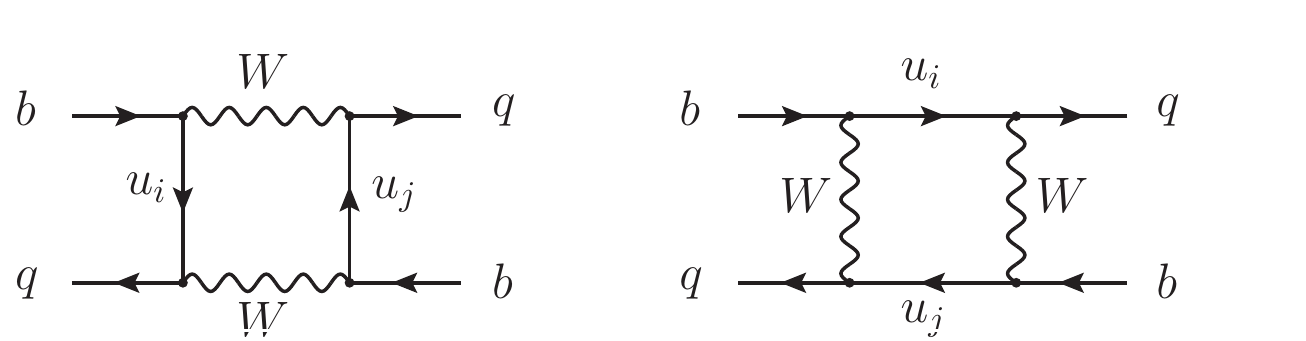}
\includegraphics[scale=0.6]{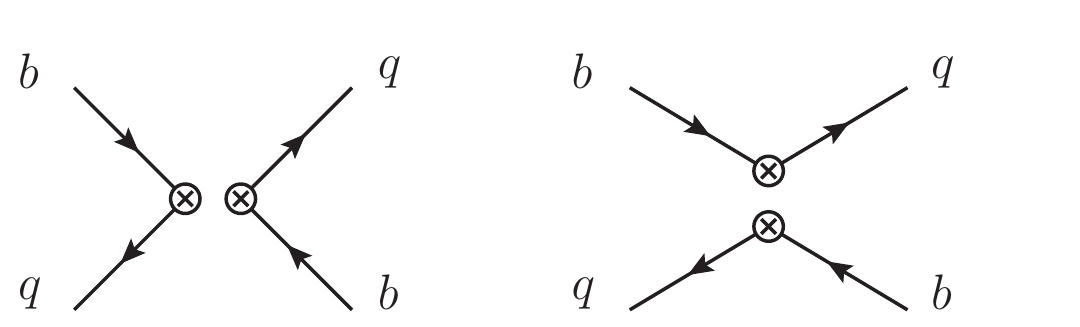}
\end{center}
\caption{Feynman diagrams for $\bar{B}_{q}\to B_{q}$ transition at LO in QCD for the full theory (left) and in the effective theory (right). Quark running in the loops are up-type and $i,j$ denote their flavor.}
\label{fig:SMmix}
\end{figure}

Since the quark content of $B$ and $\bar B$ mesons can be written as
\begin{eqnarray}
B_q\sim \bar b q\,,\hspace{0.5cm} \bar B_q \sim b \bar q\,,
\end{eqnarray}
we need two $\bar{q}$ and two $b$ field operators in order to contract the final and initial states and obtain a non-zero matrix element. On the full theory side, this necessitates a fourth order $g^4$ perturbative insertion of charged current interactions given in Eq.~(\ref{eq:SMcc}). For simplicity we shall make use of the unitary gauge for weak interactions, eliminating the would-be Goldstone fields from the theory. 
At the lowest order the mixing of $B$ mesons proceeds through the so-called box Feynman diagrams which are presented on the left side of Fig.~\ref{fig:SMmix}. 

In computation of the box diagrams the momenta of external quarks are set to zero, since the operator we are matching to contains no derivatives or masses \cite{Rothstein:2003mp}.
This leaves us with all the propagators in the loop having same momentum and greatly simplifies the integration. Under these assumptions both diagrams contribute equally and the obtained amplitude reads
\small\begin{eqnarray}
A_{\mathrm{full}}=\frac{\ii g^4}{64 \pi^2 m_W^2}
\langle \bar{b} q | \big[\bar{q}_{1L}\gamma^{\mu}b_{3L}][\bar{q}_{2L}\gamma_{\mu}b_{4L}\big]
+\big[\bar{q}_{1L}\gamma^{\mu}b_{4L}][\bar{q}_{2L}\gamma_{\mu}b_{3L}\big] |b \bar q\rangle
\sum_{i,j}^{u,c,t} V_{iq}^*V_{ib}V_{jb}V_{jq}^* F(x_i,x_j)\,,
\label{eq:SMmixAMP1}
\end{eqnarray}
\normalsize
where indices $1,\dots,4$ on the quark fields label the contractions of the external fields. $F(x_i,x_j)$ is an Inami-Lim function first presented in Ref.~\cite{Inami:1980fz} with $x_{i,j} = m_{i,j}^2/m_W^2$. When computed in the unitary gauge it is UV divergent
\begin{eqnarray}
F(x_i,x_j)=\Big(\frac{2}{\bar \epsilon}-\log\frac{m_W^2}{\mu^2}\Big)\frac{x_i+x_j-6}{4} + \cdots\,, 
\label{eq:Fxixj}
\end{eqnarray}
where we have written out only the manifestation of the UV divergence within dimensional regularization, denoting the rest with dots. What renders the amplitude (\ref{eq:SMmixAMP1}) finite are the two summations over the flavor of the up-type quarks running in the loops. In a similar fashion as in Eq.~(\ref{eq:skalprod}), we define
\begin{eqnarray}
\lambda_i^{(q)} = V_{iq}^* V_{ib}\,,
\label{eq:lambda_b}
\end{eqnarray}
where now the role of up-type and down-type quarks are interchanged and the unitarity of CKM matrix yields, in analogy to Eq.~(\ref{eq:CKMuni}) the relation 
\begin{eqnarray}
\lambda_t^{(q)}+\lambda_c^{(q)}+\lambda_u^{(q)}=0\,,
\label{eq:CKM1}
\end{eqnarray}
which tells us that we can safely drop any term in $F(x_i,x_j)$ that is not dependent on both $x_i$ and $x_j$. Consequently, the terms presented in Eq.~(\ref{eq:Fxixj}) get cancelled.

Neglecting the masses of light up quarks $x_u=x_c=0$ and making use of (\ref{eq:CKM1}) and for brevity suppressing the $(q)$ superscript in $\lambda^{(q)}_i$  we arrive at the simplified result
\begin{eqnarray}
A_{\mathrm{full}}=\frac{\ii G_F^2 m_W^2}{2 \pi^2}
\langle \bar b q|
\big[\bar{q}_{1L}\gamma^{\mu}b_{3L}][\bar{q}_{2L}\gamma_{\mu}b_{4L}\big]
+\big[\bar{q}_{1L}\gamma^{\mu}b_{4L}][\bar{q}_{2L}\gamma_{\mu}b_{3L}\big]|b \bar q\rangle 
\lambda_t^2 S_0^{\mathrm{SM}}(x_t)\,,
\label{eq:Afull}
\end{eqnarray}
where
\begin{eqnarray}
S_0^{\mathrm{SM}}(x_t)= F(x_t,x_t)-2F(x_t,0)+F(0,0)=\frac{x_t(x_t^2-11 x_t+4)}{4(x_t-1)^2}+\frac{3x_t^3\log x_t}{2(x_t-1)^3}\,.
\label{eq:s0}
\end{eqnarray}
This results explicitly confirms what we have been stating about the dominant role of the top quark in the mixing process. To complete the matching procedure we have to calculate the amplitude for the same process using effective theory~(\ref{eq:LSMmix}), where we have just one local operator to be inserted at first order of perturbation.
We can choose any of the two $b$ fields in the operator to contract the $b$ quark final state, giving us a factor of $2$. We then have a further choice of contracting the $q$ fields of the operator with the external $q$ quark fields, corresponding to the two Feynman diagrams given on the right side of Fig.~\ref{fig:SMmix}, obtaining the result
\begin{eqnarray}
A_{\mathrm{eff}}=\ii\frac{G_F^2m_W^2}{2 \pi^2}\lambda_t^{2} C_1
\langle \bar b q|
\big[\bar{q}_{1L}\gamma^{\mu}b_{3L}][\bar{q}_{2L}\gamma_{\mu}b_{4L}\big]
+\big[\bar{q}_{1L}\gamma^{\mu}b_{4L}][\bar{q}_{2L}\gamma_{\mu}b_{3L}\big]|b \bar q\rangle \,.
\label{eq:Aeff}
\end{eqnarray}
Comparing (\ref{eq:Afull}) and (\ref{eq:Aeff}) we find for the Wilson coefficient
\begin{eqnarray}
C_1^{\mathrm{SM}}(\mu_W)=S_0^{\mathrm{SM}}(x_t)\,.
\label{eq:C1SM}
\end{eqnarray}
We have demonstrated the matching at LO in QCD, however higher order perturbative QCD corrections are not negligible in such processes and can be captured in rescaling factor  $\widehat{\eta}_B$~\cite{Lenz:2010gu,Buras:1990fn}
\begin{eqnarray}
C_1^{\mathrm{SM, NLO}} = C_1^{\mathrm{SM,LO}}\widehat{\eta}_B \,,\hspace{0.5cm} \widehat{\eta}_B = 0.8393 \pm 0.0034\,.
\end{eqnarray}
Furthermore, considering NLO QCD corrections and employing RG methods the coefficient can be run down to lower energy scales $\mu_b$ at which different non-perturbative methods can be used to evaluate the matrix element of the $\mathcal O_1^q$ operator. Just for completeness we show the typical parametrization of the matrix element following Ref.~\cite{Lenz:2010gu}
\begin{eqnarray}
\langle B_q | \mathcal O^q_1 (\mu_b)|\bar{B}_q\rangle = \frac{2}{3} M^2_{B_q} f_{B_q}^2 \mathcal B_{B_q}(\mu_b)\,,
\end{eqnarray}
where $f_{B_q}$ and $\mathcal B_{B_q}$ are nonperturbative parameters, the decay constant and the bag parameter respectively.

It is apparent that considering NP to effect the charged quark currents, especially the top quark interactions, will effect the mixing amplitude. On the computational level it will lead to new contributions on the full theory side resulting in a change of the Wilson coefficient at the weak scale. 
\subsection{$|\Delta B| = 1$ transitions}
\label{sec:dB1}
In this section we consider another type of rare FCNC processes, namely the radiative decays of the neutral $B$ mesons. On quark level the transitions that we will be interested in are 
\begin{eqnarray}
b \to s \gamma \,,\hspace{0.2cm} b\to s g \,, \hspace{0.2cm} b\to s \ell^+ \ell^- \,, \hspace{0.2cm} b\to s \nu\bar{\nu}\,,\label{eq:fcnc_transitions}
\end{eqnarray}
giving rise to different decays on the hadronic level. Note that while we could again consider the light down-type quark to be either $d$ or $s$ we shall commit to the case of $s$, because the experimental sensitivities for $|\Delta B|=1$ processes that we will be studying are better when the final state quark is the $s$ quark. The results of the matching can however be applied to $d$ case by simple $s\leftrightarrow d$ change.

Very similarly to the FCNC decays of the top quark described in section \ref{sec:top_fcnc}, within the SM transitions given in Eq.~(\ref{eq:fcnc_transitions}) can not proceed through tree-level diagrams, but through two insertions of charged current interactions at one-loop level. The effective Lagrangian of the low energy effective theory, from which the heavy vector bosons and the top quark have been integrated out and is adequate for description of processes (\ref{eq:fcnc_transitions}), can be written as
\begin{eqnarray}
{\cal L}_{\mathrm{eff}}&=& \frac{4 G_F}{\sqrt{2}}\Big[ \sum_{i=1}^2 C_{i}( \lambda_u \mathcal O^{(u)}_i +  \lambda_c \mathcal O^{(c)}_i) \Big] + \frac{4 G_F}{\sqrt{2}}\lambda_t\Big[\sum_{i=3}^{10} C_{i}{\cal O}_i +  C_{\nu\bar{\nu}}{\cal O}_{\nu\bar{\nu}}\Big]\,, \label{eq:loweff1}
\end{eqnarray}
where $\lambda_i$ stands for $\lambda^{(s)}_i$ defined in Eq.~(\ref{eq:lambda_b}). The relevant operators read
\begin{align}
{\cal O}_2^c &= \big(\bar{c}_L\gamma^{\mu}b_L\big)\big(\bar{s}_L\gamma_{\mu}c_L\big)\,,& {\cal O}_9&= \frac{e^2}{(4\pi)^2}\big(\bar{s}_L\gamma^{\mu}b_L\big)\big(\bar{\ell}\gamma_{\mu}\ell\big)\,,\label{eq:ops2}\\ 
{\cal O}_7&= \frac{e m_b}{(4\pi)^2}\big(\bar{s}_L\sigma^{\mu\nu}b_R\big)F_{\mu\nu}\,,&
{\cal O}_{10}&= \frac{e^2}{(4\pi)^2}\big(\bar{s}_L\gamma^{\mu}b_L\big)\big(\bar{\ell}\gamma_{\mu}\gamma_5\ell\big)\,,\nonumber \\
\nn{\cal O}_8&= \frac{g_s m_b}{(4\pi)^2}\big(\bar{s}_L\sigma^{\mu\nu}T^ab_R\big)G^a_{\mu\nu}\,,&
{\cal O}_{\nu\bar{\nu}}&=\frac{e^2}{(4\pi)^2}\big(\bar{s}_L\gamma^{\mu}b_L\big)\big(\bar{\nu}\gamma_{\mu}(1-\gamma^5)\nu\big) \,. 
\end{align}
Here $T^a$ are $SU(3)_c$ generators in fundamental representation and $\sigma_{\mu\nu} = \ii/2 [\gamma_{\mu},\gamma_{\nu}]$. The electromagnetic and gluonic field strength tensors are defined as
\begin{eqnarray}
F_{\mu\nu} &=& \partial_{\mu}A_{\nu}-\partial_{\nu}A_{\mu}\,, \label{eq:FSdef}\\
\nonumber G_{\mu\nu}^a&=&\partial_{\mu}G^a_{\nu}-\partial_{\nu}G^a_{\mu}-g_s f_{abc}G_{\mu}^b G_{\nu}^c\,,
\end{eqnarray}
where $f_{abc}$ are the $SU(3)_c$ structure constants. Since they are not that crucial for our analysis, we omit the definition of the remaining four-quark operators $\mathcal O_{3,\dots,6}$, which can be found for example in Ref.~\cite{Buchalla:1995vs}.
\begin{figure}[h]
\begin{center}
\includegraphics[scale=0.6]{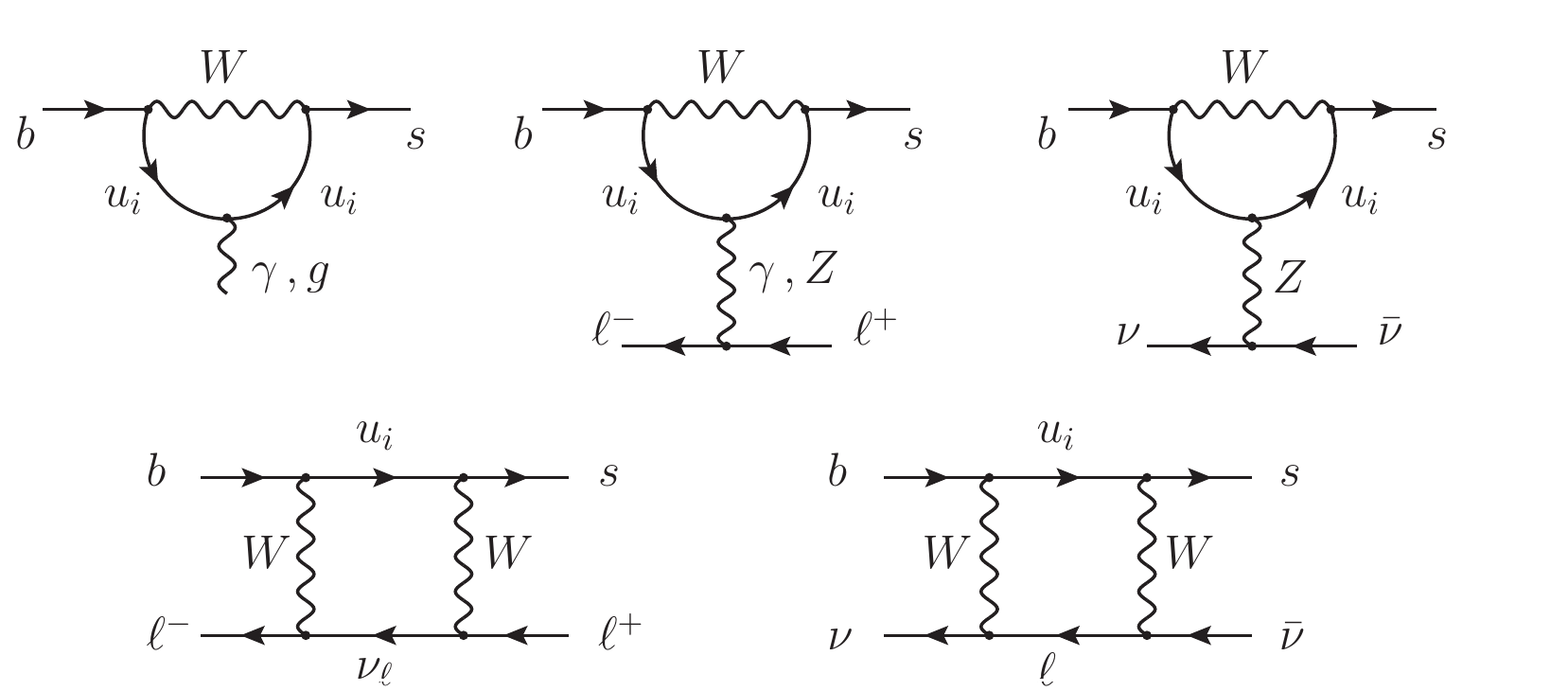}
\caption{Representative penguin and box Feynman diagrams to be calculated in the matching of SM at LO in QCD to the effective Lagrangian given in Eq.~(\ref{eq:loweff1}). Quarks running in the loops are up-type and $i$ denotes their flavor. In addition to the presented penguin diagrams one needs to consider also diagrams where the gauge bosons are emitted from the $W$ boson (except for the gluon) or the external quark legs. Diagrams with would-be Goldstone bosons which are present when working in general $R_{\xi}$ gauge for the weak interactions are not presented.}
\label{fig:sm_pingos}
\end{center}
\end{figure}

In the procedure of matching the SM to the effective theory (\ref{eq:loweff1}), again virtual top quarks turn out to play the dominant role. Representative diagrams to be computed are given in Fig.~\ref{fig:sm_pingos}. In the first diagram the photon and/or gluon are considered to be on-shell, contributing to $C_7$ and $C_8$ respectively. The remaining two penguin diagrams contribute to $C_9$, $C_{10}$ and $C_{\nu\bar\nu}$. Here the gauge bosons are off-shell coupling further to the lepton pair. We note that $C_{10}$ coefficient can only receive contributions from $Z$ mediated process due to the purely vectorial structure of the photon-lepton coupling. In addition to the penguin diagrams, $C_{9}$, $C_{10}$ and $C_{\nu\bar\nu}$ also receive contributions from box diagrams.

Calculation of the box diagrams proceeds in a similar fashion as described in section \ref{sec:dB2} for the mixing case. To compute all the Wilson coefficients coming from the penguin diagrams, we have to consider the photon and gluon to be in general off-shell and perform an expansion of the amplitudes to the second order in external momenta, neglecting throughout $m_s$ and $m_b^2$ terms. For the off-shell massive $Z$ boson the computation is somewhat simplified since its momentum can be neglected compared to its mass. 

The amplitudes for processes can be written as
\small
\begin{subequations}\label{eq:delta_B1_SM}
\begin{eqnarray}
A_{\mathrm{full}}(b\to s \ell^+\ell^-)_{\mathrm{box}}=\ii\lambda_i\frac{4G_F}{\sqrt{2}}\frac{e^2}{(4\pi)^2}\frac{1}{s_W^2}B_0(x_i)[\bar{s}_L\gamma^{\mu}b_L][
\bar\ell_L\gamma_{\mu}\ell_L]\,,\\
A_{\mathrm{full}}(b\to s \nu\bar\nu)_{\mathrm{box}}=\ii\lambda_i\frac{4G_F}{\sqrt{2}}\frac{e^2}{(4\pi)^2}\frac{1}{s_W^2}\tilde{B}_0(x_i)[\bar{s}_L\gamma^{\mu}b_L][\bar{\nu}_L\gamma_{\mu}\nu_L]\,,\\
A_{\mathrm{full}}(b\to sZ)=\ii \lambda_i\frac{4G_F}{\sqrt{2}}\frac{e}{(4\pi)^2} m_Z^2 \frac{c_W}{s_W} C_0(x_i)[\bar{s}_L\gamma^{\alpha}b_L]Z_{\alpha}(k)\,,\label{eq:tsZ}\\
A_{\mathrm{full}}(b\to s \gamma)= \ii \lambda_i\frac{4G_F}{\sqrt{2}}\frac{e}{(4\pi)^2}
\Big[D_0(x_i) [\bar{s}_L(k^2\gamma^{\alpha} - k^\alpha \gs{k})b_L] +
 m_b D_0^\prime(x_i) [\bar{s}_L\ii\sigma^{\alpha\beta}k_\beta b_R]\Big]A_{\alpha}(k)\,, \label{eq:pingo_foton}\\
A_{\mathrm{full}}(b\to s g)=\ii \lambda_i\frac{4G_F}{\sqrt{2}}\frac{g_s}{(4\pi)^2}
\Big[E_0(x_i) [\bar{s}_L(k^2\gamma^{\alpha} - k^\alpha \gs{k})T^ab_L] +
 m_b E_0^\prime(x_i) [\bar{s}_L\ii\sigma^{\alpha\beta}k_\beta T^a b_R]\Big]G_{\alpha}^a(k)\,.
\end{eqnarray}
\end{subequations}
\normalsize
The first two amplitudes are related to the box diagrams, while the rest are obtained from the penguin diagrams. We have introduced the abbreviations $c_W$ and $s_W$ which denote the cosine and sine of the Weinberg mixing angle. Further, $k$ denotes the momentum of the gauge boson, which can be neglected in Eq.~(\ref{eq:tsZ}) as argued above.
 
The loop functions $B_0,\dots,E_0^{\prime}$ depend on $x_i=m_i^2/m_W^2$, where $m_i$ is the mass of the up-type quark running in the loops of the diagrams. The subscript $0$ stands to remind that the process was calculated at LO in QCD. The analytical expressions for the loop functions are given in the Appendix~\ref{app:SM_D_B_1}. We note, that again due to the CKM unitarity relation (\ref{eq:CKM1}) all the $x_i$ independent terms encountered in the computation are dropped. 

\begin{wrapfigure}{r}{0.35\textwidth}
\begin{center}
\includegraphics[scale=0.6]{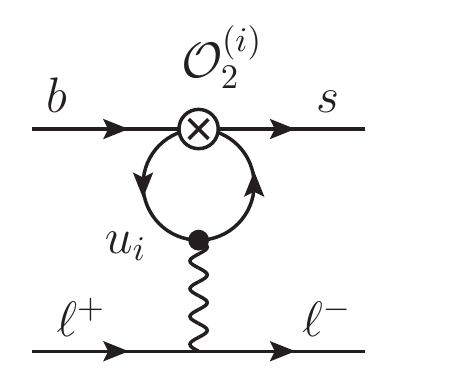}
\caption{The one-loop contribution of $\mathcal O_2$ operator to the $b\to s \ell^+ \ell^-$ on the effective theory side.}
\vspace{-2.8cm}
\label{fig:O2c}
\end{center}
\end{wrapfigure}

Functions $B_0$, $C_0$ and $D_0$ remain dependent on the choice of the gauge for weak interactions. The gauge independence is recovered once all the contributions to a particular physical final state are summed up. In particular, diagrams with off-shell photons and $Z$ bosons contribute to the leptonic final states as the bosonic field gets contracted into a propagator and further coupled to the leptons. Combining these contributions with those coming from box diagrams gives the following gauge independent combinations
\small
\begin{subequations}\label{eq:gindp}
\begin{eqnarray}
A_{\mathrm{full}}(b\to s l^+l^-) = \ii \lambda_i \frac{4G_F}{\sqrt{2}}\frac{e^2}{(4\pi)^2}
&\hspace{-0.3cm}\bigg[&\hspace{-0.3cm} \Big(\frac{2 B_0(x_i)- C_0(x_i)}{4s_W^2} 
+ C_0(x_i) + D_0(x_i)\Big)
\big[\bar s_L\gamma^{\mu}b_L\big]\big[\bar{\ell}\gamma_{\mu}\ell\big]\label{eq:bsllfull}\\
\nn&\hspace{-0.3cm}+&\hspace{-0.3cm}\frac{-2B_0(x_i)+C_0(x_i)}{4s_W^2} \big[\bar s_L\gamma^{\mu}b_L\big]\big[\bar{\ell}\gamma_{\mu}\gamma_5 \ell\big] \bigg]\,,\\
A_{\mathrm{full}}(b\to s \nu\bar\nu) = \ii \lambda_i \frac{4G_F}{\sqrt{2}}\frac{e^2}{(4\pi)^2}
&\hspace{-0.3cm}&\hspace{-0.3cm}\frac{2\tilde{B}_0(x_i) + C_0(x_i)}{4s_W^2}\,\,
\big[\bar s_L\gamma^{\mu}b_L\big]\big[\bar \nu \gamma_{\mu}(1-\gamma_5)\nu\big]\,.
\end{eqnarray}
\end{subequations}
\normalsize
To complete the matching procedure we have to compare the full theory amplitudes, with the amplitudes computed within the effective theory (\ref{eq:loweff1}). For the on-shell photon and gluon the matching is straightforward since only contributions on effective theory side are tree-level insertions of operators. Same holds for the neutrino final state and the axially coupled charged leptons
\begin{eqnarray}
C_7(\mu_W) &=& -D_0^{\prime}(x_t) /2 = -\frac{8 x_t^3+5x_t^2-7x_t}{24(x_t-1)^3} + \frac{x_t^2(3x_t-2)}{4(x_t-1)^4} \log x_t\,,\\
C_8(\mu_W) &=& -E_0^{\prime}/2(x_t) = -\frac{x_t^3-5x_t^2-2x_t}{8(x_t-1)^3} - \frac{3x_t^2\log x_t}{4(x_t-1)^4}\,,\\
C_{10}(\mu_W)&=& \frac{-2B_0(x_t)+C_0(x_t)}{4s_W^2} = \frac{1}{4s_W^2}
\Big(\frac{4x_t-x_t^2}{2(x_t-1)}-\frac{3x_t^2\log x_t}{2(x_t-1)^2}\Big)\,,\\
C_{\nu\bar\nu}(\mu_W)&=& \frac{2 \tilde{B}_0(x_t)+C_0(x_t)}{4 s_W^2}=
\frac{1}{4s_W^2}\Big(-\frac{x_t(x_t+2)}{2(x_t-1)}-\frac{3x_t^2-6x_t}{2(x_t-1)^2}\log x_t\Big)\,.
\end{eqnarray}
Since the matching was performed at LO in QCD, the expressions again apply to the high $\mu_W$ scale. For the extraction of $C_9$ coefficient we have to, in addition to the trivial tree-level contribution of ${\cal O}_9$, take into account the one-loop contribution of ${\cal O}_2$ operator presented in Fig.{\ref{fig:O2c}}.
The amplitude on the effective theory side is 
\begin{eqnarray}
A_{\mathrm{eff.}}(b\to s \ell^+ \ell^-)&=&\ii\frac{4 G_F}{\sqrt{2}}\frac{e^2}{(4\pi)^2}\big[\bar s_L\gamma^{\mu}b_L\big]\big[\bar{\ell}\gamma_{\mu}\ell\big]\bigg[\lambda_t C_9(\mu_W)\label{eq:bslleff}\\ 
&+& \sum_{i=u,c} \lambda_i C_2(\mu_W)\frac{4}{9}\Big(-\frac{2}{\epsilon} + \log\frac{m_W^2}{\mu^2} + \log x_i +1 \Big)\bigg]\,.\nonumber
\end{eqnarray}
The $2/\epsilon$ UV divergence is removed using the \MSbar renormalization\footnote{This means that $\mathcal O_2$ and $\mathcal O_9$ mix under QED renormalization.}. The constant term is characteristic to the naive dimensional regularization, which has been employed in the computation \cite{Buras:1998raa}. Because $C_2(\mu_W)=1$, we can write down, comparing (\ref{eq:bsllfull}) and (\ref{eq:bslleff}) the final expression for $C_9$ 
\begin{eqnarray}
C_9(\mu_W) &=& \frac{-18x_t^4+163 x_t^3 -259 x_t^2+108x_t}{36(x_t-1)^3}\\
\nn&+&\frac{-64x_t^4 + 76 x_t^3+30 x_t^2 -36 x_t}{36(x_t-1)^4}\log x_t\\
 &-& \frac{4}{9}+\frac{4}{9}\log x_t + -\log\frac{m_W^2}{\mu_W^2}\nn \,.
\end{eqnarray}
As in the case of $|\Delta B| = 2$ process, all of the Wilson coefficients obtain higher order perturbative QCD corrections~\cite{Misiak:2006zz, Misiak:2008ss, Misiak:2009nr}. Consideration thereof introduces renormalization mixing among many of the operators involved in $|\Delta B|=1$ processes. Application of RG methods allows us to establish the QCD running of the Wilson coefficients~\cite{Bobeth:1999mk,Gracey:2000am,Gambino:2003zm,Gorbahn:2005sa} which can then be, re-summing the large logarithms, run down to appropriately low scales, where the operator matrix elements can be evaluated thus making it possible to compare theoretical predictions with the experimental measurements of decay rates.

\section{The main strategy}\label{sec:strategy}
Having introduced the two phenomena of top quark physics that are interesting for NP searches and presenting also the importance of top quark contributions in $B$ physics, we devote the last section of the introductory chapter to introduce our main strategy of specifying and analyzing the deviations from SM physics in the top quark sector. We closely follow the reasoning presented in Ref.~\cite{Fox:2007in}, where the case of NP generating FCNC top quark decays was considered. We apply the same approach to the case of charged currents as well\footnote{We note that this concept is not linked only to NP in top quark physics which we are considering here. The indirect constraints on physics including new heavy degrees of freedom are often important and need to be considered, see for example~\cite{Dorsner:2011ai}.}.

\begin{figure}[h]
\begin{center}
\includegraphics[width=0.5 \textwidth]{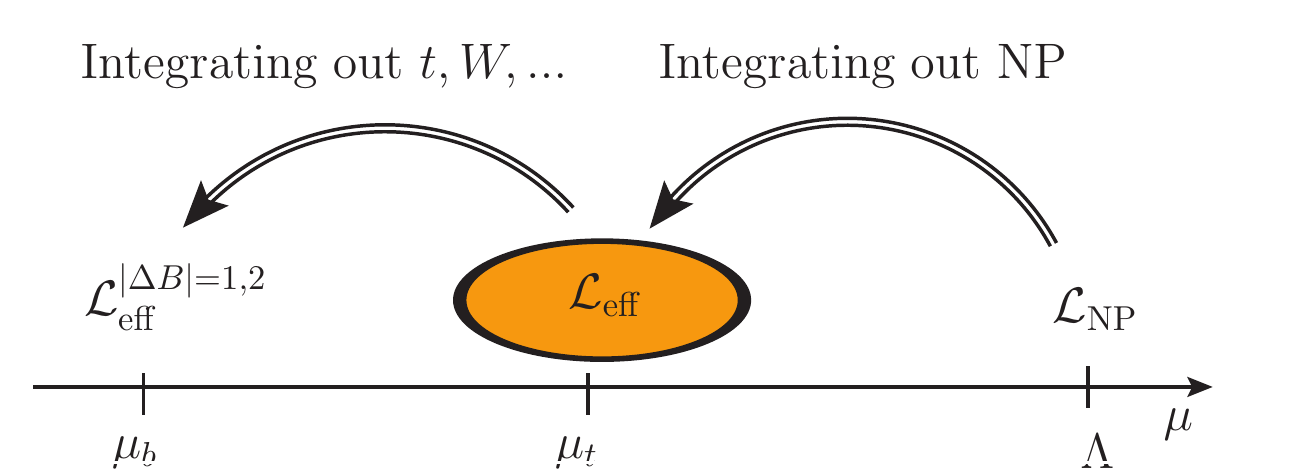}
\caption{Illustration of the effective theory approach employed in our analysis. First step presents integrating out NP particles with mass well above the electroweak scale. The second step is integrating SM degrees of freedom above the scale of $b$ quark mass to be taken when analyzing effects in $B$ physics.}
\label{fig:intout}
\end{center}
\end{figure}

The main idea is illustrated in Fig.~\ref{fig:intout} and it starts with the assumption that there exists some NP at the energy scale $\Lambda$, which is much higher than the electroweak energy scale. As this NP is integrated out, it generates operators at the electroweak scale (denoted $\mu_t$), which consist of SM fields only, and are invariant under the SM gauge group~(\ref{eq:sm_gauge_group}). Making use of the highly stressed property of OPE (\ref{eq:ope2}), that the effects of higher dimensional operators generated at high scale $\Lambda$ come suppressed with higher powers of $1/\Lambda$, we avoid committing to a particular UV completion of the SM, rather working in the framework of an effective theory, described by the Lagrangian
\begin{eqnarray}
{\cal L}_{\mathrm{eff}}={\cal L}_{\mathrm{SM}}+\frac{1}{\Lambda^2}\sum_i C_i \mathcal Q_i +\mathrm{h.c.}+ {\cal O}(1/\Lambda^3)\,,
\label{eq:lagr}
\end{eqnarray}
where ${\cal L}_{\mathrm{SM}}$ is the SM part, and $\mathcal Q_i$ are dimension-six operators with the aforementioned properties. 
The set of operators forming $\mathcal L_{\mathrm{eff}}$ are chosen such that they generate $tqZ$, $tq\gamma$, $tqg$ vertices, where $q=u,c$ or modify the SM $tWb$ vertex, depending on whether we are interested in NP in FCNC top quark decays (chapter~\ref{chap:neutral_currents}) or the main decay channel (chapter~\ref{chap:CC}). By parametrizing the appropriate  vertices in most general way and analyzing the consequences of such modifications on top quark observables, we can quantify the effects of the effective operators in $\mathcal L_{\mathrm{eff}}$ on the top quark physics side.

If however, we want to establish the effects of the same operators on the $B$ physics side as well, we need to further match our effective theory~(\ref{eq:lagr}) to the low energy effective Lagrangians responsible for $|\Delta B| = 2$ and $|\Delta B| = 1 $ processes given in Eq.~(\ref{eq:LSMmix}) and Eq.~(\ref{eq:loweff1}) respectively, as is illustrated by the second second arrow in Fig.~\ref{fig:intout}. By doing that, we gain the access to a variety of observables in $B$ physics and come across an interesting interplay of top and bottom physics.

We should note that, since the weak scale matching of NP contributions will be done at LO in QCD, there is an ambiguity of the order $\alpha_s(m_t)/4\pi$ and a residual scheme dependence, when performing the RGE evolution at next-to-leading log, which we shall be employing. However $\alpha_s$ corrections to the matching are in general model dependent and thus beyond the scope of our effective theory approach (c.f.~\cite{Becirevic:2001jj} for a more extensive discussion on this point).

\renewcommand{\op}{\mathcal O}
\chapter{NP in Top Decays: Neutral currents}\label{chap:neutral_currents}
\section{Introduction}
As we have pointed out in section \ref{sec:top_fcnc} of the introductory chapter, SM predicts highly suppressed flavor changing neutral current processes of the top quark 
\begin{eqnarray}
\nn t\to q V\,,\hspace{0.3cm} V=Z,\gamma,g\,,\hspace{0.3cm} q=c,u\,,
\end{eqnarray}
while NP beyond the SM in many cases lifts this suppression. 

For the case of $t\to q Z,\gamma$ FCNC top quark decays, the effective theory approach that we have described in section~\ref{sec:strategy} has been used in Ref.~\cite{Fox:2007in}, where the authors considered the constraints on operators that generate FCNC top quark decays from $B$ physics observables. They found that contributions of some dimension six $SU(2)_L$ gauge invariant operators are not yet constrained by $B$ physics data to such extent that the consequential top quark FCNC decays could not be observed at LHC, if the predicted sensitivities summarized in Eq.~(\ref{eq:ATC1}) are reached. On the other hand, gluonic operators governing the $t\to q g$ decays are not constrained by such indirect considerations and can, at NLO in QCD, contribute to $t\to q Z,\gamma$. In this chapter we therefore focus our attention on the top quark physics side only, analyzing the decay rates of FCNC top quark decays mediated by effective operators generating most general FCNC effective vertices. The correspondence between the Wilson coefficients of our operators and those of the $SU(2)_L$ invariant operators used in Ref.~\cite{Fox:2007in} is given in Appendix~\ref{app:tofox}.


In the first part of the chapter, which is based on our published work~\cite{Drobnak:2010wh,Drobnak:2010by}, we analyze the two-body decays  $t\to q Z,\gamma$ at NLO in QCD. In Ref.~\cite{Zhang:2008yn} it was found that $t\to q g$ decay receives almost $20\%$ enhancement from NLO QCD contributions while corrections to the $t\to q Z,\gamma$ branching ratios are much smaller. However, the authors of~\cite{Zhang:2008yn} only considered a subset of all possible FCNC operators mediating $t\to q V$ decays at leading order and furthermore neglected the mixing of the operators induced by QCD corrections. In the case of $t\to q\gamma$ decay in particular the QCD corrections generate a nontrivial photon spectrum and the correct process under study is actually $t\to q g \gamma$. Experimental signal selection for this mode is usually based on kinematical cuts, significantly affecting the spectrum. The validity of theoretical estimates based on the completely inclusive total rate should thus be reexamined. Finally, renormalization effects induced by the running of the operators from the NP scale $\Lambda$ to the top quark scale are potentially much larger than the finite matrix element corrections. Although these effects are not needed when bounding individual effective FCNC couplings from individual null measurements, they become instrumental for interpreting a possible positive signal and relating the effective description to concrete NP models.

The second part of the chapter is devoted to the study of $t \to q \ell^+ \ell^-$ decays and is based on our published work~\cite{Drobnak:2008br}. The FCNC vertex is governed by the same effective Lagrangian as in the first part, but the neutral gauge boson is further coupled through SM interaction with the pair of charged leptons. The basic goal of this analysis is the identification of possible discriminating effects of different NP models in top FCNCs, by considering different types of observables, which become attainable due to the larger phase space of the final state. 
\section{Framework}
In writing the effective Lagrangian that will generate the $tZq$, $t\gamma q$ and $tgq$ vertices of the most general form, we rely on the notation of Ref.~\cite{AguilarSaavedra:2004wm, AguilarSaavedra:2008zc}. Hermitian conjugate and chirality flipped operators are implicitly contained in the Lagrangian and contributing to the relevant decay modes
\be
{\mathcal L}_{\mathrm{eff}} = \frac{v^2}{\Lambda^2}a_L^{Z}\op_{L}^Z
+\frac{v}{\Lambda^2}\Big[b^{Z}_{LR}\op_{LR}^{Z}+b^{\gamma}_{LR}\op_{LR}^{\gamma}+b^{g}_{LR}\op_{LR}^{g}
\Big] + (L \leftrightarrow R) + \mathrm{h.c.}\,.
\label{eq:Lagr}
\ee
To explain the notation, operators considered are
\begin{align}
\op^{Z}_{L,R} &= g_Z Z_{\mu}\Big[\bar{q}_{L,R}\gamma^{\mu}t_{L,R}\Big]\,, &
\op^{Z}_{LR,RL} &= g_Z Z_{\mu\nu}\Big[\bar{q}_{L,R}\sigma^{\mu\nu}t_{R,L}\Big]\,, \label{eq:ops}\\
\nn\op^{\gamma}_{LR,RL} &= e F_{\mu\nu}\Big[\bar{q}_{L,R}\sigma^{\mu\nu}t_{R,L}\Big]\,, &
\op^{g}_{LR,RL} &= g_s G^a_{\mu\nu}\Big[\bar{q}_{L,R}\sigma^{\mu\nu}T_a t_{R,L}\Big]\,,
\end{align}
and $g_Z = 2 e/\sin 2 \theta_W$.
In addition to $F_{\mu\nu}$ and $G_{\mu\nu}^a$ that we have defined in Eq.~(\ref{eq:FSdef}), we have introduced the derivative part of the $Z$ boson field strength tensor
\begin{eqnarray}
Z_{\mu\nu} = \partial_{\mu} Z_\nu - \partial_\nu Z_\mu + \cdots \,,
\end{eqnarray}
neglecting the terms with more than one vector field, since they are not relevant for our analysis. Finally $v=246$~GeV is the electroweak condensate and $\Lambda$ is the effective scale of NP. In the remainder of the chapter, since there is no mixing between chirality flipped operators we shorten the notation, setting $a$ and $b$ to stand for either $a_L$, $b_{LR}$ or $a_R$, $b_{RL}$. 
The Feynman rules for $t\to q V$ vertices generated by operators~(\ref{eq:ops}) are given in the Appendix~\ref{app:feyn_neutral}.

Note that in principle, additional, four-fermion operators might be induced at the high scale which will also give contributions to $t\to q V$ processes, however these are necessarily $\alpha_s$ suppressed. On the other hand, such contributions can be more directly constrained via e.g. single top production measurements and we neglect their effects in the present study.  Throughout this chapter we will be neglecting the mass of the final state $(c,u)$ quark. Furthermore, when considering NLO QCD corrections we will regulate UV as well as IR divergences by working in $d=4+\epsilon$ dimensions. This kind of approach necessitates performing phase space integration in $d$ dimensions. We suggest the reader interested in dimensional regularization of IR divergences to consult the following references~\cite{Marciano:1975de,Marciano:1974tv,Muta}.
\section{Two-body $t\to q V$ decays}\label{sec:fcnc_twobody}
In this section we present the results for the NLO QCD corrections to the complete set of FCNC operators which mediate $t \to q Z,\gamma$ decays already at the leading order~(\ref{eq:ops}).

We start of by considering the virtual one-loop correction and the renormalization of UV divergences for both decay channels. We present the RGE effects linking the values of Wilson coefficients at the top quark scale to those at higher NP scale. Contributions from gluonic dipole operators are also taken into account. Next we turn our attention to the finite part of virtual corrections -- the matrix element corrections, as well as the corresponding bremsstrahlung rates. For the $t\to q \gamma$ channel we also study the relevance of kinematical cuts on the photon energy and the angle between the photon and the jet stemming from the final state quark. We present our results in analytical form and also give numerical values to estimate the significance of NLO contributions.

\subsection{Operator renormalization and RGE}\label{sec:rge}
We assume the effective $a,b$ couplings are defined near the top quark mass scale at which we evaluate virtual matrix element corrections and $\alpha_s$. A translation to a higher scale matching is governed by anomalous dimensions of the effective operators and can be performed consistently using RGE methods. To employ this mechanism we need to examine the UV divergencies generated by the NLO QCD corrections.
\begin{figure}[h]
\begin{center}
\includegraphics[scale=0.6]{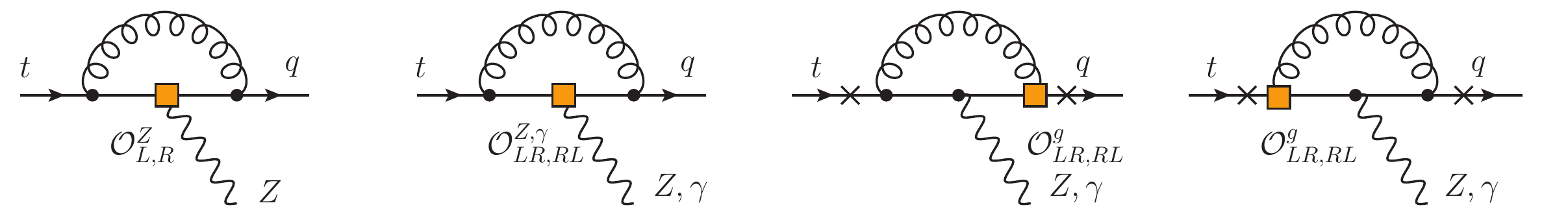}
\caption{Feynman diagrams for one-loop virtual corrections to $t\to q Z,\gamma$ decays. Squares mark the insertion of effective operators given in Eq.~(\ref{eq:ops}) and crosses the additional points from which $Z$ or $\gamma$ can be emitted. }
\label{fig:fcnc_virt}
\end{center}
\end{figure}

In addition to the diagrams presented in Fig.~\ref{fig:fcnc_virt}, the first two diagrams should be accompanied by one particle reducible diagrams with SM gluon corrections attached to the external legs. These diagrams are taken into account by setting the quark fields forming the operators $\op^Z_{L,R}$ and $\op_{LR,RL}^{\gamma,Z}$ to be renormalized $q\to \sqrt{Z_q}q$, $t\to \sqrt{Z_t} t$. Since the final state light quark $q$ is considered to be massless, the corresponding field renormalization differs from the one of the initial top quark. Using the on-shell renormalization conditions~\cite{Grozin:2005yg} we obtain
\begin{subequations}
\be
Z_t &=& 1+ 
\frac{\alpha_s}{4\pi}C_F \frac{\Gamma(1-\frac{\epsilon}{2})}{(4\pi)^{\epsilon/2}}\left(\frac{m_t}{\mu}\right)^{\epsilon}
\Big[\frac{2}{\epsUV}+\frac{4}{\epsIR}-4\Big]\,,\label{field_ren}\\
Z_q &=& 1+ 
\frac{\alpha_s}{4\pi}C_F \frac{\Gamma(1-\frac{\epsilon}{2})}{(4\pi)^{\epsilon/2}}\left(\frac{m_t}{\mu}\right)^{\epsilon}
\Big[\frac{2}{\epsUV}-\frac{2}{\epsIR}\Big]\,,
\ee
\end{subequations}
where $\mu$ is the renormalization scale parameter, $C_F=4/3$ is the color factor and separate track is kept of UV and IR divergences. Finally $\Gamma$ denotes the Euler's gamma function.
Including the tree-level LO diagrams, the quark field renormalization and the diagrams presented in Fig.~\ref{fig:fcnc_virt} yields the following amplitudes
\begin{eqnarray}
A_{t\to q\gamma} &=& \frac{v}{\Lambda^2}\Big[ b^{\gamma} \big(1+\frac{\alpha_s}{4\pi}C_F F^{\gamma}_b\big)
  +  b^g \frac{\alpha_s}{4\pi}C_F F_{bg}^{\gamma}\Big]\langle \op_{LR,RL}^{\gamma}\rangle\,,\label{eq:fcnc_amp1}\\
A_{t\to qZ} &=& \Big[\frac{v^2}{\Lambda^2} a^Z \big(1+\frac{\alpha_s}{4\pi}C_F F^Z_a\big)
  + \frac{v}{\Lambda^2}b^Z \frac{\alpha_s}{4\pi}C_F F_{ab}^Z 
  + \frac{v}{\Lambda^2}b^g \frac{\alpha_s}{4\pi}C_F F_{ag}^Z\Big]\langle \op_{L,R}^Z\rangle \label{eq:fcnc_amp2}\\
&+& \Big[\frac{v}{\Lambda^2} b^Z \big(1+\frac{\alpha_s}{4\pi}C_F F^Z_b\big)
  + \frac{v^2}{\Lambda^2}a^Z \frac{\alpha_s}{4\pi}C_F F_{ba}^Z 
  + \frac{v}{\Lambda^2}b^g \frac{\alpha_s}{4\pi}C_F F_{bg}^Z\Big]\langle \op_{LR,RL}^Z\rangle\,,\nn
\end{eqnarray}
where the complete expressions of form factors $F^x_y$ are given in the Appendix~\ref{app:form_factors_qcd}. We were able to crosscheck our expressions with those found in the literature. Namely, Eqs.~(\ref{eq:Fa}--\ref{eq:Fba}) agree with the corresponding expressions given in Ref. \cite{Ghinculov:2002pe} for the $B\to X_s \ell^+ \ell^-$ decay mediated by a virtual photon after taking into account that the dipole operator in \cite{Ghinculov:2002pe} includes a mass parameter which necessitates additional mass renormalization. On the other hand, the two form factors for the photon case (\ref{eq:Fb_gamma}, \ref{eq:Fbg_gamma}) are obtained from the corresponding $Z$ form factors in the limit where the mass of the $Z$ boson is sent to zero. To some extent we were also able to crosscheck the gluon operator induced form factors $F_{ag}^Z$ and $F_{bg}^Z$ given in Eqs.~(\ref{eq:Fag}, \ref{eq:Fbg}). Namely, we find numerical agreement of the form factor's vector component with the corresponding expressions given in Ref.~\cite{Ghinculov:2003qd}. The crosscheck is only possible in the vector part, since the SM photon coupling appearing in \cite{Ghinculov:2003qd} has no axial component.

We note that $F_b^{\gamma}$, $F_{bg}^{\gamma}$, $F_ b^{Z}$ and $F_{bg}^{Z}$ contain UV divergences that necessitate additional operator renormalization which we carry out in the $\overline{\mathrm{MS}}$ scheme obtaining the following renormalization factors
\begin{eqnarray}
Z_{b}^{\gamma}  &=& 1+\frac{\alpha_s}{4\pi}C_F\delta_{b}^{\gamma}\,,\hspace{0.5cm}
\delta_{b}^{\gamma}= - \Big(\frac{2}{\epsUV}\ +\gamma-\log(4\pi)\Big)\,,\label{renF}\\
Z_{bg}^{\gamma} &=& 1+\frac{\alpha_s}{4\pi}C_F\delta_{bg}^{\gamma}\,,\hspace{0.5cm}
\delta_{bg}^{\gamma}= -4 Q \Big(\frac{2}{\epsUV} +\gamma-\log(4\pi)\Big)\,,\\
Z_{b}^Z  &=& 1+\frac{\alpha_s}{4\pi}C_F\delta_{b}^Z\,,\hspace{0.5cm}
\delta_{b}^Z= - \Big(\frac{2}{\epsUV}\ +\gamma-\log(4\pi)\Big)\,,\label{renZ}\\
Z_{bg}^Z &=& 1+\frac{\alpha_s}{4\pi}C_F\delta_{bg}^Z\,,\hspace{0.5cm}
\delta_{bg}^Z= -2 \hat v \Big(\frac{2}{\epsUV} +\gamma-\log(4\pi)\Big)\,,
\end{eqnarray}
where $Q=2/3$ is the electric charge of the up-type quarks and $\hat v$ is defined in Eq.~(\ref{eq:some_def}) of the Appendix. The RG running is governed by the anomalous dimensions of the operators. Since operators $\mathcal O^Z_{L,R}$ do not have anomalous dimensions we assemble the remaining six operators into two vectors 
\begin{eqnarray}
\boldsymbol{\mathcal O}_{i} =  (\mathcal O^\gamma_i , \mathcal O^Z_i, \mathcal O^g_i)^T\,,\hspace{0.5cm} i = RL, LR\,,
\end{eqnarray} 
which do not mix with each other under QCD renormalization. The corresponding one-loop anomalous dimension matrix is the same for both chiralities and reads
\begin{equation}
\gamma_i = \frac{\alpha_s}{2\pi}
\left[
\begin{array}{ccc}
 C_F  & 0   & 0   \\
0  & C_F  & 0   \\
 8 C_F / 3 &   C_F (3 - 8 s^2_W) / 3  & 5C_F - 2 C_A   
\end{array}
\right]\,, \label{eq:anomal}
\end{equation}
where $C_A= 3$. We note that to compute the last entry on the diagonal of the matrix (\ref{eq:anomal}) we need to consider virtual corrections to $t\to q g$ process mediated by $\op^g_i$ operator. We have performed this calculation to crosscheck it with the well known result found in the literature (see for example Ref.~\cite{Buras:1998raa}), we however refrain from explicitly showing the details of the calculation.

Depending on the nature of new physics which generates the dipole operators at the scale $\Lambda$, the relevant $LR$ operators might explicitly include a factor of the top mass.  By redefinition of operators
$$
\widetilde{\boldsymbol{\mathcal O}}_{LR} =  (m_t/v )\boldsymbol{\mathcal O}_{LR}\,, 
$$
their running is altered by the additional mass renormalization $Z_m$ (found for example in Ref.~\cite{Buras:1998raa}), which can be taken into account by adding $6C_F$ to the diagonal entries of $\gamma_{LR}$ given in Eq.~(\ref{eq:anomal}). As we shall demonstrate, this effect is numerically not important for the interesting range of couplings and scales, which can be probed at the Tevatron and the LHC. 

We are interested in particular in the mixing of the gluonic dipole contribution into the photonic and $Z$ dipole operators. For the case with no explicit top mass effect, $LR$ and $RL$ operators receive identical corrections and the effective couplings at the top mass scale read 
\begin{subequations}
\begin{eqnarray}
 \hspace{-0.6cm}b^\gamma_{i} (\mu_t) \hspace{-0.15cm}&=& \hspace{-0.15cm} \eta ^{\kappa_1} b_i^\gamma (\Lambda )+\frac{16}{3}\left( \eta ^{\kappa_1}- \eta ^{\kappa_2}\right) b^g_i (\Lambda )\,,\\
 \hspace{-0.6cm}b^Z_{i} (\mu_t)  \hspace{-0.15cm}&=& \hspace{-0.15cm} \eta ^{\kappa_1} b_i^Z (\Lambda )   \hspace{-0.05cm}+\left[2-\frac{16}{3} s^2_W\right]\left( \eta ^{\kappa_1}- \eta ^{\kappa_2}\right) b^g_i (\Lambda )\,,
\end{eqnarray}
\end{subequations}
where $\mu_t$ is the top mass scale, $\eta = \alpha_s(\Lambda)/\alpha_s(\mu_t)$, $\kappa_1=4/3\beta_0$, $\kappa_2=2/3\beta_0$ and $\beta_0$ is part of the one-loop QCD beta function (found for example in Ref.~\cite{Buras:1998raa}). Assuming that no new colored degrees of freedom appear below the UV matching scale which would modify the QCD beta function, it evaluates to $\beta_0=7$ above the top mass scale. If we include the top mass running in the RGE of $LR$ operators, then $\kappa_{1,2}$ are modified to $\kappa_1=16/3\beta_0$, $\kappa_2=14/3\beta_0$. 

\begin{figure}[t]
\begin{center}
\includegraphics[scale=0.7]{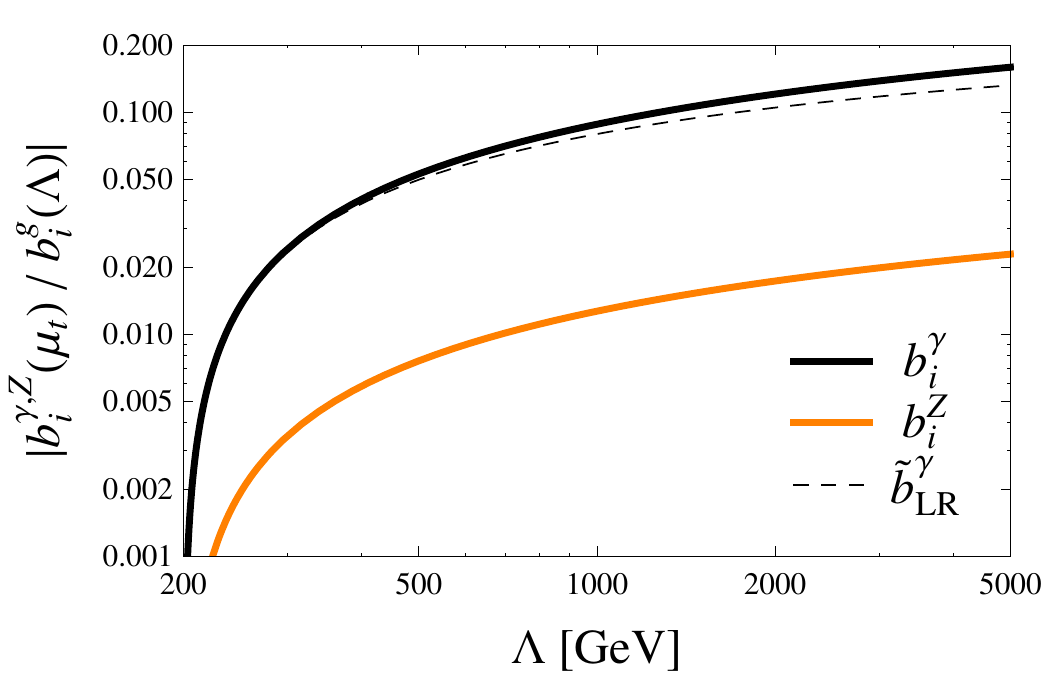}
\end{center}
\caption{The ratio of $|b_{i}^{\gamma,Z}(\mu_t)/b_i^{g}(\Lambda)|$ as a function of $\Lambda$, when $b_{i}^{\gamma,Z}(\Lambda) =0$ and the top mass scale $\mu_t\approx 200$ GeV. Solid lines represent the case with no explicit top mass effect, while the dashed line corresponds to the Wilson coefficient of $\widetilde{{\mathcal O}}_{LR}^{\gamma}$ operator. The $\widetilde{b}_{LR}^{Z}$ is not shown as its deviation from $b_{LR}^{Z}$ is unnoticeable on the plot.}
\label{fig:RGE}
\end{figure}

We illustrate the effect of the RGE running in Fig.~\ref{fig:RGE} where we plot
\begin{eqnarray}
\bigg|\frac{b_i^{\gamma,Z} (\mu_t)}{b_i^{g} (\Lambda)}\bigg|\,,\hspace{0.5cm}\text{when $b^{\gamma,Z}_i(\Lambda) = 0$}\,.
\end{eqnarray}
This shows how much $b^{\gamma,Z}_i(\mu_t)$ can be generated at the top mass scale $\mu_t\simeq 200$ GeV, due to the QCD mixing of the operators and the presence of the gluonic dipole operator at UV scale $\Lambda$.
We see that for NP matching scales above $2$ TeV the induced contributions to $b^\gamma_{i}$ are around 10\% of the $b^g_{i}$ in the UV. On the other hand, due to cancelations in the RGEs for the $b^Z_{i}$, these receive much smaller corrections (below 1 \% for the interesting range). Including the top mass renormalization reduces the induced corrections to the $\widetilde b^{\gamma,Z}_{LR}$ coupling, however for UV scales of a couple of TeV or below, this effect is negligible.

\subsection{Matrix element corrections}\label{sec:fcnc_matrix_elements}
To consistently describe rare top decays at NLO in $\alpha_s$ one has to take into account finite QCD loop corrections to the matrix elements $\bra{q \gamma} \mathcal O_i \ket{t}$ and  $\bra{q Z} \mathcal O_i \ket{t}$ evaluated at the top mass scale as well as single gluon bremsstrahlung corrections, which cancel the associated infrared and collinear divergencies in the decay rates. The total FCNC top quark decay width to $Z$ boson or a photon governed by the effective Lagrangian given in Eq.~({\ref{eq:Lagr}}) is therefore, at NLO in QCD, the sum of $\Gamma(t\to q Z,\gamma)$ and $\Gamma(t\to q g Z,\gamma)$ decay rates, where the two-body final state decay width includes the virtual QCD corrections.

Contributions due to the $O^{\gamma,Z}_{LR,RL}$ have already been computed in Ref.~\cite{Zhang:2008yn}. Here we expend the analysis to the operator basis given in Eq.~(\ref{eq:ops}), including results for $\mathcal O^{Z}_{L,R}$ current operators as well as for the admixture of the gluonic dipole operators $\mathcal O^g_{LR,RL}$. The final results that we are after can therefore be parametrized in the following way
\begin{eqnarray}
\Gamma^V &=& |a^V|^2\frac{v^4}{\Lambda^4} \Gamma_{a}^V + \frac{v^2 m_t^2}{\Lambda^4}|b^V|^2 \Gamma^V_{b}+\frac{v^3m_t}{\Lambda^4}2\mathrm{Re}\{b^{V*}a^V\} \Gamma^V_{ab} \label{eq:oso}\\
&+&
\frac{v^3m_t}{\Lambda^4} \left[2\mathrm{Re}\{a^{V*}b^g\} \Gamma^V_{ag}- 2\mathrm{Im}\{a^{V*}b^g\}\tilde{\Gamma}^V_{ag}\right]\nn \\&+& 
\frac{v^2 m_t^2}{\Lambda^4}\left[ |b^g|^2 \Gamma^V_{g}+2\mathrm{Re}\{b^{V*}b^g\} \Gamma^V_{bg} -2\mathrm{Im}\{b^{V*}b^g\}\tilde{\Gamma}^V_{bg} \right]\,,\nonumber
\end{eqnarray}
where $V= Z,\gamma$ and $a^{\gamma}=0$. Note that $\Gamma^V_{ag,bg,g}$ appearing in the second and third row of Eq.~(\ref{eq:oso}) correspond to contributions from the gluonic operator and are therefore absent in the LO result, emerging only at $\alpha_s$ order. 

\subsubsection{Tree level expressions}
At the tree-level we only have $\Gamma_{a}^Z$, $\Gamma_b^{Z,\gamma}$ and $\Gamma_{ab}^Z$ contributions, which we write in $4+\epsilon$ dimensions as 
\be
\Gamma^{\gamma(0)}_b &=& \lim_{\epsilon \to 0}m_t \alpha (1+\frac{\epsilon}{2})\Gamma(1+\frac{\epsilon}{2})\,,\\
\nn\Gamma_{a}^{Z(0)}&=&\lim_{\epsilon\to 0}\frac{m_t}{16\pi}g_Z^2(1-r_Z)^2 \Gamma(1+\frac{\epsilon}{2})(1-r_Z)^{\epsilon}\frac{1}{2r_Z}\big(1+(2+\epsilon)r_Z\big)\,,\\
\Gamma_{b}^{Z(0)}&=&\lim_{\epsilon\to 0}\frac{m_t}{16\pi}g_Z^2(1-r_Z)^2 \Gamma(1+\frac{\epsilon}{2})(1-r_Z)^{\epsilon} 2(2+\epsilon+r_Z)\,,\nonumber\\
\Gamma_{ab}^{Z(0)}&=&\lim_{\epsilon\to 0}\frac{m_t}{16\pi}g_Z^2(1-r_Z)^2 \Gamma(1+\frac{\epsilon}{2})(1-r_Z)^{\epsilon}(3+\epsilon)\,,\nonumber
\ee
where $r_Z=m_Z^2/m_t^2$.

\subsubsection{Virtual corrections}
The one-loop virtual QCD corrections to the decay amplitudes have already been presented in section~\ref{sec:rge}, where the UV divergences were renormalized. This leaves us with UV finite form factors appearing in Eqs.~(\ref{eq:fcnc_amp1}, \ref{eq:fcnc_amp2}), which however remain IR divergent and the divergences are carried over to the expressions for $t\to q V$ NLO decay widths, for which the complete expressions are given in the Appendix~\ref{app:dw1}. Here we only outline their form
\begin{eqnarray}
\Gamma^{V,\mathrm{virt}}_{a,b,ab} &=& \Gamma^{V(0)}_{a,b,ab}\Big[1 + \frac{\alpha_s}{4\pi}C_F \Gamma^{V(1)}_{a,b,ab}\Big]\,,\label{eq:FCNC_virt}\\
\nn \Gamma^{V,\mathrm{virt}}_{ag,bg} &=& \frac{\alpha_s}{4\pi}C_F \Gamma^{V(1)}_{ag,bg}\,,\\
\nn \tilde{\Gamma}^{V,\mathrm{virt}}_{ag,bg} &=& \frac{\alpha_s}{4\pi}C_F \tilde{\Gamma}^{V(1)}_{ag,bg}\,,
\end{eqnarray}
and stress that $\Gamma_{a,b,ab}^{V(1)}$ all posses IR divergences, while $\Gamma^{V(1)}_{ag,bg}$ and $\tilde{\Gamma}^{V(1)}_{ag,bg}$ are finite.

\subsubsection{Bremsstrahlung Contributions}
\begin{figure}[h]
\begin{center}
\includegraphics[scale=0.6]{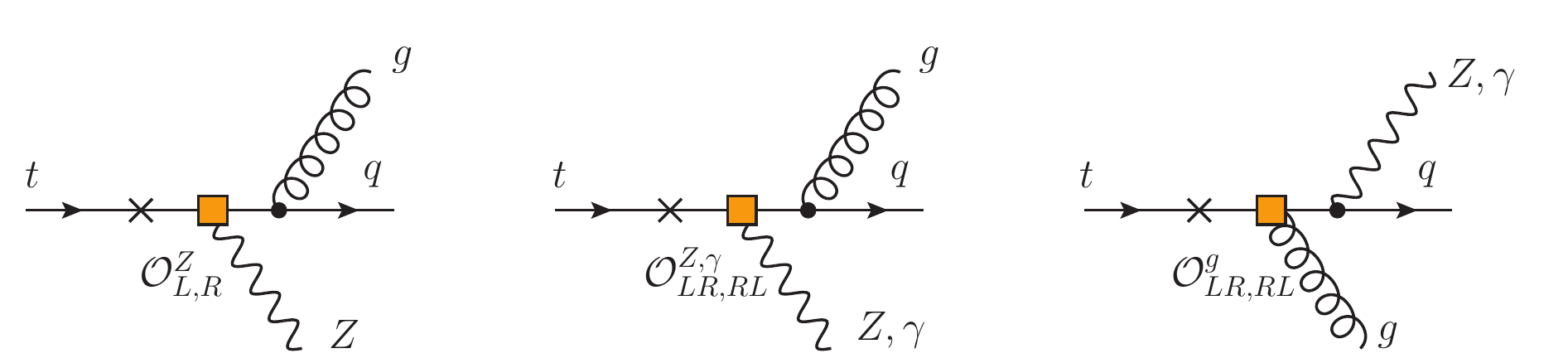}
\end{center}
\caption{Feynman diagrams for $t\to qgZ,\gamma$ bremsstrahlung process. Squares mark the insertion of effective operators given in Eq.~(\ref{eq:ops}) and crosses the additional points from which the gluon (in the first two diagrams) or $Z,\gamma$ (in the last diagram) can be emitted.}
\label{fig:fcnc_brems}
\end{figure}

The relevant Feynman diagrams contributing to $t\to q g Z,\gamma$ bremsstrahlung processes are given in Fig.~\ref{fig:fcnc_brems}. At the level of the decay width these diagrams give contributions of the same order in $\alpha_s$ as the one-loop virtual corrections presented above. Soft and collinear IR divergences emerge in the phase space integration and have to cancel the divergences present in $\Gamma^{V,\mathrm{virt}}$ once we sum the two contributions.
Computation of $\Gamma(t\to q g Z,\gamma)$ from the first two diagrams of Fig.~\ref{fig:fcnc_brems} gives contributions presented in Eqs.~(\ref{eq:bremsIR0}--\ref{eq:bremsIR3}) that indeed render the sum of three-body and two-body final state decay width IR finite
{\allowdisplaybreaks
\small
\begin{subequations}\label{GjustZ}
\be
\Gamma_{b}^{\gamma}&=&\Gamma_{b}^{\gamma(0)}\Bigg[1+\frac{\alpha_s}{4\pi}C_F\bigg[
2\log\Big(\frac{m_t^2}{\mu^2}\Big) + \frac{16}{3} - \frac{4\pi^2}{3}
\bigg]\Bigg]\,,\\
\Gamma_a^{Z}&=&\Gamma_{a}^{Z(0)}\Bigg[1+\frac{\alpha_s}{4\pi}C_F\bigg[
-4\log(1-r_Z)\log(r_Z) -2\frac{5+4r_Z}{1+2r_Z}\log(1-r_Z)\label{Ga}\\
&-&\frac{4r_Z(1+r_Z)(1-2r_Z)}{(1-r_Z)^2(1+2r_Z)}\log(r_Z)-
+\frac{5+9r_Z-6r_Z^2}{(1-r_Z)(1+2 r_Z)}-8\mathrm{Li}_2(r_Z) -\frac{4\pi^2}{3}
\bigg]\Bigg]\,,\nonumber\\
\Gamma_{b}^{Z} &=& \Gamma_{b}^{Z(0)}\Bigg[1+\frac{\alpha_s}{4\pi}C_F
\bigg[2\log\left(\frac{m_t^2}{\mu^2}\right) - 4 \log(1-r_Z)\log(r_Z)-\frac{2(8+r_Z)}{2+r_Z}\log(1-r_Z)\label{Gb}\\
&-&\frac{4r_Z(2-2r_Z-r_Z^2)}{(1-r_Z)^2(2+r_Z)}\log(r_Z)
-8\mathrm{Li}_2(r_Z)-\frac{16-11r_Z-17r_Z^2}{3(1-r_Z)(2+r_Z)} + 8 -\frac{4\pi^2}{3}\bigg]\Bigg]\,,\nonumber\\
\Gamma_{ab}^{Z} &=& \Gamma_{ab}^{Z(0)}\Bigg[1+\frac{\alpha_s}{4\pi}C_F
\bigg[
\log\left(\frac{m_t^2}{\mu^2}\right)-4\log(1-r_Z)\log(r_Z) -\frac{2(2+7r_Z)}{3r_Z}\log(1-r_Z)\label{Gab}\\
&-&\frac{4r_Z(3-2r_Z)}{3(1-r_Z)^2}\log(r_Z)+
 \frac{5-9r_Z}{3(1-r_Z)}+4-8\mathrm{Li}_2(r_Z)-\frac{4\pi^2}{3}\bigg]\Bigg]\,.\nonumber
\ee
\end{subequations}}
\normalsize
We were able to crosscheck our results given in Eqs.~(\ref{GjustZ}) with the corresponding calculation done for a virtual photon contributing to the $B\to X_s \ell^+ \ell^-$ spectrum~\cite{Asatryan:2002iy}. 
After taking into account the different dipole operator renormalization condition in~\cite{Asatryan:2002iy} (including mass renormalization) we find complete agreement with their results. $\Gamma_a^{Z}$ was also cross-checked with the corresponding calculation of the $t\to W b$ decay width at NLO in QCD~\cite{Li:1990qf}. Finally, we have compared our $\Gamma_b^Z$ expression with the results given by Zhang et al.\ in Ref.~\cite{Zhang:2008yn}. In the limit $r_Z \to 0$ our results agree with those given in \cite{Zhang:2008yn}, but we find disagreement in the $r_Z$ dependence. After our first publication of these results in~\cite{Drobnak:2010wh}, we were made aware of a new paper in preparation by the same authors, which has now been published \cite{Zhang:2010bm} and therein a corrected result for $\Gamma_b^Z$ is given that coincides with ours. 

\begin{wrapfigure}{r}{0.35\textwidth}
\begin{center}
\vspace{-0.5cm}
\includegraphics[scale=0.6]{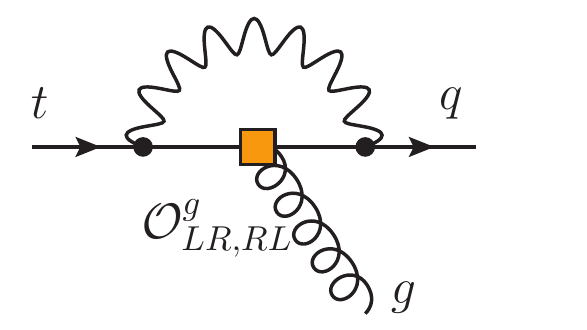}
\caption{Feynman diagram for one-loop virtual QED correction for $t\to q g$ decay governed by $\op_{LR,RL}^{g}$.}
\vspace{-0.5cm}
\label{fig:fcnc_virt_qed}
\end{center}
\end{wrapfigure}

The remaining bremsstrahlung contributions are induced by the gluonic dipole operator. What needs to be pointed out here is that while final result~(\ref{eq:oso}) for $\Gamma^Z$ is finite, $\Gamma^{\gamma}$ remains IR divergent. The divergences appear in $\Gamma_g^\gamma$ (squared contribution of third diagram of Fig.~\ref{fig:fcnc_brems}) and are not canceled by any of the virtual corrections we have considered.
To cancel them we would have to consider the decay width for $t\to q g$ governed by the gluonic operator  and include the one-loop virtual QED corrections. The corresponding Feynman diagram is shown in Fig.~\ref{fig:fcnc_virt_qed}.

The $t\to q \gamma g$ decay process involves three (one almost) massless particles in the final state. Virtual matrix element corrections contribute only at the soft gluon endpoint ($E_g = 0$) and result in non-vanishing  $b^\gamma b^g$ interference contributions. They involve IR divergencies which are in term canceled by the real gluon emission contributions. These also produce non-vanishing $|b^g|^2$ contributions, and  create a non-trivial photon spectrum involving both soft and collinear divergencies. The later appear whenever a photon or a gluon is emitted collinear to the light quark jet.  An analogous situation is encountered in the $B\to X_s\gamma$ decay measured at the $B$-factories. However, there the photon energy in the $B$ meson frame can be reconstructed and a hard cut ($E_\gamma^{\mathrm{cut}}$) on it removes the soft photon divergence. The cut also ensures that the $B\to X_s g$ process contributing at the end-point $E_\gamma = 0$ is suppressed. On the other hand, in present calculations the collinear divergencies are simply regulated by a non-zero strange quark mass, resulting in a moderate $\log(m_s/m_b)$ contributions to the rate.
The situation at the Tevatron and the LHC is considerably different.  The initial top quark boost is not known and the reconstruction of the decay is based on triggering on isolated hard photons with a very loose cut on the photon energy (a typical value being $E_\gamma>10 $\,GeV in the lab frame \cite{Aad:2009wy}). Isolation criteria are usually specified in terms of a jet veto cone $\Delta R = \sqrt {\Delta \eta^2 + \Delta \phi^2}$ where $\Delta\eta$ is the difference in pseudorapidity and $\Delta\phi$ the difference in azimuthal angle between the photon and nearest charged track. Typical values are $\Delta R > (0.2-0.4)$ \cite{Carvalho:2007yi}. 

Rather than including QED corrected $t\to qg$ rate, we render the $\Gamma_{g}^{\gamma}$ finite by modeling the non-trivial cuts in the top quark frame with a cut on the projection of the photon direction onto the direction of any of the two jets ($\delta r_j= 1- {\bf p}_\gamma \cdot {\bf p}_j / E_\gamma E_j$), where $j=g,q$ labels the gluon and light quark jet respectively. The effects of the different cuts on the decay Dalitz plot are shown in the left graph of Fig.~\ref{fig:foton_cuts_1}. 
Since at this order there are no photon collinear divergencies associated with the gluon jet, the $\delta r_g$ cut around the gluon jet has a numerically negligible effect on the rate. On the other hand the corresponding cut on the charm jet - photon separation does not completely remove the divergencies in the spectrum. However, they become integrable. The combined effect is that the contribution due to the gluonic dipole operator can be enhanced compared to the case of $B\to X_s \gamma$. 

We present the full analytical formulae for the $t\to q \gamma g$ and $t\to q Z g$ decay rates including the effects of kinematical cuts for the former channel in Appendix~\ref{app:dw2}. 

\subsubsection{Numerical analysis}
\begin{figure}[h]
\begin{center}
\includegraphics[height=6.5cm]{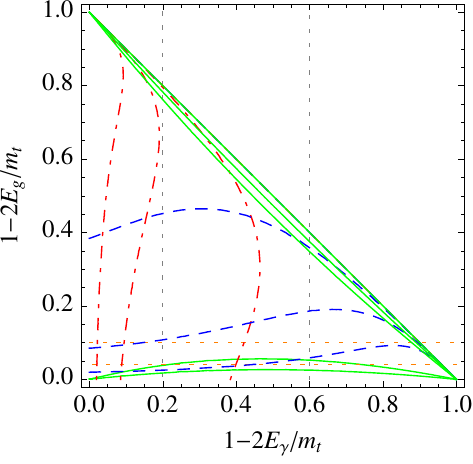}\hspace{0.8cm}
\includegraphics[height=6.5cm]{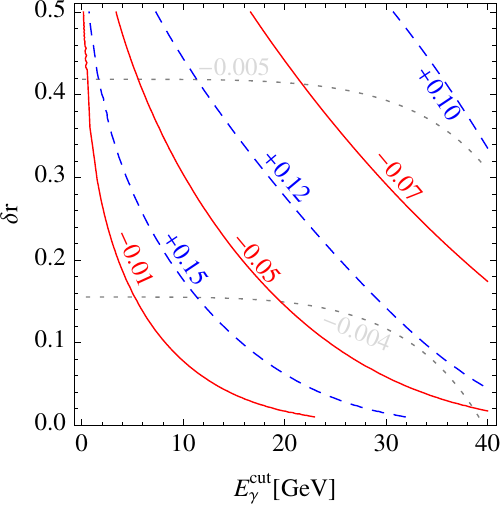}
\end{center}
\caption{{\bf Left}: The $t\to q \gamma g$ Dalitz plot. Contours of constant photon and gluon infrared and collinear divergent contributions are drawn in red (dot-dashed) and blue (dashed) lines respectively. The collinear divergencies appear at the horizontal and vertical boundaries of the phase-space, while the IR divergencies sit in the top and right corners. The cuts on the photon energy correspond to vertical lines, the cuts on the gluonic jet energy to horizontal lines. Full green lines correspond to cuts on the jet veto cone around the photon.
{\bf Right}: Relative size of $\alpha_s$ corrections to the $\mathrm{Br}(t\to q \gamma)$ at representative ranges of $\delta r_c\equiv\delta r$ and $E^{\mathrm{cut}}_\gamma$. Contours of constant correction values are plotted for $b^g=0$ (gray, dotted), $b^g=b^\gamma$ (red) and $b^g = - b^\gamma$ (blue, dashed).}
\label{fig:foton_cuts_1}
\end{figure}
\begin{figure}[h!]
\begin{center}
\includegraphics[scale=0.8]{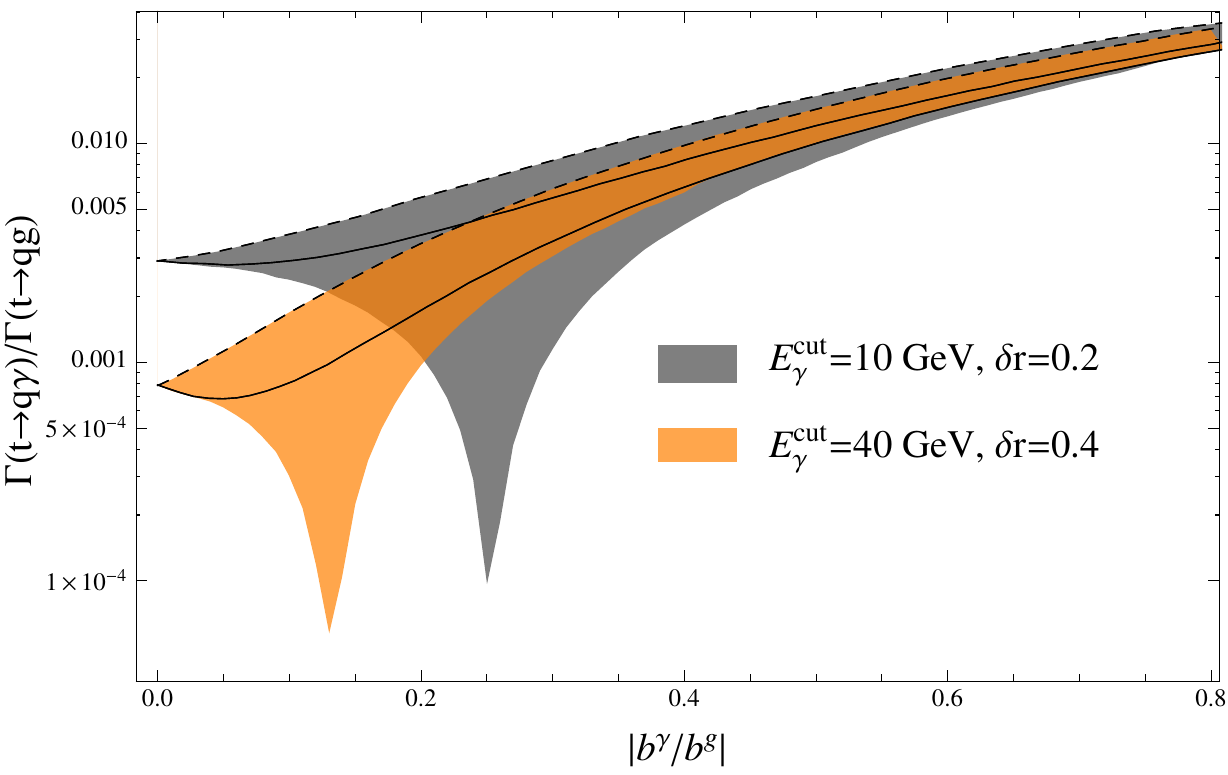}
\end{center}
\caption{\label{fig:tcg-tcg}  The ratio of radiative rates $\Gamma(t\to q \gamma) / \Gamma(t\to q g )$ versus the absolute ratio of the relevant effective FCNC couplings $|b^\gamma/b^g|$. Two representative choices of experimental kinematic cuts are shown. The two regions represent the possible spread due to the unknown relative phase between $b^\gamma$ and $b^g$ couplings, while the lines correspond to maximal positive (full) and negative (dashed) interference $b^\gamma b^g$. See text for details.}
\end{figure}

In all the numerical analysis of this section we use the following values for the parameters
\begin{align}
m_W &= 80.4\,\, \mathrm{GeV}\,, & m_t &=172.3\,\, \mathrm{GeV}\,,  &m_Z &= 91.2 \,\, \mathrm{GeV}\,,\label{eq:numV}\\
 \mu &= m_t\,,  & \alpha_s(m_t)&=0.107\,, & \sin^2 \theta_W &= 0.231\,.\nn
\end{align}
Turning first to $t\to q \gamma$ decay we show in the right graph of Fig.~\ref{fig:foton_cuts_1} the $b^g$ induced correction to the tree-level $\mathrm{Br}(t\to q \gamma)$ for representative ranges of $\delta r$ and $E^{\mathrm{cut}}_\gamma$.
We observe, that the contribution of $b^g$ can be of the order of $10-15\%$ of the total measured rate, depending on the relative sizes and phases of $\mathcal O_{LR,RL}^{g,\gamma}$ and on the particular experimental cuts employed. Consequently, a bound on $\mathrm{Br}(t\to q \gamma)$ can, depending on the experimental cuts, probe both $b^{g,\gamma}$ couplings. In order to illustrate our point, we plot  the ratio of radiative rates $\Gamma(t\to q \gamma) / \Gamma(t\to q g )$, both computed at NLO in QCD versus the ratio of the relevant effective FCNC dipole couplings $|b^\gamma/b^g|$ in Fig.~\ref{fig:tcg-tcg}. The NLO $\Gamma(t\to q g)$ result is taken from Ref.~\cite{Zhang:2008yn}. 
We show the correlation for two representative choices of experimental kinematic cuts for the $t\to q \gamma$ decay. The vertical spread of the bands is due to the variation of the relative phase between $b^\gamma$ and $b^g$ couplings. We also display the two interesting limits where the $b^\gamma b^g$ interference is maximal positive (zero relative phase) and negative (relative phase $\pi$). We see that apart from the narrow region around $|b^\gamma/b^g| \sim 0.2$, where the two contributions may be fine-tuned and conspire to diminish the total $t\to q \gamma$ rate, the two radiative rates are well correlated. In particular, depending on the kinematical cuts employed, there is a natural lower bound on ratio of decay rates, valid outside of the fine-tuned region. Finally, for  $|b^\gamma/b^g|>0.6$ the correlation becomes practically insensitive to the particular experimental cuts employed and also the unknown relative phase between $b^\gamma$ and $b^g$ couplings. 

For the $t\to qZ$ decay channel we present some numerical values to estimate the significance of QCD corrections. In particular we parametrize the decay width given in Eq.~(\ref{eq:oso}) as
\small
\be
\Gamma^Z\hspace{-0.2cm}&=&\hspace{-0.2cm}\frac{m_t}{16\pi}g_Z^2\Bigg\{
\frac{v^4}{\Lambda^4}|a^Z|^2  \Big[x_{a} +\frac{\alpha_s}{4\pi}C_F y_{a} \Big]+
\frac{v^2m_t^2}{\Lambda^4}|b^Z|^2  \Big[ x_{b}+\frac{\alpha_s}{4\pi}C_F y_{b}\Big]2
+\frac{v^3 m_t}{\Lambda^4}2\mathrm{Re}\{b^{Z*} a^Z\}\Big[x_{ab} +\frac{\alpha_s}{4\pi}C_F y_{ab}\Big]\nn\\
&+&|b^g|^2 \frac{v^2m_t^2}{\Lambda^4}\frac{\alpha_s}{4\pi}C_F y_{g}
+\frac{v^3 m_t}{\Lambda^4}\Big[
2\mathrm{Re}\{a^{Z*} b^{g}\}\frac{\alpha_s}{4\pi}C_F y_{ag}
-2\mathrm{Im}\{a^{Z*} b^{g}\}\frac{\alpha_s}{4\pi}C_F \tilde{y}_{ag}\Big] \nonumber\\
&+&\frac{v^2m_t^2}{\Lambda^4}\Big[2\mathrm{Re}\{b^{Z*}b^g\}\frac{\alpha_s}{4\pi}C_F y_{bg}-2\mathrm{Im}\{b^{Z*}b^g\}\frac{\alpha_s}{4\pi}C_F \tilde{y}_{bg} \Big] \Bigg\}\,.\label{eq:grdi}
\ee
\normalsize
Here $x_i$ stand for the tree-level contributions, while $y_i\,, \tilde{y}_i$ denote the corresponding QCD corrections. Numerical values of the coefficients are given in Tab.~\ref{table:num}. We see that corrections due to the gluon dipole operator are an order of magnitude smaller (except $y_g$, which is even more suppressed) than corrections to the $Z$ operators themselves and have opposite sign. 
\begin{table}[h]
\begin{center}
\begin{tabular}{llll}
\hline\hline
$x_{b}=2.36$ & $x_{a}=1.44$ & $x_{ab}=1.55$   \\
$y_{b}=-17.90$ & $y_{a}=-10.68$ & $y_{ab}=-10.52$ &$y_{g}=0.0103$\\
 $y_{bg}=3.41$ & $y_{ag}=2.80$ & $\tilde{y}_{bg}=2.29$ &  $\tilde{y}_{ag} = 1.50$\\
 \hline\hline
\end{tabular}
\caption{\label{table:num}Numerical values of coefficient functions appearing in Eq.~(\ref{eq:grdi}).} 
\end{center}
\end{table}

Next we investigate the relative change of the decay rates and branching ratios when going from  LO to NLO in QCD.
\begin{table}[!h]
\begin{center}
\begin{tabular}{l|l|l|l||l|l}\hline\hline
	&$b^Z=b^g=0$&$a^Z=b^g=0$&$a^Z=b^Z, b^g=0$& $b^Z=0, a^Z =b^g$ &$a^Z=0, b^Z =b^g$\\
\hline
$\Gamma^{\mathrm{NLO}}/\Gamma^{\mathrm{LO}}$&$0.92$& $0.91$& $0.92$ & $0.95$&$0.94$\\
$\Brfrac$&$1.001$&$0.999$&$1.003$ &$1.032$&$1.022$\\
\hline\hline
\end{tabular}
\caption{Numerical values of $\Gamma^{\mathrm{NLO}}/\Gamma^{\mathrm{LO}}$ and $\mathrm{Br}^{\mathrm{NLO}}(t\to q Z)/\mathrm{Br}^{\mathrm{LO}}(t\to q Z)$ for certain values and relations between Wilson coefficients.}
\label{brsZ}
\end{center}
\end{table}
The results are presented in Tab.~\ref{brsZ}. We see that the change in the decay width is of the order 10\%. There is a severe cancellation between the QCD corrections to $\Gamma(t\to c Z)$ and the main decay channel $\Gamma(t\to b W)$. This cancellation causes the change of the branching ratio to be only at the per-mille level when $b^g$ is set to zero. In the case when only operators $\mathcal O_{L,R}^Z$ are considered this cancellation is anticipated since the NLO correction to $\Gamma^Z_{a}$ is of the same form as the correction to the rate of the main decay channel. If we treat $b$ quarks as massless, exact cancellation is avoided only due to the difference in the masses of $Z$ and $W$ bosons. It is more surprising that similar cancellation is obtained also when only the dipole $Z$ operator is considered. However, setting $b^g=a^Z$ or $b^g=b^Z$, the impact of QCD corrections is increased by an order of magnitude and reaches a few percent.

\subsection{Summary}
To summarize, we have presented a study of $t\to q Z,\gamma$ decays mediated by the effective operators given in Eq.~(\ref{eq:ops}) at NLO in QCD. We found that QCD corrections can induce sizable mixing of the relevant operators, both through their renormalization scale running as well as in the form of finite matrix element corrections. These effects are found to be relatively small for the $t\to q Z$ decay, but can be of the order 10\% in the $t\to q \gamma$ channel, depending on the kinematical cuts employed. The accurate interpretation of experimental bounds on radiative top processes in terms of effective FCNC operators requires the knowledge of the experimental cuts involved and can be used to probe $\mathcal O^g_{LR,RL}$ contributions indirectly. 


\newpage
\section{Three-body $t\to q \ell^+ \ell^-$ decays }\label{sec:three_body}

This section is devoted to the study of $t \to q \ell^+ \ell^-$ with the basic goal of identifying discriminating effects of different NP models in top FCNCs which can be approached by the experimental study. Exploring the three-body decay channel brings about two main advantages. First one is the larger phase-space which offers more observables to be considered -- in particular the angular asymmetries among the final state lepton and jet directions. The second advantage is that the three-body final state that we are to consider is common to both $Z$ and $\gamma$ mediated FCNC decays. Since some BSM models predict observable FCNC top quark decays in both $Z$ and $\gamma$ channels the interference effects in the common three-body channel is something worth exploring.

Since the standard forward-backward asymmetry for the leptons vanishes in the photon mediated decays, we consider another asymmetry which we call the left-right asymmetry and is associated with the lepton angular distribution in the lepton-quark rest frame (see section \ref{sec:3body_obs}). This asymmetry is nontrivial also in the photon mediated decays. We explore the ranges of values for these two asymmetries in $t\rightarrow q\ell^+\ell^-$ decays mediated by both $Z$ boson and the photon. We also consider the interference effects as we expect them to significantly affect the ranges of the asymmetries. Our results can serve as a starting point for more elaborate investigations of experimental sensitivity to the proposed observables including QCD corrections, proper jet fragmentation and showering and the impact of experimental cuts and detector effects.

\subsection{Effective Lagrangian}
\begin{wrapfigure}{r}{0.3 \textwidth}
\begin{center}
\vspace{-0.8cm}
\includegraphics[scale=0.6]{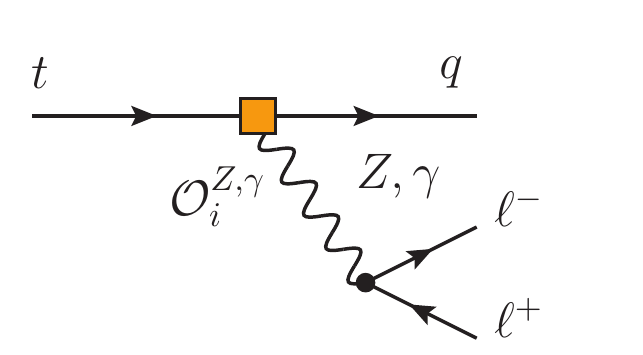}
\caption{$t\to q \ell^+ \ell^-$ Feynman diagram, where the FCNC operators $\op_{L,R}^Z$, $\op_{LR,RL}^{Z,\gamma}$, are given in Eq.~(\ref{eq:ops}).}\label{fig:fcnc_three}
\vspace{-0.3cm}
\end{center}
\end{wrapfigure}
The effective Lagrangian governing the FCNC top quark decay is considered to be the same as given in Eq.~(\ref{eq:Lagr}). We shall assume the FCNC mediating gauge boson ($Z$ boson or the photon) to further couple with a pair of charged leptons. Note that the gluonic operator will not play a role in the analysis of this section since the gluon does not couple to leptons. For completeness we present here the part of SM Lagrangian that couples charged leptons with the photon and the $Z$ boson 
\begin{eqnarray}
\mathcal L_{\ell} = g_Z Z_{\mu} \left[ c_R\, \bar \ell_R \gamma^{\mu} \ell_R + c_L \,\bar \ell_L \gamma^{\mu} \ell_L\right] + e A_{\mu} \bar \ell \gamma^{\mu} \ell\,, \label{eq:Lell}
\end{eqnarray}
where the $Z$ couplings are $c_R = \sin^2\theta_W$ and $c_L =- (\cos 2\theta_W /2)$. The Feynman diagram of the process that we shall be studying is given in Fig.~{\ref{fig:fcnc_three}}. Since top quark is massive enough for the produced $Z$ boson to be on-shell we shall use the standard Breit-Wigner formula to accommodate for its finite decay width by replacing $(s-m_Z^2) \to (s - m_Z^2 +\ii m_Z \Gamma_Z)$ in the denominator of the propagator, where $s$ denotes the square of the intermediate $Z$ boson momentum and $\Gamma_Z$ is its total decay width.

\subsection{Observables}\label{sec:3body_obs}
We consider scenarios where detection of a NP signal in the FCNC decay channel $t\to q \ell^+\ell^-$ could be most easily complemented by other observables in the same decay mode. This would allow distinguishing between different possible effective amplitude contributions and thus different underlying NP models. 
We neglect kinematical effects of lepton masses and the light quark jet invariant mass, as these are expected to yield immeasurably small effects in the kinematical phase-space set by the large top quark mass. 

We start with the double-differential decay rate ${\dd \Gamma}/{(\dd u \dd s)}$, where $s=m_{\ell^+\ell^-}^2$ is the invariant mass of the lepton pair and
$u = m_{j \ell^+}^2$ is the invariant mass of the final state quark (jet) and the lepton of positive charge $\ell^+$. Integrating this decay rate over one of the kinematical variables, we obtain
the partially integrated decay rate distributions ($\dd\Gamma / \dd u$, $\dd\Gamma / \dd s$), while the full decay rate ($\Gamma$) is obtained after completely integrating these distributions. The branching ratio is obtained by normalizing the decay width to the width of the main decay channel~(\ref{eq:SM_MDC}).

\begin{figure}[h]
\begin{center}
\includegraphics[scale=0.7]{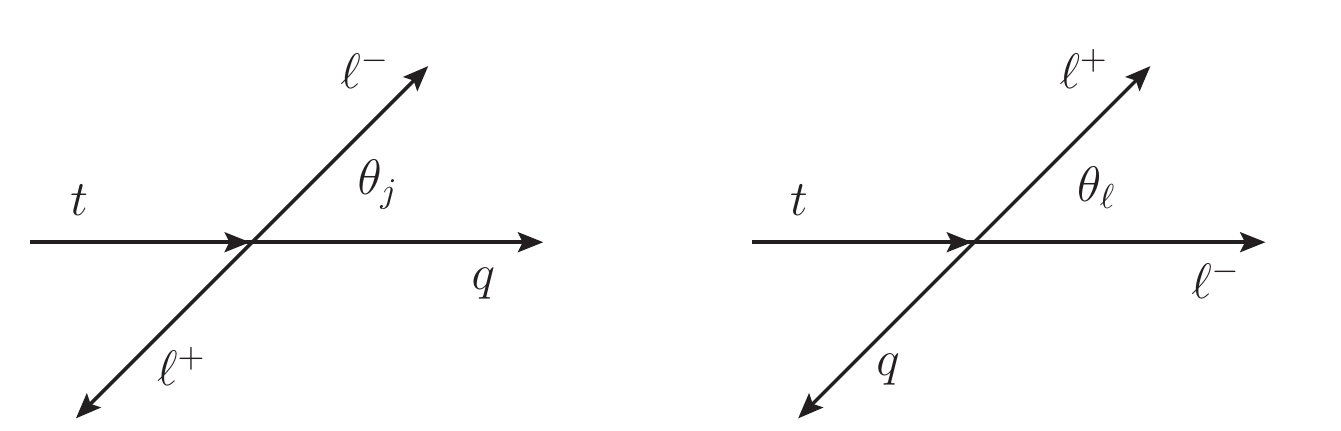}
\caption{\label{fig:angles} Definition of two angles relevant to our analysis. The arrows represent the three momenta of particles. Left diagram corresponds to the lepton pair rest-frame and the right diagram to the rest-frame of the positively charged lepton and the light quark. }
\end{center}
\end{figure}

The differential decay rate distribution can also be decomposed in terms of two independent angles, as defined in Fig.~\ref{fig:angles}. 
In the $\ell^+ \ell^-$ rest-frame $z_j=\cos\theta_{j}$ measures the relative direction between the negatively charged lepton and the light quark jet. Conversely, in the rest-frame of the positively charged lepton and the quark jet, we can define $z_{\ell}=\cos\theta_{\ell}$ to measure the relative directions between the two leptons. In terms of these variables, we can define two asymmetries ($i=j,\ell$) as
\begin{equation}
A_i = \frac{\Gamma_{z_i>0}-\Gamma_{z_i<0}}{\Gamma_{z_i>0} + \Gamma_{z_i<0}}\,,
\label{eq:As}
\end{equation}
where we have denoted $\Gamma_{z_i\lessgtr 0}$ as the integrated decay rates with an upper or lower cut on one of the $z_i$ variables. We can then identify $A_{j}\equiv A_{\mathrm{FB}}$ as the commonly known {\sl forward-backward asymmetry} (FBA) and in addition define $A_{\ell}\equiv A_{\mathrm{LR}}$ as the {\sl left-right asymmetry} (LRA). The two angles and the asymmetries they define are related via a simple permutation of final state momentum labels between the quark jet and the positively charged lepton, and consequently via a $u\leftrightarrow s$ interchange.

Since the asymmetries as defined in Eq.~(\ref{eq:As}) are normalized to the decay rate, they represent independent observables with no spurious correlations to the branching ratio. On the other hand, correlations among the two asymmetries are of course present and indicative of the particular NP operator structures contributing to the decay.

\subsection{Signatures}
Next we study the signatures of various possible contributions to the $t\to q \ell^+\ell^-$ decay using the integrated FBA and LRA observables defined in the previous section. 
Before exploring individual mediation cases a general remark is in order. Since all the effective operators of our basis~(\ref{eq:ops}) come suppressed with an undetermined NP cut-off scale, the actual values of the effective couplings ($a$, $b$) are unphysical (can always be shifted with a different choice of the cutoff scale). The total decay rate determines the overall magnitude of the physical product of the couplings with the cut-off scale. On the other hand relative sizes or ratios of couplings (independent of the cut-off) determine the magnitude of the asymmetries. The extremal cases are then naturally represented when certain (combinations of) couplings are set to zero -- often the case in concrete NP model implementations.

\subsubsection{Photon mediation}
As pointed out in section~\ref{sec:fcnc_matrix_elements}, the direct detection of energetic photons is considered to be the prime strategy in the search for photon mediated FCNC top quark decays. However the $t\to q\ell^+\ell^-$ channel, where the photon is coupled to the charged lepton pair can serve as an additional handle. Due to the infrared pole in the di-lepton invariant mass distribution we introduce a low $\sy=m_{\ell}^2/m_t^2$ cut denoted $\sy_{\mathrm{min}}\equiv\epsilon/m_t^2$ and present the total decay width as its function
\begin{eqnarray}
\Gamma^{\gamma} = \frac{m_t}{16\pi^3}\frac{g_Z^4 v^4 }{\Lambda^4}B_{\gamma}f_{\gamma}(\sy_{\mathrm{min}})\,, \hspace{0.5cm}
B_{\gamma} = \frac{m_t^2}{v^2}\frac{e^4}{g_Z^4}\frac{|b_{LR}^{\gamma}|^2+|b_{RL}^{\gamma}|^2}{2}\,.\label{eq:gamma_gamma}
\end{eqnarray}

\begin{wrapfigure}{r}{0.5\textwidth}
\begin{center}
\vspace{-0.7cm}
\includegraphics[scale=0.6]{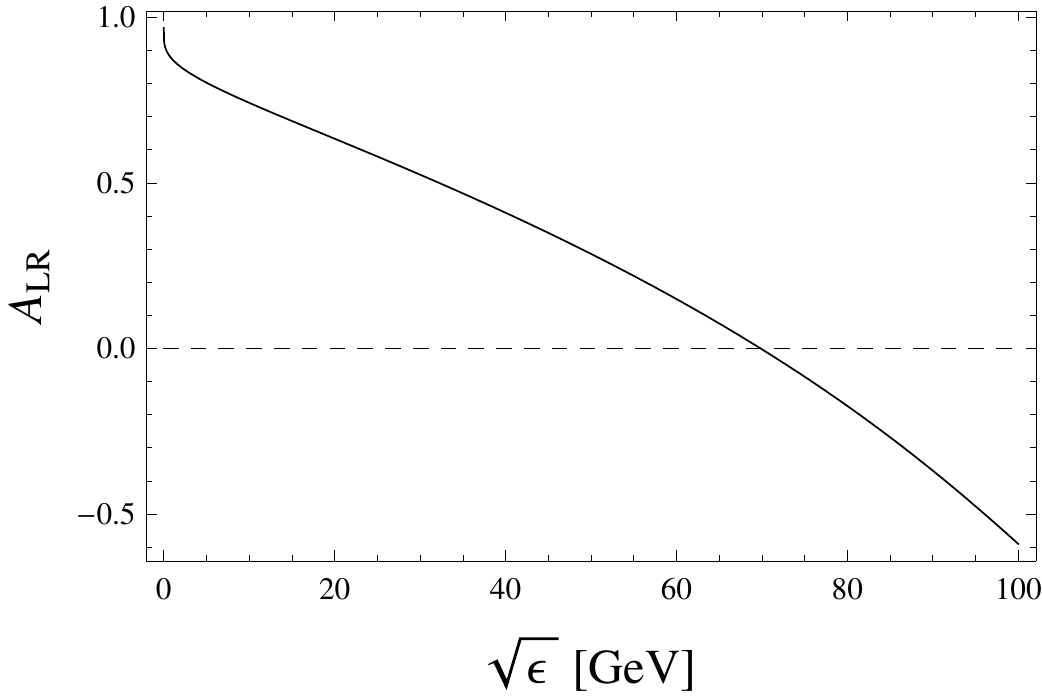}
\vspace{-0.3cm}
\caption{\label{fig:lra_photon} The dependence of the photon-mediated LRA on the low di-lepton invariant mass cut $\epsilon$. }
\end{center}
\vspace{-1.0cm}
\end{wrapfigure}

The physical cut is of course at $\epsilon=4m_{\ell}^2$. We have also define an auxiliary variable $B_{\gamma}$ summarizing the relevant NP parameter dependencies. Function $f_{\gamma}$ depends only on the di-lepton invariant mass cutoff $\sy_{\mathrm{min}}$ and is presented in Eq.~(\ref{fg}) of the Appendix.

While the FBA vanishes identically, due to the purely vectorial coupling of the photon to the leptons, the LRA can be written in the following way
\begin{eqnarray}
A_{\mathrm{LR}}^\gamma&=&\frac{g_{\gamma}(\sy_{\mathrm{min}})}{f_{\gamma}(\sy_{\mathrm{min}})}\,,\label{eq:def_g}
\end{eqnarray}
where function $g_{\gamma}$ is presented in Eq.~(\ref{gg}) of the Appendix and also depends only on the cut of the di-lepton invariant mass.  Consequently LRA does not depend on the effective dipole couplings in any way, however showing a non-trivial dependence on the low $\sy$ cut, which we plot in Fig.~\ref{fig:lra_photon}. We see that the value of the integrated LRA is highly sensitive to the cut having value $1$ in the limiting case of $\sqrt{\epsilon}\to 0$, decreasing with the cut set higher and even exhibiting a change of sign between $60$ and $80$ GeV.

\subsubsection{$Z$ mediation}
Current search strategies for $t\to q Z$ decays actually consider $t\to q \ell^+\ell^-$ decay channel, where the charged lepton pair is to be identified as the decay product of the $Z$ boson. This is achieved by imposing a cut on the invariant lepton mass around the $Z$ mass to reduce backgrounds. As long as such cuts are loose compared to the width of the $Z$, we do not expect them to affect our observables. 
The decay width and the two asymmetries can be written as
\begin{subequations}\label{eq:fcnc_Z_1}
\begin{eqnarray}
\Gamma^Z &=&  \frac{m_t}{16\pi^3}\frac{g_Z^4v^4}{\Lambda^4}\Big[f_A A + f_B B + f_C C\Big]\,, \label{eq:xx}\\
A_{\mathrm{FB}}^Z &=& f_{\alpha\beta\gamma}\frac{\alpha -4\beta+4\gamma}{f_A A + f_B B + f_C C}\,, \label{eq:xx1}\\
A_{\mathrm{LR}}^Z &=&\frac{ g_A A + g_B B + g_C C + g_{\alpha\beta\gamma}[\alpha-4\beta+4\gamma]}{f_A A + f_B B + f_C C}\,,\label{eq:xx2}
\end{eqnarray}
\end{subequations}
where the parts depending on the NP generated FCNC top quark couplings have been separated from the parts that depend just on the SM parameters and the phase space integration. The NP dependent parameters introduced are
\begin{align}
 A&= \frac{|a_R^Z|^2+|a_L^Z|^2}{2} L_+\,,&
\alpha &= \frac{|a_R^Z|^2-|a_L^Z|^2}{2}L_-\,,\\
\nonumber B&= \frac{m_t^2}{v^2}\frac{|b_{LR}^Z|^2+|b_{RL}^Z|^2}{2}  L_+\,,&
\nonumber  \beta &= \frac{m_t^2}{v^2}\frac{|b_{LR}^Z|^2-|b_{RL}^Z|^2}{2}L_-\,,\\
\nonumber C&=-\frac{m_t}{v}\frac{\mathrm{Re}\{b^Z_{LR}a_L^{Z*}+b_{RL}^Za_{R}^{Z*}\}}{2} L_+\,,&
\nonumber \gamma&=\frac{m_t}{v}\frac{ \mathrm{Re}\{b_{LR}^Za_L^{Z*}-b_{RL}^{Z}a_{R}^{Z*}\}}{2}L_-\,,
\end{align}
where $L_{\pm} = \frac{1}{2}\sin^4\theta_W\pm\frac{1}{8}\cos^22\theta_W$ are the factors coming from the charged lepton couplings to the $Z$ boson governed by the Lagrangian given in Eq.~(\ref{eq:Lell}).

On the other hand the remaining parameters $f_i$ and $g_i$ do not depend on the effective FCNC couplings  and are presented in the form of integrals in Eqs.~(\ref{eq:fA}, \ref{eq:gZ}) of the Appendix.
\begin{figure}[h]
\begin{center}
\includegraphics[scale=0.8]{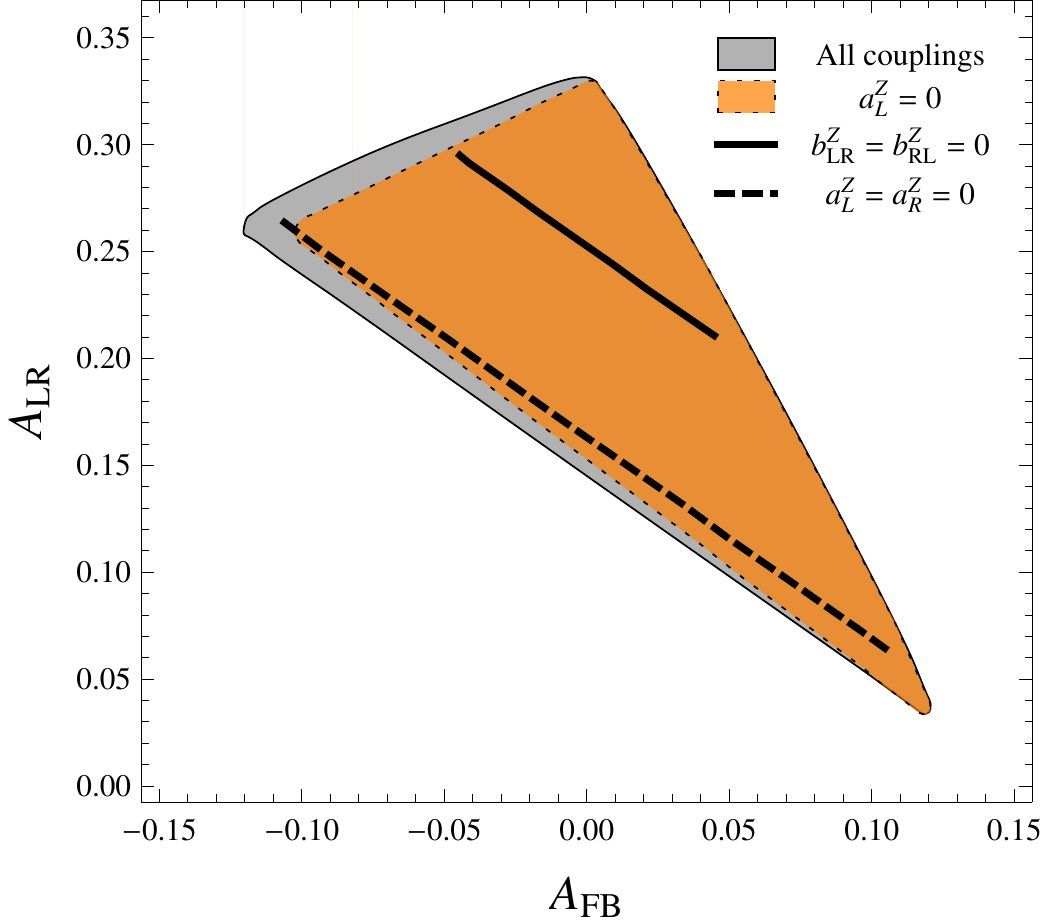}
\caption{\label{fig:assym_Z} The correlation of FBA and LRA in $Z$ mediated decay. The gray area (solid border) represents decays with all possible current and dipole $Z$ FCNC couplings. The orange area (dotted border) corresponds to decays with $a_L^Z$ set to zero, while the solid and dashed lines represent decays with only current and only dipole couplings respectively. The following numerical values have been used $m_{Z,t} = 91.2\,,171.2$ GeV, $\Gamma_Z = 2.5$ GeV and $\sin^2\theta_W = 0.231$.}
\end{center}
\end{figure}

Performing a random sweep over the values of FCNC couplings that give the same value of the total FCNC decay width (\ref{eq:xx}), we explore the possible ranges and correlations between the two asymmetries (\ref{eq:xx1}, \ref{eq:xx2}) in Fig.~\ref{fig:assym_Z}. On the same plot we also project the limits, where only dipole or only current interactions of the  $Z$ contribute. In Ref.~\cite{Fox:2007in} strong indirect limits were reported on the left-handed FCNC couplings of the $Z$ coming from low energy observables. Therefore we also superimpose the possible predictions for the two asymmetries when these couplings are set to zero.

We observe that the LRA can be used to distinguish between dipole and current FCNC couplings of the $Z$, while the FBA can distinguish the chiralities of the couplings.


\subsubsection{Interference of photon and $Z$ mediation}
Several NP models predict comparable decay rates for $t\to q Z,\gamma$. This may in turn lead to a situation, where an experimental search using a common final state may be more promising than dedicated  searches in each channel separately. In addition, the asymmetries in $t\to q\ell^+\ell^-$ may shed additional light on the specific couplings involved. The decay rate in this case depends again on the di-lepton invariant mass cutoff 
\begin{eqnarray}\label{eq:FCNC_width2}
\Gamma^{\gamma+Z}(\sy_{\mathrm{min}}) =  \Gamma^{\gamma}(\sy_{\mathrm{min}}) + \Gamma^{Z}(\sy_{\mathrm{min}}) + \Gamma^{\mathrm{int}}(\sy_{\mathrm{min}})\,,
\end{eqnarray}
where the pure photon contribution $\Gamma^{\gamma}$ is given in Eq.~(\ref{eq:gamma_gamma}) while the pure $Z$ and interference contributions can be written as 
\begin{subequations}\label{eq:new_fs}
\begin{eqnarray}
\Gamma^{Z}(\sy_{\mathrm{min}}) &=&\frac{m_t}{16\pi^3}\frac{v^4g_Z^4}{\Lambda^4}\Big[f_A^{\epsilon}A+f_B^{\epsilon}B+f_C^{\epsilon}C\Big]\,,\\
\Gamma^{\mathrm{int}}(\sy_{\mathrm{min}}) &=&\frac{m_t}{16\pi^3}\frac{v^4g_Z^4}{\Lambda^4}\Big[f_{W_{12}} (W_1+W_2) +f_{W_{34}}( W_3+ W_4)\Big]\,.
\end{eqnarray}
\end{subequations}
Here $W_1,\dots,W_4$ are the newly introduced NP dependent constants containing both $Z$ and $\gamma$ effective FCNC couplings 
\begin{align}
W_1&= \frac{m_t^2}{v^2}\frac{e^2}{g_Z^2}\frac{1}{2}\mathrm{Re}\{b_{LR}^{\gamma*}b_{LR}^{Z}c_L+b_{RL}^{\gamma*}b_{RL}^{Z}c_R \} \,, &
 W_2&= \frac{m_t^2}{v^2}\frac{e^2}{g_Z^2}\frac{1}{2}\mathrm{Re}\{b_{LR}^{\gamma*}b_{LR}^{Z}c_R+b_{RL}^{\gamma*}b_{RL}^{Z}c_L \}\,,& \\
\nonumber W_3&= \frac{m_t}{v}\frac{e^2}{g_Z^2}\frac{1}{2}\mathrm{Re}\{-b_{LR}^{\gamma*}a_L^{Z}c_L - b_{RL}^{\gamma*}a_{R}^{Z}c_R \}\,, &
\nonumber W_4&= \frac{m_t}{v}\frac{e^2}{g_Z^2}\frac{1}{2}\mathrm{Re}\{-b_{LR}^{\gamma*}a_L^{Z}c_R - b_{RL}^{\gamma*}a_{R}^{Z}c_L \}\,.
\end{align}
The two asymmetries can be expressed as the following fractions
\begin{eqnarray}
 A_{\mathrm{FB}}^{\gamma+Z} &=&  
 \frac
 {f_{\alpha\beta\gamma}^{\epsilon} (\alpha-4\beta+4\gamma)+ f_W \big(2(W_2-W_1)+ W_4-W_3\big)}
 {f_{\gamma}B_{\gamma}  + f_A^{\epsilon}A+f_B^{\epsilon}B+f_C^{\epsilon}C +f_{W_{12}} (W_1+W_2) +f_{W_{34}}( W_3+ W_4)}
 \,,\\
 A_{\mathrm{LR}}^{\gamma+Z} &=& \frac
 {g_{\gamma}B_{\gamma}+g_A^{\epsilon} A + g_B^{\epsilon} B+g_C^{\epsilon} C+g_{\alpha\beta\gamma}(\alpha-4\beta+4\gamma)+\sum_{i=1}^4 g_{W_i} W_i }
 {f_{\gamma}B_{\gamma}  + f_A^{\epsilon}A+f_B^{\epsilon}B+f_C^{\epsilon}C +f_{W_{12}} (W_1+W_2) +f_{W_{34}}( W_3+ W_4)}\,.
\end{eqnarray}
The newly introduced functions $f_i$ and $g_i$ now depend on the $Z$ boson parameters as well as the di-lepton invariant mass cutoff $\epsilon$. They are presented in Eqs.~(\ref{eq:fcnc_int_1}) of the Appendix. 
\begin{figure}[h]
\begin{center}
\includegraphics[scale=0.7 ]{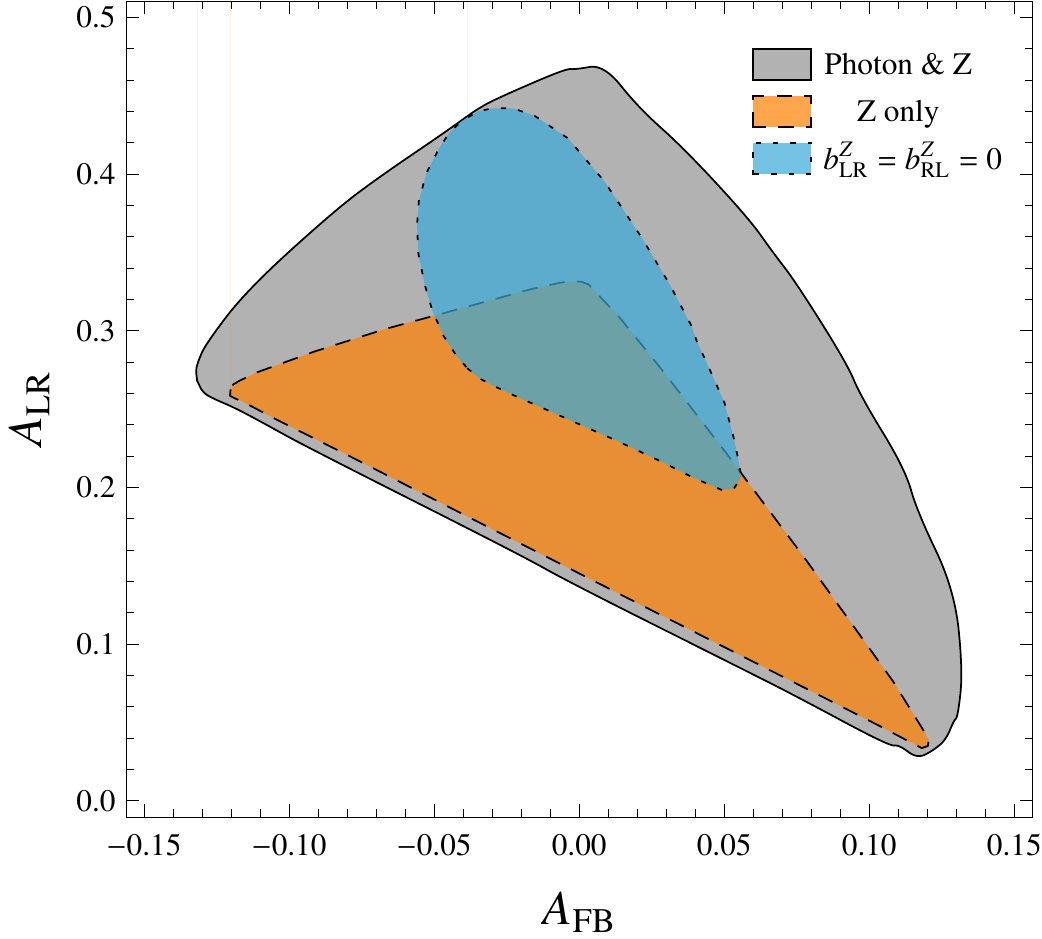}
\caption{\label{fig:assym_Zg} The correlation of FBA and LRA in $Z$ and $\gamma$ mediated decay. The gray area (solid border) represents decays with all possible $Z$ and $\gamma$ FCNC couplings. The blue (dotted border) area corresponds to decays with only current $Z$ FCNC couplings. For comparison, the orange (dashed border) area represents $Z$ mediated decays. Numerical values used are the same as in Fig.~\ref{fig:assym_Z} with addition of $\sqrt{\epsilon} = 40 $ GeV.}
\end{center}
\end{figure}

Performing a random sweep across both $\gamma$ and $Z$ NP FCNC couplings that give the same decay width~(\ref{eq:FCNC_width2}) we explore the possible correlation between the FBA and the LRA in this scenario. We present the results with a fixed cut on $\sqrt{\epsilon}$ set to $40$~GeV in Fig.~\ref{fig:assym_Zg}. We also present possible points for the case when only the current FCNC $Z$ couplings contribute. We observe that in principle interference effects can produce a larger LRA compared to the case of pure $Z$ mediation.

\begin{wraptable}{r}{0.3\textwidth}
\begin{tabular}{c|c|c}
 & only $Z$ & $Z$ and $\gamma$ \\\hline\hline
$A_{\mathrm{FB}}$ &0.045 & 0.035\\
$A_{\mathrm{LR}}$ &0.206 & 0.226\\\hline\hline
\end{tabular}
\caption{Values of FBA and LRA for highest allowed coefficients given by Fox et al. in Ref.~\cite{Fox:2007in}, $\sqrt{\epsilon}= 40$ GeV.}
\label{tab:asym_tabela}
\end{wraptable}

In Ref.~\cite{Fox:2007in} upper bounds on coefficients accompanying the operators responsible for FCNC $t\to c Z$ and $t\to c \gamma$ are presented. Using transcription formulae presented in Eqs.~(\ref{eq:fcnc_trans}) of the Appendix, we can evaluate FBA and LRA associated with these upper bounds. The numerical values are presented in Tab.~\ref{tab:asym_tabela}.

These values serve just for illustration that nonzero values of asymmetries can indeed be obtained. They do not represent any kind of upper bounds for asymmetries. There is no reason to think that the highest allowed values of coefficients (\ref{eq:fcnc_trans}) are to give the largest possible asymmetries which are complicated functions of these coefficients.

\subsection{Summary}

We have considered the top quark decay $t\to q\ell^+\ell^-$ as a probe of BSM physics manifested in FCNC top quark transitions. In addition to the branching ratio, we have defined two angular asymmetries which can serve to further discriminate between different NP scenarios. Comparing possible contributions to the decay mode via SM $\gamma$ and $Z$ mediation we can draw the following conclusions: large values of FBA ($|A_{\mathrm{FB}}|\gg 0.1$) cannot be accounted for in decay modes mediated by $Z,\gamma$ bosons as long as we assume these bosons to have SM couplings to the charged leptons. We have shown in Fig.~\ref{fig:assym_Z}, that the $A_{\mathrm{FB}}\in[-0.12,0.12]$.

A measured point in $(A_{\mathrm{FB}},A_{\mathrm{LR}})$ plane could exclude models with only current or only dipole FCNC couplings of $Z$  if it were located off the solid and dashed black lines in Fig.~\ref{fig:assym_Z}. Treating the $Z$ and photon as indistinguishable mediators expands the allowed LRA region  to larger positive values.

Current experimental sensitivity studies look at the two-body decay modes $t\to c Z,\gamma$ ~\cite{Carvalho:2007yi}. Our analysis may be applicable to the potential measurement of $t\to c Z$ at ATLAS since they will be identifying the $Z$ boson through its decay to a lepton pair. Angular asymmetries of this pair and the remaining hard jet could provide additional information on the $tcZ$ FCNC vertex. The
$t\to c\gamma$ decay is generically characterized by a single high $p_T$ photon. Current search strategies for this FCNC include the detection of this photon, and not its eventual decay to a lepton pair. 

In order to fully explore our decay mode, one would need to relax or modify certain criteria used by current search strategies to reduce SM backgrounds. In addition, the reconstruction of the LRA might require top quark charge tagging. In principle our results are applicable also to the purely hadronic decay modes, where the two leptons are replaced by $b$-tagged jets for example, however in this case the asymmetries are compromised by the lack of knowledge of the sign of the $b$ quark charges.



\chapter{NP in Top Decays: Charged currents}
\renewcommand{\op}{\mathcal Q}
\label{chap:CC}
\section{Introduction}
In this chapter we analyze the possible deviations from the SM coupling strength and structure of the {\sl charged quark currents}. Following the strategy outlined in section~\ref{sec:strategy} of the introductory chapter we explore on one hand the implications of such deviations on the low energy $B$ physics and on the other hand delve into the possible consequences to be observed in top quark decays, particularly focusing on the main decay channel and the $tWb$ interaction.
Results presented in this chapter are based on our published work ~\cite{Drobnak:2011aa,Drobnak:2011wj,Drobnak:2010ej}. 

Unlike in the case of FCNC top quark decays, the complete analysis of low energy constraints for model independent anomalous charged quark currents has not been performed yet. In particular analysis of anomalous $tWd,s$ couplings on $B$ meson mixing has been attempted in Ref.~\cite{Lee:2008xs,Lee:2010hv}, using however a subset of all possible effective $tWd,s$ operators. Furthermore effects in $b\to s \gamma$ transition have been analyzed in Ref.~\cite{Grzadkowski:2008mf}. Again, only a subset of the operators that we shall be considering has been used and no other $\Delta B=1$ transitions have been ruminated.

We therefore first set out to upgrade the analysis of indirect implications available in the literature
focusing on observables in $\Delta B=2$ and $\Delta B=1$ processes presented in sections \ref{sec:dB2} and \ref{sec:dB1} respectively. The experimentally well measured low energy observables which mostly agree with the SM predictions provide constraints on the anomalous charged quark current structures. Since the top quark with its possible anomalous interactions is in these processes a virtual particle we label such constraints on top quark physics as indirect.

Only then do we turn to the direct top quark physics where the structure of $tWb$ vertex can be probed  by analyzing the main decay channel of on-shell produced top quarks at Tevatron and LHC. The derived indirect constraints give us an idea of how much room there is left for deviation away from SM predictions given that the NP can be parametrized in the form of anomalous $tWb$ couplings.

We briefly comment that there is another class of observables in which virtual top quarks play an important role called {\sl electroweak precision observables} (EWPO). Namely, anomalous $tWb$ vertices can effect the $S,T,U$ oblique parameters and the $Z\to b\bar{b}$ decays (see for example \cite{Peskin:1991sw,ALEPH:2005ab}). Following our publications of the results that are presented in this work, the analysis of anomalous top quark couplings on EPWO was given in Refs.~\cite{Zhang:2012cd,Greiner:2011tt}. Although the operator basis parametrizing NP used by the authors of Refs.~\cite{Zhang:2012cd,Greiner:2011tt} does not coincide with the one used in this work, especially in regards to flavor structure, there are some common operators for which we can confront the obtained indirect bounds to find they are comparable and compatible.

This chapter is structured as follows. The first section is devoted to the formulation of the effective theory. Specifying our framework in terms of effective operator basis we proceed to analyze the effect in $B$ physics, giving detailed study of $\Delta B = 2$ transitions, namely $B$ meson mixing and $\Delta B = 1$ transitions comprised of FCNC $B$ meson decays. In the last section we focus on the top quark physics and the effects of anomalous $tWb$ couplings on helicity fractions in its main decay channel. Here we perform our analysis at NLO in QCD, which seems sensible, since as pointed out in section~\ref{sec:hfSM}, NLO QCD corrections prove crucial in consideration of helicity fractions within the SM. We show that the direct bounds are in a sense complementary to the indirect bounds and reveal a nice interplay between top and bottom physics. 

\section{Effective Lagrangian}

Following the strategy outlined in section~\ref{sec:strategy}, our first objective is to specify the operator basis  $\mathcal Q_i$ of our interest, namely the determination of gauge and flavor structure of the dimension six operators appearing in the effective Lagrangian (\ref{eq:lagr}). To analyze the effects in $B$ physics, we will have to further perform the second step illustrated in Fig.~\ref{fig:intout}. Much like in the SM case presented in sections \ref{sec:dB2} and \ref{sec:dB1}, we will have to integrate out the SM degrees of freedom with masses greater than that of the $b$ quark, matching $\mathcal L_{\mathrm{eff}}$ to low energy effective Lagrangians (\ref{eq:LSMmix}, \ref{eq:loweff1}). By doing that we shall gain access to different $B$ physics observables of interest and the possibility to see how predictions are affected by NP.

\subsection{Gauge structure}
Our operator basis consists of all dimension-six operators invariant under the SM gauge group that generate charged current quark interactions with the $W$. The possible gauge structures we can use are \cite{Buchmuller:1985jz}
\begin{subequations}\label{eq:GaugeStr}
\begin{eqnarray}
&&\big[\bar{u}\gamma^{\mu}d\big](\phi_u^{\dagger}\ii D_{\mu}\phi_d)\,,\\
&&\big[\bar{Q}\gamma^{\mu} Q\big](\phi_d^{\dagger}\ii D_{\mu}\phi_d)\,,\hspace{0.5cm}\big[\bar{Q}\gamma^{\mu}\tau^a Q\big](\phi^{\dagger}\tau^a\ii D_{\mu}\phi)\,,\label{eq:GaugeStr_LL}\\
&&\big[\bar{Q}\sigma^{\mu\nu}\tau^a u\big]\phi_u W_{\mu\nu}^a\,,\\
&&\big[\bar{Q}\sigma^{\mu\nu}\tau^a d\big]\phi_d W_{\mu\nu}^a\,.
\end{eqnarray}
\end{subequations}
Here $Q$ stands for the quark $SU(2)_L$ doublet $u$, $d$ are the up- and down-type quark $SU(2)_L$ singlets respectively, $\tau^a$ are the $SU(2)_L$ Pauli matrices. In addition we have the covariant derivative and field strength definitions
\begin{eqnarray}
D_{\mu}&=&\partial_{\mu}+\ii \frac{g}{2}W_{\mu}^a\tau^a +\ii \frac{g'}{2}B_{\mu} Y\,,  \\
W^a_{\mu\nu}&=&\partial_{\mu}W_{\nu}^a-\partial_{\nu}W_{\mu}^a - g\epsilon_{abc}W_{\mu}^b W_{\nu}^c\,,\nn
\end{eqnarray}
and finally $\phi_{u,d}$ are the up- and down-type Higgs fields (in the SM $\phi_u\equiv \tilde{\phi} =\ii \tau^2 \phi_d^*$). The operators in Eq.~(\ref{eq:GaugeStr}) are written with quark fields in the interaction basis being flavor universal.

\subsection{Flavor structure}

On the flavor side, we restrict the structure of the operators to be consistent with {\sl Minimal Flavor Violation} (MFV) hypothesis \cite{Buras:2003jf,D'Ambrosio:2002ex,Grossman:2007bd}, which postulates that even in the presence of NP operators, Yukawa matrices present in the SM remain the sole source of flavor violation. 

The way to implement this concept is to make the operators formally invariant under the SM flavor group~(\ref{eq:SM_G_flav}) and insist that the only $\mathcal G^{\rm SM}$ symmetry breaking spurionic fields in the theory are the up and down quark Yukawa matrices $Y_{u,d}$, introduced in Eq.~(\ref{eq:intro_yuk}), formally transforming as $(3,\bar 3,1)$ and $(3,1,\bar 3)$ respectively.
 
The four distinct quark bilinears appearing in Eq.~(\ref{eq:GaugeStr}) have different transformation properties under $\mathcal G^{\rm SM}$: $\bar u d$, $\bar Q Q$, $\bar Q u$ and $\bar Q d$ transforming as $(1,\bar 3, 3)$, $({1\oplus8},1,1)$, $(\bar 3,3,1)$ and $(\bar 3, 1, 3)$ respectively. From them we can construct the most general $\mathcal G^{\rm SM}$ invariant structures as 
\begin{equation}
\bar u Y_u^\dagger \mathcal A_{ud} Y_d d\,, ~~~ \bar Q \mathcal A_{QQ} Q\,, ~~~ \bar Q \mathcal A_{Qu} Y_u u\,,~~~ \bar Q \mathcal A_{Qd} Y_d d\,,
\label{eq:flav}
\end{equation}
where $\mathcal A_{xy}$ are arbitrary polynomials of $Y_{u} Y_{u}^\dagger$ and/or $Y_{d}Y_d^\dagger$,  transforming as $({1\oplus8},1,1)$.

In order to identify the relevant flavor structures in terms of physical parameters, we can without the loss of generality consider $Y_{u,d}$ condensate values in the down basis in which $\langle Y_d \rangle$ is diagonal~(\ref{eq:down_basis})
\begin{eqnarray}
\langle Y_d \rangle \simeq \mathrm{diag}(0,0,m_b)/v_d\,,\hspace{0.5cm} \langle Y_u \rangle \simeq V_{}^\dagger \mathrm{diag}(0,0,m_t)/v_u\,.
\end{eqnarray}
Here $V$ is the SM CKM matrix~(\ref{eq:CKMmat}) and we have introduced separate up- and down-type Higgs condensates $v_{u,d}$. We have also neglected the masses of first two generation quarks, which is the approximation we will be using throughout this chapter.

Further, once we assume electroweak symmetry breaking (see Eq.~(\ref{eq:breaking}) and text around it) we rewrite the quark fields in the mass eigenbasis. Making the flavor indices and chirality explicit these fields are
\begin{eqnarray}
Q_i=(V^*_{ki} u_{Lk},d_{Li})^{\mathrm{T}}\,, \hspace{0.5cm} u_{iR}\,,\hspace{0.5cm} d_{iR}\,.
\end{eqnarray}

We consider first the simplest case of linear MFV where $\mathcal A_{xy}$ is such that powers of $Y_d$ in (\ref{eq:flav}) do not exceed 1, and powers of $Y_u$ do not exceed 2. Obtained flavor structures are given in first two columns of Tab.~\ref{tab:FlavStr}. Following~\cite{Kagan:2009bn}, the generalization of the above discussion to MFV scenarios where large bottom Yukawa effects can be important is straight forward. We implement it by raising the highest allowed power of $Y_d$ appearing in (\ref{eq:flav}) to 2. This gives us additional flavor structures presented in the last two columns of Tab.~\ref{tab:FlavStr}. We can see from the form of $Y_u Y_u^\dagger$ and $Y_d Y_d^\dagger$ with flavor indices explicitly written out
\begin{eqnarray}
(Y_u Y_u^\dagger)_{ij} = \frac{m_t^2}{v_u^2}V^*_{ti}V_{tj}\,,\hspace{0.5cm}(Y_d Y_d^\dagger)_{ij} = \frac{m_b^2}{v_d^2}\delta_{3i}\delta_{3j}\,,
\end{eqnarray}
that apart from these two forms, the only additional flavor structure that can be obtained is
\begin{eqnarray}
(Y_u Y_u^\dagger Y_d Y_d^{\dagger})_{ij} &=& \frac{m_t^2}{v_u^2}\frac{m_b^2}{v_d^2}V^*_{ti}\delta_{3j}\,.
\end{eqnarray}
This means that our list given in Tab.~\ref{tab:FlavStr} is exhausting as forms of $\mathcal A$ more complex than presented therein generate no new flavor structures, rather just give higher powered overall factors.
\begin{table}[h]
\setlength{\doublerulesep}{0.05cm}
\begin{center}
\begin{tabular}{c|cc|c c}\hline\hline
& $\mathcal A=\mathds{1}$ & $\mathcal A= Y_u Y_u^{\dagger}$ & $\mathcal A = Y_d Y_d^{\dagger} $& $\mathcal A = Y_u Y_u^\dagger Y_d Y_d^{\dagger} $ \\\hline
$\bar u Y_u^\dagger \mathcal A_{ud} Y_d d$&$\frac{m_t}{v_u}\frac{m_b}{v_d} \bar t_R V_{tb}  b_R$& & &\\
$\bar Q \mathcal A_{QQ} Q$				&$\bar Q_i Q_i$& $\frac{m_t^2}{v_u^2} \bar Q_i V^*_{ti} V_{tj} Q_j$ & $\frac{m_b^2}{v_d^2}\bar Q_3 Q_3$&$\frac{m_b^2}{v_d^2}\frac{m_t^2}{v_u^2}\bar Q_i V_{ti}^* V_{tb}Q_3$\\
$\bar Q \mathcal A_{Qu} Y_u u$			&$\frac{m_t}{v_u}\bar Q_i V^{*}_{ti}t_R $& &$\frac{m_b^2}{v_d^2}\frac{m_t}{v_u}\bar Q_3 V_{tb}^* t_R$ &\\
$\bar Q \mathcal A_{Qd} Y_d d$			&$\frac{m_b}{v_d}\bar Q_3 b_R$& $\frac{m_b}{v_d}\frac{m_t^2}{v_u^2}\bar Q_i V_{ti}^* b_R$&&\\\hline\hline
\end{tabular}
\end{center}
\caption{MFV consistent flavor structures of the quark bilinear parts of dimension-six operators that generate charged current quark interactions with the $W$. First two columns represent the simplest case of linear MFV with the highest allowed powers of $Y_u$ and $Y_d$ set to 2 and 1 respectively while the last two columns represent MFV scenarios where large bottom Yukawa effects can be important, so the highest allowed $Y_d$ power is raised to 2.}
\label{tab:FlavStr}
\end{table}

\subsection{Final operator basis}
\label{sec:FinalBasis}
Before writing down the final set of operators that we will be analyzing we first put the obtained structures under further scrutiny and make some modifications.
\begin{itemize}
\item Since $\bar Q_i Q_i$ is completely flavor universal, when coupled to the $W$ it would modify the effective Fermi constant as extracted from charged quark currents compared to the muon lifetime. Existing tight constraints on such deviations~\cite{Antonelli:2009ws} do not allow for significant effects in $B$ meson or top quark phenomenology and we do not consider this structure in our analysis.
\item On the other hand, $\bar Q_i V^*_{ti} V_{tj} Q_j$ potentially leads to large tree-level FCNCs in the down quark sector when coupled to the $Z$. Explicitly, both the singlet and triplet operators (\ref{eq:GaugeStr_LL}) include the following term 
\begin{eqnarray}
\frac{\ii g}{2c_W}V_{ti}^*V_{tj}\big[\bar{d}_{Li}\gamma^{\mu} d_{Lj}\big]Z_{\mu}\varphi_0^2\,,
\end{eqnarray}
where $\varphi_0$ is the lower $SU(2)_L$ component of $\phi_d$, having zero electric charge. After it acquires a VEV, such terms generate the aforementioned FCNC couplings. We therefore consider as our operator the linear combination of the singlet and the triplet, choosing the relative negative sign between them therby getting rid of the FCNCs in the bottom sector.
\item Similarly, $\bar Q_i V^*_{ti} V_{tb} b_R$ includes the following term
\begin{eqnarray}
-V_{ti}^*\big[\bar{d}_{Li}\sigma^{\mu\nu}b_R\big] \varphi_0(s_W F_{\mu\nu}+c_W Z_{\mu\nu})\,,
\end{eqnarray}
which, once $\varphi_0$ acquires a VEV, generates tree-level FCNC $b\to s \gamma,Z$ transitions. These are already tightly constrained by $B\to X_s\gamma$ and $B\to X_s \ell^+ \ell^-$~\cite{Hurth:2008jc}. Since all the charged current mediating $SU(2)_L$ invariant operators of dimension six or less containing such a flavor structure do necessarily involve either the $Z$ or the photon, we drop this structure from our subsequent analysis.
\end{itemize}
Taking these considerations into account, we obtain the set of seven effective dimension six operators invariant under the SM gauge group and consistent with MFV hypotheses that involve charged quark currents
\begin{subequations}
\label{eq:ops1}
\begin{eqnarray}
 \mathcal Q_{RR}&=& V_{tb} [\bar{t}_R\gamma^{\mu}b_R] \big(\phi_u^\dagger\ii D_{\mu}\phi_d\big) \,, \\
 \mathcal Q_{LL}&=&[\bar Q^{\prime}_3\tau^a\gamma^{\mu}Q'_3] \big(\phi_d^\dagger\tau^a\ii D_{\mu}\phi_d\big)-[\bar Q'_3\gamma^{\mu}Q'_3]\big(\phi_d^\dagger\ii D_{\mu}\phi_d\big),\\
 \mathcal Q'_{LL}&=&[\bar Q_3\tau^a\gamma^{\mu}Q_3] \big(\phi_d^\dagger\tau^a\ii D_{\mu}\phi_d\big) -[\bar Q_3\gamma^{\mu}Q_3]\big(\phi_d^\dagger\ii D_{\mu}\phi_d\big),\\
 \mathcal Q^{\prime\prime}_{LL}&=&[\bar Q'_3\tau^a\gamma^{\mu}Q_3] \big(\phi_d^\dagger\tau^a\ii D_{\mu}\phi_d\big)-[\bar Q'_3\gamma^{\mu}Q_3]\big(\phi_d^\dagger\ii D_{\mu}\phi_d\big),\\
 \mathcal Q_{LRt} &=& [\bar Q'_3 \tau^a\sigma^{\mu\nu} t_R]{\phi_u}W_{\mu\nu}^a \,,\\
\mathcal Q'_{LRt} &=& [\bar Q_3 \tau^a\sigma^{\mu\nu} t_R]{\phi_u}W_{\mu\nu}^a \,,\\
\mathcal Q_{LRb} &=& [\bar Q_3 \tau^a\sigma^{\mu\nu} b_R]\phi_d W_{\mu\nu}^a \,,
\end{eqnarray}
\end{subequations}
where we have introduced 
\begin{eqnarray}
\bar Q'_3 =\bar Q_i V^*_{ti}= (\bar{t}_L,V_{ti}^*\bar{d}_{iL})^T\,.
\end{eqnarray}
We note, that the final set of operators coincides with those considered in the $B\to X_s \gamma$ analysis of anomalous $tWb$ couplings~\cite{Grzadkowski:2008mf} expanded by the three primed operators which originate from the structures given by the last two columns of Tab.~\ref{tab:FlavStr}, corresponding to higher order down-Yukawa MFV\footnote{In the final form of the operators we do not explicitly write out the $m_t/v_u$ and $m_b/v_d$ factors present in Tab.~\ref{tab:FlavStr}, technically shifting them to the Wilson coefficients.}. 

Furthermore we do not make the operators hermitian, hence effects of operators $\mathcal Q_i^{\dagger}$ are accompanied by $C_i^{*}$ and will be kept track of separately.

Notice that starting with the most general MFV construction we are led to a set of operators, where largest deviations in charged quark currents are expected to involve the third generation (a notable exception being the flavor universal $\bar Q_i Q_i$ structure present already in the SM, which we have dropped from our analysis).

As a consequence of our effective theory approach, operators $\op_i$ given in Eq.~(\ref{eq:ops1}) modify not only the $tWb$ vertex but withhold a much richer flavor and gauge structure. Since our aim is to analyze the effects of these operators in $B$ physics we make two notes regarding the additional effects $\op_i$ might cause that we shall not be pursuing.
\begin{itemize}
\item $\mathcal Q_{LL}$ and $\mathcal Q_{LRt}$ also modify $tWs$ and $tWd$ vertices. Consequently, they also contribute to $K^0-\bar K^0$ mixing at one-loop. However, their contributions to neutral kaon and as well as $B$ meson oscillations turn out to be universal and purely real (see discussion below Eq.~(\ref{eq:kappas})) so they cannot increase the $\epsilon$ and $\epsilon^{\pr}$ predictions.
\item $\mathcal Q_{LRb}$ and $\mathcal Q'_{LL}$ also modify $uWb$ and $cWb$ vertices.  This could interfere with $V_{cb}$ and $V_{ub}$ extraction from semileptonic $B$ decays. Since these quantities are crucial for the reconstruction of the CKM matrix in MFV models, a consistent analysis of these operators would  require a modified CKM unitarity fit, which is beyond the scope of this work.
\end{itemize}

The Feynman rules for all the vertices generated by $\op_i$ that are relevant for our analysis are presented in the Appendix~\ref{app:feyn_charged}. Since we shall be working in the general $R_{\xi}$ gauge, beneficing the check of the $\xi$ dependence cancelation in the final results, we will have to consider also the would-be Goldstone bosons in our calculations.
\section{$|\Delta B|=2$ transitions}
\label{sec:mixing}

Recently, possible NP effects in the $B_{q}-\bar B_{q}$, mixing amplitudes ($q=d,s$) have received considerable attention (c.f.~\cite{Lenz:2010gu} and references within). In particular within the SM, the $B_d-\bar B_d$ mass difference and the time-dependent CP asymmetry in $B_d\to J/\psi K_s$ are strongly correlated with the branching ratio $\mathrm{Br}(B^+\to \tau^+ \nu)$. The most recent global analysis point to a disagreement of this correlation with direct measurements at the level of 2.9 standard deviations~\cite{Lenz:2010gu}. 

Similarly in the $B_s$ sector, the measured CP-asymmetries by the Tevatron experiments, namely in $B_s \to J/\psi\phi$ and in di-muonic inclusive decays when combined, deviate from the SM prediction for the CP violating phase $\phi_s$ in $B_s-\bar B_s$ mixing by $3.3$ standard deviations~\cite{Lenz:2010gu}. This indication of NP effect is however weakened by the latest LHCb result of the $\phi_s$ phase inferred from combined analysis of $B_s\to J/\psi \phi$ and $B_s\to J/\psi f(980)$ channels \cite{lhcb:phi_s}, showing agreement with SM prediction therefore eliminating the possibility of large NP contributions.

In this section we analyze the effects of our operators (\ref{eq:ops1}) on the  $B_{q}-\bar B_{q}$ mixing amplitudes. We first perform the matching to the low energy Lagrangian, where we consider only diagrams with one $\op_i$ operator insertions, resulting in first order corrections in $1/\Lambda^2$ expansion. Consideration of only single NP operator insertion is a good approximation given the small size of observed deviations in the CP-conserving $B_{q}$ mixing observables from SM predictions. However, we have also computed higher order insertions and checked explicitly that they do not change our conclusions of the numerical analysis presented in section \ref{sec:BBnumerics}.

To obtain the constraints on NP contributions we rely on the recent global CKM and $B_{q}$ mixing fits given in Refs.~\cite{Lenz:2010gu} and \cite{Lenz:2012az}. 

\subsection{Matching}
In order to study the effects of our operators (\ref{eq:ops1}) on the matrix elements relevant in $B_{q}-\bar B_{q}$ mixing, we normalize them, following~\cite{Lenz:2010gu}, to the SM values given in Eq.~(\ref{eq:mixnorm}) by writing
\begin{equation}
M_{12}^{q}=\frac{1}{2m_{B_q}}\langle B_{q}^0|{\cal H}_{\mathrm{eff}}|\bar{B}_{q}^0\rangle_{\mathrm{disp}}= M_{12}^{q,\mathrm{SM}}\Delta_{q}\,,\label{mat}
\end{equation}
where the deviation of parameter $\Delta_{q}$ from $1$ quantifies NP contributions. Proceeding in a similar fashion as in section \ref{sec:SMmix}, where we have analyzed the mixing amplitudes in the SM, we now match our effective theory to the low energy effective theory relevant for $|\Delta B|=2$ transitions which is governed by the Lagrangian 
\begin{eqnarray}
{\cal L}_{\mathrm{eff}} =-\frac{G_F^2 m_W^2}{4 \pi^2}\big(V_{tb}V^*_{tq}\big)^2 \sum_{i=1}^5 C_i(\mu) {\cal O}_i^{q}\,.
\label{eq:lagrBB}
\end{eqnarray}
Compared to the low energy effective Lagrangian (\ref{eq:LSMmix}), to which we were matching the pure SM contributions, effective Lagrangian of (\ref{eq:lagrBB}) contains four additional operators~\cite{Becirevic:2001xt}
\begin{eqnarray}
\mathcal O_2^d = \big[\bar{d}_R^{\alpha}b_L^{\alpha}\big]\big[\bar{d}_R^{\beta}b_L^{\beta}\big]\,,\hspace{0.5cm}
\mathcal O_3^d =\big[\bar{d}_R^{\alpha}b_L^{\beta}\big]\big[\bar{d}_R^{\beta}b_L^{\alpha}\big]\,,\\
\nn\mathcal O_4^d= \big[\bar{d}_R^{\alpha}b_L^{\alpha}\big]\big[\bar{d}_L^{\beta}b_R^{\beta}\big]\,,\hspace{0.5cm}
\mathcal O_5^d =\big[\bar{d}_R^{\alpha}b_L^{\beta}\big]\big[\bar{d}_L^{\beta}b_R^{\alpha}\big]\,,
\end{eqnarray}
which need to be included since non-SM chirality structures are present in our operator basis. We have explicitly written out the $\alpha,\beta$ color indices.

\begin{figure}[h]
\begin{center}
\includegraphics[scale= 0.6]{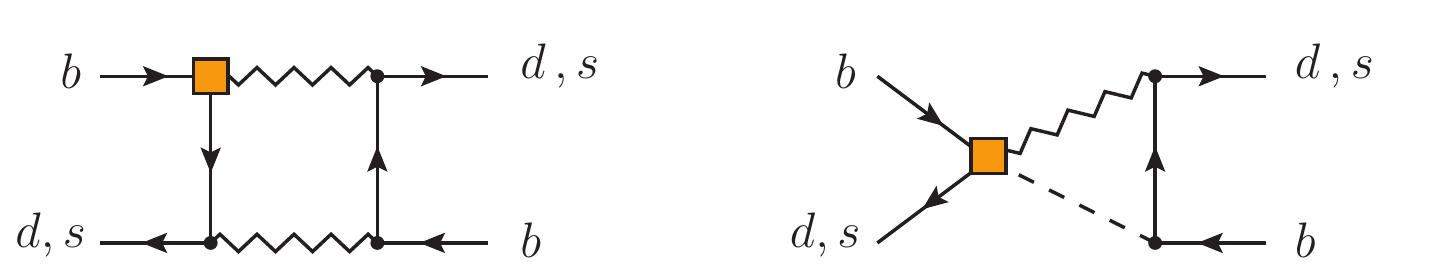}
\caption{Feynman diagrams for $\bar{B}_q\to B_q$ transitions with one insertion of $\mathcal Q_i$ operators, labeled with a square. The zigzag lines represent $W$ gauge bosons or would-be Goldstone scalars $\phi$. Quarks running in the loop are up-type quarks. {\bf Left}: Box diagram to which all $\mathcal Q_i$ contribute. {\bf Right}: Triangle diagrams generated only by $\mathcal Q_{LL}^{(\prime\prime)}$ operators. }
\label{fig:NPmix}
\end{center}
\end{figure}

In the matching procedure the $W$ boson and the top quark are integrated out by computing the box diagrams such as the one depicted in Fig.~\ref{fig:NPmix}, which now contain anomalous couplings, from insertions of operators $\op_i$. The box diagrams with anomalous couplings appearing in the bottom-right corner instead the top-left and the crossed diagrams with internal quark and boson lines exchanged are completely symmetric and need not be computed separately. 

We note that working in general $R_{\xi}$ gauge for weak interactions brings about new anomalous interactions of Would-be goldstone bosons generated by $\mathcal Q^{({\prime},{\prime}{\prime})}_{LL}$ and $\mathcal Q_{RR}$ operators. What is more, in the general $R_{\xi}$ gauge, operators $\op_{LL}$ and $\op_{LL}^{\pr\pr}$ contribute to mixing amplitudes also through the triangle diagrams shown on the righthand side of  Fig.~\ref{fig:NPmix}. 

\begin{figure}[h]
\begin{center}
\includegraphics[scale=0.6]{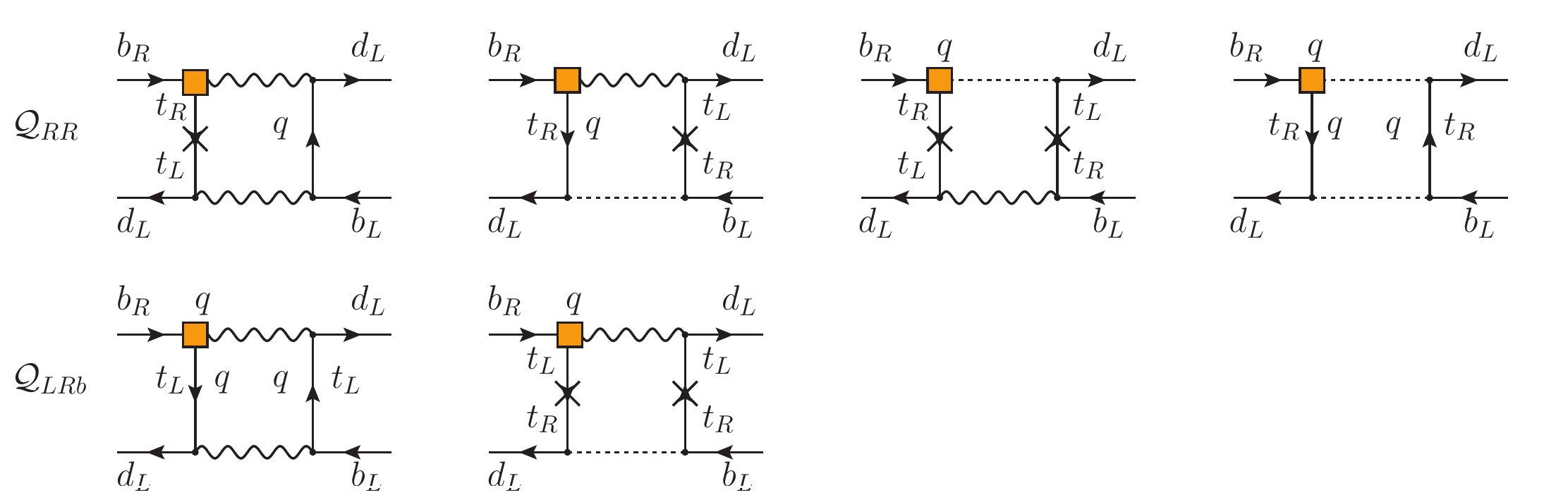}
\end{center}
\caption{Schematic consideration of $\op_{RR}$ and $\op_{LRb}$ insertions in the mixing diagrams. Crosses and $q$s on the top quark propagators mark which part of the propagator is needed due to specific chirality demands of the vertices. $q$ at the anomalous vertices mark that the Feynman rules for those vertices include the loop momentum $q$. }
\label{fig:chirality}
\end{figure}

By simple consideration of the chirality structure of the diagrams we find that one insertion of operators $\mathcal Q_{RR}$ and $\mathcal Q_{LRb}$ give contributions suppressed by the down quark masses. This is illustrated in Fig.~\ref{fig:chirality}, which shows explicitly the chiralities of quark fields and the number of times the loop integration momentum $q$ appears in the mixing diagrams. Because we neglect the down quark masses, no chirality flip of these fields is possible, therefore their chirality is determined by the interaction vertices. The up (top) quarks in the loop however can experience a helicity flip or instead pick up the loop integration momentum $q$ from the propagator. What is more some anomalous couplings also include momentum $q$ in the Feynman rule. If we sum the number of times that $q$ appears in the diagram to be odd, such diagram gives a zero contribution due to the symmetry of the integration over $\dd^4 q$ and the fact that the only momentum appearing in the diagram is $q$ (see section~\ref{sec:dB2})
\begin{eqnarray}
\int \frac{\dd^4 q}{(2\pi)^4} \frac{q^{2n} q^\mu}{\prod_{i}[q^2-m^2_i]} = 0 \,,\hspace{0.5cm} n=0,1,2,\dots \,.
\end{eqnarray}
This is what happens in all one-insertion diagrams for $\op_{RR}$ and $\op_{LRb}$, which are presented in Fig.~\ref{fig:chirality}, meaning that these two operators contribute only upon two insertions and will therefore not be considered further in this section. In turn the only relevant four quark operator of the effective Lagrangian (\ref{eq:lagrBB}) for the matching is $\mathcal O_1^q$ and we can write
\begin{eqnarray}\label{eq:C1_and_NP}
C_1 = C_1^{\mathrm{SM}} + \delta C_1\,,
\end{eqnarray}
where $C_1^{\mathrm{SM}}$ is given in Eq.~(\ref{eq:C1SM}) and $\delta C_1$, which is identical for $q=d$ and $q=s$ case is
\begin{eqnarray}
\delta C_1 =
\mathrm{Re}[\kappa_{LL}] S_0^{LL}(x_t)
+\mathrm{Re}[\kappa_{LRt}]S_0^{LRt}(x_t)
+\kappa_{LL}^{\prime(\prime\prime)}S_{0}^{LL\prime(\prime\prime)}(x_t)
+\kappa_{LRt}^{\prime}S_{0}^{LRt\prime}(x_t)\,,
\label{LOWils}
\end{eqnarray}
where $x_t=m_t^2/m_W^2$ and we have defined 
\begin{eqnarray}
\kappa_{LL}^{(\prime,\prime\prime)}=\frac{C_{LL}^{(\prime,\prime\prime)}}{\Lambda^2\sqrt{2}G_F}\,,\hspace{0.5cm} 
\kappa_{LRt}^{(\prime)}=\frac{C_{LRt}^{(\prime)}}{\Lambda^2 G_F}\,.
\label{eq:kappas}
\end{eqnarray}
The $S_0^i(x_t)$ loop functions are given in Eqs.~(\ref{eq:S0s}) of the Appendix. Their gauge independence has been checked by the cancelation of all $\xi$-dependent terms. There are two features of the obtained results that we want to point out. 
\begin{itemize}
\item First one is that, $S_0^{LL}$ and $S_0^{LL\prime\prime}$ contributions turn out to be UV-divergent. The divergences originate from the triangle diagrams with two would-be Goldstone bosons running in the loop\footnote{Had we been working in the unitary gauge, these diagrams would not exist. The UV divergences would however still emerge from the box diagrams, since one $\op_{LL}$ or $\op_{LL}^{\prime\prime}$ insertions demand one of the up-type quarks to be the top quark. This means that rather than having two summations in Eq.~(\ref{eq:SMmixAMP1}) as we did in the SM case, we have just one, which leaves the results UV divergent.}. We renormalize them using the $\overline{\mathrm{MS}}$ prescription, leading to remnant renormalization scale dependent terms of the form  $x_t\log m_W^2/\mu^2$ . Because of this ultraviolet renormalization, it would be inconsistent to assume that no other operators but those in Eq.~(\ref{eq:ops1}) comprise the dimension-six part of the Lagrangian (\ref{eq:lagr}). In particular, on dimensional grounds it is easy to verify that the appropriate MFV consistent counter-terms are generated by the four-quark operators of the form
\begin{eqnarray}
\op_{4Q}=\big[\bar Q \mathcal A_{QQ} \gamma^{\mu}Q\big]\big[\bar Q \mathcal A^\prime_{QQ} \gamma_{\mu}Q\big]\,, \hspace{0.3cm} \mathcal A_{QQ} = Y_u Y_u^{\dagger}\,,\hspace{0.3cm} \mathcal A_{QQ}^\prime = \bigg\{\hspace{-0.2cm} \begin{array}{c l }
Y_u Y_u^\dagger\,,& \text{for } \op_{LL}\,\\
Y_u Y_u^\dagger Y_d Y_d^\dagger\,,&\text{for } \op_{LL}^{\pr\pr}\,
\end{array}\,,
\end{eqnarray}
giving the change in Wilson coefficient
\begin{eqnarray}
\delta C_1 = \frac{C_{4Q}}{\Lambda^2\sqrt{2}G_F}8\pi^2\,x_t\,.
\end{eqnarray}
Appearance of $x_t$ in the expression above is crucial to match the $x_t$ factor  accompanying the UV divergences in $\op_{LL}$ and $\op_{LL}^{\pr\pr}$ contributions to $\delta C_1$.

Generic tree-level contributions of this kind to $\delta C_1$ have been analyzed in detail in~\cite{Ligeti:2010ia} although not in the context of  radiative corrections but as standalone dimension-six $\Delta F=2$ effective operators adhering to MFV -- we will not consider them further. It is however important to keep in mind that our derived bounds on $\kappa_i$ presented in the next section assume that the dominant NP effects at the $\mu\simeq m_t$ scale are represented by a single $\mathcal Q_i$ insertion.

\item The second feature to be pointed out is that only real parts of $\kappa_{LL}$ and $\kappa_{LRt}$ enter Eq.~(\ref{LOWils}) and thus cannot introduce a new CP violating phase. On a computational level, this is due to the fact that these operators always contribute to the mixing amplitudes in hermitian conjugate pairs. In particular, a box diagram with an insertion of operator $\op_{LL}$ or $\op_{LRt}$ in the upper-left corner is accompanied by a diagram with insertion of $\op_{LL}^\dagger$ and $\op_{LRt}^\dagger$ in the upper-right corner, since as was pointed out in section \ref{sec:FinalBasis}, these operators also effect $tWd$ and $tWs$ vertices. Both diagrams give the same result but one with $\kappa_{LL,LRt}$ and the other with $\kappa_{LL,LRt}^*$ pre-factors respectively, resulting in the appearance of $\mathrm{Re}[\kappa_{LL,LRt}]$ in the sum of both contributions. Similarly, the triangular diagrams generated by $\op_{LL}$ are also generated by $\op_{LL}^{\dagger}$.

This inability to introduce new phases can also be understood more generally already at the operator level. Namely as shown in Ref.~\cite{Blum:2009sk}, a necessary condition for new flavor violating structures $\mathcal Y_x$ to introduce new sources of CP violation in quark transitions is that 
\begin{eqnarray}
{\rm Tr}(\mathcal Y_x[\langle Y_u Y_u^\dagger \rangle , \langle Y_d Y_d^\dagger \rangle])\neq 0\,.
\label{eq:Perez1}
\end{eqnarray}
In MFV models (where $\mathcal Y_x$ is built out of $Y_u$ and $Y_d$ ) this condition can only be met if $\mathcal Y_x$ contains products of both $Y_u$ and $Y_d$. In our analysis this is true for all operators except $\mathcal Q_{LL}$ and $\mathcal Q_{LRt}$.
\end{itemize}
\subsection{Semi-numerical formula}
\label{sec:BBnumerics}
In order to evaluate the hadronic matrix elements of the operators, we have to evolve the Wilson coefficient~(\ref{eq:C1_and_NP}) from the matching scale at the top quark mass to the low energy scale at the bottom quark mass. Next-to-leading log (NLL) running for the SM Wilson coefficient $C_1^{\mathrm{SM}}$ in the $\overline{\mathrm{MS}}$ (NDR) scheme is 
\cite{Buras:2001ra}
\begin{eqnarray}
C_1^{\mathrm{SM}}(m_b)&=& 0.840\, C_1^{\mathrm{SM}}(m_t)\,.
\end{eqnarray}
Because we are relying on results with consistent $\overline{\mathrm{MS}}$ renormalization procedures, we need to use the $\overline{\mathrm{MS}}$ quark masses $m_t\equiv\overline{m}_t(\overline{m}_t)$, $m_b\equiv\overline{m}_b(\overline{m}_b)$.
Following the reasoning outlined in the last paragraph of section~\ref{sec:strategy} we assume the same running also for the $\delta C_1$ part.
Under this assumption the effects of running between the change $\delta C_1$ and the SM contribution cancel out and the parameter for quantification of NP in mixing amplitudes can then be written as
\begin{eqnarray}
\nn \Delta &=& 1+\frac{\delta C_1}{C_1^{\mathrm{SM}}} = 1 + \sum_i \frac{S_0^{i}(x_t)}{S_0^{\mathrm{SM}}(x_t)}\\
&=&1- 2.57\, \mathrm{Re}[\kappa_{LL}]-1.54\,\mathrm{Re}[\kappa_{LRt}]
+2.00\, \kappa_{LL}^{\prime}-1.29\, \kappa_{LL}^{\prime\prime} - 0.77\,\kappa_{LRt}^{\prime}\,,
\label{eq:seminum1}
\end{eqnarray}
where the parameters $\kappa_i$ are understood to be evaluated at the high matching scale $m_t$. In order to be consistent with the global analysis of Ref.~\cite{Lenz:2010gu}, on which we shall rely in the next section, we have used the numerical values for masses and other parameters as specified therein.

\subsection{Bounds on NP contributions}
Let us first assume that the anomalous couplings $\kappa_i$ are real. Using Eq.~(\ref{eq:seminum1}), we consider  one $\kappa_i(\mu=m_t)$ at the time to be non-zero. The assumption of real $\kappa_i$  makes our NP fall under the ``scenario II'' of \cite{Lenz:2010gu}, for which the global analysis gives
\begin{eqnarray}
\Delta = 0.90
\Big[\hspace{-0.2cm}\begin{array}{c}
{\scriptstyle +0.07} \vspace{-0.1cm}\\
{\scriptstyle -0.07} 
\end{array}\hspace{-0.2cm}\Big]
\Big[\hspace{-0.2cm}\begin{array}{c}
{\scriptstyle +0.31} \vspace{-0.1cm}\\
{\scriptstyle -0.10} 
\end{array}\hspace{-0.2cm}\Big]\,,
\end{eqnarray}
here the bracketed intervals represent the $1\sigma$ and $2\sigma$ C.L. intervals around the central value, which we can use to obtain $95\%$ C.L. bounds on $\kappa_i$. We present our results in Tab.~\ref{tab:MixingBounds}.
\begin{table}[h!]
\begin{center}
\begin{minipage}{0.2\textwidth}
\begin{tabular}{c|c}\hline\hline
& $95 \%$ C.L. \\\hline
$\kappa_{LL}$ 			&$\begin{array}{r} 0.08\\-0.09\end{array}$\\\hline
$\kappa_{LL}^{\pr}$ 		&$\begin{array}{r} 0.11\\-0.09\end{array}$\\\hline
$\kappa_{LL}^{\pr\pr}$	&$\begin{array}{r} 0.18\\-0.18\end{array}$\\\hline
$\kappa_{LRt}$			&$\begin{array}{r} 0.13\\-0.14\end{array}$\\\hline
$\kappa_{LRt}^{\pr}$		&$\begin{array}{r} 0.29\\-0.29\end{array}$\\ \hline\hline
\end{tabular}
\end{minipage}
\begin{minipage}{0.5\textwidth}\begin{center}
\includegraphics[scale=0.88]{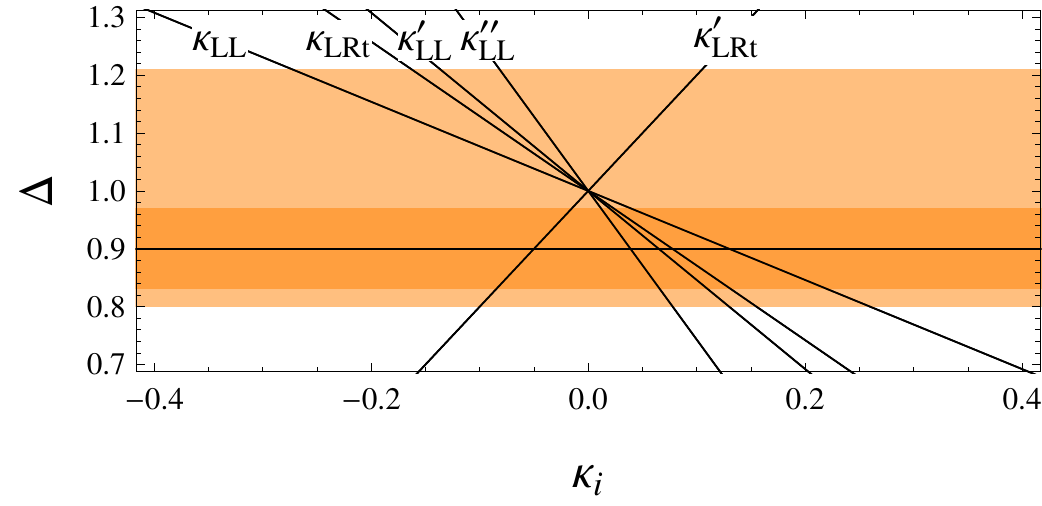}
\end{center}
\end{minipage}
\caption{$95\%$ C.L. allowed intervals for $\kappa_i$ which are considered to be real and analyzed with only one being different than zero at a time. Accompanying graph shows $\Delta$ as function of one $\kappa_i$. The horizontal line represents the central fitted value and the orange bands the $1\sigma$ and $2\sigma$ regions for $\Delta$ as obtained in the global analysis of \cite{Lenz:2010gu}. $95\%$ C.L. limits on the parameters $\kappa_i$, are obtained by looking at where the functions cross the outer orange regions. }
\label{tab:MixingBounds}
\end{center}
\end{table}
Compared to existing $B\to X_s \gamma$ constraints for couplings $\kappa_{LL}$ and $\kappa_{LRt}$ given in Ref.~\cite{Grzadkowski:2008mf}, we find our bounds on $\kappa_{LL}$ to be comparable, while bounds on $\kappa_{LRt}$ are improved. 

Relaxing the assumption of real $\kappa_i$, contributions of the primed operators can introduce new CP violating phases if the anomalous couplings have nonzero imaginary components. The analysis of such general complex contributions to $\Delta$ fall under the ``scenario III'' of \cite{Lenz:2010gu}.
\begin{figure}[h!]
\begin{center}
\includegraphics[scale=0.8]{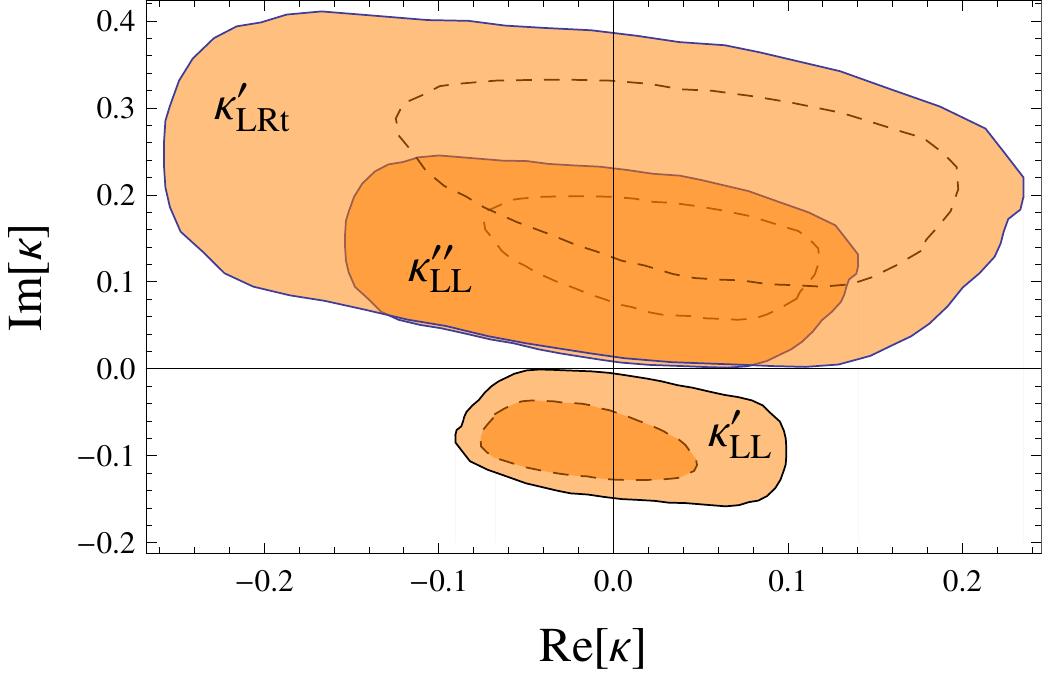}
\caption{68\% (dashed) and 95\% (solid) C.L. allowed regions for $\kappa_{LL}^{\pr(\pr\pr)}$ and $\kappa_{LRt}^{\pr}$ in the complex plain obtained from Eq.~(\ref{eq:seminum1}) and the results of Ref.~\cite{Lenz:2012az}. }
\label{fig:phis}
\end{center}
\end{figure}
As we have already mentioned at the beginning of this chapter, the best fitted values for NP when new phases are considered given in \cite{Lenz:2010gu} are highly influenced by the extraction of the $\phi_s$, the mixing angle in the $B_s$ system, from Tevatron data. One can imagine the fits to change quite significantly on the side of imaginary component of $\Delta$, if the newest measurement of $\phi_s^{\mathrm{exp.}}$ from LHCb~\cite{lhcb:phi_s} is taken into consideration since it is in agreement with the SM prediction. This is confirmed by comparing the graphical results from Ref.~\cite{Lenz:2010gu} and the recent update given in Ref.~\cite{Lenz:2012az}, which we use to present the $68\%$ and $95\%$ C.L. regions in the complex plane for $\kappa_{LL}^{\pr(\pr\pr)}$ and $\kappa_{LRt}^\pr$ shown in Fig.~\ref{fig:phis}. We can still observe the preference of non-zero imaginary components by the global fits, significance of which is however reduced compared to the situation prior to the LHCb measurement~\cite{Drobnak:2011wj}.

\subsection{Summary}
To summarize, we have matched our effective Lagrangian (\ref{eq:lagr}) to the low energy effective Lagrangian (\ref{eq:lagrBB}) and analyzed the impact of the effective operators $\op_i$ on $|\Delta B| = 2$ mixing to first order in $1/\Lambda^2$, allowing one $\op_i$ insertion in mixing diagrams.

We have shown that operators $\op_{RR}$ and $\op_{LRb}$ do not contribute upon one insertion, and while operators $\op_{LL}$ and $\op_{LRt}$ can not contribute any new phases to the mixing, the primed operators $\op_{LL}^{\pr (\pr\pr)}$ and $\op_{LRt}^{\pr}$ can. 

Effects of $\op_i$ turn out to be the same for $B_d$ and $B_s$ systems, so that they can be parametrized in terms of one parameter $\Delta$. Following the global analysis of \cite{Lenz:2010gu,Lenz:2012az} we were able to put constraints on $\kappa_i$. In particular, first assuming the Wilson coefficients $\kappa_i$ to be real, were able to obtain for them the $95\%$ C.L. allowed intervals given in Tab.~\ref{tab:MixingBounds}, which compared to the $b\to s \gamma$ constraint given in Ref.~\cite{Grzadkowski:2008mf} prove to be competitive for $\kappa_{LL}$ and improved for $\kappa_{LRt}$. For the three primed operators, which can contribute new phases, we have obtained the 95\% C.L. allowed regions in the corresponding complex plains, presented in Fig.~{\ref{fig:phis}}. 
\section{$|\Delta B| = 1$ transitions}
In this section we turn our analysis of the rare $|\Delta B|=1$ processes introduced in section~\ref{sec:dB1}. After performing the one-loop matching of our operator basis~(\ref{eq:ops1}) onto the low energy effective Lagrangian~(\ref{eq:loweff1}), we obtain corrections to the relevant Wilson coefficients. We proceed by calculating the effects in the inclusive $B\to X_s\gamma$ and $B \to  X_s \ell^+ \ell^-$. In order to derive bounds on both the real and imaginary parts of the appropriate Wilson coefficients we include the experimental results not only for the decay rates but also for the CP asymmetry in $B \to X_s \gamma$. After performing a global fit of the Wilson coefficients, we derive predictions for several rare $B$ meson processes: $B_s \to \mu^+ \mu^-$, the forward-backward asymmetry in $B \to K^* \ell^+ \ell^-$ and the  branching ratios for $B \to K^{(*)} \nu \bar \nu$.

\subsection{Matching}
The procedure of matching closely resembles the one described for the SM case in section~\ref{sec:dB1}. We are again interested in NP contributions to the observables at order $1/\Lambda^2$ and thus only consider single operator insertions. 

\begin{figure}[h]
\begin{center}
\includegraphics[scale= 0.6]{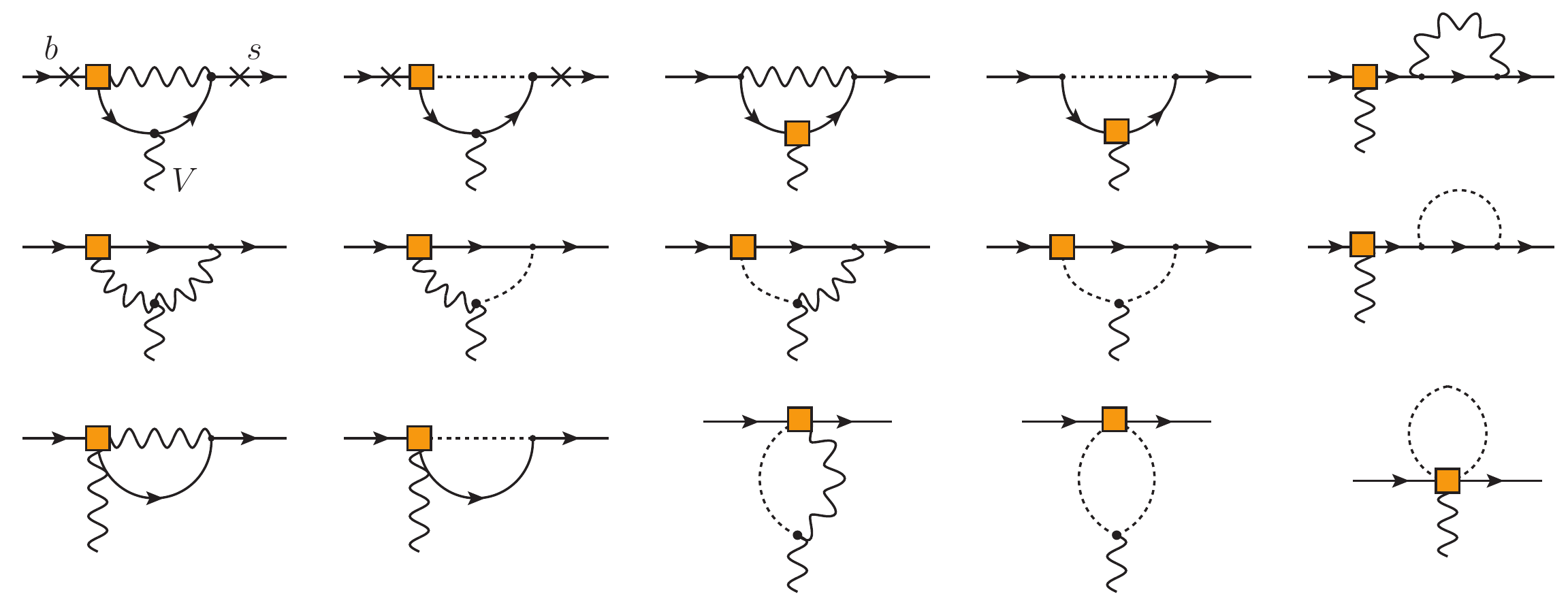}
\caption{Types of Feynman diagrams encountered when computing $b\to s V$ transitions, where $V$ stands for $\gamma, Z, g$. Dotted lines represent would-be Goldstone bosons, crosses mark additional points where $V$ can be emitted in one-particle-reducible diagrams and square represents an anomalous coupling. Gluon emission is only possible from quark lines and with the SM coupling. Quarks running in the loops are up-type.}
\label{fig:feyns11}
\end{center}
\end{figure}

\begin{figure}[h]
\begin{center}
\includegraphics[scale= 0.6]{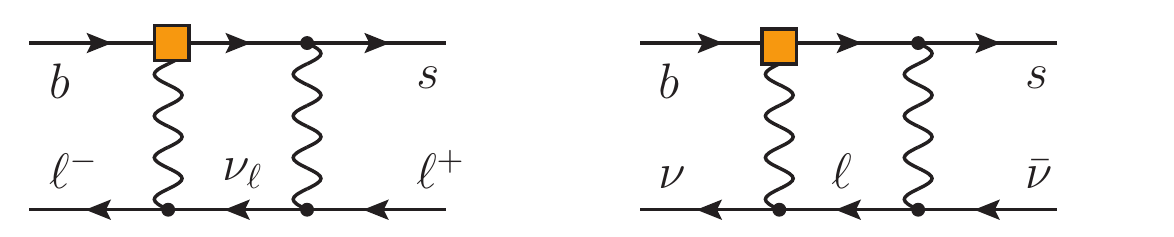}
\caption{Box Feynman diagrams contributing to $b\to s \ell^+ \ell^-$ and $b\to s \nu\bar{\nu}$ transitions. Diagrams with would-be Goldstone bosons are absent, since the leptons are treated as massless. Quark running in the loop is up-type.}
\label{fig:feyns3}
\end{center}
\end{figure}

Generic penguin and box diagrams with anomalous couplings are shown in Fig.~\ref{fig:feyns11} and Fig.~\ref{fig:feyns3}, where again due to the commitment to $R_{\xi}$ gauge, we are faced with diagrams with would-be Goldstone bosons. Exact diagrams for a specific ${\cal Q}_i$ can be reconstructed using Feynman rules given in the Appendix~\ref{app:feyn_charged}.

As the result of the matching procedure we obtain deviations from the SM values for the Wilson coefficients, $C_i = C_i^{\rm SM} + \delta C_i$, which we parametrize as
\begin{eqnarray}
\delta C_i(\mu) &=&\sum_{j}\kappa_j(\mu) f_i^{(j)}(x_t,\mu) + \kappa_j^{*}(\mu) \tilde{f}^{(j)}_i (x_t,\mu)\,,\label{eq:fs}
\end{eqnarray}
where index $j$ runs over the operator basis~(\ref{eq:ops1}), $\mu$ is the matching scale and $\kappa_j$ are rescaled Wilson coefficients defined as\footnote{For easier reading we repeat the definition of $\kappa_{LL}^{(\pr,\pr\pr)}$ and $\kappa_{LRt}^{(\pr)}$ previously defined in Eq.~(\ref{eq:kappas})}
\begin{eqnarray}
\kappa_{LL}^{(\prime,\prime\prime)}=\frac{C_{LL}^{(\prime,\prime\prime)}}{\Lambda^2\sqrt{2}G_F}\,,\hspace{0.3cm}
\kappa_{RR}=\frac{C_{RR}}{\Lambda^2 2\sqrt{2} G_F}\,,\hspace{0.3cm}
\kappa_{LRb}=\frac{C_{LRb}}{\Lambda^2 G_F}\,,\hspace{0.3cm}
\kappa_{LRt}^{(\prime)}=\frac{C_{LRt}^{(\prime)}}{\Lambda^2 G_F}\,.
\label{eq:kappas2}
\end{eqnarray}
Separate track is kept of $\op_i$ and $\op_i^\dagger$ contributions that are quantified by functions $f_i^{(j)}$ and $\tilde{f}_i^{(j)}$ respectively of which the analytic expressions are given in the Appendix~\ref{app:SM_D_B_1}.

We note, that the matching procedure for operator ${\cal Q}_{LL}^{\pr}$ stands out compared to the other operators. The charged current structure of this operator resembles that of the SM operator $\bar{Q}_i\gamma^{\mu}\tau^a Q_i W^a_\mu$. Consequently, when considering $\op_{LL}^{\pr}$, Wilson coefficients $C_{1,\dots,8}$ are changed in a trivial way $C_i(\mu) = (1+\kappa_{LL}^{\pr}(\mu))C_i^{\mathrm{SM}}(\mu)$. The change of the remaining Wilson coefficients $C_{9,10,\nu\bar{\nu}}$, matching of which involves the $Z$ boson, is however not of this form.

As in the case of $|\Delta B| =2$ process, some of the diagrams in Fig.~\ref{fig:feyns11} are UV divergent. We remove these divergences using the $\overline{\mathrm{MS}}$ prescription leading to remnant $\log (m_W^2/\mu^2)$ terms. We shall quantify the matching scale dependence of our results and consequently their sensitivity to the UV completion of the effective theory by varying the scale between $\mu=2m_W$ and $\mu=m_W$. Since UV renormalization is necessary, again our operator basis needs to be extended to include operators that can serve as the appropriate counter-terms. Within the employed MFV framework examples of these operators read
\begin{eqnarray}
{\cal Q}_1^{\mathrm{c.t.}}=\big[\bar{Q}\sigma^{\mu\nu}\mathcal A_{Qd} Y_d\tau^a d\big]\phi_d W^a_{\mu\nu}\,,\hspace{0.5cm}
{\cal Q}_2^{\mathrm{c.t.}}=\big[\bar{Q}\gamma^{\mu}\mathcal A_{QQ}Q\big]\big[\bar{\ell}\gamma_{\mu}\ell\big]\,, \label{eq:counterterms} \\
{\cal Q}_3^{\mathrm{c.t.}}=\big[\bar{Q}\gamma^{\mu}\tau^a \mathcal A^{\pr}_{QQ}Q\big]\big[\phi_d^{\dagger}\tau^a\ii D_{\mu}\phi_d\big] + \big[\bar{Q}\gamma^{\mu}\mathcal A^{\pr}_{QQ}Q\big]\big[\phi_d^{\dagger}\ii D_{\mu}\phi_d\big]\,.\nn
\end{eqnarray}
The operator ${\cal Q}_1^{\mathrm{c.t.}}$ produces a counter-term for divergences in $\delta C_7$ , while ${\cal Q}_{2,3}^{\mathrm{c.t.}}$ provide counter-terms for divergent parts of $\delta C_{9,10,\nu\bar{\nu}}$. The operator ${\cal Q}_3^{\mathrm{c.t.}}$ generates a tree-level $bZs$ vertex. 
The sets of flavor  matrices needed to match the structures of divergencies generated by the various operators in \eqref{eq:ops1} are
\begin{eqnarray}
\mathcal A_{Qd}&=&Y_u Y_u^{\dagger}\,, \\
\nn\mathcal A_{QQ}&=&Y_u Y_u^{\dagger}\,,\, Y_u Y_u^{\dagger} Y_d Y_d^{\dagger}\,,\nonumber \\
\nn\mathcal A_{QQ}^{\pr}&=&Y_u Y_u^{\dagger}\,,\, (Y_u Y_u^{\dagger})^2\,,\,Y_u Y_u^{\dagger}Y_d Y_d^{\dagger}\,,\, (Y_u Y_u^{\dagger})^2Y_d Y_d^{\dagger}\,,\,
Y_u Y_u^{\dagger} Y_d Y_d^{\dagger}Y_u Y_u^{\dagger}\,.
\end{eqnarray}
Just as in the analysis of NP in $|\Delta B|=2$ processes, we will selectively set contributions of certain operators to be nonzero in our numerical analysis. Consequently in the following, we will drop the implicit (tree-level) contributions of the operators in~\eqref{eq:counterterms} to $\delta C_i$, as these have been already investigated and constrained in the existing literature~\cite{Hurth:2008jc}.

All $f_i^{(j)}, \tilde{f}_i^{(j)}$ are found to be $\xi$ independent and a crosscheck with results from Ref.~\cite{Grzadkowski:2008mf} is possible for some of them. We confirm their original results for all the operators except $\mathcal Q_{LRb}$, while an updated version of~\cite{Grzadkowski:2008mf} confirms our result also for this operator. 

\begin{table}[h]
\begin{tabular}{c|c|cc|cc|cc|cc|cc|cc}\hline\hline
&SM&$\kappa_{LL}$&$\kappa_{LL}^*$&$\kappa_{LL}^{\pr}$&$\kappa_{LL}^{\pr*}$&$\kappa_{LL}^{\pr\pr}$&$\kappa_{LL}^{\pr\pr*}$&$\kappa_{RR}$&$\kappa_{LRb}$&$\kappa_{LRt}$&$\kappa_{LRt}^*$&$\kappa_{LRt}^{\pr}$&$\kappa_{LRt}^{\pr*}$\\\hline
$f_7$&-0.19 &0.45& 0.45& -0.19& 0& 0.45& 0& -45.3& 85.5& -0.13& -0.17& -0.15& -0.17\\
$f_8$&-0.095&0.24& 0.24& -0.095& 0& 0.48& 0& -20.2& 54.5& 0.15& 0.05& 0& 0.05\\
$f_9$&1.34&-1.11& -1.11& 1.35& 0.09& -1.11& 0.009& 0& 0& 0.64& 0.64& 0.009& 0.64\\
$f_{10}$&-4.16&1.48& 1.48& -4.28& -0.12& 1.48& -0.12& 0& 0& -2.41& -2.41& 0& -2.41 \\
$f_{\nu\bar{\nu}}$&-6.52&2.38& 2.38& -6.63& -0.12& 2.38& -0.12& 0& 0& -4.25& -4.25& 0& -4.25\\ \hline\hline
\end{tabular}
\caption{Numerical values of functions $f_i^{(j)}$ and $\tilde{f}_i^{(j)}$ at $\mu=2 m_W$. Numerical values used for the input parameters are $\overline{m}_t(2m_W)=165.0$~GeV, $s_W^2=0.231$,  $m_W= 80.4$~GeV, $\overline{m}_b(2m_W)= 2.9$~GeV, $|V_{tb}|^2=1$. All $f_i$ values correspond to matching at LO in QCD.}
\label{tab:cs}
\end{table}

To quantify the effects of our seven operators on Wilson coefficients (\ref{eq:fs}) we present the numerical values of $f_i^{(j)}$ evaluated at $\mu=2m_W$ in Tab.~\ref{tab:cs}. We see that the contributions of the operator ${\cal Q}_{LL}$ and ${\cal Q}_{LL}^{\dagger}$ are identical in all cases, which means that $\kappa_{LL}$ can not induce new CP violating phases in the Wilson coefficients. Likewise, $\mathcal Q_{LRt}$ contributions to $C_{9,10,\nu\bar\nu}$ are Hermitian but this operator can induce a new CP violating phase in $C_{7,8}$. 
Finally at order $1/\Lambda^2$, operators ${\cal Q}_{RR}$ and ${\cal Q}_{LRb}$ which contain right-handed down quarks only contribute to $C_{7,8}$. These contributions are however very significant, since they appear enhanced as $m_t/m_b$ (\ref{eq:rr}, \ref{eq:lrb}) due to the lifting of the chiral suppression, as already pointed out in Ref.~\cite{Grzadkowski:2008mf}.

\subsection{Bounds on anomalous couplings}\label{sec:bounds}

Having computed $\delta C_i$ in terms of $\kappa_{j}$, we turn our attention to observables affected by such shifts. In particular at order $1/\Lambda^2$, the presently most constraining observables -- the decay rates for $B\to X_s\gamma$ and $B\to X_s \ell^+\ell^-$  are mostly sensitive to the real parts of $\kappa_i$~\cite{Huber:2005ig}.  While in general both $B\to X_{d,s} \gamma$ channels are complementary in their sensitivity to flavor violating NP contributions~\cite{Crivellin:2011ba}, within MFV such effects are to a very good approximation universal and the smaller theoretical and experimental uncertainties in the later mode make it favorable for our analysis. In order to bound imaginary parts of $\kappa_i$, we consider the CP asymmetry in $B\to X_s \gamma$. Finally, we compare and combine these bounds with the ones obtained from $B_{q} - \bar B_{q}$ oscillation observables given in Tab.~\ref{tab:MixingBounds}.

\subsubsection{Real parts}
We consider the inclusive $B\to X_s\gamma$ and $B\to X_s \ell^+\ell^-$ branching ratios, for which the presently most precise experimental values have been compiled in~\cite{Asner:2010qj, Huber:2007vv}
\begin{align}
&\mathrm{Br}[\bar{B}\to X_s \gamma]_{E_{\gamma}>1.6~\mathrm{GeV}}=(3.55 \pm 0.26)\times 10^{-4}\,, \\
&\mathrm{Br}[\bar{B}\to X_s \mu^+ \mu^-]_{\mathrm{low}\, q^2}\hspace{0.2cm}=(1.60\pm 0.50)\times 10^{-6}\,.\nn
\end{align}

Because the SM contributions to $C_{i}(\mu_b)$ and the corresponding operator matrix elements are mostly real~\cite{Huber:2005ig}, the linear terms in $\delta C_i$, which stem from SM--NP interference contributions contribute mostly as $\mathrm{Re}[\delta C_i]$. 
These are the only terms contributing at order $1/\Lambda^2$.
Therefore, the bounds derived from these two observables are mostly sensitive to the real parts of $\kappa_j$.  Using results of~\cite{Huber:2005ig}, we have explicitly verified that the small $\mathrm{Im}[\delta C_i]$ contributions to ${\mathrm{Br}} [\bar B\to X_s \ell^+\ell^-]$ have negligible effect for all operators except $\mathcal Q_{RR,LRb}$. However, even for these operators ${\rm Im}[\kappa_i]$ are much more severely constrained by $\mathcal A_{X_s \gamma}$, discussed in the next section. Also, using known NLO $B\to X_s\gamma$ formulae~\cite{Chetyrkin:1996vx}, we have verified that ${\rm Im}[\delta C_i]$ contributions to this decay rate are negligible. To analyze the effects of $\delta C_i$ on the two branching ratios, we therefore neglect the small $\mathrm{Im}[\delta C_i]$ contributions and employ the semi-numerical formulae given in Ref.~\cite{DescotesGenon:2011yn} with a few modifications that we specify below.
\begin{itemize}
\item In~\cite{DescotesGenon:2011yn} all predictions are given in terms of $\delta C_i$ at the scale $\mu_b=4.8$~GeV. Since we wish to check how our results depend on the matching scale $\mu$, we express $\delta C_i(\mu_b)$ using NNLO QCD running~\cite{Bobeth:1999mk,Gracey:2000am,Gambino:2003zm,Gorbahn:2005sa} as
\begin{eqnarray}
\delta C_7(\mu_b) &=& 0.627\,\delta C_7(m_W)\,,\\
\delta C_7(\mu_b) &=& 0.579\,\delta C_7(2m_W)\,.\nn
\end{eqnarray}
On the other hand, $C_{9,10}$ are only affected by EW running and their change with scale from $2m_W$ to $m_W$ is negligible. 
\item The authors of Ref.~\cite{DescotesGenon:2011yn} assumed $\delta C_8 =0$, which is not the case in our analysis. However, LO $C_7$ and $C_8$ (thus also $\delta C_7$ and $\delta C_8$) enter both observables in approximately the same combination (conventionally denoted as $C_7^{eff}$, c.f.~\cite{Buras:1993xp}). Employing the known SM NNLO matching and RGE running formulae~\cite{Bobeth:1999mk,Gracey:2000am,Gambino:2003zm,Gorbahn:2005sa} we can correct for this with a simple substitution in the expressions of Ref.~\cite{DescotesGenon:2011yn} for the branching ratios
$\delta C_7 \to \delta C_7 + 0.24\, \delta C_8$, where we have neglected the small difference between the matching conditions at $\mu = 2m_W$ and $\mu=m_W$. We have verified that in this way we reproduce approximately the known $\delta C_8$ dependencies in $B\to X_s\gamma$~\cite{Freitas:2008vh} and $B\to X_s \ell^+\ell^-$~\cite{Huber:2005ig}. 
\item As pointed out in the previous section, ${\cal Q}_{LL}^{\pr}$ is to be treated differently than the other operators. Its effects in $\mathcal O_{1,\dots,8}$ can be seen as a shift in the CKM factor appearing $V_{tb}V_{ts}^*\to(1+\kappa_{LL}^{\pr})V_{tb}V_{ts}^*$. Consequently SM predictions for these contributions simply get multiplied by the factor of $|1+\kappa_{LL}^{\pr}|^2$ and only $\delta C_{9,10,\nu\bar{\nu}}$ need to be considered separatly.
\end{itemize}
Taking all this into account and considering only one operator ${\cal Q}_i$ to contribute at a time, we obtain the 95\% C.L. bounds on ${\rm Re} [\kappa_i]$ shown in Tab.~\ref{tab:bounds}.
\begin{table}[h!]
\hspace{-1cm}
\begin{center}
\begin{tabular}{c|ccc|c|c}\hline\hline
&$B-\bar{B}$&$B\to X_s\gamma$&$B\to X_s \mu^{+}\mu^-$ & combined & $C_i(2m_W)\sim 1$ \\
\hline
\LINE{$\kappa_{LL}$}{$\bs{0.08}{-0.09}$}
{$\bs{0.03}{-0.12}$}
{$\bs{0.48}{-0.49}$}
{$\bs{0.04}{-0.09}\Big(\bs{0.03}{-0.10}\Big)$}& $\Lambda> 0.82\,\, \mathrm{TeV}$\\\hline
\LINE{$\kappa_{LL}^{\pr}$}{$\bs{0.11}{-0.11}$}
{$\bs{0.17}{-0.04}$}
{$\bs{0.31}{-0.30}$}
{$\bs{0.11}{-0.06}\Big(\bs{0.10}{-0.06}\Big)$}& $\Lambda> 0.74\,\, \mathrm{TeV}$\\\hline
\LINE{$\kappa_{LL}^{\pr\pr}$}{$\bs{0.18}{-0.18}$}
{$\bs{0.06}{-0.22}$}
{$\bs{1.02}{-1.04}$}
{$\bs{0.08}{-0.17}\Big(\bs{0.05}{-0.15}\Big)$}& $\Lambda> 0.60\,\, \mathrm{TeV}$\\\hline
\LINE{$\kappa_{RR}$}{}
{$\bs{0.003}{-0.0006}$}
{$\bs{0.68}{-0.66}$}
{$\bs{0.003}{-0.0006}\Big(\bs{0.002}{-0.0006}\Big)$}& $\Lambda> 3.18\,\, \mathrm{TeV}$\\\hline
\LINE{$\kappa_{LRb}$}{}
{$\bs{0.0003}{-0.001}$}
{$\bs{0.34}{-0.35}$}
{$\bs{0.0003}{-0.001}\Big(\bs{0.003}{-0.01}\Big)$}& $\Lambda> 9.26\,\, \mathrm{TeV}$\\\hline
\LINE{$\kappa_{LRt}$}{$\bs{0.13}{-0.14}$}
{$\bs{0.51}{-0.13}$}
{$\bs{0.38}{-0.37}$}
{$\bs{0.13}{-0.07}\Big(\bs{0.12}{-0.14}\Big)$}& $\Lambda> 0.81\,\, \mathrm{TeV}$\\\hline
\LINE{$\kappa_{LRt}^{\pr}$}{$\bs{0.29}{-0.29}$}
{$\bs{0.41}{-0.11}$}
{$\bs{0.75}{-0.73}$}
{$\bs{0.27}{-0.07}\Big(\bs{0.25}{-0.06}\Big)$}& $\Lambda> 0.56\,\, \mathrm{TeV}$\\\hline\hline
\end{tabular}
\end{center}
\hspace{1.4cm}
\begin{center}
\includegraphics[scale= 1.0]{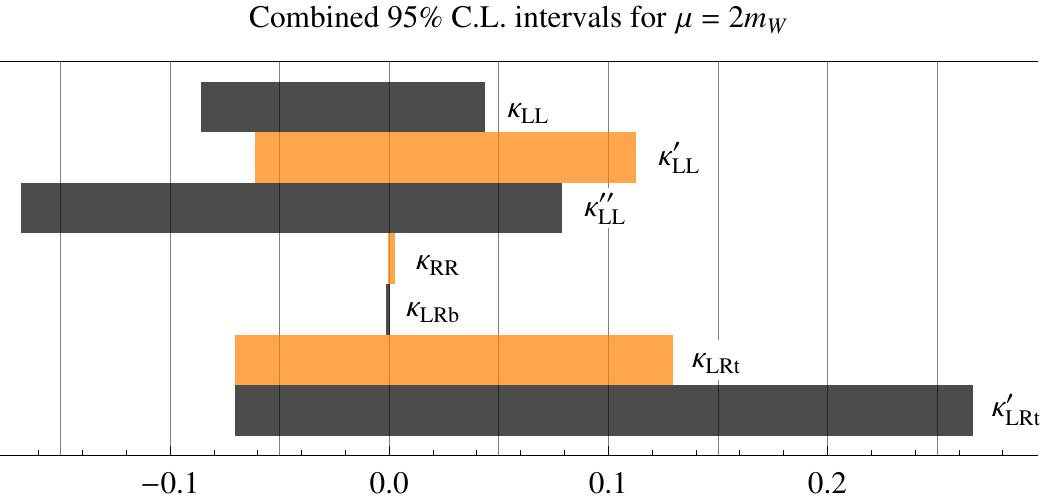}
\end{center}
\caption{Lower and upper $95\%$ C.L. bounds on real parts of individual anomalous couplings $\kappa_j$ for $\mu = 2 m_W$ and bracketed for $\mu=m_W$, where $\overline{m}_t(m_W)=173.8$~GeV and $\overline{m}_b(m_W)=3.06$~GeV have been used. *\,The $B\to X_s \ell^+\ell^-$ bounds on $\mathrm{Re}[\kappa_{RR,LRb}]$ are valid in the $\mathrm{Im}[\kappa_{RR,LRb}]=0$ limit; see text for details. Last column shows estimated lower bounds for NP scale $\Lambda$ assuming the combined bounds and Wilson coefficients $C_i$ to be of the order one. Accompanying plot serves for visual comparison of the presently allowed intervals.}
\label{tab:bounds}
\end{table}

The first column shows bounds obtained from $B_q-\bar{B}_q$ mixing as analyzed in section~\ref{sec:mixing}, while the ``combined'' column corresponds to combined bounds from all three observables. For the later we also present the results when the matching scale is set to $\mu=m_W$ to check the scale dependence of our results. We can see that the bounds obtained change significantly only in the case of $\kappa_{LRb}$ where lowering the scale to $\mu=m_W$ loosens the bounds by almost an order of magnitude. We have also checked that the $B\to X_s \gamma$ bounds agree nicely with those obtained in Ref.~\cite{Grzadkowski:2008mf}. In the last column we present the lower bound on NP scale $\Lambda$ obtained from the combined bounds and under the assumption that Wilson coefficients are of the order $1$ at scale $2m_W$.

The bounds on $\kappa_{RR}$ and $\kappa_{LRb}$ indeed turn out to be an order of magnitude more stringent then for the rest of the coefficients. This was anticipated by the numerical values given in Tab.~{\ref{tab:cs}} where very large effects of operators $\op_{RR}$ and $\op_{LRb}$ on $C_7$ and $C_8$ were observed.

\subsubsection{Imaginary parts}
We have shown in section \ref{sec:mixing} that imaginary parts of primed Wilson coefficients~(\ref{eq:kappas}) can affect the CP violating phase in $B_{q}-\bar B_{q}$ mixing and nonzero values were found to be favored by  the global fit of~\cite{Lenz:2012az}. To constrain imaginary parts of the remaining four operators, which do not contribute with new phases in $B_{q}-\bar B_{q}$ mixing, we consider the direct CP asymmetry in $B\to X_s \gamma$ for which the current world average experimental value reads \cite{Asner:2010qj}
\begin{eqnarray}
A_{X_s \gamma}=\frac{\Gamma(\bar{B}\to X_s\gamma)-\Gamma(B\to X_{\bar{s}}\gamma)}{\Gamma(\bar{B}\to X_s\gamma)+\Gamma(B\to X_{\bar{s}}\gamma)} = -0.012 \pm 0.028\,. \label{eq:cpnow}
\end{eqnarray} 

Based on the recent analysis of this observable in Ref.~\cite{Benzke:2010tq} we obtain the following semi-numerical formula
\begin{eqnarray}
A_{X_s \gamma}&=&0.006+0.039(\tilde{\Lambda}_{17}^u-\tilde{\Lambda}_{17}^c) \\
\nonumber&+&\Big[0.008+0.051 (\tilde{\Lambda}_{17}^u-\tilde{\Lambda}_{17}^c)\Big]\mathrm{Re}[\delta C_7(2m_W)]\\
\nonumber &+&\Big[0.012(\tilde{\Lambda}_{17}^u- \tilde{\Lambda}_{17}^c)+0.002\Big]\mathrm{Re}[\delta C_8(2m_W)]\\
\nonumber&+&\Big[-0.256+0.264 \tilde{\Lambda}_{78}-0.023 \tilde{\Lambda}_{17}^u-2.799 \tilde{\Lambda}_{17}^c\Big]\mathrm{Im}[\delta C_7(2m_W)]\\
\nonumber&+&\Big[-0.668 \tilde{\Lambda}_{17}^c-0.005 \tilde{\Lambda}_{17}^u-0.563 \tilde{\Lambda}_{78}+0.135\Big]\mathrm{Im}[\delta C_8(2m_W)]\,.
\end{eqnarray}

\begin{figure}
\begin{center}\vspace{-0.55cm}
\includegraphics[scale=0.6]{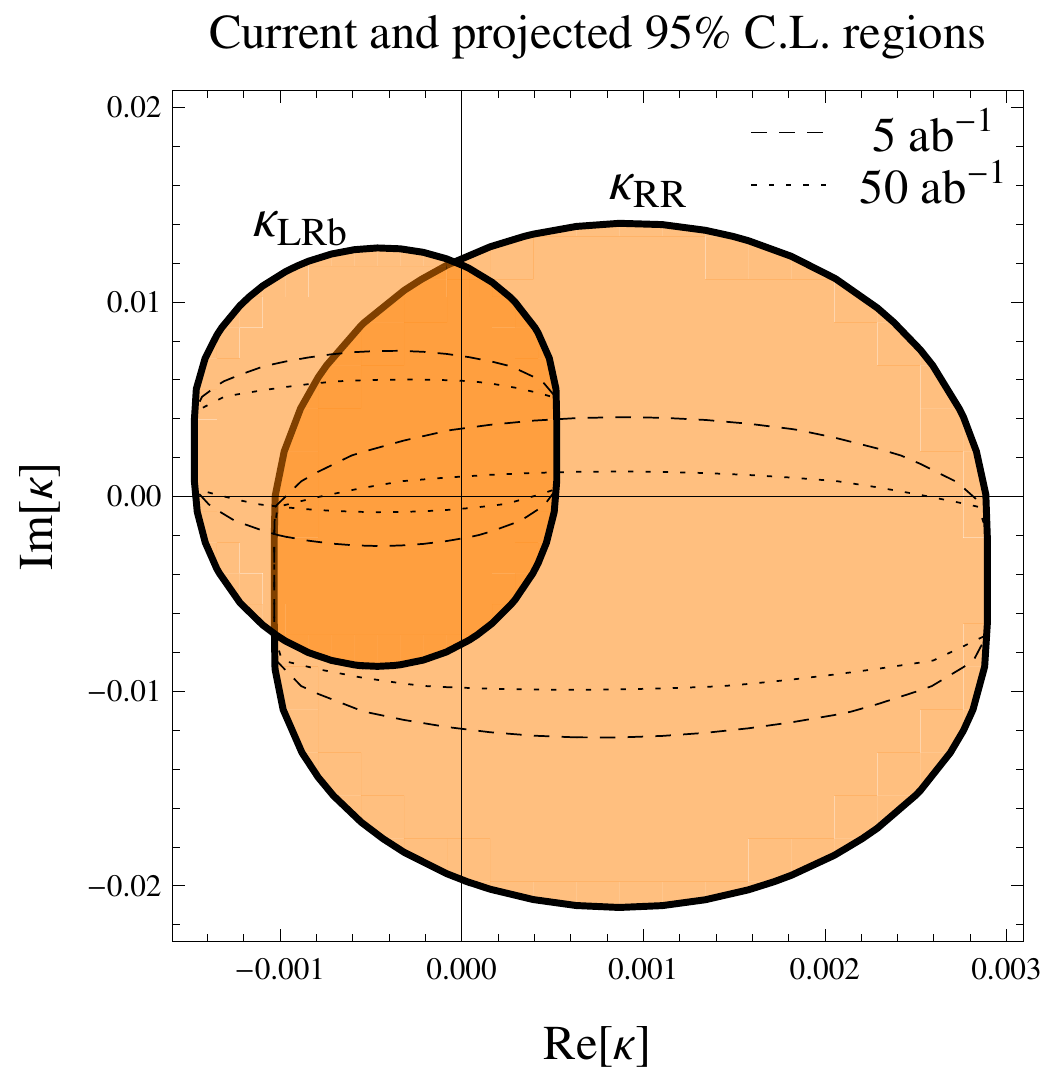}
\caption{$95\%$ C.L. allowed regions in the $\kappa_{RR}(2m_W)$ and $\kappa_{LRb}(2m_W)$ complex plain (solid). The constraints are dominated by ${\rm Br}[{B}\to X_s \gamma]$ and $A_{X_s\gamma}$. Dashed and dotted lines represent bounds obtained under the projected sensitivity of Super-Belle \cite{Browder:2008em,Aushev:2010bq} measurement with 5 and 50 $\mathrm{ab}^{-1}$ of integrated luminosity.  }
\label{fig:CP}
\end{center}
\end{figure}

The estimated intervals for hadronic parameters $\tilde{\Lambda}_{17}^u$, $\tilde{\Lambda}_{17}^c$ and $\tilde{\Lambda}_{78}$  as specified in Ref.~\cite{Benzke:2010tq} dominate the theoretical uncertainty making it sufficient to use a LO QCD analysis in the perturbative regime. Thus, in addition to the numerical parameters specified in~\cite{Benzke:2010tq}, we have used the LO QCD running for $\delta C_{7,8}$ in this observable.
Performing a combined analysis of all considered bounds on the real and imaginary parts of individual $\kappa_i$ in which we marginalize the hadronic parameters entering $A_{X_s\gamma}$ within the allowed intervals, we can obtain the allowed regions in the complex plain of $({\rm Re}[\kappa_i],{\rm Im}[\kappa_i])$ shown in Fig.~\ref{fig:CP}.  

As already argued, the imaginary part of $\kappa_{LL}$ does not contribute to the $\delta C_i$ and thus remains unconstrained. It also turns out that due to the large hadronic uncertainties, the imaginary parts of $\kappa^{(\prime)}_{LRt}$, $\kappa^{\pr(\pr\pr)}_{LL}$ remain largely unconstrained by $A_{X_s \gamma}$ and $B_{q}-\bar B_{q}$ mixing observables still provide the strongest constraints (except for ${\rm Im}[\kappa_{LRt}]$ which again remains unconstrained). On the contrary, constraints on the imaginary parts of $\kappa_{RR}$ and $\kappa_{LRb}$ reach per-cent level, as can be seen in Fig.~\ref{fig:CP}, where we also illustrate the projected bounds for Super-Belle, assuming the measured central value to be the same as given in Eq.~(\ref{eq:cpnow}) and using the estimated Super-Belle accuracy given in Ref.~\cite{Browder:2008em,Aushev:2010bq}.
Finally we note that in absence of the long-distance effects on NP contributions considered in~\cite{Benzke:2010tq}, $A_{X_s \gamma}$ would exhibit an even greater sensitivity to the imaginary parts of $\delta C_{7,8}$~\cite{Barbieri:2011fc}, thus we consider our derived bounds as conservative.
\subsection{Predictions}
Having derived bounds on anomalous $\kappa_j$ couplings, it is interesting to study to what extent these can still affect other rare $B$ decay observables.  Analyzing one operator at a time we set the matching scale to $\mu=2m_W$ and consider $\kappa_j$ to be real.

We turn once more to the semi-numerical formulae given in Ref.~\cite{DescotesGenon:2011yn}, and first consider  the branching ratio $\mathrm{Br}[\bar B_s\to\mu^+\mu^-]$ for which CDF's latest analysis yields~\cite{Aaltonen:2011fi}
\begin{eqnarray}
4.6\times 10^{-9}<\mathrm{Br}[\bar B_s\to\mu^+\mu^-]<3.9\times 10^{-8}\,,\hspace{0.5cm} \text{at $90\%$ C.L.}\,,
\end{eqnarray}
while the LHCb collaboration reports on the upper limit \cite{Aaij:2012ac}
\begin{eqnarray}
\mathrm{Br}[\bar B_s\to\mu^+\mu^-] < 4.5\times 10^{-9}\,,\hspace{0.5cm} \text{at $95\%$ C.L.}\,,
\end{eqnarray}
and a CMS, Atlas and LHCb combined limit has recently been made public \cite{CMS:Bsmumu}
\begin{eqnarray}
\mathrm{Br}[\bar B_s\to\mu^+\mu^-] < 4.2\times 10^{-9}\,,\hspace{0.5cm} \text{at $95\%$ C.L.}\,.
\end{eqnarray}
In addition we explore the differential forward-backward asymmetry $A_{\mathrm{FB}}(q^2)$ in the $\bar{B}_d\to \bar{K}^*\ell^+\ell^-$ decay, for which the latest measurement of LHCb has recently been published in Ref.~\cite{Aaij:2011aa}.
Finally, following Ref.~\cite{Altmannshofer:2009ma} we analyze the allowed effects of $\kappa_j$ on the branching ratios $\mathrm{Br}[B\to K^{(*)}\nu\bar{\nu}]$, which are expected to become experimentally accessible at the super-B factories~\cite{O'Leary:2010af}.

\begin{figure}[h]
\begin{center}
\includegraphics[scale= 0.513]{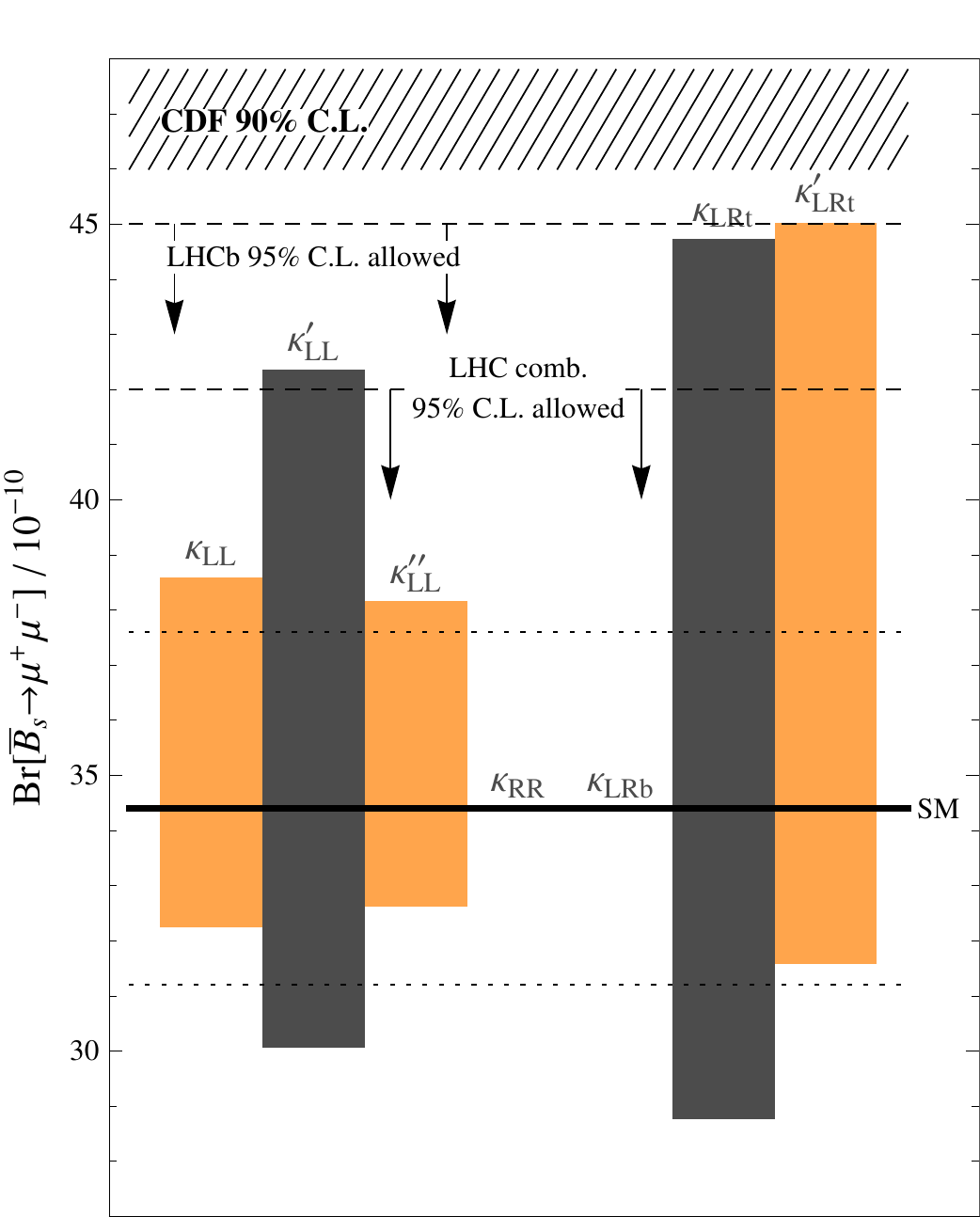}\hspace{0.8cm}
\includegraphics[scale=0.48]{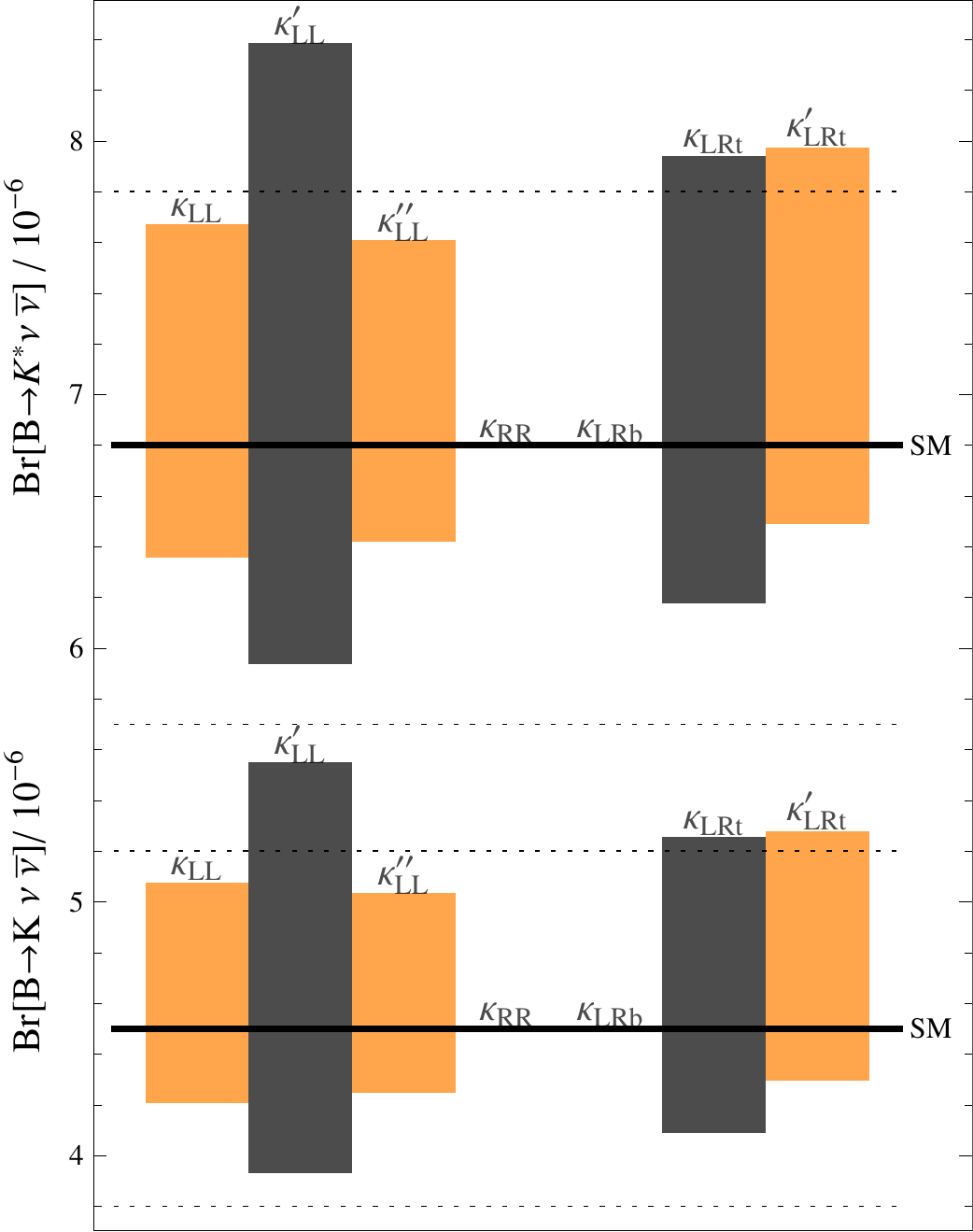}
\caption{Ranges of values for branching ratios obtained as anomalous couplings are varied within the $95\%$ C.L. intervals given in Tab.~\ref{tab:bounds}. We also show the SM predictions (black) with $1\sigma$ theoretical uncertainty band (dotted) and for the muonic decay channel the lower end of the experimental $90\%$ C.L. interval from \cite{Aaltonen:2011fi}, the 95\% C.L. upper bound from LHCb \cite{Aaij:2012ac} and the latest combined LHC upper bound \cite{CMS:Bsmumu}.}
\label{fig:predict1}
\end{center}
\end{figure}

\begin{figure}[h!]
\begin{center}
\includegraphics[scale= 0.6]{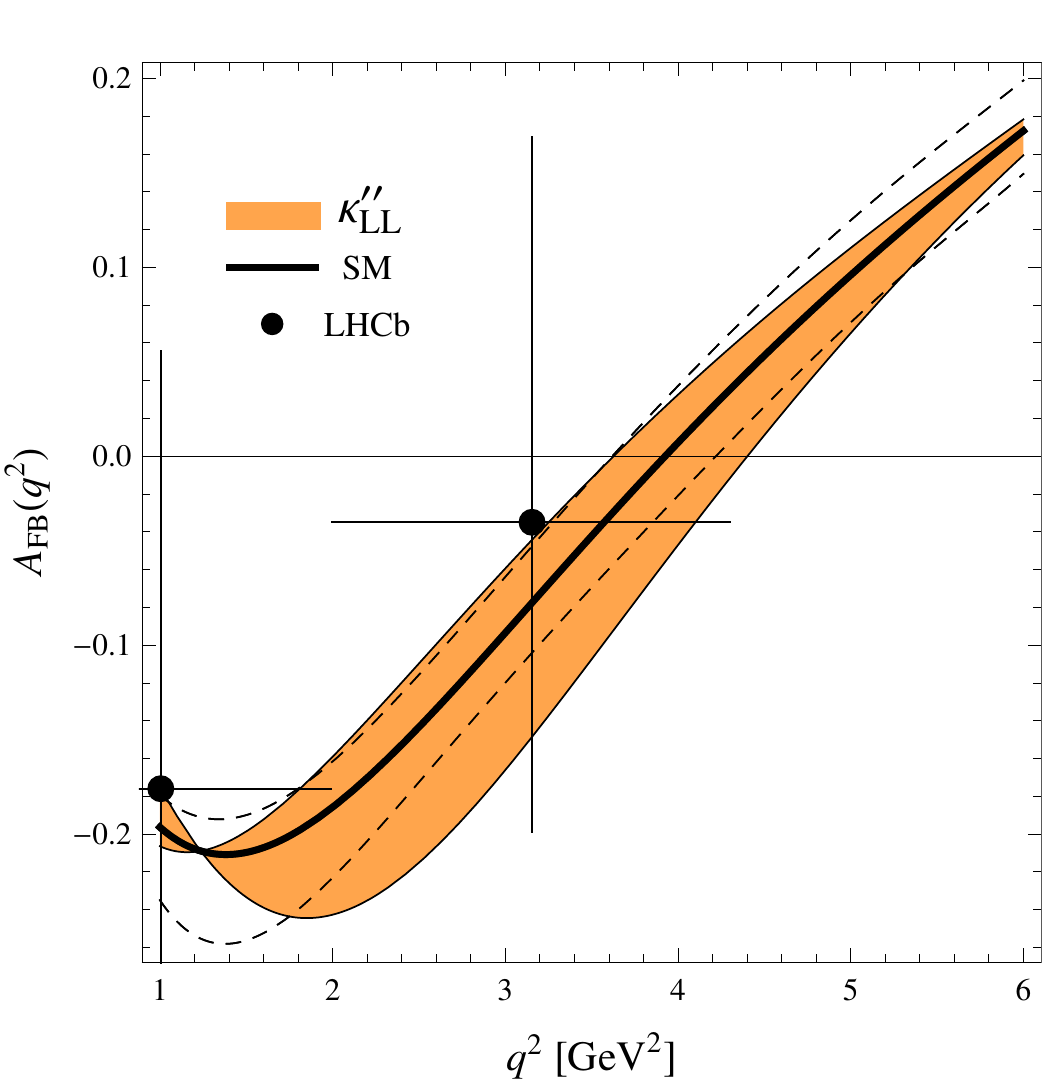}\hspace{0.8cm}
\includegraphics[scale= 0.6]{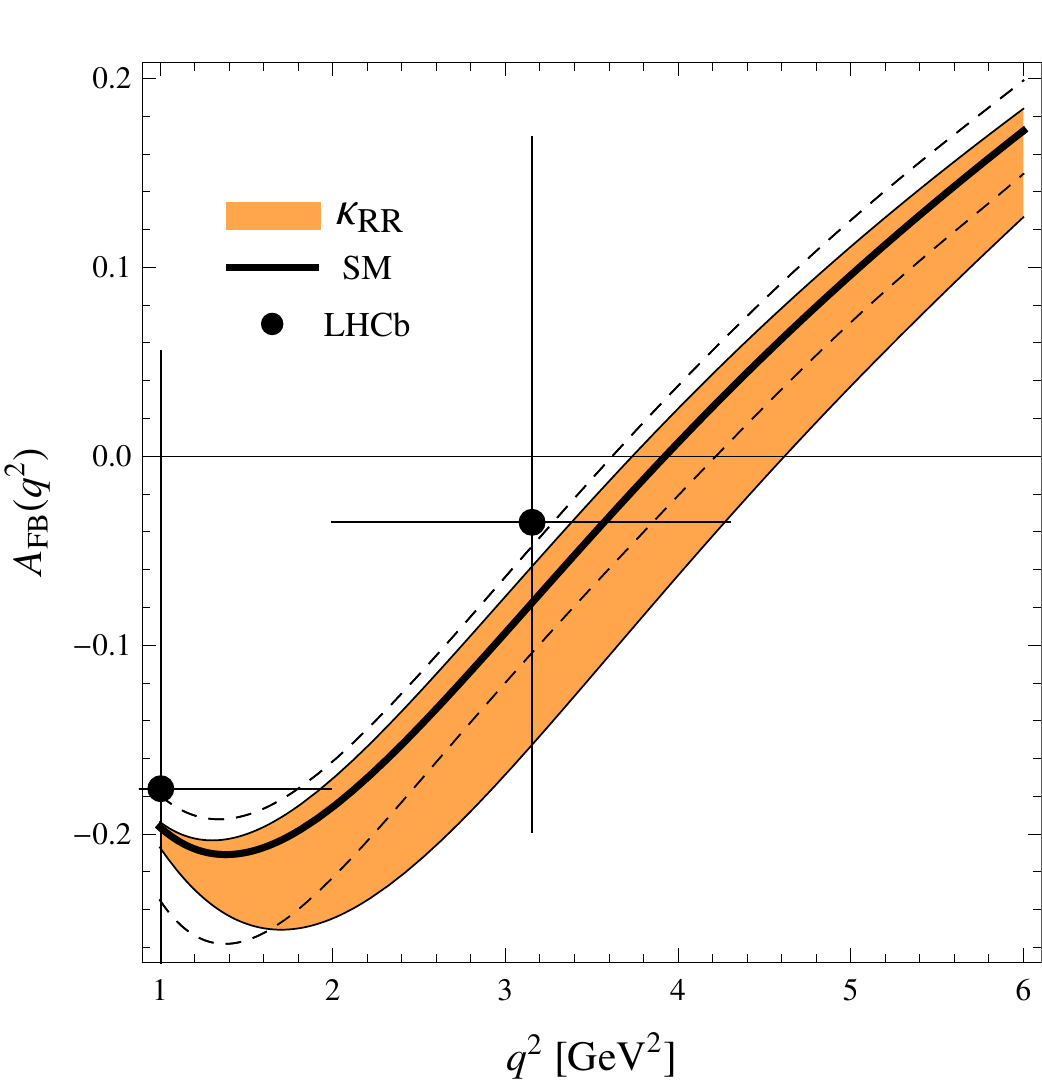}
\caption{$A_{\mathrm{FB}}(q^2)$ band obtained when varying real parts of $\kappa_{LL}^{\pr\pr}$ (left) and $\kappa_{RR}$ (right) within the 95\% C.L. interval given in Tab.~\ref{tab:bounds}. Also presented are the SM predicted central value (black) with $1\sigma$ theoretical uncertainty band (dashed) and the latest measured points with experimental errors given in Ref.~\cite{Aaij:2011aa}.  }
\label{fig:predict2}
\end{center}
\end{figure}

We present our findings in Figs.~\ref{fig:predict1} and~\ref{fig:predict2}. The effects of anomalous couplings $\kappa_j$ on all branching ratios are similar. There is a slight tenancy of anomalous $\kappa_j$ couplings to increase the predictions compared to the SM values at the level of the present theoretical uncertainties, with the exception of $\kappa_{RR}$ and $\kappa_{LRb}$ of which effects are negligible. In particular, none of the contributions can accommodate the recent CDF measurement of ${\rm Br}[\bar B_s \to \mu^+\mu^-]$ at the $90\%$ C.L.\,, while the latest measurements from LHC are already starting to constrain $\kappa'_{LL}$ and $\kappa^{(\prime)}_{LRt}$ contributions and could in future become an important factor to be included in the combined bound analysis. Furthermore we find that the forward-backward asymmetry $A_{\mathrm{FB}}(q^2)$ can still be somewhat effected by $\kappa_{LL}^{\pr\pr}$ and $\kappa_{RR}$, for which we present the bands obtained when varied within the $95\%$ C.L. intervals in Fig.~\ref{fig:predict2}. While not sensitive at the moment, in the near future, improved measurements by the LHCb experiment could possibly probe such effects. On the other hand, the contributions of other anomalous couplings all fall within the theoretical uncertainty bands around the SM predicted curve.

\subsection{Summary}
We have investigated contributions of  anomalous charged quark currents in flavor changing neutral current mediated $|\Delta B|=1$ processes within an effective field theory framework assuming minimal flavor violation. We have determined the indirect bounds on the real and imaginary parts of the anomalous couplings. 
In particular, we are able for the first time to constrain the imaginary parts of $\kappa_{RR}$ and $\kappa_{LRb}$ already at order $1/\Lambda^2$. Taking into account the obtained bounds on real parts of $\kappa_i$ we have predicted the magnitude of effects that the operators considered might have on the branching ratio of the $B_s \to \mu^+ \mu^-$ decay, the forward-backward asymmetry in $B \to K^* \ell^+ \ell^-$, as well as the branching ratios of $B \to K^{(*)} \nu \bar \nu$ decays. The better knowledge of these (especially the potential further lowering of the $\mathrm{Br}[B_s\to \mu^+\mu^-]$ upper limit) and other recently proposed~\cite{Bobeth:2007dw} observables in the future could further constrain some of the anomalous couplings.

\section{Helicity fractions at NLO in QCD}
\label{sec:hel_nlo}

Having exhausted the implications of charged quark current operators (\ref{eq:ops1}) on the $B$ meson mixing, radiative and rare semileptonic decays, we turn in this section, to the study of how the non-SM $tWb$ interactions induced by these operators influence the $W$ gauge boson helicity fractions in unpolarized top quark decays at NLO in QCD. We aim to confront these effects with the indirect bounds obtained for the couplings from $B$ physics to see how much deviation from SM predictions in helicity fractions may still be compatible with the low energy observations and further examine if perhaps LHC and Tevatron measurements might turnout to provide more stringent constraints on such NP.
\subsection{Framework}
Since in this section we shall be dealing with $t\to W b$ decays exclusively, the $tWb$ vertex and its deviation from the SM value and structure is the only charged quark interaction of interest. Following \cite{AguilarSaavedra:2008zc}, we can obtain the most general parametrization thereof by considering the following effective Lagrangian
\begin{eqnarray}
\mathcal L_{\mathrm{eff}} = -\frac{g}{\sqrt{2}}\bar{b}\Big[\gamma^{\mu} \big(a_L P_L +a_R P_R\big)
-(b_{RL} P_L + b_{LR} P_R)\frac{2\ii \sigma^{\mu\nu}}{m_t}q_{\nu}  \Big]t W_{\mu}\,,\label{eq:effsimple}
\end{eqnarray}
where $q$ is the momentum of the $W$ boson and $P_{R,L}=1/2(1\pm\gamma^5)$ are the chirality projectors. Note that $a_L$ includes also the SM contribution $a_L = V_{tb} + \delta a_L$. Extraction of the modified Feynman rule for $tWb$ vertex is straight forward and is given in the Appendix~\ref{app:feyn_charged}. All of the operators (\ref{eq:ops1}) considered in the previous two sections generate these couplings with the following correspondence
\begin{eqnarray}
\delta a_L =  V_{tb}^* \kappa_{LL}^{(\pr,\pr\pr)*}\,,\hspace{0.3cm} a_R = V_{tb}^{*}\kappa_{RR}^*\,,\hspace{0.3cm} b_{LR} = -\frac{m_t}{2 m_W}V_{tb}^{*}\kappa_{LRt}^{(\pr)}\,,\hspace{0.3cm} b_{RL} = -\frac{m_t}{2 m_W} V_{tb}^* \kappa_{LRb}^*\,,\label{eq:translation}
\end{eqnarray}
where $\kappa_j$ have been defined in Eq.~(\ref{eq:kappas}).

We shall parametrize the main decay channel of the top quark in the following way
\begin{eqnarray}
\Gamma_{t\to W b}&=&\frac{m_t}{16\pi}\frac{g^2}{2}\sum_{i}\Gamma^i\,,
\end{eqnarray}
where $i= L,+,-$ stands for longitudinal, transverse-plus and transverse-minus as introduced in section~\ref{sec:hfSM}.

The $\Gamma^i$ decay rates have already been studied to quite some extent in the existing literature. The tree-level analysis of the effective interactions~(\ref{eq:effsimple}) has been conducted in Ref.~\cite{AguilarSaavedra:2006fy}. QCD corrections, however, have been studied only for the chirality conserving (SM) operators. As we have argued in chapter~\ref{sec:hfSM} QCD corrections are especially important for the observable $\mathcal F_+$ since they allow to lift the helicity  suppression present at the LO in the SM. Helicity suppression in this observable is also exhibited in the presence non SM dipole structure $b_{LR}$ of $tWb$ vertex, which is especially interesting since it is least constrained by indirect bounds coming from $B$ physics presented in Tab.~\ref{tab:bounds}. It might therefore have the potential to modify the $t\to W b$ decay properties in an observable way.

\subsection{Computation}
\begin{figure}[h!]
\begin{center}
\includegraphics[scale=0.6]{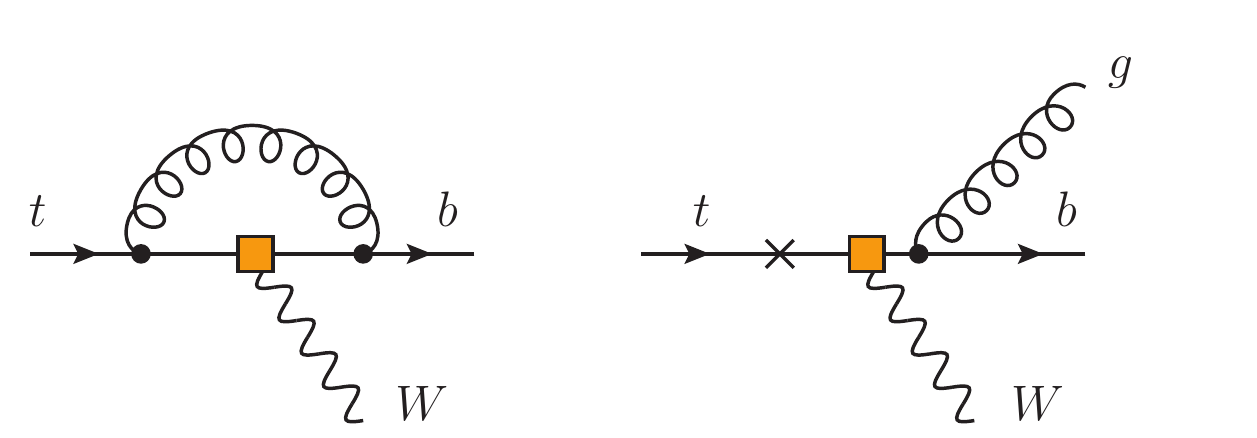}
\end{center}
\caption{Feynman diagrams for next-to-leading order QCD contributions. Square marks the insertion of the generally parametrized $tWb$ interaction specified in Eq.~(\ref{eq:effsimple}) and cross the additional point from which the gluon can be emitted. }
\label{fig:feyndiags}
\end{figure}

We compute the ${\cal O}(\alpha_s)$ corrections to the polarized rates $\Gamma^i$ in the $m_b=0$ limit using the most general $tWb$ interaction vertex extracted from Eq.~(\ref{eq:effsimple}). The appropriate Feynman one-loop and bremsstrahlung diagrams to be considered are presented in Fig.~\ref{fig:feyndiags}. We regulate UV and IR divergences by working in $d=4+\epsilon$ dimensions. The renormalization procedure and the fusing of virtual and bremsstrahlung contributions to attain the cancelation of IR divergences closely resembles the procedure (for the $t\to qZ$ case) described in chapter~\ref{sec:rge}, where we have analyzed the NLO QCD corrections for FCNC top quark decays. There are however some differences in the computation worth pointing out. 

First, since there are no contributions of gluonic NP operators, there are no QCD mixing effects to be considered. Furthermore, we need to make use of the covariant projector technique introduced in section~\ref{sec:hfSM} of the introductory chapter and summarized in Tab.~\ref{tab:projectors}, to project out the desired helicities of the $W$ boson. This makes the analysis more involved. What is more, since one of the projectors includes an explicit $\epsilon_{\alpha\beta\gamma\delta}$ tensor one can envision encountering problems with $\gamma^5$ in $d$ dimensions if naive dimensional regularization, whereby $\gamma^5$ is assumed to anti-commute with other $\gamma^{\mu}$ matrices in $d=4+\epsilon$ dimensions as well, is used. To avoid conceivable problems we use the prescriptions based on 't Hooft--Veltman $\gamma^5$ regularization that have been derived by S.A. Larin and are given in Ref.~\cite{Larin:1993tq,Ball:2004rg}
\begin{eqnarray}
\gamma_{\mu}\gamma_5 &\to& (1-4a_s)\frac{\ii}{3!}\epsilon_{\mu\nu_1\nu_2\nu_3}\gamma^{\nu_1}\gamma^{\nu_2}\gamma^{\nu_3}\,,\\
\sigma_{\mu\nu}\gamma_5&\to&-\frac{\ii}{2}\epsilon_{\mu\nu\alpha\beta}\sigma^{\alpha\beta}\,,
\end{eqnarray}
where $a_s = C_F \alpha_s/(4\pi)$ is needed because since the anticommutativity of $\gamma^5$ is violated the standard properties of the axial current and Ward identities are also violated and need to be restored by additional renormalization (see Ref.~\cite{Larin:1993tq} for details).

\subsection{The decay rates}
In the $m_b=0$ limit, which we are employing, there is no mixing between chirality flipped operators and the decay rates can be written as
\begin{eqnarray}
\Gamma^{(L,+,-)}&=& |a_L|^2 \Gamma^{(L,+,-)}_a + |b_{LR}|^2 \Gamma^{(L,+,-)}_{b} \label{e3}
+ 2\mathrm{Re}\{a_L b_{LR}^*\} \Gamma^{(L,+,-)}_{ab} + \langle L \leftrightarrow R,+\leftrightarrow -\rangle\,,\label{eq:form}
\end{eqnarray}
where considering the $a_R$, $b_{RL}$ pair can be accommodated by changing the role of transverse plus and transverse minus decay widths $\Gamma^+ \leftrightarrow \Gamma^-$ as indicated by the bracketed term in Eq.~(\ref{eq:form}). Analytical formulae for $\Gamma^{i}_{a,b,ab}$ functions are given in the Appendix~\ref{app:gamma_main}. We have crosschecked $\Gamma^i_{a}$ with the corresponding expressions given in \cite{Fischer:2000kx} and found agreement between the results.

\begin{table}[h!]
\begin{center}
\begin{tabular}{l|c|c|c|c||c|c|c|c}\hline\hline
			& $L$ 	& $+$ 	& $-$& unpolarized &  &$L$ & $+$ &$-$ \\\hline
$\Gamma_a^{i,\mathrm{LO}}$&$\frac{(1-x^2)^2}{2x^2}$& $0$ & $(1-x^2)^2$&$\frac{(1-x^2)^2(1+2x^2)}{2x^2}$
&$\Gamma^{i,\mathrm{NLO}}_a/\Gamma_a^{i,\mathrm{LO}}$  &$0.90$&$3.50$&$0.93$\\
$\Gamma_b^{i,\mathrm{LO}}$& $2x^2(1-x^2)^2$ &$0$&$4(1-x^2)^2$&$2(1-x^2)^2(2+x^2)$
&$\Gamma^{i,\mathrm{NLO}}_b/\Gamma_b^{i,\mathrm{LO}}$  &$0.96$&$4.71$&$0.91$\\
$\Gamma_{ab}^{i,\mathrm{LO}}$&$(1-x^2)^2$&$0$&$2(1-x^2)^2$&$3(1-x^2)^2$
&$\Gamma^{i,\mathrm{NLO}}_{ab}/\Gamma_{ab}^{i,\mathrm{LO}}$ &$0.93$&$3.75$&$0.92$\\\hline\hline
\end{tabular}
\caption{{\bf Left}: Tree-level decay widths for different $W$ helicities and their sum, which gives the unpolarized width. All results are in the $m_b=0$ limit and we have defined $x=m_W/m_t$. {\bf Right}: Numerical values for $\Gamma^\mathrm{NLO}/\Gamma^{\mathrm{LO}}$ with the following input parameters $m_t=173.0$ GeV, $m_W = 80.4$ GeV, $\alpha_s(m_t) = 0.108$. Scale $\mu$ appearing in NLO expressions is set to $\mu=m_t$. In addition $m_b=4.8$ GeV. These values are used throughout the section for all numerical analysis. }
\label{tab:NP_hel}
\end{center}
\end{table}

The LO (${\cal O}(\alpha_s^0)$) contributions to decay rates $\Gamma_{a,b,ab}^{i,\mathrm{LO}}$ are obtained with a tree-level 
calculation and are given on the left side of Tab.~\ref{tab:NP_hel}. Our results coincide with those given in \cite{AguilarSaavedra:2006fy}, if the mass $m_b$ is set to zero. 

The change of $\Gamma_{a,b,ab}^{i}$ going form LO to NLO QCD is presented on the right side of Tab.~\ref{tab:NP_hel}. Since in the $m_b=0$ limit $\Gamma_{a,b,ab}^{+,\mathrm{LO}}$ vanish, we use the full $m_b$ dependence of the LO rate when dealing with $W$ transverse-plus helicity. Effectively we neglect the ${\cal O}(\alpha_s m_b)$ contributions. In Ref.~\cite{Fischer:2001gp} it has been shown, that these sub-leading contributions can scale as $\alpha_s (m_b/m_W)^2 \log (m_b/m_t)^2 $ leading to a relative effect of a couple of percent compared to the size of $\mathcal O (\alpha_s)$ corrections in the $m_b=0$ limit.

\subsection{Effects on ${\cal F}_+$}
\begin{figure}[h!]
\begin{center}
\includegraphics[scale=0.6]{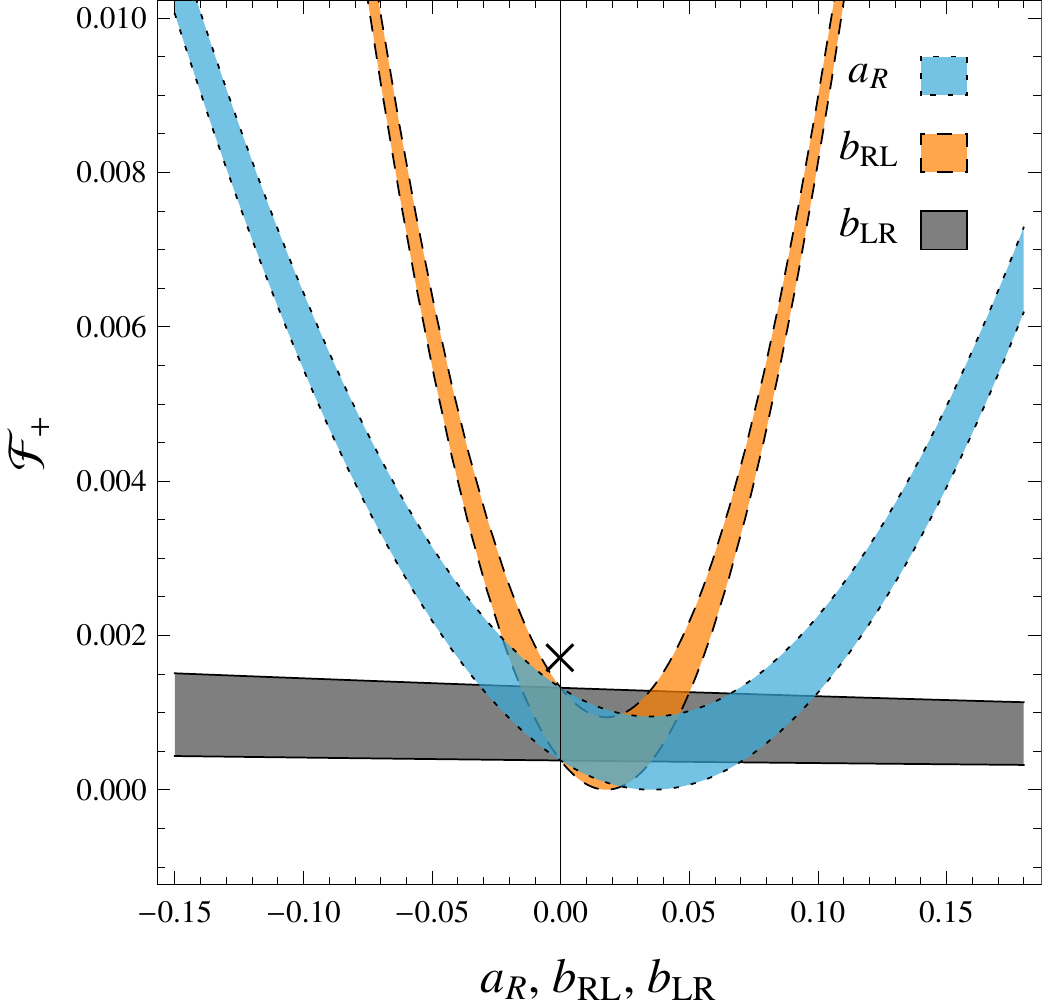}\hspace{1cm}
\includegraphics[scale=0.6]{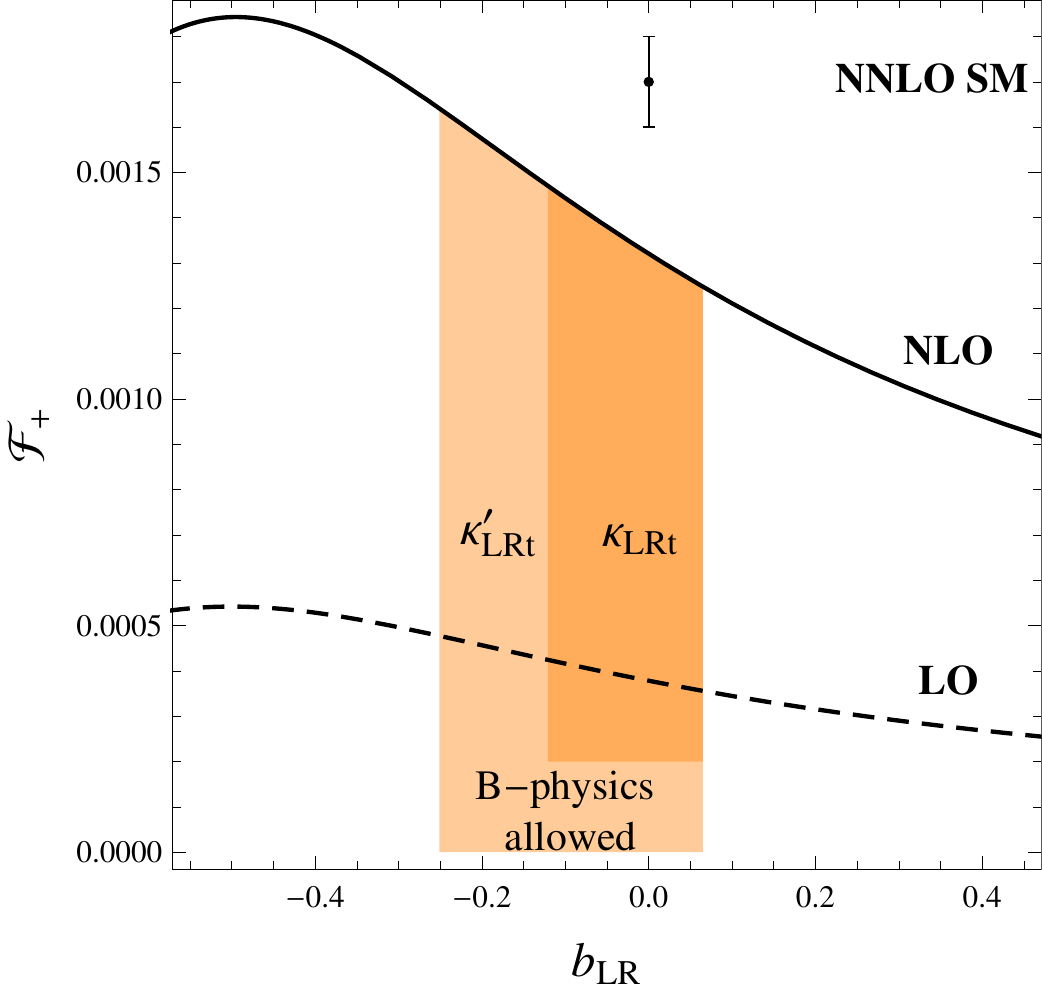}
\end{center}
\caption{Dependence of $\mathcal F_+$ on anomalous couplings which are considered to be real and non-zero one at the time. {\bf Left}: Value of ${\cal F}_+$ for $a_R$ (blue, dotted), $b_{RL}$ (orange, dashed) and $b_{LR}$ (black, solid). Lower and upper lines correspond to LO and NLO results respectively and cross marks the SM NNLO prediction. {\bf Right}: Value of ${\cal F}_+$ for $b_{LR}$. Dashed line corresponds to LO results, while the solid line represents the NLO results. We also present the SM NNLO value along with its error bars given in Eq.~(\ref{eq:e22b}) and the $95\%$ C.L. allowed intervals for $b_{LR}$ given in Eq.~(\ref{eq:ind_translated}).}
\label{fig:F+}
\end{figure}

In Fig.~\ref{fig:F+} we present how each separate anomalous coupling, assumed to be real, affects $\mathcal F_+$. Deviation of the left-handed current coupling from the SM value $\delta a_L$ can not be probed with helicity fractions as long as no interference effects with other NP is considered, since its effects simply factor out in the decay widths. The impact of going from LO to NLO in QCD is presented in terms of the bands where the lower line corresponds to LO while the upper line presents the inflated NLO result. 

A relatively narrow range in the size of anomalous couplings is shown since using Eq.~(\ref{eq:translation}) we translate the indirect constraints on anomalous couplings given in Tab.~\ref{tab:bounds} to find the $95\%$ C.L. allowed intervals are quite narrow
\begin{eqnarray}
 -0.0006 \le  a_R \le 0.003\,,\hspace{0.5cm}
 -0.0004 \le b_{RL} \le 0.0016\,,\hspace{0.5cm}
 -0.14 (-0.29) \le b_{LR} \le 0.08\,.\label{eq:ind_translated}
\end{eqnarray}
Two separate lower bounds on $b_{LR}$, which is substantially less constrained the other two anomalous couplings, correspond to which operator $\mathcal Q_{LRt}$ or $\mathcal Q_{LRt}^\prime$ we assume the $b_{LR}$ to be generated from. The graph on the righthand side of Fig.~\ref{fig:F+} shows the $\mathcal F_+$ dependence on $b_{LR}$ in more detail, along with intervals given in Eq.~(\ref{eq:ind_translated}). We see that the increase is substantial when going to NLO in QCD, but still leaves ${\cal F}_+$ at the $1-2$ per-mille level.

Since the indirect constraints on non-zero values of $a_R$ and $b_{RL}$ are very stringent they can not produce large $\mathcal F_+$, both giving the maximal values of $\mathcal F_{+} =0.00133$, which
is within $1\%$ of the SM prediction $\mathcal F_+^{\mathrm{SM,NLO}}=0.00132 $.

\subsection{Effects on ${\cal F}_L$}
\begin{figure}[h!]
\begin{center}
\includegraphics[scale=0.6]{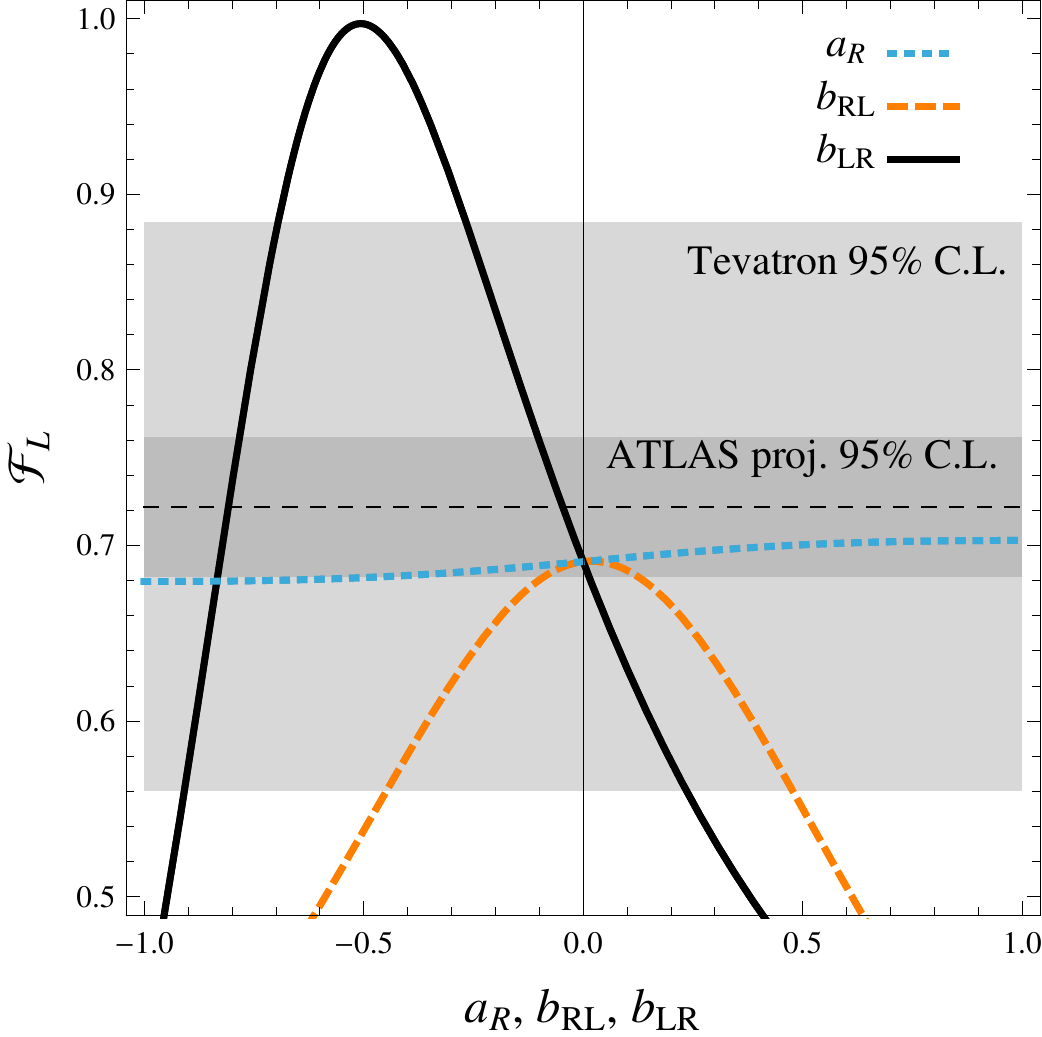}\hspace{1cm}
\includegraphics[scale=0.6]{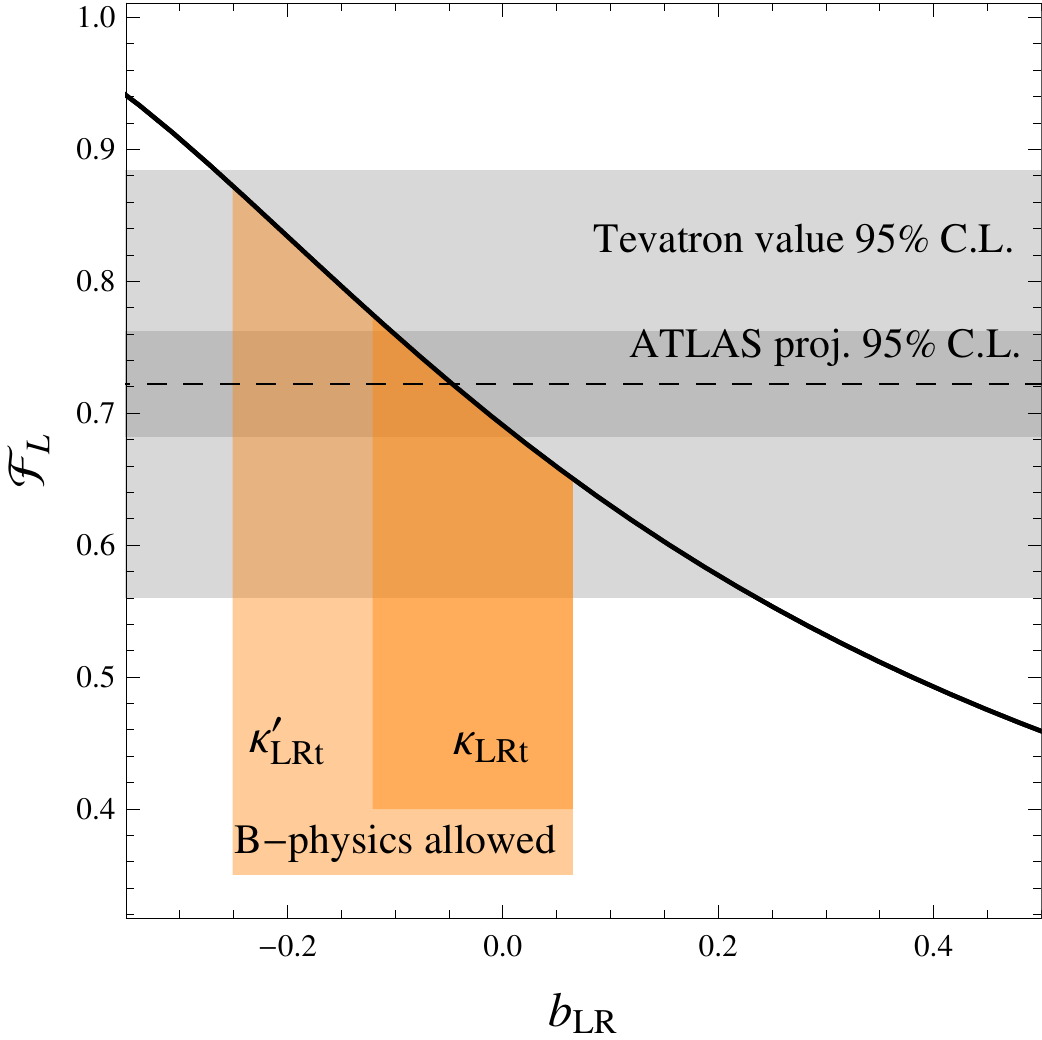}
\end{center}
\caption{Dependence of $\mathcal F_L$ on anomalous couplings which are considered to be real and non-zero one at the time. Also shown is the central measured Tevatron value (dashed) and the $95\%$ C.L. interval as well as the expected ATLAS $95\%$ C.L. interval put on top of the Tevatron central value. {\bf Left}: Value of ${\cal F}_L$ for $a_R$ (blue, dotted), $b_{RL}$ (orange, dashed) and $b_{LR}$ (black, solid) at NLO in QCD. {\bf Right}: Dependance on $b_{LR}$ and the $95\%$ C.L. allowed intervals for $b_{LR}$ given in Eq.~(\ref{eq:ind_translated}). }
\label{fig:FL}
\end{figure}

Analyzing a single real NP contribution at the time, leading QCD corrections decrease ${\cal F}_L$ by approximately $1\%$ in all cases. In in Fig.~\ref{fig:FL} we show the $\mathcal F_L$ NLO dependance on the anomalous couplings. Possible effects of $a_R$ and $b_{RL}$ are again severely constrained by indirect $B$ physics considerations. On the other hand, we find that the most recent combined Tevatron measurement of $\mathcal F_L$ given Eq.~(\ref{eq:hel_exp}) allows to put competitive bounds on $b_{LR}$ compared to the indirect constraints given in Eq.~(\ref{eq:ind_translated}). A detailed plot of ${\cal F}_L$ dependance on $b_{LR}$ is given on the right graph of Fig.~\ref{fig:FL}.

We see that at present the indirect constraints are a bit better, however if the projected sensitivity is reached, the direct bounds from $\mathcal F_L$ could turn out to be more stringent.

\subsection{Comparison with direct constraints}
As we have shown, there is an interesting interplay between direct and indirect constrains when considering the anomalous charged quark currents of the top quark. In this subsection we would like to stress this point further, going a bit beyond the scope of this work, by including another important top quark process that is influenced by anomalous $tWb$ vertices, namely the single top quark production which proceeds through weak interactions.

Not going into details of the subject, we only comment that measurement of the single top quark production cross section~\cite{Group:2009qk} and its agreement with the SM predicted value serve to constrain NP contributions affecting the cross section~\cite{Tait:2000sh}. Having an additional sensitive observable at disposal one can consider pairs of NP operators altering the $tWb$ vertex contributing simultaneously and obtain $95\%$ C.L. allowed regions in the corresponding NP parameter planes. This was performed in Ref.~\cite{AguilarSaavedra:2011ct} using the single top production cross section and the helicity fractions as the constraining observables.

\begin{figure}[h!]
\begin{center}
\includegraphics[scale= 0.6]{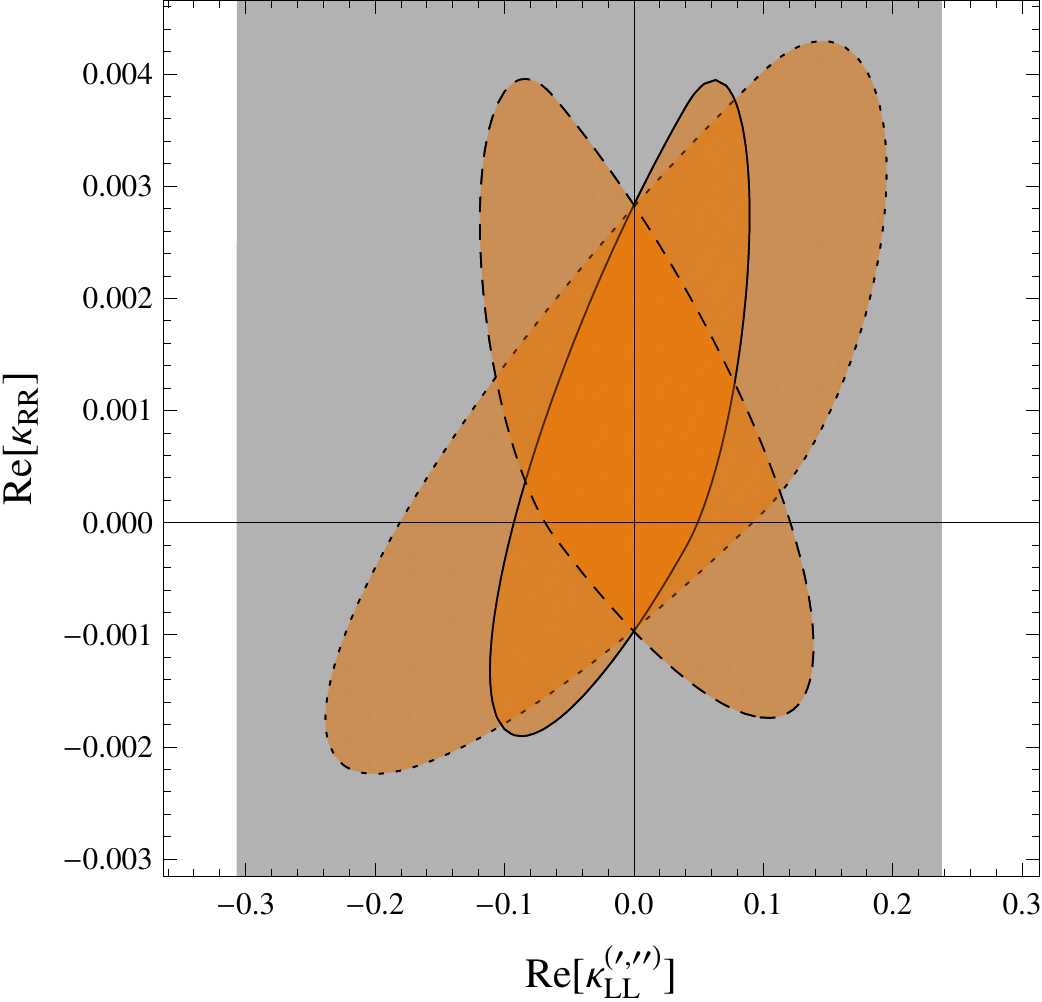}\hspace{0.8cm}
\includegraphics[scale=0.6]{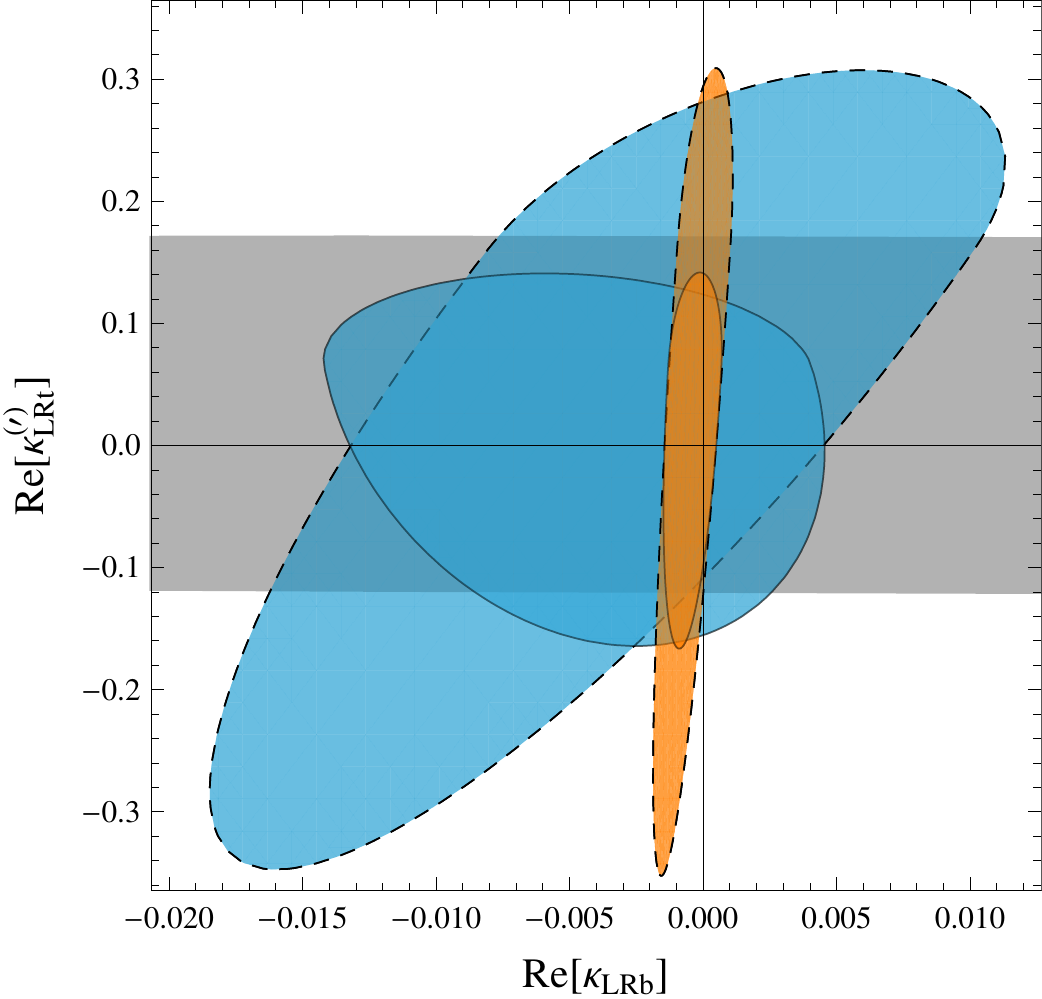}
\caption{95\% C.L. allowed regions in different $(\kappa_i,\kappa_j)$ planes. The gray bands represent the allowed regions from direct Tevatron constraints given in Ref.~\cite{AguilarSaavedra:2011ct} and $\kappa_i$ are assumed to be real. {\bf Left}: $\kappa_{RR}$ - $\kappa_{LL}$ (solid), $\kappa_{LL}^{\prime}$ (dashed), $\kappa_{LL}^{\pr\pr}$ (dotted) plane. Matching scale is set to $\mu=2 m_W$. {\bf Right}: $\kappa_{LRb}$ - $\kappa_{LRt}$ (solid), $\kappa_{LRt}^{\prime}$ (dashed) plane. Matching scale is set to $\mu=2 m_W$ (narrow regions) and $\mu=m_W$ (wider regions).}
\label{fig:2d1}
\end{center}
\end{figure}

We compare the regions presented there with those that we can obtain using our indirect constraints from $B$ physics in Fig.~\ref{fig:2d1}. The comparison nicely summarizes the interplay of direct and indirect constraints on anomalous $tWb$ interactions and shows that they are in a way complementary. On both plots the gray area represents the $95\%$ C.L. allowed regions obtained in Ref.~\cite{AguilarSaavedra:2011ct}. They appear as bands because the direct constraints in $\kappa_{RR}$ and $\kappa_{LRb}$ directions are much weaker than the indirect. On the other hand we can see that Tevatron constraints on $\kappa_{LL}^{(\pr,\pr\pr)}$ and $\kappa_{LRt}^{(\pr)}$ are comparable and in some cases more stringent than the indirect. Having analyzed the helicity fraction constrains in detail we can deduce that the single top production contributions to the direct constraints presented in Fig.~\ref{fig:2d1} are significant, improving the direct bounds considerably. 

\subsection{Summary}
We have analyzed the decay of an unpolarized top quark to a bottom quark and a polarized $W$ boson as mediated by the most general effective $tWb$ vertex at ${\cal O}(\alpha_s)$. We have shown, that within this approach the helicity fraction ${\cal F}_+$  can reach maximum values of the order of $2$ per-mille in the presence of a non-SM $b_{LR}$ contributions. Leading QCD effects increase the contributions $b_{LR}$ substantially owing to the helicity suppression of the LO result.

Indirect constraints coming from $B$ physics already severely restrict the contributions of anomalous $tWb$ couplings. In particular, considering only real contributions of a single anomalous coupling at a time, all considered anomalous couplings except $b_{LR}$ are constrained to yield $\mathcal F_+$ within $2\%$ of the SM prediction.  Even in the presence of the much less constrained $b_{LR}$ contributions, a potential determination of ${\cal F}_+ $ significantly deviating from the SM prediction, at the projected sensitivity of the LHC experiments~\cite{AguilarSaavedra:2007rs}, could not be explained within such framework. Based on the existing SM calculations of higher order QCD and electroweak corrections \cite{Czarnecki:2010gb, Do:2002ky}, we do not expect such corrections to significantly affect our conclusions.

Finally, with increased precision of the ${\cal F}_L $ and the single top quark production cross-section measurements at the Tevatron and the LHC the direct bounds on $b_{LR}$ and $\delta a_L$ are expected to regnant over the indirect.

\chapter{Concluding remarks}
\vspace{-0.2cm}
As we are approaching the end of Tevatron era and have well entered into the exciting era of the LHC, the hunt for BSM physics is in full effect. Searching for new particles is by no means the only way that LHC can produce new answers and questions. One of the area that we hope it will shed some light on, is the flavor problem of the SM. There is no question that this is where top quark, with its high mass, plays an outstanding role. Since LHC can be considered a true top quark factory, top quark physics is for the first time being probed at high precision. Precise determination of top quarks parameters and interactions could serve as a window to observations of physics beyond SM. In this work we have considered different aspects of top quark decays and how NP, which we have parametrized using effective theory approach, could affect them.

On one hand we have investigated the effects of perturbative NLO QCD corrections on different decay rates of the top quark, something to be considered when dealing with quarks and being confronted with measurements of ever increasing precision. In particular, we have investigated the branching ratios of $t\to q \gamma,Z$ decays and different kinematical asymmetries in subsequent three body decays $t\to q \ell^+ \ell^-$ as well as the main decay channel of the top quark $t\to W b$, paying special attention to helicity fractions as observables sensitive to the structure of the $tWb$ vertex.

On the other hand, we tried to accent the importance of considering the effects in well measured observables of meson physics, whenever deviations from SM in top quark physics are present. The dominant role that top quark plays in rare processes of meson physics, where it appears as a virtual particle, should always be kept in mind. While the analysis of indirect constraints for operators generating FCNC top quark decays has already been performed and can be found in literature, a comprehensive analysis of indirect constraints on operators generating anomalous charged currents with top quark has not and is an essential result of our work. As we have shown, the precise measurements of different ``top quark sensitive'' observables in $|\Delta B| = 2$ and $|\Delta B|=1$ processes put constraints on NP. The significance of some indirect constraints is not expected to be met by the direct constraints from the LHC data.

Whether any deviation from SM predictions in top quark decays is to be observed or not, the future measurements are expected to play an important role in the flavor aspects of constructing and constraining BSM models.

\cleardoublepage
\phantomsection
\addcontentsline{toc}{chapter}{Acknowledgments}
\chapter*{Acknowledgments}
I would like to sincerely thank my advisors Svjetlana Fajfer and Jernej F. Kamenik for all their help and guidance throughout my graduate study years. Their receptiveness and willingness to help is something I have come to greatly appreciate. Special thanks go to Jure Zupan for his hospitality at University of Cincinnati and his valuable insights about the world of physics.
\newpage
\thispagestyle{plain}

\cleardoublepage
\phantomsection
\thispagestyle{plain}
\addcontentsline{toc}{chapter}{List of publications}
\chapter*{List of publications}
\small
  \begin{itemize}
  \item J.~Drobnak, S.~Fajfer and J.~F.~Kamenik,\newline
  ``Signatures of NP models in top FCNC decay $t\to c(u) \ell^+ \ell^-$'',\newline
  JHEP {\bf 0903} (2009) 077, 
  [arXiv:0812.0294 [hep-ph]].
  
  \item J.~Drobnak, S.~Fajfer and J.~F.~Kamenik,\newline
  ``Flavor Changing Neutral Coupling Mediated Radiative Top Quark Decays at Next-to-Leading Order in QCD'',\newline
  Phys.\ Rev.\ Lett.\  {\bf 104} (2010) 252001,
  [arXiv:1004.0620 [hep-ph]].
  
  \item J.~Drobnak, S.~Fajfer and J.~F.~Kamenik,\newline
  ``QCD Corrections to Flavor Changing Neutral Coupling Mediated Rare Top Quark Decays'',\newline
  Phys.\ Rev.\ D {\bf 82} (2010) 073016,
  [arXiv:1007.2551 [hep-ph]].
 
  \item J.~Drobnak, S.~Fajfer and J.~F.~Kamenik,\newline
  ``New physics in $t\to b W$ decay at next-to-leading order in QCD'',\newline
  Phys.\ Rev.\ D {\bf 82} (2010) 114008,
  [arXiv:1010.2402 [hep-ph]].
  
  \item J.~Drobnak, S.~Fajfer and J.~F.~Kamenik,\newline
  ``Interplay of $t\to b W$ Decay and $B_q$ Meson Mixing in Minimal Flavor Violating Models'',\newline
  Phys.\ Lett.\ B {\bf 701} (2011) 234,
  [arXiv:1102.4347 [hep-ph]].
  
  \item I.~Dorsner, J.~Drobnak, S.~Fajfer, J.~F.~Kamenik and N.~Kosnik,\newline
  ``Limits on scalar leptoquark interactions and consequences for GUTs'',\newline
  JHEP {\bf 1111} (2011) 002,
  [arXiv:1107.5393 [hep-ph]].
  
  \item J.~Drobnak, S.~Fajfer and J.~F.~Kamenik,\newline
  ``Probing anomalous $tWb$ interactions with rare $B$ decays'',\newline
  Nucl.\ Phys.\ B {\bf 855} (2012) 82,
  [arXiv:1109.2357 [hep-ph]].
  
  \item N.~Kosnik, I.~Dorsner, J.~Drobnak, S.~Fajfer and J.~F.~Kamenik,\newline
  ``Scalar diquark in $t \bar{t}$ production and constraints on Yukawa sector of grand unified theories'',\newline
  PoS EPS {\bf -HEP2011} (2011) 380,
  [arXiv:1111.0477 [hep-ph]].
  
  \item J.~Drobnak, J.~F.~Kamenik and J.~Zupan,\newline
  ``Flipping $t \bar{t}$ asymmetries at the Tevatron and the LHC'',\newline
  Accepted for publication in Phys.\ Rev.\ D,
  arXiv:1205.4721 [hep-ph].
\end{itemize}
\normalsize
\newpage
\thispagestyle{plain}

\bibliographystyle{h-physrev}
\cleardoublepage
\phantomsection
\thispagestyle{plain}
\addcontentsline{toc}{chapter}{Bibliography}
\bibliography{reference}

\begin{appendices}

\chapter{Feynman rules}\label{app:feyn_rules}
In this Appendix we present the Feynman rules involving different types of anomalous couplings generated by effective Lagrangians defined various sections of chapters~\ref{chap:neutral_currents} and~\ref{chap:CC}.
\section{Neutral currents}\label{app:feyn_neutral}
In Tab.~\ref{tab:feyns_fcnc} we present Feynman rules for FCNC top quark transitions governed by the Lagrangian given in Eq.~(\ref{eq:Lagr}).
\begin{table}[h!]
\begin{tabular}{m{1.6cm}l}
\includegraphics[scale= 0.5]{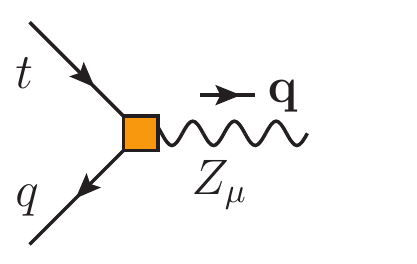}
 &$\ii g_Z \frac{v^2}{\Lambda^2}\Big[\gamma^{\mu}a^{Z}_{R,L} -\frac{2\ii \sigma^{\mu\nu}q_{\nu}}{v}b^Z_{LR,RL}\Big]P_{R,L}$ \\
 \includegraphics[scale= 0.5]{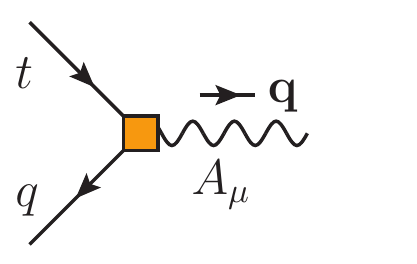}
 &$-\ii e \frac{v^2}{\Lambda^2}\frac{2\ii \sigma^{\mu\nu}q_{\nu}}{v}b^\gamma_{LR,RL}\,P_{R,L}$ \\
\includegraphics[scale= 0.5]{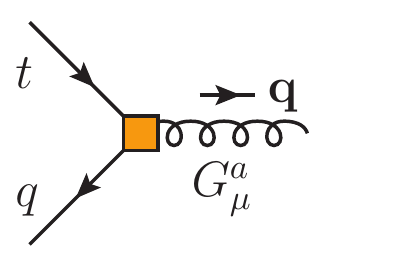}
 &$-\ii g_s \frac{v^2}{\Lambda^2}\frac{2\ii \sigma^{\mu\nu}q_{\nu}}{v}b^g_{LR,RL}T^a\,P_{R,L}$ \\\
\end{tabular}
\caption{Feynman rules for $tVq$ FCNC vertices, $q_{\mu}$ is the momentum of the outgoing gauge boson and $P_{R,L}=(1\pm \gamma^5)/2$ are the chirality projectors.}
\label{tab:feyns_fcnc}
\end{table}
\vfill

\section{Charged currents}\label{app:feyn_charged}
Feynman rules for vertices generated by operators given in Eq.~(\ref{eq:ops1}) that turn out to be relevant for our analysis are shown in Tab.~\ref{tab:feyns_cc}. We use the following abbreviations, labeling flavor with $i,j$ 
\small
\begin{eqnarray}
v_R &=& \kappa_{RR} \delta_{3i}\delta_{3j}\,,\hspace{0.5cm} \tilde{v}_R=\frac{c_W}{s_W}v_R\,,\\ 
\nn v_L &=& \kappa_{LL}\delta_{3i}+\kappa_{LL}^{\pr}\delta_{3j}+\kappa_{LL}^{\pr\pr}\delta_{3i}\delta_{3j}\,,\\
\nn \tilde{v}_L&=&\frac{c_W^2-s_W^2}{2 c_W s_W} v_L-\frac{1}{2c_Ws_W}\Big(\kappa_{LL}^*\delta_{3i}+\kappa_{LL}^{\pr*}\delta_{3j}+\kappa_{LL}^{\pr\pr*}\frac{V_{ib}V_{tj}}{V_{ij}}\Big)\,,\\
\nn g_R &=& -\kappa_{LRb}\,,\\
\nn g_L &=& -\kappa_{LRt}^* \delta_{3i}-\kappa_{LRt}^{\pr*}\delta_{3i}\delta_{3j}\,,
\end{eqnarray}
\normalsize
\vspace{-0.4cm}
\begin{table}[h!]
\begin{tabular}{m{1.6cm}l|m{1.6cm}l}
\includegraphics[scale= 0.5]{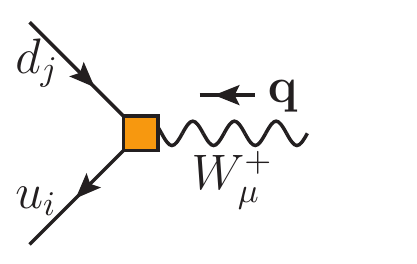}
 &$-\frac{\ii g}{\sqrt{2}}V_{ij}\Big[\gamma^{\mu}v_{R,L} +\frac{\ii \sigma^{\mu\nu}q_{\nu}}{m_W}g_{R,L}\Big]P_{R,L}$ &
\includegraphics[scale= 0.5]{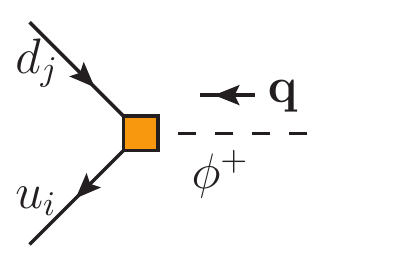}
 &$-\frac{\ii g}{\sqrt{2}}V_{ij}\frac{\gs{q}}{m_W}(-v_{R,L}) P_{R,L} $\\
 \includegraphics[scale= 0.5]{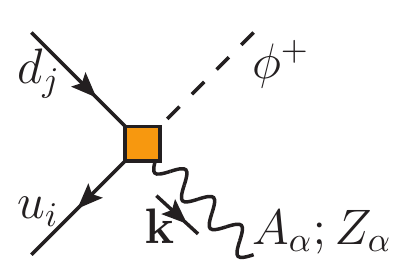}
 &$\begin{array}{l}-\frac{\ii g}{\sqrt{2}} V_{ij}\frac{e}{m_W}\Big[\{v_{R,L};\tilde{v}_{R,L} \}\gamma^{\mu}\\
 +\{1;\frac{c_W}{s_W}\}(-g_{R,L})\frac{\ii \sigma^{\alpha\mu}k_{\mu}}{2 m_W}\Big]P_{R,L}\end{array}$&
\includegraphics[scale= 0.5]{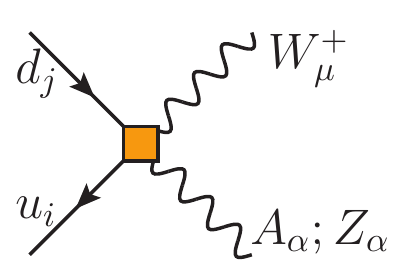}
 &$-\frac{\ii g}{\sqrt{2}}V_{ij} e\Big[\{1;\frac{c_W}{s_W}\}(-g_{R,L})\frac{\ii \sigma^{\mu\alpha}}{m_W}\Big]P_{R,L}$\\\hline
 \includegraphics[scale= 0.5]{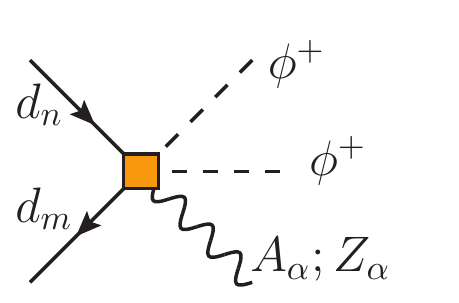}
 &$\begin{array}{l}\ii\big(\frac{g}{\sqrt{2}}\big)^2 \frac{e}{m_W^2}V_{tm}^*V_{tn}\{1;\frac{c_W^2-s_W^2}{2c_W s_W}\}\gamma^{\alpha}P_L\\
 \times(\kappa_{LL}+\delta_{3n}\kappa_{LL}^{\pr\pr})\end{array}$&
\includegraphics[scale= 0.5]{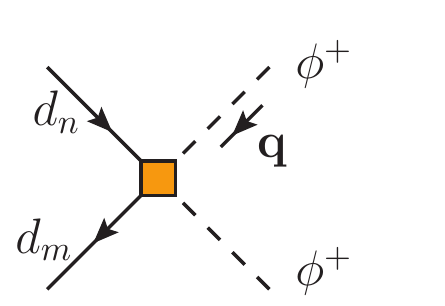}
 &$\begin{array}{l}-\ii\big(\frac{g}{\sqrt{2}}\big)^2V_{tm}^* V_{tn}\frac{ \gs{q}}{m_W^2}{P}_L\\\times(\kappa_{LL}+\delta_{3n}\kappa_{LL}^{\prime\prime})\end{array}$\\
\includegraphics[scale= 0.5]{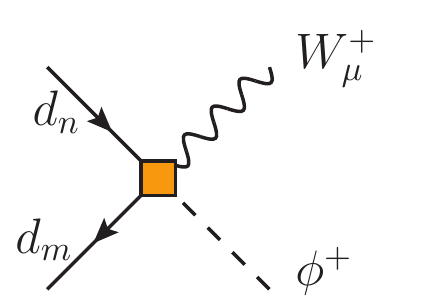}
 &$\begin{array}{l}\ii \big(\frac{g}{\sqrt{2}}\big)^2V_{tm}^* V_{tn}\frac{1}{m_W}\gamma^{\mu}{P}_L\\ \times(\kappa_{LL}+\delta_{3n}\kappa_{LL}^{\prime\prime})\end{array}$&
 \includegraphics[scale= 0.5]{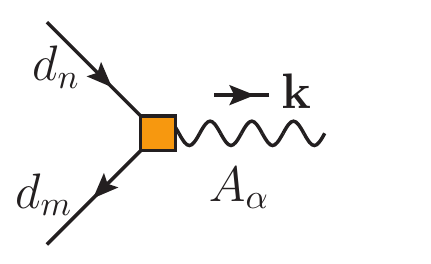}
 &$\ii e\,\kappa_{LRb}\delta_{3m}\delta_{3n}\frac{\ii \sigma^{\alpha\mu}k_{\mu}}{2m_W}P_R$\\\hline
\includegraphics[scale= 0.5]{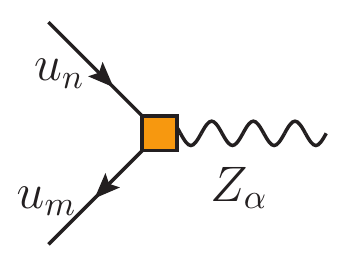}
 &$\begin{array}{l}-\ii e \frac{1}{2s_Wc_W}\gamma^{\alpha}P_L\\
 \times\big(\delta_{3m}\delta_{3n}\kappa_{LL}+V_{mb}V_{nb}^*(\kappa_{LL}^{\pr}+\delta_{3m}\kappa_{LL}^{\pr\pr})\big)\end{array}$&
\includegraphics[scale= 0.5]{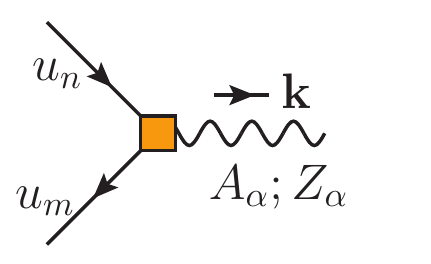}
 &$\begin{array}{l}-\ii e\{1;\frac{c_W}{s_W}\} \frac{\ii \sigma^{\alpha\nu}k_{\nu}}{2 m_W}P_R \\ 
 \times\big(\delta_{3m}\delta_{3n} \kappa_{LRt}+V_{mb}\delta_{3n}\kappa_{LRt}^{\pr}\big)\end{array}$ \\
\end{tabular}
\caption{ Feynman rules for the anomalous vertices. Indicies $i,j$ and $m,n$ label quark flavor.}
\label{tab:feyns_cc}
\end{table}

Feynman rule for $tWb$ vertex obtained from Eq.~(\ref{eq:effsimple}) is given in Tab.~\ref{tab:feyn_main1}.
\begin{table}[h!]
\begin{tabular}{m{1.6cm}l}
\includegraphics[scale= 0.5]{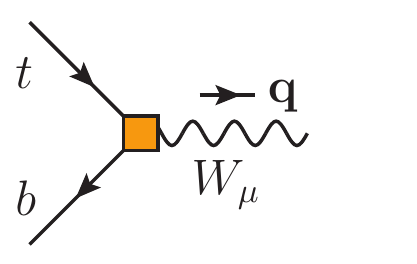}
 &$-\ii \frac{g}{\sqrt{2}}\Big[\gamma^{\mu}a_{R,L} -\frac{2\ii \sigma^{\mu\nu}q_{\nu}}{m_t}b_{LR,RL}\Big]P_{R,L}$ \\
\end{tabular}
\caption{Feynman rule for the general parametrization of $tWb$ vertex, $q_\mu$ is the momentum of the outgoing gauge boson.}
\label{tab:feyn_main1}
\end{table}

\chapter{Loop functions}
In this Appendix we present analytic expressions for various loop functions obtained in the calculation of one-loop amplitudes in FCNC processes.
\section{FCNC top decay form factors}\label{app:form_factors_qcd}
Here we present the form factors defined in Eqs.~(\ref{eq:fcnc_amp1}, \ref{eq:fcnc_amp1}). Expressions are given in $d=4+\epsilon$ dimensions regularizing both UV as well as IR divergences denoted $\epsilon_{\mathrm{UV}}$ and $\epsilon_{\mathrm{IR}}$ respectively.
Further we define 
$$C_{\epsilon}=(m_t/\mu)^{\epsilon}\Gamma(1-\epsilon/2)/(4\pi)^{\epsilon/2}\,,$$
and for the UV divergent form factors add the counter terms denoted by $\delta$.
The form factors are
\small
\be
F_{b}^{\gamma}&=&C_{\epsilon}
\Bigg[-\frac{4}{\epsIR^2}+\frac{5}{\epsIR}+\frac{2}{\epsUV} -6 \Bigg] + \delta_{b}^{\gamma}\,, \label{eq:Fb_gamma}\\
F_{bg}^{\gamma}&=&Q C_{\epsilon}
\Bigg[\frac{8}{\epsUV}-11+\frac{2}{3}\pi^2-2\pi\ii \Bigg]+ \delta_{bg}^{\gamma}\,,\label{eq:Fbg_gamma}\\
F_{a}^Z&=&C_{\epsilon}\left[
-\frac{4}{\epsIR^2}+\frac{5-4\log(1-r_Z)}{\epsIR}-2\log^2(1-r_Z)+3\log(1-r_Z)-2\mathrm{Li}_2(r_Z)-6\right]\,,\label{eq:Fa}\\
F_{b}^Z&=&C_{\epsilon}
\Big[
-\frac{4}{\epsIR^2}+\frac{5-4\log(1-r_Z)}{\epsIR}+\frac{2}{\epsUV} -2 \log^2(1-r_Z)+4\log(1-r_Z)\label{eq:Fb}\\
&-&2\mathrm{Li}_2(r_Z)-6 \Big]+\delta_{b}^Z\,,\nn\\
F_{ab}^Z&=&-4 m_t \log(1-r_Z)\,,\label{eq:Fab}\\
F_{ba}^Z&=&-\frac{1}{m_t}\frac{1}{2r_Z}\log(1-r_Z)\,,\label{eq:Fba}
\ee
\be
F_{ag}^Z&=&m_t\bigg[\hat v+\hat a \label{eq:Fag}\\
\nonumber &+& (\hat v-\hat a)\Big\llbracket\frac{r_Z(4-r_Z)(1+r_Z)}{(1-r_Z)^3}f_1 -\frac{2r_Z(4-r_Z)}{(1-r_Z)^4}f_2-\frac{1-7r_Z+3r_Z^2}{(1-r_Z)^2}+\frac{2 r_Z}{1-r_Z}\log r_Z\Big\rrbracket\bigg]\,,\\
F_{bg}^Z&=&C_{\epsilon}\bigg[
2 \hat v \frac{2}{\epsUV}+(\hat v+\hat a)(f_1-2)\label{eq:Fbg}\\
&+&(\hat v-\hat a)\Big\llbracket-\frac{r_Z}{1-r_Z}\log(r_Z) -\ii  \pi- \frac{7/2-4r_Z+2r_Z^2}{(1-r_Z)^2} -\frac{(1+r_Z)(2+r_Z)}{2(1-r_Z)^3}f_1+\frac{2+r_Z}{(1-r_Z)^4}f_2\Big\rrbracket\bigg]+ \delta_{bg}^Z \,.\nonumber
\ee
\normalsize
where we have defined
\begin{eqnarray}\label{eq:def_fcnc}
r_Z = m_Z^2/m_t^2\,,\hspace{0.5cm} \hat{v} = T_3 - 2 \sin \theta_W Q\,,\hspace{0.5cm} \hat{a} = T_3\,. \label{eq:some_def}
\end{eqnarray}
For the up-type quarks $Q = 2/3$ and $T_3 = 1/2$. Further we have introduced auxiliary functions $f_1$ and $f_2$ for shorter notation
\small
\be
f_1&=&2\sqrt{\frac{4-r_Z}{r_Z}}\arctan\Big(\sqrt{\frac{r_Z}{4-r_Z}}\Big)\,,\\
f_2&=&-2\mathrm{Li}_2(r_Z-1)+2\arctan\Big(\frac{1-r_Z}{3-r_Z}\sqrt{\frac{4-r_Z}{r_Z}}\Big)
\arctan\Big(\frac{r_Z}{2-r_Z}\sqrt{\frac{4-r_Z}{r_Z}}\Big) \nonumber\\
 &+&2\mathrm{Re}\Big\{\mathrm{Li}_2\Big((1-r_Z)^2
\big(1-\frac{r_Z}{2}\frac{2-r_Z}{1-r_Z}(1+\ii\sqrt{\frac{4-r_Z}{r_Z}})\big)\Big)
-\mathrm{Li}_2\Big(\frac{1-r_Z}{2}\big(2-r_Z-\ii \sqrt{(4-r_Z)r_Z}\big)\Big)\Big\}\,.\nonumber
\ee
\normalsize
\section{$|\Delta B|=2$ loop functions}\label{app:NP_D_B_2}
Here we present SM as well as NP loop functions for $|\Delta B|=2$ processes defined in Eqs.~(\ref{LOWils}). Functions $S_{0}^{LL (\pr)}$ are UV divergent and the forms presented here are $\overline{\mathrm{MS}}$ renormalized.
\begin{subequations}\label{eq:S0s}
\begin{eqnarray}
S_0^{\mathrm{SM}}(x_t)\hspace{-0.2cm}&=&\hspace{-0.2cm}\frac{1}{2}S_0^{LL\pr}(x_t)=\frac{x_t(x_t^2-11 x_t+4)}{4(x_t-1)^2}+\frac{3x_t^3\log x_t}{2(x_t-1)^3}\,,\\
S_{0\overline{\mathrm{MS}}}^{LL}(x_t)\hspace{-0.2cm}&=&\hspace{-0.2cm}2S_{0\overline{\mathrm{MS}}}^{LL\pr\pr}(x_t)=-\frac{x_t \left(x_t^2+10 x_t+1\right)}{2 \left(x_t-1\right)^2} + x_t \log \frac{m_W^2}{\mu^2}\\
&&\hspace{1.7cm}+\frac{x_t \left(x_t^3-3 x_t^2+12 x_t-4\right) \log x_t}{\left(x_t-1\right)^3}\,,\nn\\
S_0^{LRt}(x_t)\hspace{-0.2cm}&=&\hspace{-0.2cm}2S_0^{LRt\pr}(x_t)=3\sqrt{x_t}\bigg[-\frac{x_t(x_t+1)}{(x_t-1)^2}+\frac{2x_t^2\log x_t}{(x_t-1)^3}\bigg]\,.
\end{eqnarray}
\end{subequations}
\section{$|\Delta B|=1$ loop functions}\label{app:SM_D_B_1}
Below we present loop functions obtained in the calculation of $|\Delta B|=1$ processes within the SM in the general $R_{\xi}$ gauge defined in Eqs.~(\ref{eq:delta_B1_SM}).
\begin{eqnarray}
B_0(x)&=& 
\frac{x}{2(x-1)}-\frac{x \log x}{2(x-1)^2} - \frac{1}{2}\,\phi(x,\xi)\,,\\
\tilde{B}_0(x)&=&
-\frac{2 x}{x-1}+\frac{2 x \log x}{(x-1)^2}+ \frac{1}{2}\, \phi(x,\xi)\,,\\
C_0(x)&=&-\frac{x(x^2-7x+6)}{2(x-1)^2}-\frac{x(3x+2)}{2(x-1)^2}\log x -\phi(x,\xi)\,,\\
D_0(x)&=&\frac{4}{9} \log x + \frac{x^2(19x-25)}{36(x-1)^3} +\frac{x^2(-5x^2+2x+6)}{18(x-1)^4}\log x +  \phi(x,\xi)\,,\\
D_0^\prime(x)&=& \frac{8x^3+5x^2-7x}{12(x-1)^3}-\frac{x^2(3x-2)}{2(x-1)^4}\log x\,,\\ 
E_0(x)&=&\frac{2}{3}\log x -\frac{x(x^2+11x-18)}{12(x-1)^3}-\frac{x^2(4x^2-16x+15)\log x}{6(x-1)^4}\,,\\
E_0^\prime(x)&=&\frac{x(x^2-5x-2)}{4(x-1)^3} + \frac{3x^2\log x}{2(x-1)^4}\,.
\end{eqnarray}
Function
\begin{eqnarray}
\phi(x,\xi)=
\frac{x^2(\xi-1)(x\xi+7x-8\xi)\log x}{4(x-1)^2(x-\xi)^2}
-\frac{x(x\xi-7x+6\xi)}{4(x-1)(x-\xi)}
-\frac{x\xi(6x+\xi^2-7\xi)\log\xi}{4(\xi-1)(x-\xi)^2}\,,
\end{eqnarray}
captures all the $\xi$ dependance and has the property $\lim_{x\to0} \phi(x,\xi) = \lim_{\xi \to 1} \phi(x,\xi) = 0$. It is obvious that the following linear combinations are gauge independent
\begin{eqnarray}
2B_0(x)-C_0 (x)\,,\hspace{0.5cm}2\tilde{B}_0(x)+C_0 (x)\,,\hspace{0.5cm}  C_0(x) + D_0(x) \,.
\end{eqnarray}

Next we present analytical expressions for functions $f_i^{(j)}$ and $\tilde{f}_i^{(j)}$ defined in Eq.~(\ref{eq:fs}). For shorter notation we further decompose
\begin{eqnarray*}
f_{9}^{(j)}=g^{(j)}-\frac{1}{4s_W^2} h^{(j)}\,,\hspace{0.5cm}f_{10}^{(j)}=\frac{1}{4 s_W^2}h^{(j)}\,,\hspace{0.5cm} f^{(j)}_{\nu\bar{\nu}}=\frac{1}{4s_W^2}k^{(j)}\,.
\end{eqnarray*}
Functions containing explicit $\mu$ dependance have been renormalized using $\overline{\mathrm{MS}}$ scheme. Below we give all nonzero contributions
{\allowdisplaybreaks
\small
\begin{eqnarray}
f_7^{(LL)} &=&\tilde{f}_7^{(LL)}=f_7^{(LL\pr\pr)}= \frac{22 x^3-153x^2+159x-46}{72 (x-1)^3}+\frac{3x^3-2x^2}{4(x-1)^4}\log x\,,\\
f_7^{(LL\pr)}&=&-\frac{8x^3+5x^2-7x}{24(x-1)^3}+\frac{3x^3-2x^2}{4(x-1)^4}\log x \,,\\
f_7^{(RR)}&=&\frac{m_t}{m_b}\Big[\frac{-5x^2+31x-20}{12 (x-1)^2}+\frac{2x-3x^2}{2(x-1)^3}\log x\Big]\label{eq:rr}\,,\\
f_7^{(LRb)}&=&\frac{m_W}{m_b}\Big[-\frac{x}{2}\log \frac{m_W^2}{\mu^2}+\frac{6x^3-31x^2+19x}{12(x-1)^2}+\frac{-3x^4+16x^3-12x^2+2x}{6(x-1)^3}\log x\Big] \label{eq:lrb}\,,\\
f_7^{(LRt)}&=&\frac{m_t}{m_W}\Big[\frac{1}{8}\log\frac{m_W^2}{\mu^2}+\frac{-9x^3+63x^2-61x+19}{48(x-1)^3}+\frac{3x^4-12x^3-9x^2+20x-8}{24(x-1)^4}\log x\Big]\,,\\
\tilde{f}_7^{(LRt)}&=&\tilde{f}_7^{(LRt\prime)}=\frac{m_t}{m_W}\Big[\frac{-3x^3+17x^2-4x-4}{24(x-1)^3}+\frac{2x-3x^2}{4(x-1)^4}\log x\Big]\,,\\
f_7^{(LRt\prime)}&=&\frac{m_t}{m_W}|V_{tb}|^2\Big[\frac{-x^2-x}{8(x-1)^2}+\frac{x^2\log x}{4(x-1)^3}\Big]\,,\\
f_8^{(LL)}&=&\tilde{f}_8^{(LL)}=f_8^{(LL\pr\pr)}=\frac{5 x^3-9 x^2+30 x-8}{24 (x-1)^3}-\frac{3 x^2 \log x}{4 (x-1)^4}\,,\\
f_8^{(LL\pr)}&=&\frac{-x^3+5x^2+2x}{8 (x-1)^3}-\frac{3 x^2 \log x}{4 (x-1)^4}\,,\\
f_8^{(RR)}&=&\frac{m_t}{m_b}\Big[\frac{-x^2-x-4}{4 (x-1)^2}+\frac{3 x \log x}{2 (x-1)^3}\Big]\,,\\
f_8^{(LRb)}&=&\frac{m_W}{m_b}\Big[\frac{x^2+5 x}{4 (x-1)^2}+\frac{2 x^3-6 x^2+x}{2 (x-1)^3}\log x\Big]\,,\\
f_8^{(LRt)}&=&\frac{m_t}{m_W}\Big[\frac{3 x^2-13 x+4}{8 (x-1)^3}+\frac{5 x-2 }{4 (x-1)^4}\log x\Big]\,,\\
\tilde{f}_8^{(LRt)}&=&\tilde{f}_8^{(LRt\pr)}=\frac{m_t}{m_W}\Big[\frac{x^2-5 x-2}{8 (x-1)^3}+\frac{3 x \log (x)}{4 (x-1)^4}\Big]\,,\\
g^{(LL)}&=&\tilde{g}^{(LL)}=(-x-\frac{4}{3})\log\frac{m_W^2}{\mu^2}+\frac{250x^3-384x^2+39x+77}{108 (x-1)^3}\\
&+&\frac{-18x^5+48x^4-102x^3+135x^2-68x+8}{18(x-1)^4}\log x \,,\\
g^{(LL\pr)}&=&(\frac{4}{9}-\frac{x}{2})\log\frac{m_W^2}{\mu^2}+\frac{125x^3-253x^2+138x -16}{36(x-1)^3}\\
&+&\frac{-9x^5+12x^4-48x^3+99x^2-59x+8}{18(x-1)^4}\log x - |V_{tb}|^2\frac{x}{2}\,,\nonumber\\
\tilde{g}^{(LL\pr)}&=&\tilde{h}^{(LL\pr)}=\tilde{f}^{(LL\pr)}_{\nu\bar{\nu}}=-\frac{x}{2}\log\frac{m_W^2}{\mu^2}+\frac{x}{2}(1-\log x -|V_{tb}|^2)\,,\\
g^{(LL\pr\pr)}&=&-\big(\frac{4}{3}+\frac{x}{2}+|V_{tb}|^2\frac{x}{2} \big)\log\frac{m_W^2}{\mu^2}
+ \frac{250x^3-384x^2+39x + 77}{108(x-1)^3}\\
&+&\frac{-9x^5+12x^4-48x^3+99x^2-59x+8}{18(x-1)^4}\log x - |V_{tb}|^2\frac{x}{2}\log x \,,\nonumber \\
\tilde{g}^{(LL\pr\pr)}&=&\tilde{h}^{(LL\pr\pr)}=\tilde{f}^{(LL\pr\pr)}_{\nu\bar{\nu}}=|V_{tb}|^2\Big(-\frac{x}{2}\log\frac{m_W^2}{\mu^2}-\frac{x}{2}\log x \Big)\,,\\
g^{(LRt)}&=&\tilde{g}^{(LRt)}=\frac{m_t}{m_W}\Big[\frac{-99x^3+136x^2+25x-50}{72(x-1)^3}+\frac{24x^3-45x^2+17x+2}{12(x-1)^4}\log x\Big]\,,\\
g^{(LRt\pr)}&=&\frac{m_t}{m_W}|V_{tb}|^2\Big[\frac{x^2+3x-2}{8(x-1)^2}+\frac{x-2x^2}{4(x-1)^3}\log x\Big]\,,\\
\tilde{g}^{(LRt\pr)}&=&\frac{m_t}{m_W}\bigg[\frac{-54x^3+59x^2+35x-34}{36(x-1)^3}+\frac{15x^3-27x^2+10x+1}{6(x-1)^4}\log x\\ 
&+& |V_{tb}|^2\Big[\frac{x^2+3x-2}{8(x-1)^2}+\frac{x-2x^2}{4(x-1)^3}\log x\Big]\bigg]\,,\\
h^{(LL)}&=&\tilde{h}^{(LL)}=-(x+\frac{3}{2})\log\frac{m_W^2}{\mu^2}+\frac{11x-5}{4(x-1)}+\frac{-2x^3+x^2-2x}{2(x-1)^2}\log x\,,\\
h^{(LL\pr)}&=&-\frac{x}{2}\log\frac{m_W^2}{\mu^2}+\frac{3x}{2(x-1)}-\frac{x^3+x^2+x}{2(x-1)^2}\log x-|V_{tb}|^2\frac{x}{2}\,, \\
h^{(LL\pr\pr)}&=&-\big(\frac{3}{2}+\frac{x}{2}+|V_{tb}|^2\frac{x}{2}\big)\log\frac{m_W^2}{\mu^2}+\frac{11x-5}{4(x-1)}-\frac{x^3+x^2+x}{2(x-1)^2}\log x-|V_{tb}|^2 \frac{x}{2}\log x\,,\\
h^{(LRt)}&=&\tilde{h}^{(LRt)}=\tilde{h}^{(LRt\pr)}=\frac{m_t}{m_W}\Big[-\frac{3x}{2(x-1)}+\frac{3x\log x}{2(x-1)^2}\Big]\,,\\
k^{(LL)}&=&\tilde{k}^{(LL)}=h^{(LL)}-\frac{3}{(x-1)}+\frac{3x\log x}{(x-1)^2}\,,\\
k^{(LL\pr)}&=&h^{(LL\pr)}-\frac{3x}{(x-1)}+\frac{3x\log x}{(x-1)^2}\,,\\
k^{(LL\pr\pr)}&=&h^{(LL\pr\pr)}-\frac{3}{(x-1)}+\frac{3x\log x}{(x-1)^2}\,,\\
k^{(LRt)}&=&\tilde{k}^{(LRt)}=\tilde{k}^{(LRt\pr)}=h^{(LRt)}+\frac{3}{x-1}-\frac{3x\log x}{(x-1)^2}\,.
\end{eqnarray}
\normalsize
}
\chapter{Decay Widths}\label{app:allwidths}
In this Appendix we present various decay widths obtained in calculating rates at NLO in QCD. For the FCNC processes covered in chapter~\ref{chap:neutral_currents} we separately present the decay widths for $t\to q Z$, $t\to q \gamma$ including the virtual NLO QCD corrections and the decay widths for bremsstrahlung processes of $t\to q g Z$, $t\to q g \gamma$. For the main decay channel analysis given in chapter~\ref{chap:CC} we present only the combined virtual and bremsstrahlung decay rates.
\section{FCNC decays}
\subsection{Virtual corrections}\label{app:dw1}
Here we present the decay widths defined in Eq.~(\ref{eq:FCNC_virt}) with $C_F = 4/3$
{\allowdisplaybreaks
\small
\be
\Gamma_{b}^{\gamma,\mathrm{virt.}}&=&\Gamma_{b}^{\gamma(0)}
\bigg[1+\frac{\alpha_s}{4\pi}C_F\Big[-\frac{8}{\epsIR^2}+\frac{6}{\epsIR}-7-\frac{\pi^2}{3}+2\log\left(\frac{m_t^2}{\mu^2}\right)\Big]\bigg]\,,\label{eq:Gamma_virt_first}\\
\Gamma_{bg}^{\gamma,\mathrm{virt.}}&=&\Gamma_{b}^{\gamma(0)}
\frac{\alpha_s}{4\pi}C_F Q\Big[-11+\frac{2\pi^2}{3}+4\log\left(\frac{m_t^2}{\mu^2}\right)\Big]\,,\\
\tilde{\Gamma}_{bg}^{\gamma,\mathrm{virt.}}&=&\Gamma_{b}^{\gamma(0)}
\frac{\alpha_s}{4\pi}C_F Q\Big[-2\pi\Big]\,,\\
\Gamma_{a}^{Z,\mathrm{virt.}}&=&\Gamma_{a}^{Z(0)}\bigg[1+\frac{\alpha_s}{4\pi} C_F\Big\llbracket-\frac{8}{\epsIR^2}+\frac{-16 \log(1-r_Z)+\frac{4}{1+2 r_Z}+6}{\epsIR}\label{virtIR1} \\*
&-& 16 \log^2(1-r_Z)+\frac{2(5+8r_Z)}{1+2 r_Z}\log(1-r_Z)-\frac{\pi^2}{3}-\frac{2(6+7r_Z)}{1+2r_Z} - 4\mathrm{Li}_2(r_Z)\Big\rrbracket\bigg]\,,\nonumber\\
\Gamma_{b}^{Z,\mathrm{virt.}}&=&\Gamma_{b}^{Z(0)}\bigg[1+\frac{\alpha_s}{4\pi}C_F\Big\llbracket-\frac{8}{\epsIR^2}+\frac{-16 \log(1-r_Z)-\frac{8}{2+ r_Z}+10}{\epsIR} \label{virtIR2}\\*
&-& 16 \log^2(1-r_Z)
+\frac{2(4+9r_Z)}{2+r_Z}\log(1-r_Z)-\frac{\pi^2}{3}+2\log\left(\frac{m_t^2}{\mu^2}\right)	 - \frac{2(7+6r_Z)}{2+r_Z}-4\mathrm{Li}_2(r_Z)\Big\rrbracket\bigg]\,,\nonumber \\
\Gamma_{ab}^{Z,\mathrm{virt.}}&=&\Gamma_{ab}^{Z(0)}\bigg[1+\frac{\alpha_s}{4\pi}C_F\Big\llbracket-\frac{8}{\epsIR^2}+\frac{-16 \log(1-r_Z)+\frac{22}{3}}{\epsIR} \label{virtIR3}\\*
&-&16 \log^2(1-r_Z)-\frac{2(2-15 r_Z)}{3 r_Z}\log(1-r_Z)-\frac{\pi^2}{3}-\frac{26}{3}-4\mathrm{Li}_2(r_Z)\Big\rrbracket\bigg]\,,\nonumber \\
\Gamma_{ag}^{Z,\mathrm{virt.}}&=&\Gamma_{ab}^{Z(0)}\frac{\alpha_s}{4\pi}C_F \bigg[
2 \hat v \log\left(\frac{m_t^2}{\mu^2}\right)+(\hat v-\hat a)\Big[\frac{1}{3}\log(r_Z)+\frac{2f_2}{3(1-r_Z)^2}\Big]\\*
&+&\frac{2}{3}f_1\frac{\hat a(2-r_Z)+\hat v(1-2 r_Z)}{1-r_Z}+\frac{\hat a}{3} (4+\frac{1}{r_Z})-\frac{14\hat v}{3}\bigg]\,,\nonumber\\
\Gamma_{bg}^{Z,\mathrm{virt.}}&=&\Gamma_{b}^{Z(0)}\frac{\alpha_s}{4\pi}C_F\bigg[
2 \hat v \log\left(\frac{m_t^2}{\mu^2}\right) + (\hat v-\hat a)\Big[\frac{r_Z}{2+r_Z}\log(r_Z)+ \frac{4 f_2}{(1-r_Z)^2(2+r_Z)}\Big]\\*
&+& f_1\frac{\hat a (4+r_Z-r_Z^2) - \hat v(3+r_Z)r_Z}{(1-r_Z)(2+r_Z)}
- \hat v\frac{11+4r_Z}{2+r_Z}+ \hat a \frac{6}{2+r_Z}\bigg]\,,\nonumber\\
\tilde{\Gamma}_{ag}^{Z,\mathrm{virt.}}&=&\Gamma_{ab}^{(0)} \frac{\alpha_s}{4\pi}C_F (\hat v-\hat a)(-\pi)\,,\\
\tilde{\Gamma}_{bg}^{Z,\mathrm{virt.}}&=&\Gamma_{b}^{(0)}\frac{\alpha_s}{4\pi}C_F (\hat v-\hat a)(-\pi)\,.\label{eq:gamma_virt_last}
\ee
\normalsize
}

\subsection{Bremsstrahlung}\label{app:dw2}
Below we give the $t\to qgZ,\gamma$ bremsstrahlung decay rates, where for the photon channel Eqs.~(\ref{eq:GbbF1}, \ref{eq:GbbF2}, \ref{eq:GbbF3}) we include the kinematical cuts. For the purpose of shorter notation we define $\hat{x} \equiv \delta r_c$ and $\hat{y} \equiv 2 E_{\gamma}^{\mathrm{cut}}/m_t$.
{\allowdisplaybreaks
\small
\be
\Gamma_b^{\gamma,\mathrm{brems.}}&=&\Gamma_b^{\gamma(0)}\frac{\alpha_s}{4\pi}C_F\Big[
\frac{8}{\epsIR^2}-\frac{6}{\epsIR} + \frac{37}{3} - \pi^2\Big]\,, \label{eq:bremsIR0}\\
\Gamma_{bg}^{\gamma,\mathrm{brems.}}&=&\Gamma_b^{\gamma(0)}\frac{\alpha_s}{4\pi}C_F\Big[
-\frac{25}{3} + \frac{2}{3}\pi^2\Big]\,,\\
\Gamma_{a}^{Z,\mathrm{brems.}}&=& \Gamma_a^{Z(0)}\frac{\alpha_s}{4\pi}C_F\bigg[
\frac{8}{\epsIR^2}+\frac{16\log(1-r_Z)-\frac{4}{1+2r_Z}-6}{\epsIR} + 16 \log^2(1-r_Z)- 4\log (r_Z)\log(1-r_Z)\label{eq:bremsIR1}\\*
&-& 4\frac{5 + 6 r_Z}{1+2r_Z}\log(1-r_Z)-\frac{4(1-r_Z-2r_Z^2)r_Z}{(1-r_Z)^2(1+2r_Z)} \log(r_Z)- \pi^2
-4\mathrm{Li}_2(r_Z) +\frac{7+r_Z}{(1-r_Z)(1+2r_Z)}+10\bigg]\,,\nonumber\\
\Gamma_{b}^{Z,\mathrm{brems.}}&=& \Gamma_b^{Z(0)}\frac{\alpha_s}{4\pi}C_F\bigg[
\frac{8}{\epsIR^2}+\frac{16\log(1-r_Z)+\frac{8}{2+r_Z}-10}{\epsIR} + 16\log^2(1-r_Z)-4\log(r_Z)\log(1-r_Z)\label{eq:bremsIR2}\\*
&-&4\frac{6+5r_Z}{2+r_Z}\log(1-r_Z)-\frac{4(2-2r_Z-r_Z^2)r_Z}{(1-r_Z)^2(2+r_Z)}\log(r_Z)
-\pi^2-4\mathrm{Li}_2(r_Z)-\frac{4-8r_Z}{(1-r_Z)(2+r_Z)}+\frac{43}{3}\bigg]\,,\nonumber\\
\Gamma_{ab}^{Z,\mathrm{brems.}}&=& \Gamma_{ab}^{Z(0)}\frac{\alpha_s}{4\pi}C_F\bigg[
\frac{8}{\epsIR^2}+\frac{16\log(1-r_Z)-\frac{22}{3}}{\epsIR} + 16\log^2(1-r_Z)-4\log(r_Z)\log(1-r_Z)\label{eq:bremsIR3}\\*
&-&\frac{44}{3}\log(1-r_Z)-\frac{4(3-2r_Z)r_Z}{3(1-r_Z)^2}\log(r_Z)
-\pi^2-4\mathrm{Li}_2(r_Z)-\frac{4}{3(1-r_Z)}+\frac{47}{3}\bigg]\,,\nonumber\\
\Gamma_{ag}^{Z,\mathrm{brems.}}&=&\frac{\Gamma^{Z(0)}_{ab}}{3(1-r_Z)^2} \frac{\alpha_s}{4\pi}C_F\Bigg[
2\hat v\bigg\llbracket \frac{1}{4}(3 -4 r_Z+r_Z^2)+\log(r_Z)(1-r_Z-r_Z^2)-\mathrm{Li}_2(1-r_Z)\label{Bremsag}\\*
&+&r_Z\sqrt{(4-r_Z)r_Z}\Big(\arctan\Big(\sqrt{\frac{r_Z}{4-r_Z}}\Big)+\arctan\Big(\frac{r_Z-2}{\sqrt{(4-r_Z) r_Z}}\Big)\Big)\nonumber \\*
&+&2\mathrm{Re}\Big\{\mathrm{Li}_2\Big(\frac{1}{2}(1-r_Z)(2-r_Z-\ii\sqrt{(4-r_Z)r_Z}\Big)\Big\}
\bigg\rrbracket \nonumber\\*
&+&2\hat a\bigg\llbracket \frac{1}{4}(3 -8 r_Z+5r_Z^2)+\frac{1}{2}\log(r_Z)(-2-7r_Z+2r_Z^2)+\mathrm{Li}_2(1-r_Z)\nonumber\\*
&+&(3-r_Z)\sqrt{(4-r_Z)r_Z}\Big(\arctan\Big(\sqrt{\frac{r_Z}{4-r_Z}}\Big)+\arctan\Big(\frac{r_Z-2}{\sqrt{(4-r_Z) r_Z}}\Big)\Big)\nonumber \\*
&-&2\mathrm{Re}\Big\{\mathrm{Li}_2\Big(\frac{1}{2}(1-r_Z)(2-r_Z-\ii\sqrt{(4-r_Z)r_Z}\Big)\Big\}
\bigg\rrbracket\Bigg]\,,\nonumber\\
\Gamma_{bg}^{Z,\mathrm{brems.}}&=& \frac{\Gamma_b^{Z(0)}}{(1-r_Z)^2 2(2+r_Z)} \frac{\alpha_s}{4\pi}C_F\Bigg[
2 \hat v\bigg\llbracket \frac{1}{3}(1-r_Z)(-25+2r_Z-r_Z^2)-4r_Z\log(r_Z)(1+r_Z)\label{Bremsbg}\\*
&-&4(1-r_Z)\sqrt{(4-r_Z)r_Z}\Big(\arctan\Big(\sqrt{\frac{r_Z}{4-r_Z}}\Big)+\arctan\Big(\frac{r_Z-2}{\sqrt{(4-r_Z) r_Z}}\Big)\Big)-4\mathrm{Li}_2(1-r_Z)\nonumber \\*
&+&8 \mathrm{Re}\Big\{\mathrm{Li}_2\Big(\frac{1}{2}(1-r_Z)(2-r_Z-\ii\sqrt{(4-r_Z)r_Z}\Big)\Big\}
\bigg\rrbracket \nonumber\\*
&+&2\hat a\bigg\llbracket 9-r_Z(2+7r_Z)+r_Z\log(r_Z)(8+5r_Z)+4\mathrm{Li}_2(1-r_Z) \nonumber\\*
&+&2(2-r_Z)\sqrt{(4-r_Z)r_Z}\Big(\arctan\Big(\sqrt{\frac{r_Z}{4-r_Z}}\Big)+\arctan\Big(\frac{r_Z-2}{\sqrt{(4-r_Z) r_Z}}\Big)\Big) \nonumber \\*
&-&8 \mathrm{Re}\Big\{\mathrm{Li}_2\Big(\frac{1}{2}(1-r_Z)(2-r_Z-\ii\sqrt{(4-r_Z)r_Z}\Big)\Big\}
\bigg\rrbracket\Bigg]\,,\nonumber\\
\Gamma_{g}^{Z} &=& \frac{\Gamma_b^{Z(0)}}{(1-r_Z)^2 2(2+r_Z)}\frac{\alpha_s}{4\pi}C_F\Bigg[
\frac{\hat v^2}{6} \bigg\llbracket(1-r_Z)(77-r_Z-4 r_Z^2)+ 3\log(r_Z)(10-4r_Z-9r_Z^2) \label{Bremsg}\\*
&+&6 \sqrt{\frac{r_Z}{4-r_Z}}(20+10r_Z-3r_Z^2)
   \Big(\arctan\Big(\sqrt{\frac{r_Z}{4-r_Z}}\Big)+\arctan\Big(\frac{r_Z-2}{\sqrt{(4-r_Z) r_Z}}\Big)\Big)
+12\log^2(r_Z)\nonumber \\*
&+&48 \mathrm{Re}\Big\{\mathrm{Li}_2\Big(\frac{1}{2}+\frac{\ii}{2} 
   \sqrt{\frac{4-r_Z}{r_Z}}\Big)-\mathrm{Li}_2\Big(\frac{r_Z}{2}+\frac{\ii}{2} \sqrt{(4-r_Z) r_Z}\Big)\Big\}\bigg\rrbracket \nonumber\\*
&+&\frac{\hat a^2}{6}\bigg\llbracket\frac{(1-r_Z)}{r_Z}(1-70r_Z+38r_Z^2-5r_Z^3)+3\log(r_Z)(2+46r_Z-9r_Z^2+4\log(r_Z)) \nonumber\\*
&-&6(20-3r_Z)\sqrt{(4-r_Z)r_Z}\Big(\arctan\Big(\sqrt{\frac{r_Z}{4-r_Z}}\Big)+\arctan\Big(\frac{r_Z-2}{\sqrt{(4-r_Z) r_Z}}\Big)\Big)\nonumber \\*
&+&48 \mathrm{Re}\Big\{\mathrm{Li}_2\Big(\frac{1}{2}+\frac{\ii}{2} 
   \sqrt{\frac{4-r_Z}{r_Z}}\Big)-\mathrm{Li}_2\Big(\frac{r_Z}{2}+\frac{\ii}{2} \sqrt{(4-r_Z) r_Z}\Big)\Big\}\bigg\rrbracket\nonumber \\*
&+&\hat a \hat v \bigg\llbracket - 7 + 22 r_Z -15 r_Z^2 -\log(r_Z)(6-5r_Z^2+4\log(r_Z))\nonumber\\*
&+&2(2+r_Z)\sqrt{(4-r_Z)r_Z}\Big(\arctan\Big(\sqrt{\frac{r_Z}{4-r_Z}}\Big)+\arctan\Big(\frac{r_Z-2}{\sqrt{(4-r_Z) r_Z}}\Big)\Big)\nonumber \\*
&-&16 \mathrm{Re}\Big\{\mathrm{Li}_2\Big(\frac{1}{2}+\frac{\ii}{2} 
   \sqrt{\frac{4-r_Z}{r_Z}}\Big)-\mathrm{Li}_2\Big(\frac{r_Z}{2}+\frac{\ii}{2} \sqrt{(4-r_Z) r_Z}\Big)\Big\}\bigg\rrbracket \Bigg]\,,\nonumber \\ 
\Gamma_{b}^{\gamma,\mathrm{brems.}}&=&\Gamma_{b}^{\gamma(0)} \frac{\alpha_s}{4\pi}C_F\Bigg[
\frac{8}{\epsIR^2}-\frac{6}{\epsIR}+1-\pi^2-2\frac{\uc(1-\uc)(2\uc-1)}{2-\uc \xc}+\uc 
+\frac{4}{\xc}(2-\uc)(1-\uc)\label{eq:GbbF1}\\*
&-&16\frac{1-\uc}{\xc^2}
-2\log^2(1-\uc)+(\uc^2+2\uc-10)\log(1-\uc)-\frac{2\xc^2-24 \xc+32}{\xc^3}\log\Big(\frac{2-\xc}{2-\uc\xc}\Big)\nonumber\\*
&-&6\log\Big(\frac{2-\xc}{2-\xc\uc(2-\uc)}\Big)-\Big(\frac{2}{\xc}+\uc^2+2\uc\Big)
\log\Big(\frac{2-\xc\uc(2-\uc)}{2-\uc\xc}\Big)\nonumber\\*
&+&12\sqrt{2/\xc-1}\arctan\Big(\frac{1-\uc}{\sqrt{2/\xc-1}}\Big)
+4\mathrm{Li}_2\Big(\xc \frac{1-\uc}{\xc -2}\Big)-2\mathrm{Li}_2\Big(\xc\frac{(1-\uc)^2}{\xc-2}\Big)
\Bigg]\,,\nonumber\\
\Gamma_{bg}^{\gamma,\mathrm{brems.}}&=&
\Gamma_{b}^{\gamma(0)} \frac{\alpha_s}{4\pi}C_F Q\Bigg[
-\frac{(1-\uc)(2-\xc)(\uc\xc^2-2\uc\xc-2\xc+8)}{\xc^2(2-\uc\xc)}+\frac{2\pi^2}{3}\label{eq:GbbF2}\\*
&-&4(1-\uc)\log\Big(\frac{2-\xc\uc(2-\uc)}{(1-\uc)(2-\uc\xc)}\Big)
-4\log(\uc)\log\Big(\frac{2-\xc\uc(2-\uc)}{2}\Big)+2\log\Big(\frac{\xc}{2}\Big)\log\Big(\frac{2-\xc}{2-\xc\uc(2-\uc)}\Big)\nonumber \\*
&-&\frac{4}{\xc^3}(\xc^2-4\xc+4)\log\Big(\frac{2-\xc}{2-\xc\uc}\Big)
+4\Big(\mathrm{Li}_2\Big(\frac{\xc}{2}\Big)-\mathrm{Li}_2(\uc)-\mathrm{Li}_2\Big(\frac{\xc\uc}{2}\Big)\Big)\nonumber\\*
&-&8\arctan\Big(\frac{1-\uc}{\sqrt{2/\xc-1}}\Big)\Big(\sqrt{2/\xc-1}- \arctan(\sqrt{2/\xc-1}\Big) \nonumber\\*
&+&8\mathrm{Re}\Big\{\mathrm{Li}\Big(\frac{1}{2}\big(2-\xc-\ii\sqrt{(2-\xc)\xc}\big)\Big)-
\mathrm{Li}_2\Big(\frac{1}{2}\big(2-\xc\uc-\ii \uc\sqrt{(2-\xc)\xc}\big)\Big)\Big\}\Bigg] \nonumber\,,\\
\Gamma_{g}^{\gamma}&=&
\Gamma_{b}^{\gamma(0)} \frac{\alpha_s}{4\pi}C_F Q^2\Bigg[-\frac{(1-\uc)(2-\xc)(3\uc\xc^2-4\xc\uc-8\xc+16)}{\xc^2(2-\xc\uc)}+\frac{2\pi^2}{3}\label{eq:GbbF3}\\*
&+&\big(4-2\xc+4\log\Big(\frac{\xc}{2}\Big)\big)\log(\uc) 
+(3-\uc)(1-\uc)\log\Big(\xc\frac{1-\uc}{2-\xc \uc}\Big)\nonumber\\*
&+&\frac{2}{\xc^3}(2-\xc)(\xc^3-\xc^2+6\xc-8)\log\Big(\frac{2-\xc}{2-\xc\uc}\Big)
+4\Big(\mathrm{Li}_2\Big(\frac{\xc\uc}{2}\Big)-\mathrm{Li}_2\Big(\frac{\xc}{2}\Big)-\mathrm{Li}_2(\uc)\Big)
\Bigg]\,.\nonumber
\ee
\normalsize
}

\section{Main decay channel}\label{app:gamma_main}
Here we present analytical formulae for all nine $\Gamma^{L,+,-}_{a,b,ab}$ appearing in Eq.~(\ref{e3}) to ${\cal 
O}(\alpha_s)$ order and in the $m_b=0$ limit. Note that in this section $x= m_W/m_t$ 
{\allowdisplaybreaks
\small
\begin{eqnarray}
\Gamma_{a}^L &=&\frac{(1-x^2)^2}{2 x^2}+\frac{\alpha_s}{4\pi}C_F\Bigg[ \frac{(1-x^2)(5+47 x^2-4x^4)}{2 x^2}-\frac{2\pi^2}{3}\frac{1+5x^2+2x^4}{x^2}-\frac{3(1-x^2)^2}{x^2}\log(1-x^2)\\*
&-&\frac{2(1-x)^2(2-x+6x^2+x^3)}{x^2}\log(x)\log(1-x)- \frac{2(1+x)^2(2+x+6x^2-x^3)}{x^2}\log(x)\log(1+x)\nn\\*
&-&\frac{2(1-x)^2(4+3x+8x^2+x^3)}{x^2}\mathrm{Li}_2(x)-\frac{2(1+x)^2(4-3x+8x^2-x^3)}{x^2}\mathrm{Li}_2(-x)+16(1+2x^2)\log(x)\Bigg]\,,\nn\\
\Gamma_b^L&=&2x^2(1-x^2)^2+\frac{\alpha_s}{4\pi}C_F\Bigg[-2x^2(1-x^2)(21-x^2)+\frac{2\pi^2}{3}4x^2(1+x^2)(3-x^2)+4x^2(1-x^2)^2\log\Big(\frac{m_t^2}{\mu^2}\Big)\nn\\*
&-&16x^2(3+3x^2-x^4)\log(x)-4(1-x^2)^2(2+x^2)\log(1-x^2)-8x(1-x)^2(3+3x^2+2x^3)\log(x)\log(1-x)\nn\\*
&+&8x(1+x)^2(3+3x^2-2x^3)\log(x)\log(1+x)-8x(1-x)^2(3+2x+7x^2+4x^3)\mathrm{Li}_2(x)\nn\\*
&+&8x(1+x)^2(3-2x+7x^2-4x^3)\mathrm{Li}_2(-x)\Bigg]\,,\\
\Gamma_{ab}^L&=&(1-x^2)^2+\frac{\alpha_s}{4\pi}C_F\Bigg[-(1-x^2)(1+11x^2)-\frac{2\pi^2}{3}(1-7x^2+2x^4)+(1-x^2)^2 \log\Big(\frac{m_t^2}{\mu^2}\Big)\nn\\*
&-&\frac{2(1-x^2)^2(1+2x^2)}{x^2}\log(1-x^2)-4x^2(7-x^2)\log(x)-4(1-x)^2(1+5x+2x^2)\log(x)\log(1-x)\nn\\*
&-&4(1+x)^2(1-5x+2x^2)\log(x)\log(1+x)-4(1-x)^2(3+9x+4x^2)\mathrm{Li}_2(x)\nn\\*
&-&4(1+x)^2(3-9x+4x^2)\mathrm{Li}_2(-x)\Bigg]\,,\nn\\
\Gamma_{a}^+ &=&\frac{\alpha_s}{4\pi}C_F\Bigg[-\frac{1}{2}(1-x)(25+5x+9x^2+x^3)+\frac{\pi^2}{3}(7+6x^2-2x^4)-2(5-7x^2+2x^4)\log(1+x)\\*
&-&2(5+7x^2-2x^4)\log(x)-\frac{(1-x)^2(5+7x^2+4x^3)}{x}\log(x)\log(1-x)-\frac{(1-x)^2(5+7x^2+4x^3)}{x}\mathrm{Li}_2(x)\nn\\*
&+&\frac{(1+x)^2(5+7x^2-4x^3)}{x}\log(x)\log(1+x)+\frac{5+10x+12x^2+30x^3-x^4-12x^5}{x}\mathrm{Li}_2(-x)\Bigg]\,,\nn\\
\Gamma_{b}^+ &=&\frac{\alpha_s}{4\pi}C_F\Bigg[ \frac{4}{3}x(1-x)(30+3x+7x^2-2x^3-2x^4)-4\pi^2 x^4-8(5-9x^2+4x^4)\log(1+x)\\*
&+&8x^2(1+5x^2)\log(x)-4(1-x)^2(4+5x+6x^2+x^3)\log(x)\log(1-x)\nn\\*
&-&4(1+x)^2(4-5x+6x^2-x^3)\log(x)\log(1+x)-4(1-x)^2(4+5x+6x^2+x^3)\mathrm{Li}_2(x)\nn\\*
&-&4(4+3x-16x^2+6x^3+16x^4-x^5)\mathrm{Li}_2(-x)\Bigg]\,,\nn\\
\Gamma_{ab}^+ &=&\frac{\alpha_s}{4\pi}C_F\Bigg[
2x(1-x)(15-11x)+\frac{2\pi^2}{3}x^2(5-2x^2)-2(13-16x^2+3x^4)\log(1+x)\\*
&-&2(1-x)^2(5+7x+4x^2)\log(x)\log(1-x)-2(1+x)^2(5-7x+4x^2)\log(x)\log(1+x)\nn\\*
&+&2x^2(1+3x^2)\log(x)-2(1-x)^2(5+7x+4x^2)\mathrm{Li}_2(x)-2(3+3x-31x^2+x^3+12x^4)\mathrm{Li}_2(-x)\Bigg]\,,\nn\\
\Gamma_{a}^- &=&(1-x^2)^2+\frac{\alpha_s}{4\pi}C_F\Bigg[-\frac{1}{2}(1-x)(13+33x-7x^2+x^3)+\frac{\pi^2}{3}(3+4x^2-2x^4)\\*
\nn&-&2(5+7x^2-2x^4)\log(x)-\frac{2(1-x^2)^2(1+2x^2)}{x^2}\log(1-x)-\frac{2(1-x^2)(1-4x^2)}{x^2}\log(1+x)\\*
\nn&-&\frac{(1-x)^2(5+7x^2+4x^3)}{x}\log(x)\log(1-x)+\frac{(1+x)^2(5+7x^2-4x^3)}{x}\log(x)\log(1+x)\\*
\nn&-&\frac{(1-x)^2(5+3x)(1+x+4x^2)}{x}\mathrm{Li}_2(x)+\frac{5+2x+12x^2+6x^3-x^4-4x^5}{x}\mathrm{Li}_2(-x)\Bigg]\,,\\
\Gamma_{b}^- &=&4(1-x^2)^2+\frac{\alpha_s}{4\pi}C_F\Bigg[\frac{4}{3}(1-x)(16-14x+22x^2+18x^3-3x^4-3x^5)\\*
\nn&-&\frac{\pi^2}{3}4(4+x^4) +8x^2(1+5x^2)\log(x)-24(1-x^2)^2\log(1-x)+8(1-x^2)(2-x^2)\log(1+x)\\*
\nn&-&4(1-x)^2(4+5x+6x^2+x^3)\log(x)\log(1-x)-4(1+x)^2(4-5x+6x^2-x^3)\log(x)\log(1+x)\\*
&-&4(1-x)^2(12+21x+14x^2+x^3)\mathrm{Li}_2(x)-4(12+3x+6 x^3-x^5)\mathrm{Li}_2(-x)+8(1-x^2)^2\log\Big(\frac{m_t^2}{\mu^2}\Big)\Bigg]\,,\nn\\
\Gamma_{ab}^- &=&2(1-x^2)^2+\frac{\alpha_s}{4\pi}C_F\Bigg[2(1-x)(9-6x+6x^2-5x^3)-\frac{2\pi^2}{3}(5+2x^4)+2x^2(1+3x^2)\log(x)\\*
\nn&-&\frac{2(1-x^2)^2(1+5x^2)}{x^2}\log(1-x)-\frac{2(1-x^2)(1-9x^2-2x^4)}{x^2}\log(1+x)\\*
\nn&-&2(1-x)^2(5+7x+4x^2)\log(x)\log(1-x)-2(1+x)^2(5-7x+4x^2)\log(x)\log(1+x)\\*
&-&2(1-x)^2(13+23x+12x^2)\mathrm{Li}_2(x)-2(15+3x+5x^2+x^3+4x^4)\mathrm{Li}_2(-x)+2(1-x^2)^2\log\Big(\frac{m_t^2}{\mu^2}\Big)\Bigg]\,.\nn
\end{eqnarray}
\normalsize
}

\chapter{Three-body FCNC top decays}\label{app:allTB}
In this Appendix we present some details of the $t\to q\ell^+ \ell^-$ analysis presented in section~\ref{sec:three_body} of chapter~{\ref{chap:neutral_currents}}.
\section{Analytical formulae}\label{app:tqll}
Below we give the complete analytic formulae for the partial differential decay rate distributions
in terms of our chosen kinematical variables and the expression for functions appearing in FBA and LRA. Mostly they are given in unevaluated integral form, as analytic integration, though possible in most cases, yields very long expressions.

\subsection{Photon mediation}
The double and single differential decay widths are given as
\begin{eqnarray}
\frac{\dd\Gamma^\gamma}{\dd \uy\dd\sy}&=&\frac{m_t}{16\pi^3}\frac{g_Z^4v^4}{\Lambda^4}B_{\gamma}\times\frac{1}{\sy}\Big[\sy(2\uy-1) + 2\uy^2-2\uy+1\Big]\,,\\
\frac{\dd\Gamma^\gamma}{\dd\hat s} &=& \frac{m_t}{16\pi^3}\frac{g_Z^4v^4}{\Lambda^4} B_{\gamma} \frac{(1-\sy)^2(\sy+2)}{3\sy}\,.\nn
\end{eqnarray}
Functions $f_{\gamma}$ and $g_{\gamma}$ defined in Eqs.~(\ref{eq:gamma_gamma}, \ref{eq:def_g}) are 
\begin{eqnarray}
f_{\gamma}(x) &=& \frac{1}{9} \Big[-x^3 + 9x-6\log(x)-8\Big]\,,\label{fg}\\
g_{\gamma}(x) &=&-\frac{13}{18}+3x-2x^2+x^3-\frac{2}{3}\log(4x)\,.\label{gg}
\end{eqnarray}

\subsection{$Z$ mediation}\label{app:Zmediation}
We define
$$
\hat{m}_Z = \frac{m_Z^2}{m_t^2}\,,\hspace{0.5cm}\gamma_Z = \frac{\Gamma_Z}{m_Z}\,.
$$
The double and single differential decay widths are given as
\begin{eqnarray}
\frac{\dd \Gamma^Z}{\dd \sy\dd \uy}&=&\frac{m_t}{16\pi^3}\frac{g_Z^4v^4}{\Lambda^4}\frac{1}{(\sy-\hat{m}_Z)^2+\sy^2\gamma_Z^2}
 \Big[\frac{ A + \alpha}{4}(1- \sy - \uy )( \sy + \uy) +
 +\frac{ A - \alpha}{4} (1- \uy)\uy  \\
\nonumber &+&(B+\beta)\uy\sy(\uy+\sy)+(B-\beta)\sy (1-\sy-\uy) (1-\uy)+(C+\gamma)\sy (1-\uy-\sy)+(C-\gamma)\uy\sy\Big]\,,\\
\frac{\dd\Gamma^Z}{\dd\hat s} &=&  \frac{m_t}{16\pi^3}\frac{g_Z^4v^4}{\Lambda^4}\frac{(\sy-1)^2}{(\sy-\hat{m}_Z)^2 + \sy^2\gamma_Z^2}
\Big[\frac{A}{12}(2\sy+1) + \frac{B}{3}\sy(\sy+2) + C \sy\Big]\,.
\end{eqnarray}
For the sake of shorter notation we first define
\begin{eqnarray}
\nonumber r_1 &=& \frac{(1-\sy)^2}{(\sy - \hat{m}_Z)^2 + \sy^2\gamma_Z^2}\,,\\
\nonumber r_2 &=& \frac{\frac{1}{8}(1-\uy)^2}{[(1-z)(1-\uy)-2\hat{m}_Z]^2 + \gamma_Z^2(1-z)^2(1-\uy)^2	}\,,
\end{eqnarray}
then present $f_i$ functions defined in Eqs.~(\ref{eq:fcnc_Z_1}) in the from of the following integrals
\begin{align}
f_A&=\int_{0}^{1} \dd \sy \,r_1\, \frac{1}{12}(1+2\sy)\,,&\label{eq:fA}
f_B&=\int_{0}^{1} \dd \sy \,r_1\, \frac{1}{3}(2\sy+\sy^2)\,,&\\
\nn f_C&=\int_{0}^{1} \dd \sy \,r_1\, \sy\,,&
f_{\alpha\beta\gamma} &= -\frac{1}{8} f_C\,.
\end{align}
The $g_i$ functions present in LRA expressions defined in Eqs.~(\ref{eq:fcnc_Z_1}) are more complicated due to the fact that 
the angular variable appears in the resonant factor of the matrix element. So for the sake of brevity
we define additional functions $G_i$ in which the $\uy$ integration is performed
\begin{eqnarray}
g_i&=&\int_0^1 \dd z \,G_i - \int_{-1}^{0}\dd z  \,G_i\,,\hspace{0.5cm} i = A,B,C,\alpha\beta\gamma\,,\label{eq:gZ}\\
\nonumber G_A &=&\int_{0}^{1} \dd \uy\, r_2\, (1+5\uy+2\uy z-z^2+\uy z^2)\,,\\
\nonumber G_B &=&\int_{0}^{1} \dd \uy\, r_2\, 4(1-\uy+2\uy^2-2\uy z-z^2+3\uy z^2-2\uy^2z^2)\,,\\
\nonumber G_C&=&\int_{0}^{1} \dd \uy\, r_2\, 4(1 + \uy - 2\uy z - z^2 +\uy z^2)\,,\\ 
\nonumber G_{\alpha\beta\gamma} &=&\int_{0}^{1} \dd \uy\, r_2\, (1 - 3 \uy + 2\uy  z - z^2 + \uy  z^2)\,.
\end{eqnarray}


\subsection{Interference between $Z$ and photon mediation}\label{app:fcnc_gammaZ}

The interference contribution between the $Z$ and the photon to the double differential decay rate is
\begin{eqnarray}
\nonumber\frac{\dd \Gamma^{\mathrm{int}}}{\dd \sy\dd\uy}&=&\frac{m_t}{16\pi^3}\frac{v^4g_Z^4}{\Lambda^4}\mathrm{Re}\Bigg\{\frac{\sy-\hat{m}_Z-\ii\sy\gamma_Z}{(\sy-\hat{m}_Z)^2+\sy^2\gamma_Z^2}\times\\
\nonumber &&
\Big[2W_1(1-\sy-\uy)(1-\uy)+2W_2\uy(\uy+\sy)+ W_3(1-\sy-\uy) + W_4\uy\Big]\Bigg\}\,.
\end{eqnarray}
In all further computations we neglect the imaginary part in the propagator's numerator $\gamma_Z \sim 0.02$. This means that $\mathrm{Re}$ acts only on the model dependent constants $W_1,\dots,W_4$.
$f_i^{\epsilon}$ and $g_i^{\epsilon}$ are the same as $f_i$ and $g_i$, except that the integration limits are altered due to the di-lepton invariant mass cutoff $\epsilon$. In $f_X$ the $\sy$ integration is now in the $[\epsilon/m_t^2,1]$ region, in $g_X$ the intervals for $z$ are $[0,1-2\epsilon/m_t^2]$ and $[-1,0]$, and for the $\uy$ in $G_X$ functions $\uy\in[0,1-\frac{2\epsilon}{1-z}]$. We further define
\begin{eqnarray}
\nonumber r_3 &=& \frac{[(1-\uy)(1-z)-2\hat{m}_Z](1-\uy)}{[(1-z)(1-u)-2\hat{m}_Z]^2 + \gamma_Z^2(1-z)^2(1-u)^2}\,.
\end{eqnarray}
The new $f_i$ and $g_i$ functions defined in Eqs.~(\ref{eq:new_fs}) are
\begin{subequations}\label{eq:fcnc_int_1}
\begin{eqnarray}
f_{W_{12}} &=&\int_{\epsilon/m_t^2}^1 \dd \sy\, r_1\,(\hat{s}-\hat{m}_Z) \frac{1}{3} (\sy +2)\,,\label{fW12}\\
f_{W_{34}} &=&\int_{\epsilon/m_t^2}^1 \dd \sy\, r_1\,(\hat{s}-\hat{m}_Z) \frac{1}{2}\,,\label{fW34}\\
f_{W} &=&\frac{1}{2} f_{W_{34}}\,,\label{fW}\\
g_i&=&\int_0^{1-2\epsilon/m_t^2} \dd z \,G_i - \int_{-1}^{0}\dd z  \,G_i\,,\label{gZg}
\end{eqnarray}
\end{subequations}
\begin{align}
 G_{W_1}&=\int_{0}^{1-\frac{2\epsilon/m_t^2}{1-z}} \dd \uy\, r_3\, (1-\uy)^2(1+z)\,,&
 G_{W_2}&=\int_{0}^{1-\frac{2\epsilon/m_t^2}{1-z}} \dd \uy\, r_3\, \uy (1+\uy-z+z\uy)\,,\\ 
\nonumber G_{W_3}&=\int_{0}^{1-\frac{2\epsilon/m_t^2}{1-z}} \dd \uy\, r_3\, \frac{1}{2} (1+\uy+z-z\uy)\,,&
 G_{W_4}&=\int_{0}^{1-\frac{2\epsilon/m_t^2}{1-z}} \dd \uy\, r_3\, \uy \,.
\end{align}

\vfill \hfill
\section{Matching to the parametrization of Fox et al.}\label{app:tofox}
Here we present the conversion of $\Leff$ presented in Ref.~\cite{Fox:2007in} to the form in Eq.~(\ref{eq:Lagr}).
Fox et al. give a complete set of dimension six operators that generate a $tcZ$ or  $tc\gamma$ vertex 
\begin{align*}
O^u_{LL} &=\ii \Big[\bar{Q}_3\tilde{H}\Big] \Big[(\slashed{D}\tilde{H})^{\dagger}Q_2\Big]
-\ii\Big[\bar{Q}_3(\slashed{D}\tilde{H})\Big] \Big[\tilde{H}^{\dagger}Q_2\Big] \,,&
O_{LL}^h &= \ii \Big[\bar{Q}_3\gamma^{\mu}Q_2\Big]\Big[H^{\dagger} (D_{\mu}H) - (D_{\mu}H)^{\dagger} H \Big] \,,\\
O_{RL}^w &=g_2\Big[\bar{Q}_2\sigma^{\mu\nu}\sigma^a \tilde{H}\Big]t_R W^a_{\mu\nu} \,,&
O_{RL}^b &= g_1\Big[\bar{Q}_2\sigma^{\mu\nu}\tilde{H}\Big]t_R B_{\mu\nu}\,,\\
O_{LR}^w & = g_2\Big[\bar{Q}_3\sigma^{\mu\nu}\sigma^a \tilde{H}\Big]c_R W_{\mu\nu}^a \,,&
O_{LR}^b &= g_1\Big[\bar{Q}_3\sigma^{\mu\nu} \tilde{H} \Big]c_R B_{\mu\nu} \,,\\
O^u_{RR} &=\ii \bar{t}_R\gamma^{\mu}c_R \Big[H^{\dagger} (D_{\mu}H) - (D_{\mu}H)^{\dagger} H \Big] \,,&
\end{align*}
for notational details see Ref.~\cite{Fox:2007in}. We have suppressed the addition of h.c. for every operator. Keeping only FCNC parts and the VEV of the Higgs field we obtain
\begin{align*}
O^{u}_{LL} &= \frac{v^2}{2}[g A_{\mu}^3  - g' B_{\mu}]\,\,\Big[ \bar{ t}_L \gamma^{\mu}c_L\Big]  \,,&
O_{LL}^h &= \frac{v^2}{2}[g A_{\mu}^3  - g' B_{\mu}]\,\,\Big[\bar{t}_L \gamma^{\mu} c_L + \bar{b}_L \gamma^{\mu} s_L\Big]  \,, \\
O_{RL}^w &=g\frac{v}{\sqrt{2}}W_{\mu\nu}^3\,\,\Big[\bar{c}_L \sigma^{\mu\nu} t_R\Big]  \,,&
O_{RL}^b &= g'\frac{v}{\sqrt{2}} B_{\mu\nu}\,\,\Big[\bar{c}_L \sigma^{\mu\nu} t_R\Big] \,, \\
O_{LR}^w &= g\frac{v}{\sqrt{2}}W_{\mu\nu}^3\,\,\Big[\bar{t}_L \sigma^{\mu\nu} c_R\Big]  \,, &
O_{LR}^b &=g'\frac{v}{\sqrt{2}} B_{\mu\nu}\,\,\Big[\bar{t}_L \sigma^{\mu\nu} c_R\Big] \,, \\
O_{RR}^u &= \frac{v^2}{2}[g A_{\mu}^3  - g' B_{\mu}]\,\,\Big[\bar{t}_R\gamma^{\mu}c_R \Big]  \,.&
\end{align*}
Finally our coupling constants can be expressed as
\begin{subequations}\label{eq:fcnc_trans}
\begin{eqnarray}
a_L^Z &=& \frac{1}{2}\Big[C^u_{LL}+ C_{LL}^h\Big]\,,\label{eq:fox1}\\
a_R^Z &=& \frac{C^u_{RR}}{2} \,, \label{eq:fox2}\\
b_{LR}^Z &=& \frac{C_{RL}^w \cos^2\theta_W - C_{RL}^b \sin^2\theta_W}{\sqrt{2} } \,,\label{eq:fox3}\\
b_{RL}^Z &=& \frac{C_{LR}^w \cos^2\theta_W -C_{LR}^b \sin^2\theta_W}{\sqrt{2}}\,,\label{eq:fox4}\\
b_{LR}^{\gamma} &=& \frac{C_{RL}^w + C_{RL}^b }{\sqrt{2}}\,,\label{eq:fox5}\\
b_{RL}^{\gamma} &=& \frac{C_{LR}^w  + C_{LR}^b }{\sqrt{2}}\,.\label{eq:fox6}
\end{eqnarray}
\end{subequations}

\end{appendices}

\renewcommand{\figurename}{Slika}
\renewcommand{\tablename}{Tabela}
\renewcommand{\chaptername}{Poglavje}
\chapter{Raz\v{s}irjen povzetek}
\section{Uvod}
Glavni cilj teoreti\v{c}ne fizike osnovnih delcev je razumevanje, opisovanje in napovedovanje pojavov, ki se odvijajo na najmanj\v{s}ih eksperimentalno dosegljivih razdaljah. Glavno matemati\v{c}no orodje pri tem je kvantna teorija polja. 

Stalni napredki na teoreti\v{c}nem in eksperimentalnem podro\v{c}ju so privedli do oblikovanja teorije znane pod imenom Standardni Model (SM), katerega glavni koncepti so bili zasnovani v \v{s}estdesetih letih prej\v{s}njega stoletja~\cite{Weinberg:1967tq}. 

Glavni karakteristiki SM sta enostavnost in izjemna prediktivna mo\v{c}. Zadnja desetletja je zaznamovalo testiranje narazli\v{c}nej\v{s}ih napovedi te teorije z bogato paleto sofisticiranih eksperimentov, najodmevnej\v{s}i od katerih je {\sl Veliki Hadronski Trkalnik} (LHC).

Kot renormalizabilno teorijo osnovnih delcev in interakcij SM zaznamuje umeritvena grupa, pod katero je invariantna, in mehanizem  zloma elektro-\v{s}ibke simetrije s pomo\v{c}jo Higgsovega bozona
\begin{eqnarray}
SU(3)_c \times SU(2)_L \times U(1)_{Y} \xrightarrow{\langle \phi \rangle} SU(3)_c \times U(1)_Q\,, \hspace{0.5cm} Q = Y + T_3\,,
\end{eqnarray}
kar povsem dolo\v{c}a vsebino vektorskih in skalarnih polj. Fermionska polja, ki jih delimo na kvarke in leptone, pa se v SM pojavijo v treh ponovitvah (dru\v{z}inah) istih reprezentacij umeritvene grupe, za katere re\v{c}emo, da imajo razli\v{c}ne okuse. Baza v kateri so umeritvene interakcije diagonalne se razlikuje od baze masnih lastnih stanj, ki izvirajo iz interakcijskih \v{c}lenov tipa Yukawa. Pravimo, da je Yukawa sektor SM edini vir fizike okusa, ki v kvarkovskem sektorju privede do nabitih tokov, ki spreminjajo okus
\begin{eqnarray}
\mathcal L_{\mathrm{cc}} =- \frac{g}{\sqrt{2}}
\big[\bar{u}_{iL}\gamma^{\mu}d_{jL}\big]V_{ij}W_{\mu}^+ - \frac{g}{\sqrt{2}} 
\big[\bar{d}_{jL}\gamma^{\mu}u_{iL}\big]V^*_{ij}W_{\mu}^-\,.
\label{eq:SMcc!}
\end{eqnarray}
Me\v{s}anje med dru\v{z}inami v nabitih tokovih opisuje unitarna matrika {\sl Cabbibo-Kobayashi-Maskawa} (CKM)
\begin{eqnarray}
V = \left(\begin{array}{ccc}
V_{ud}& V_{us}& V_{ub}\\
V_{cd}& V_{cs}& V_{cb}\\
V_{td}& V_{ts}& V_{tb}\\
\end{array}\right)\,.
\label{eq:CKMmat!}
\end{eqnarray}

Kljub veliki uspe\v{s}nosti SM kot teorije velja splo\v{s}no sprejeto prepri\v{c}anje, da SM ni dokon\v{c}na teorija osnovnih delcev in njihovih interakcij. Nenazadnje ne opisuje kvantne gravitacije, ki postane pomembna pri energijah reda Planck-ove skale $\Lambda_{\mathrm{P}}\sim 10^{16} $ GeV. Tudi del\v{c}na vsebina SM ne zadostuje za opis vsega, kar smo do sedaj opazili v naravi, saj SM nikakor ne pojasni obstoja temne snovi in energije~\cite{Olive:2003iq,Trimble:1987ee}. Odkritje nevtrinskih oscilacij nedvomno potrjuje obstoj nevtrinskih mas, ki jih v SM ne poznamo. Poleg tega sta tu \v{s}e dve veliki konceptualni uganki, ki nam dajeta misliti, da mora kon\v{c}na teorija biti \v{s}e bolj dovr\v{s}ena. Hierarhi\v{c}ni problem~\cite{Martin:1997ns,Wells:2009kq} izpostavlja te\v{z}ko razumljivo veliko razliko med elektro-\v{s}ibko in Planckovo skalo. Problem fizike okusa pa izra\v{z}a, da SM ne zna razlo\v{z}iti parametrov fizike okusa in njihove o\v{c}itno hierarhi\v{c}ne ureditve.

V dobi LHC lahko prvi\v{c} preu\v{c}ujemo fiziko kvarka top z veliko preciznostjo, saj LHC lahko smatramo za pravo tovarno kvarkov top. Kvark top, ki izstopa s svojo veliko maso $m_t = 173.2 \pm 0.9$ GeV~\cite{Lancaster:2011wr} in veliko razpadno \v{s}irino glavnega razpadnega kanala,
\begin{eqnarray}
\Gamma (t\to W b ) =|V_{tb}|^2\frac{m_t}{16 \pi}\frac{g^2}{2}\frac{(1-x^2)^2(1+2x^2)}{2x^2} \sim 1.5 \,\,\mathrm{GeV}\,,\label{eq:SM_MDC!}
\end{eqnarray}
ki nam omogo\v{c}a, da kvark top v razpadih obravnavamo kot prost delec, postaja vse bolj zanimiv in dostopen za iskanje {\sl nove fizike} (NF) onkraj SM. Osrednje vpra\v{s}anje, ki ga bomo posku\v{s}ali raziskati v tem delu, je, na kak\v{s}en na\v{c}in se NF lahko manifestira in opazi v razpadih kvarka top. Glavno vodilo so redki razpadi, za katere SM napoveduje zelo majhno verjetnost. Posledi\v{c}no so odstopanja od napovedi SM lahko jasno opazljiva.
 
Za iskanje NF so zanimivi razpadi kvarka top, ki potekajo preko {\sl nevtralnih tokov, ki spreminjajo okus} (FCNC). Ti razpadi v okviru SM niso mogo\v{c}i na drevesnem redu, zato so njihova razvejitvena razmerja neopazljivo majhna~\cite{Eilam:1990zc, AguilarSaavedra:2004wm}
\begin{eqnarray}
\mathrm{Br}[t\to c \gamma]\sim 10^{-14}\,,\hspace{0.5cm}
\mathrm{Br}[t\to c Z]\sim 10^{-14}\,,\hspace{0.5cm}
\mathrm{Br}[t\to c g]\sim 10^{-12}\,.
\end{eqnarray}
Potencialna detekcija tak\v{s}nih razpadov bi nedvomno pomenila prisotnost NF.

Po drugi strani lahko odstopanje od napovedi SM i\v{s}\v{c}emo tudi v glavnem razpadnem kanalu kvarka top. Ta poteka preko nabite \v{s}ibke interakcije na nivoju drevesnega reda~(\ref{eq:SMcc!}) in velja za eksperimentalno signaturo kvarka top. V kolikor bi struktura nabitih kvarkovskih tokov odstopala od strukture, ki jo poznamo v SM, bi se to poznalo na {\sl su\v{c}nostni dele\v{z}ih} (eng. {\sl helicity fractions}) bozona $W$, ki nastane pri razpadu. Definiramo jih tako, da razpadno \v{s}irino glavnega razpadnega kanala razdelimo na tri dele
\begin{eqnarray}
\Gamma(t\to W b) = \Gamma_L + \Gamma_+ + \Gamma_- \,,
\end{eqnarray}
kjer $L$ ozna\v{c}uje longitudinalno, $+$ in $-$ pa pozitivno in negativno transverzalno stanje su\v{c}nosti bozona $W$. Su\v{c}nostne dele\v{z}e nato vpeljemo kot $\mathcal F_{L,+,-} = \Gamma_{L,+,-}/\Gamma$. Ker SM napoveduje zelo majhno vrednost $\mathcal F_+$ \cite{Fischer:2001gp,Czarnecki:2010gb,Do:2002ky,Fischer:2000kx}
\begin{eqnarray}
\mathcal F_L^{\rm SM} =  0.687(5)\,,\hspace{0.5cm} \mathcal F_+^{\rm SM} = 0.0017(1) \label{eq:e22b!}\,,
\end{eqnarray}
so te opazljivke zanimive za iskanje NF. Zgoraj navedene vrednosti vklju\v{c}ujejo kvantne popravke vi\v{s}jega reda, najpomembnej\v{s}i od katerih so popravki {\sl kvantne kromodinamike} (QCD). 
\begin{SCfigure}[3.5][h!]
\includegraphics[width=0.2\textwidth]{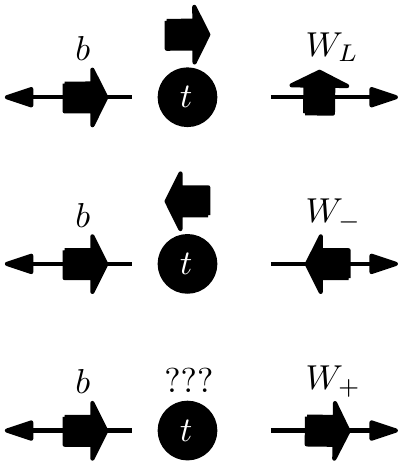}
\caption{Ilustracija razpada kvarka top v mirovnem sistemu po glavnem razpadnem kanalu v limiti $m_b=0$. \v{S}iroke pu\v{s}\v{c}ice predstavljajo tretjo komponento spina, tanke pu\v{s}\v{c}ice pa smer gibanja. V brezmasni limiti su\v{c}nost in ro\v{c}nost polj sovpadata. Ker je \v{s}ibka interakcija v SM izklju\v{c}no levo-ro\v{c}na, je su\v{c}nost kvarka $b$ v omenjeni limiti vedno negativna. Tretja slika prikazuje situacijo, ki je zaradi ohranitve spina prepovedana, saj ima kvark top spin $1/2$. To nam or\v{s}e, zakaj je napovedana vrednost za $\mathcal F_+$ v SM majhna. Ta enostavna slike se podre, ko opustimo brezmasno limito ali v proces vklju\v{c}imo kvantne popravke vi\v{s}jega reda.}
\label{fig:illust!}
\end{SCfigure}
Intuitivna razlaga za majhnost $\mathcal F_+$ je podana na sliki Fig.~\ref{fig:illust!}. 
V primeru, da se izmerjene vrednosti su\v{c}nostnih dele\v{z}ev ujemajo z napovedmi SM, meritve slu\v{z}ijo omejevanju NF, v primeru, da bi se izmerjene vrednosti bistveno razlikovale od napovedanih (zlasti v primeru signifikantno neni\v{c}elnega $\mathcal F_+$), pa bi to lahko pomenilo odkritje NF v nabitih tokovih s kvarkom top.

Pri obravnavi NF v nabitih in nevtralnih tokovih, ki vsebujejo kvark top, ne moremo mimo dejstva, da kvark top igra zelo pomembno vlogo tudi v fiziki ni\v{z}jih energij, kjer se pojavlja kot virtualen delec. Zlasti v teoreti\v{c}nih napovedih za redke procese mezonov $B$ in $K$ lahko v primeru NF v fiziki kvarka top pri\v{c}akujemo spremembe. V primeru, da se meritve ujemajo z napovedmi SM, lahko slu\v{z}ijo za postavitev indirektnih omejitev na prispevke NF. Po drugi strani, v kolikor bodo natan\v{c}no merjene v prihodnje, lahko NF postavlja zanimive napovedi.

Z ozirom na to v primeru obravnave NF v nabitih tokovih podrobno analiziramo implikacije NF na opazljivke v procesih me\v{s}anja nevtralnih mezonov $B$ ($|\Delta B|=2$ procesi) in njihovih razpadih ($|\Delta B|=1$ procesi), ki jih opi\v{s}emo z naslednjimi efektivnimi Lagrangeovimi funkcijami, v katerih ni polj, ki bi imela mase ve\v{c}je od mase kvarka $b$
\begin{subequations}\label{eq:x!}
\begin{eqnarray}
\mathcal L_q^{|\Delta B|=2}&=&- \frac{G_F^2 m_W^2}{4\pi^2}(V_{tq}^*V_{tb})^2 C_1(\mu)\mathcal O_1^q \,,\\
{\cal L}_{\mathrm{eff}}^{|\Delta B|=1}&=& \frac{4 G_F}{\sqrt{2}}\Big[ \sum_{i=1}^2 C_{i}( \lambda_u \mathcal O^{(u)}_i +  \lambda_c \mathcal O^{(c)}_i) \Big] + \frac{4 G_F}{\sqrt{2}}\lambda_t\Big[\sum_{i=3}^{10} C_{i}{\cal O}_i +  C_{\nu\bar{\nu}}{\cal O}_{\nu\bar{\nu}}\Big]\,. \label{eq:loweff1!}
\end{eqnarray}
\end{subequations}
Izraze za Wilsonove koeficiente $C_i$ na elektro-\v{s}ibki skali izra\v{c}unamo s postopkom ujemanja polne teorije, ki vsebuje vse prostostne stopnje, na efektivne teorije, ki jih opisujeta zgoraj navedeni Lagrangeovi funkciji. S pomo\v{c}jo ena\v{c}b renormalizacijske grupe, ki izvirajo iz anomalnih dimenzij efektivnih operatorjev, nato spustimo skalo do reda mase kvarka $b$, kjer je z razli\v{c}nimi neperturbativnimi mogo\v{c}e izra\v{c}unati matri\v{c}ne elemente efektivnih operatorjev in s tem napovedati amplitude za razli\v{c}ne procese.

V kolikor NF posega v proces ujemanja na visokih skalah, se ves njen vpliv na procese v fiziki $B$ manifestira kot sprememba Wilsonovih koeficientov na visoki skali ujemanja.

Tudi pri obravnavi NF v nevtralnih in nabitih kvarkovskih tokovih se poslu\v{z}imo metod efektivnih teorij. S pomo\v{c}jo efektivnih operatorjev vi\v{s}jih dimenzij, kljub nepoznavanju fizike onkraj SM, lahko sistemati\v{c}no parametriziramo u\v{c}inke NF na omenjene tokove
\begin{eqnarray}
{\cal L}_{\mathrm{eff}}={\cal L}_{\mathrm{SM}}+\frac{1}{\Lambda^2}\sum_i C_i \mathcal Q_i +\mathrm{h.c.}+ {\cal O}(1/\Lambda^3)\,,
\label{eq:lagr!}
\end{eqnarray}
kjer ${\cal L}_{\mathrm{SM}}$ predstavlja SM del, $\mathcal Q_i$ pa so operatorji dimenzije 6, invariantni na operacije umeritvene grupe SM ter so sestavljeni le iz polj, ki jih vsebuje SM. Pri tem smo uporabili zelo mo\v{c}en in pomemben koncept efektivnih teorij, ki nam zagotavlja, da vi\v{s}je dimenzionalne operatorje spremljajo vi\v{s}je negativne potence skale nove fizike in so zato njihovi prispevki vedno manj\v{s}i.
\begin{figure}[h]
\begin{center}
\includegraphics[width=0.5 \textwidth]{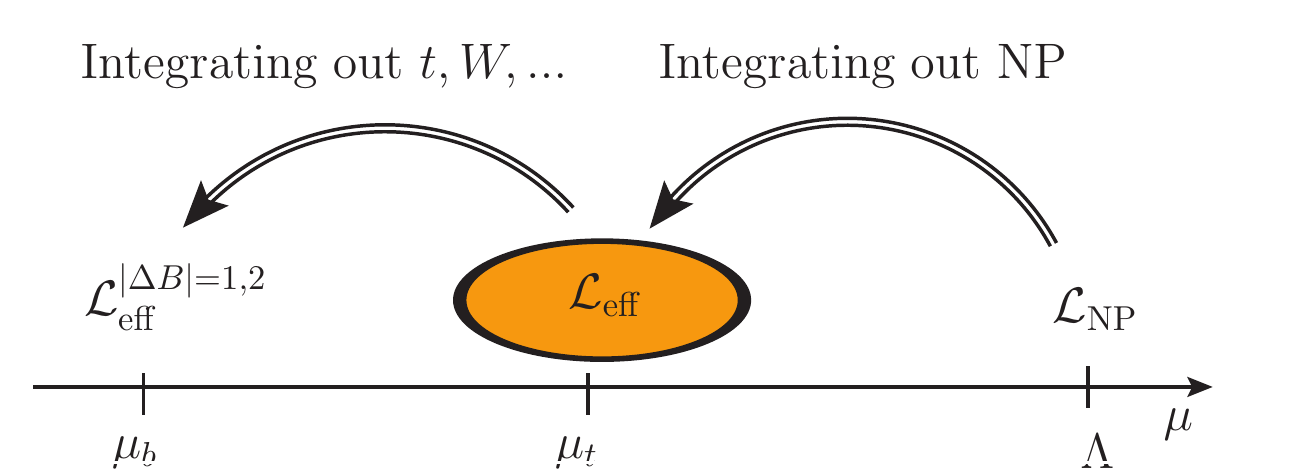}
\caption{Shemati\v{c}en prikaz na\v{s}ega pristopa k obravnavi NF. Prvi korak predstavlja dolo\v{c}itev baze operatorjev v En.~(\ref{eq:lagr!}) s \v{c}imer parametriziramo vpliv NF, katere skala $\Lambda$ je dale\v{c} nad elektro-\v{s}ibko skalo $\mu_t$, na fiziko kvarka top. Drugi korak pa predstavlja nadaljnje ujemanje efektivne teorije~(\ref{eq:lagr!}) z efektivnima teorijama~(\ref{eq:x!}), ki nam omogo\v{c}a obravnavo vplivov NF na opazljivke fizike mezonov $B$.}
\label{fig:intout!}
\end{center}
\end{figure}

Sl.~\ref{fig:intout!} shemati\v{c}no prikazuje na\v{s}o strategijo analize NF. Prvi korak ponazarja na\v{s}e nepoznavanje NF manifestirane na skali $\Lambda$, ki je precej vi\v{s}ja od elektro-\v{s}ibke skale, katere vplive lahko parametriziramo z efektivno Lagrangeovo funkcijo oblike~(\ref{eq:lagr!}) z izbiro ustrezne baze operatorjev, glede na to kak\v{s}ne spremembe v fiziki kvarka top \v{z}elimo obravnavati. \v{C}e \v{z}elimo nadalje obravnavati tudi vplive NF na fiziko mezonov $B$, moramo izvr\v{s}iti \v{s}e drugi korak - ujemanje Lagrangeove funkcije~(\ref{eq:lagr!}) z Lagrangeovimi funkcijami ~(\ref{eq:x!}).


\section{NF v razpadih kvarka top: Nevtralni tokovi}
\subsection{Uvod}
Razvejitvena razmerja za procese 
\begin{eqnarray}
\nn t\to q V\,,\hspace{0.3cm} V=Z,\gamma,g\,,\hspace{0.3cm} q=c,u\,,
\end{eqnarray}
ki so v okviru SM neopazljivo majhna, lahko v okviru rez\v{s}irjenih teorij postanejo ob\v{c}utno ve\v{c}ja~\cite{AguilarSaavedra:2004wm,Yang:2008sb,deDivitiis:1997sh,delAguila:1998tp} in potencialno opazljiva na LHC, saj ATLAS ocenjuje sposobnost odkritja omenjenih razpadov v kolikor bi bila razvejitvena razmerja vsaj reda $\sim 10^{-5}$~\cite{Carvalho:2007yi}.

V tem poglavju s pomo\v{c}jo efektivnih operatorjev analiziramo dvo-del\v{c}ne razpade $t \to q Z,\gamma$ in tro-del\v{c}ne razpade $t\to q \ell^+ \ell^-$. Pri dvodel\v{c}nih razpadih upo\v{s}tevamo popravke prvega reda v QCD in analiziramo tako posledice me\v{s}anja efektivnih operatorojev pod renormalizacijo, kot tudi efekte kon\v{c}nih popravkov, vklju\v{c}no s tako imenovanimi ``bremsstrahlung'' procesi. Pri obravnavi tro-del\v{c}nih razpadov pa se osredoto\v{c}imo na iskanje opazljivk, ki bi lahko pomagale pri diskriminaciji med razli\v{c}nimi oblikami NF, ki poveljuje razpadom FCNC. To nam omogo\v{c}a bogate\v{s}i fazni prostor tro-del\v{c}nega kon\v{c}nega stanja.

Pri parametrizaciji NF, ki generira $tZq$, $t\gamma q$ in $tgq$ vozli\v{s}\v{c}a sledimo Ref.~\cite{AguilarSaavedra:2004wm, AguilarSaavedra:2008zc}
\be
{\mathcal L}_{\mathrm{eff}} = \frac{v^2}{\Lambda^2}a_L^{Z}\op_{L}^Z
+\frac{v}{\Lambda^2}\Big[b^{Z}_{LR}\op_{LR}^{Z}+b^{\gamma}_{LR}\op_{LR}^{\gamma}+b^{g}_{LR}\op_{LR}^{g}
\Big] + (L \leftrightarrow R) + \mathrm{h.c.}\,,
\label{eq:Lagr!}
\ee
kjer so operatorji definirani kot
\begin{align}
\op^{Z}_{L,R} &= g_Z Z_{\mu}\Big[\bar{q}_{L,R}\gamma^{\mu}t_{L,R}\Big]\,, &
\op^{Z}_{LR,RL} &= g_Z Z_{\mu\nu}\Big[\bar{q}_{L,R}\sigma^{\mu\nu}t_{R,L}\Big]\,, \label{eq:ops!}\\
\nn\op^{\gamma}_{LR,RL} &= e F_{\mu\nu}\Big[\bar{q}_{L,R}\sigma^{\mu\nu}t_{R,L}\Big]\,, &
\op^{g}_{LR,RL} &= g_s G^a_{\mu\nu}\Big[\bar{q}_{L,R}\sigma^{\mu\nu}T_a t_{R,L}\Big]\,.
\end{align}
Pri ra\v{c}unanju popravkov QCD se poslu\v{z}imo dimezijske regularizacije, s pomo\v{c}jo katere regulariziramo tako UV kot IR divergence.

Analiza u\v{c}inkov teh operatorjev v fiziki mezonov je bila opravljena v Ref.~\cite{Fox:2007in} in je v tem delu ne ponavljamo ali nadgrajujemo in s tem presko\v{c}imo oba koraka shemati\v{c}no prikazana na Sl.~\ref{fig:intout!} ter se osredoto\v{c}imo le na obravnavo pojavov s kvarki top na masni lupini. Iz analize~\cite{Fox:2007in} povzamemo, da obstajajo operatorji, ki generirajo efektivna FCNC vozli\v{s}\v{c}a s kvarkom top, za katere indirektne omejitve na njihove prispevke ne izklju\v{c}ujejo potencialne opazljivosti FCNC razpadov kvarkov top.

\subsection{Dvo-del\v{c}ni razpadi}
\begin{figure}[h]
\begin{center}
\includegraphics[scale=0.6]{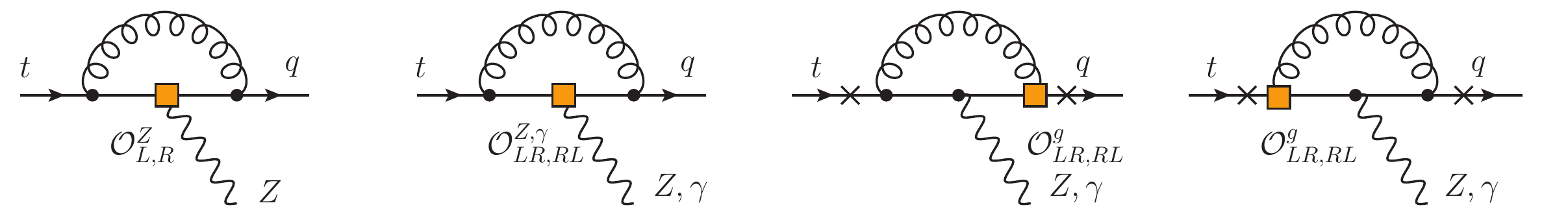}
\caption{Feynmanovi diagrami virtualnih popravkov QCD za razpade $t\to q Z,\gamma$. Kvadratki ozna\v{c}ujejo delovanje efektivnih opreratorjev podanih v En.~(\ref{eq:ops!}), kri\v{z}ci pa dodatne to\v{c}ke iz katerih se lahko izsevajo bozoni $Z$ in $\gamma$.}
\label{fig:fcnc_virt!}
\end{center}
\end{figure}

Najprej se osredoto\v{c}imo na virtualne popravke QCD. Feynmanovi diagrami za obravnavo popravkov na nivoju ene zanke so prikazani na Sl.~\ref{fig:fcnc_virt!}.
V prvi vrsti ti popravki privedejo do me\v{s}anja operatorjev pod renormalizacijo. Z uporabo metod efektivnih teorij lahko izpeljemo ena\v{c}be renormalizacijske grupe, ki nam povezuje vrednosti Wilsonovih koeficientov ovrednotenih na razli\v{c}nih skalah.
Ker operatorji $\mathcal O^Z_{L,R}$ nimajo anomalnih dimenzij, lahko preostalih 6 operatorjev zberemo v dva vektorja,
\begin{eqnarray}
\boldsymbol{\mathcal O}_{i} =  (\mathcal O^\gamma_i , \mathcal O^Z_i, O^g_i)^T\,,\hspace{0.5cm} i = RL, LR\,,
\end{eqnarray} 
ki se med seboj ne me\v{s}ata. Matrika anomalnih dimenzij na nivoju ene zanke je 
\begin{equation}
\gamma_i = \frac{\alpha_s}{2\pi}
\left[
\begin{array}{ccc}
 C_F  & 0   & 0   \\
0  & C_F  & 0   \\
 8 C_F / 3 &   C_F (3 - 8 s^2_W) / 3  & 5C_F - 2 C_A   
\end{array}
\right]\,. \label{eq:anomal!}
\end{equation}

Pogosto zaradi strukture NF, ki generira operator $LR$, ta lahko vsebuje ekspliciten faktor mase kvarka top. Da analiziramo, ali to privede do opaznih sprememb v analizi renormalizacijske grupe, definiramo nov operator
$$
\widetilde{\boldsymbol{\mathcal O}}_{LR} =  (m_t/v )\boldsymbol{\mathcal O}_{LR}\,.
$$

Me\v{s}anje gluonskega dipolnega operatorja s fotonsikm in operatojem z dipolno $Z$ sklopitvijo povzemajo spodnje ena\v{c}be
\begin{subequations}\label{eq:rge!}
\begin{eqnarray}
 b^\gamma_{i} (\mu_t) &=& \eta ^{\kappa_1} b_i^\gamma (\Lambda )+\frac{16}{3}\left( \eta ^{\kappa_1}- \eta ^{\kappa_2}\right) b^g_i (\Lambda )\,,\\
 b^Z_{i} (\mu_t)  &=&  \eta ^{\kappa_1} b_i^Z (\Lambda )  +\left[2-\frac{16}{3} s^2_W\right]\left( \eta ^{\kappa_1}- \eta ^{\kappa_2}\right) b^g_i (\Lambda )\,,
\end{eqnarray}
\end{subequations}
kjer $\mu_t$ predstavlja skalo mase kvarka top, $\eta = \alpha_s(\Lambda)/\alpha_s(\mu_t)$, $\kappa_1=4/3\beta_0$, $\kappa_2=2/3\beta_0$ in $\beta_0$ je del beta funkcije QCD na nivoju ene zanke~\cite{Buras:1998raa}.

\begin{figure}
\begin{center}
\includegraphics[scale=0.7]{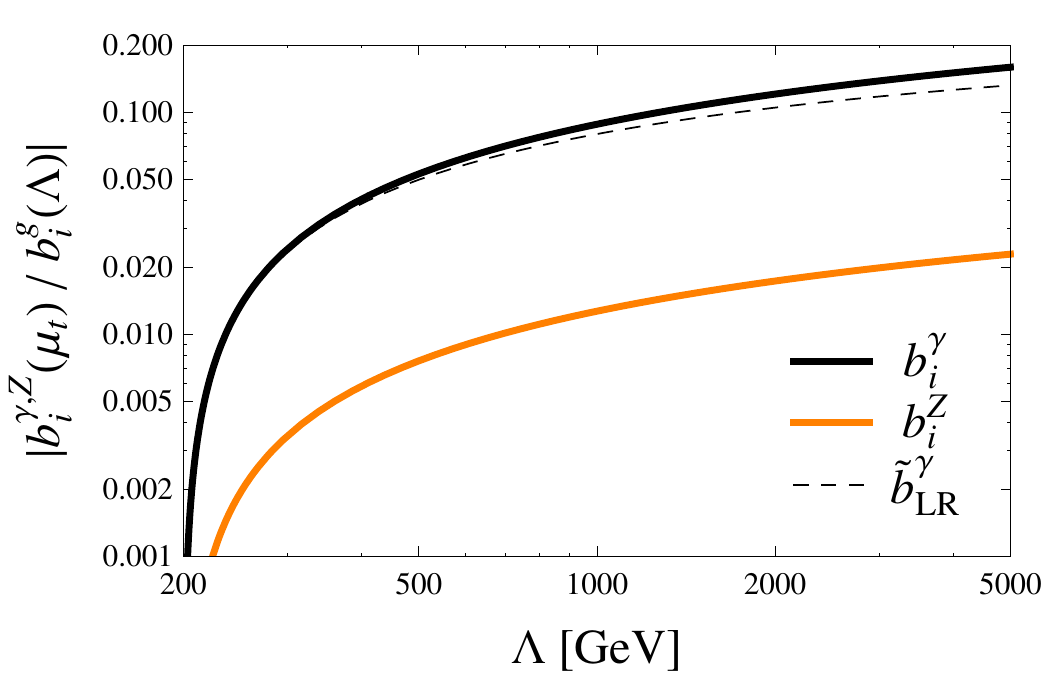}
\end{center}
\caption{Razmerje $|b_{i}^{\gamma,Z}(\mu_t)/b_i^{g}(\Lambda)|$ kot funkcija $\Lambda$ ob predpostavki $b_{i}^{\gamma,Z}(\Lambda) =0$ in  $\mu_t\approx 200$ GeV. Polne \v{c}rte se nana\v{s}ajo na operatorje brez eksplicitne mase kvarka top, prekinjena \v{c}rta pa se nana\v{s}a na Wilsonov coefficient operatorja $\widetilde{{\mathcal O}}_{LR}^{\gamma}$. $\widetilde{b}_{LR}^{Z}$ ni prikazan, saj na grafu odstopanje od $b_{LR}^{Z}$ ni opazno.}
\label{fig:RGE!}
\end{figure}

Ob predpostavki, da pod UV skalo nimamo dodatnih barvnih prostostnih stopenj, ki bi modificirale beta funkcijo, imamo $\beta_0=7$ za skale nad $\mu_t$. \v{C}e za operatorje $LR$ vzamemo redefinirano obliko $\widetilde{\boldsymbol{\mathcal O}}_{LR}$, se $\kappa_{1,2}$ spremenita v $\kappa_1=16/3\beta_0$, $\kappa_2=14/3\beta_0$. 

Posledice spreminjanja renormalizacijske skale ponazarja Sl.~\ref{fig:RGE!}, kjer prikazujemo 
\begin{eqnarray}
\bigg|\frac{b_i^{\gamma,Z} (\mu_t)}{b_i^{g} (\Lambda)}\bigg|\,,\hspace{0.5cm}\text{ko $b^{\gamma,Z}_i(\Lambda) = 0$}\,,
\end{eqnarray}
ki nam pove kolik\v{s}na $b^{\gamma,Z}_i(\mu_t)$ lahko generirano na skali mase kvarka top $\mu_t\simeq 200$ GeV, izklju\v{c}no z me\v{s}anjem operatorjev pod renomalizacijo QCD in prisotnostjo gluonskega operatorja na visoki skali $\Lambda$. 

Opazimo lahko, da inducirani prispevki $b^\gamma_{i}$ v primeru energijske skale NF  okoli $\Lambda \sim 2$ TeV, zna\v{s}ajo 10\% vredosti $b^g_{i}$ koeficienta generiranega na skali $\Lambda$. Po drugi strani so, zaradi od\v{s}tevanja podobnih prispevkov v ena\v{c}bah En.~(\ref{eq:rge!}), inducirani prispevki k $b^Z_{i}$ mnogo manj\v{s}i (pod $1\%$ na prikazanem razponu skale $\Lambda$). Vklju\v{c}itev eksplicitnega faktorja mase kvarka top v operatorje teh zaklju\v{c}kov ne spremeni. 

\begin{figure}[h]
\begin{center}
\includegraphics[scale=0.6]{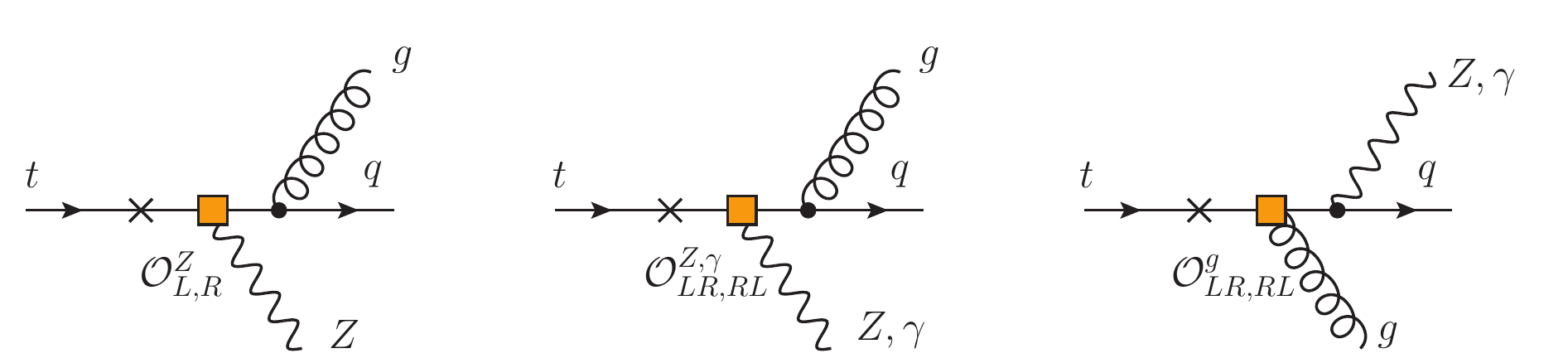}
\end{center}
\caption{Feynmanovi diagrami bremsstrahlung procesov $t\to qgZ,\gamma$. Kvadratki ozna\v{c}ujejo delovanje operatorja NF, kri\v{z}ci pa ozna\v{c}ujejo dodatne to\v{c}ke iz katerih se lahko izseva gluon (v prvih dveh diagramih) ali $Z,\gamma$ (v zadnjem diagramu).}
\label{fig:fcnc_brems!}
\end{figure}

Po analizi renormalizacijskih lastnosti operatorjev NF nam preostanejo \v{s}e $\alpha_s$ popravki matri\v{c}nih elementov $\bra{q \gamma} \mathcal O_i \ket{t}$ in  $\bra{q Z} \mathcal O_i \ket{t}$, ki jih ovrednotimo na skali mase kvarka top, in bremsstrahlung popravki, katerih Feynmanovi diagrami so prikazani na Sl.~\ref{fig:fcnc_brems!}.
Razpadne \v{s}irine teh diagramov so istega reda v $\alpha_s$ kot razpadne \v{s}irine dvo-del\v{c}nih kon\v{c}nih stanj. Se\v{s}tevek obeh prispevkov poskrbi, da je rezultat IR kon\v{c}en. Razpadno \v{s}irino s popravki reda $\alpha_s$ parametriziramo kot
\begin{eqnarray}
\Gamma^V &=& |a^V|^2\frac{v^4}{\Lambda^4} \Gamma_{a}^V + \frac{v^2 m_t^2}{\Lambda^4}|b^V|^2 \Gamma^V_{b}+\frac{v^3m_t}{\Lambda^4}2\mathrm{Re}\{b^{V*}a^V\} \Gamma^V_{ab} \label{eq:oso!}\\
&+&
\frac{v^3m_t}{\Lambda^4} \left[2\mathrm{Re}\{a^{V*}b^g\} \Gamma^V_{ag}- 2\mathrm{Im}\{a^{V*}b^g\}\tilde{\Gamma}^V_{ag}\right]\nn \\&+& 
\frac{v^2 m_t^2}{\Lambda^4}\left[ |b^g|^2 \Gamma^V_{g}+2\mathrm{Re}\{b^{V*}b^g\} \Gamma^V_{bg} -2\mathrm{Im}\{b^{V*}b^g\}\tilde{\Gamma}^V_{bg} \right]\,,\nonumber
\end{eqnarray}
kjer $V= Z,\gamma$ in $a^{\gamma}=0$. $\Gamma^V_{ag,bg,g}$ v drugi in tretji vrstici En.~(\ref{eq:oso!}) povzemajo prispevke gluonskega operatorja in so zato odsotni v prvem redu ($\alpha_s^0$) in se pojavijo \v{s}ele na redu $\alpha_s$. Analiti\v{c}ni izrazi vseh razpadnih \v{s}irin so podani v dodatku~\ref{app:allwidths}.

Kromodinamski popravki procesa s fotonom, ki nima mase, v kon\v{c}nem stanju so nekoliko bolj kompleksni kot v primeru bozona $Z$. Rezultat, ki ga dobimo iz obravnave predstavljenih diagramov virtualnih in bremsstrahulg korekcij, je IR divergenten. To divergenco lahko odstranimo, \v{c}e v obravanavo vklju\v{c}imo dodaten diagram za proces $t\to q g$ s fotonskim popravkom ene zanke. Ker pa eksperimentalno iskanje procesa $t\to q \gamma$ vselej vklju\v{c}uje detekcijo izoliranega fotona, k \v{c}emer omenjeni diagram ne prispeva, razpadno \v{s}irino raje regulariziramo z vpeljavo reza, ki zagotovi, da sta smeri fotona in lahkega kvarka ali gluona dovolj narazen $\nolinebreak{\delta r_j= 1- {\bf p}_\gamma \cdot {\bf p}_j / E_\gamma E_j}$, kjer $j=g,q$. Izka\v{z}e se, da je odvisnost rezultata od reza $\delta_q$ znatna in lahko privede do pove\v{c}anja prispevka gluonskega operatorja. Poleg tega vpeljemo \v{s}e eksperimentalno motiviran rez na energijo fotona $E_{\gamma}^{\mathrm{cut}}$. Razpadne \v{s}irine za FCNC razpade kvarka top v foton predstavimo kot funkcije definiranih rezov.

\begin{figure}[h!]
\begin{center}
\includegraphics[height=5.5cm]{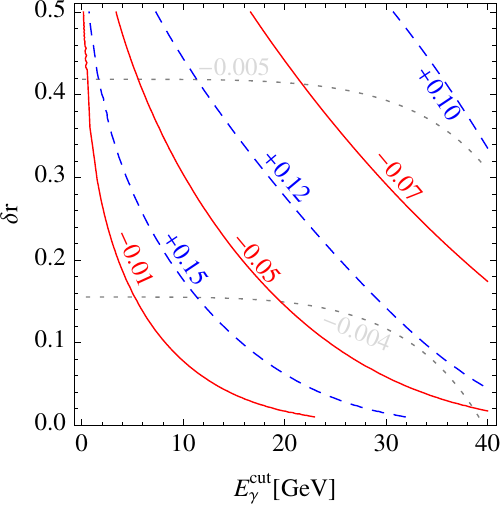}
\end{center}
\caption{Relativna velikost $\alpha_s$ popravkov k . Prikazana sta reprezentativna intervala za $\delta r_c\equiv\delta r$ and $E^{\mathrm{cut}}_\gamma$. Konture z dolo\v{c}eno velikostjo popravkov so narisane za $b^g=0$ (sivo, pike), $b^g=b^\gamma$ (rde\v{c}e) and $b^g = - b^\gamma$ (modro, \v{c}rte).}
\label{fig:foton_cuts!}
\end{figure}

\begin{figure}[h!]
\begin{center}
\includegraphics[scale=0.8]{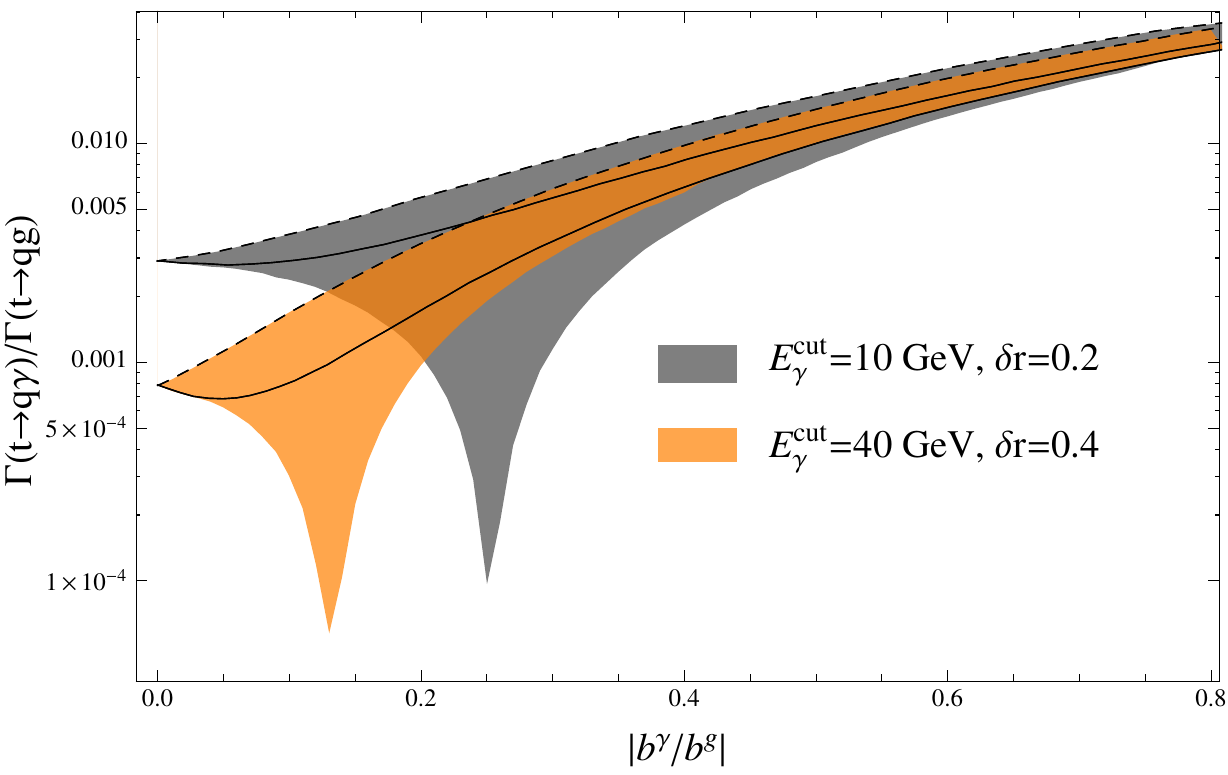}
\end{center}
\caption{\label{fig:tcg-tcg!} Razmerje razpadnih \v{s}irin $\Gamma(t\to q \gamma) / \Gamma(t\to q g )$ v odvisnosti od absolutne vrednosti razmerja relevantnih FCNC Wilsonovih koeficientov $|b^\gamma/b^g|$. Prikazani so rezultati za dve reprezentativni izbiri kinemati\v{c}nih rezov. Obmo\v{c}ja so kon\v{c}nih razse\v{z}nosti v vertikalni smeri, zaradi neznane relativne faze med $b^\gamma$ in $b^g$. Prikazane \v{c}rte so za maksimalno pozitivno (polni) in maksimalno negativno (\v{c}rtkano) interferenco $b^\gamma b^g$.}
\end{figure}

Numeri\v{c}no analizo prikazuje Sl.~\ref{fig:foton_cuts!}, kjer so narisane konture konstantne relativne velikosti $\alpha_s$ popravkov k $\mathrm{Br}(t\to q \gamma)$ v ravnini rezov. Opazimo, da so prispevki gluonskega operatorja lahko reda $10-15\%$ celotne razpadne \v{s}irine, odvisno od relativne faze in velikosti Wilsonovih koeficientov operatorjev $\mathcal O_{LR,RL}^{g,\gamma}$. To pomeni, da eksperimentalna meja na $\mathrm{Br}(t\to q \gamma)$ lahko dejansko omejuje tako $b^{\gamma}$ kot $b^{g}$. To opa\v{z}anje lahko dodatno podkrepimo z analizo razmerja $\Gamma(t\to q \gamma) / \Gamma(t\to q g )$, kjer obe razpadni \v{s}irini izra\v{c}unamo do reda $\alpha_s$\footnote{$\Gamma(t\to q g)$ s kromodinamskimi popravki je vzeta iz Ref.~\cite{Zhang:2008yn}. }, v odvisnosti od razmerja relevantnih FCNC Wilsonovih koeficientov $|b^\gamma/b^g|$, kar prikazuje Sl.~\ref{fig:tcg-tcg!} za dve reprezentativni izbiri kinemati\v{c}nih rezov. Vertikalna dimenzija se oblikuje ob spreminjanju relativne faze med koeficientoma $b^\gamma$ in $b^g$.

V primeru razpada $t \to q Z$ so popravki QCD precej manj dramati\v{c}ni. Signifikanco popravkov povzamemo v Tab.~\ref{tab:brsZ!}, kjer navajamo relativno spremembo v razpadnih \v{s}irinah in razvejitvenih razmerjih ob prehodu iz reda $\alpha_s^0$ na red $\alpha_s$.
\begin{table}[!h]
\begin{center}
\begin{tabular}{l|l|l|l||l|l}\hline\hline
	&$b^Z=b^g=0$&$a^Z=b^g=0$&$a^Z=b^Z, b^g=0$& $b^Z=0, a^Z =b^g$ &$a^Z=0, b^Z =b^g$\\
\hline
$\Gamma^{\mathrm{NLO}}/\Gamma^{\mathrm{LO}}$&$0.92$& $0.91$& $0.92$ & $0.95$&$0.94$\\
$\Brfrac$&$1.001$&$0.999$&$1.003$ &$1.032$&$1.022$\\
\hline\hline
\end{tabular}
\caption{Numeri\v{c}ne vrednosti razmerji $\Gamma^{\mathrm{NLO}}/\Gamma^{\mathrm{LO}}$ in $\mathrm{Br}^{\mathrm{NLO}}(t\to q Z)/\mathrm{Br}^{\mathrm{LO}}(t\to q Z)$ za dolo\v{c}ene vrednosti FCNC Wilsonovih koeficientov. }
\label{tab:brsZ!}
\end{center}
\end{table}

Opazimo, da relativna sprememba razpadnih \v{s}irin lahko dose\v{z}e $10\%$, sprememba v razvejitvenih razmerjih pa je mnogo manj\v{s}a. Razlog za to je skoraj\v{s}nje popolno izni\v{c}enje $\alpha_s$ prispevkov k razpadni \v{s}irini $t\to q Z$ in razpadni \v{s}irini glavnega razpadnega kanala $t\to Wb$, na katerega so razvejitvena razmerja normirana. V kolikor bi tak rezultat pri\v{c}akovali za prispevke $a^Z$, je podoben rezultat za $b^Z$ netrivialen. Vidimo tudi, da pri dolo\v{c}enih faznih odnosih med $b^Z$ in $b^g$ sprememba lahko naraste na par procentov.

\subsection{Tro-del\v{c}ni razpadi}

V tem poglavju se osredoto\v{c}imo na razpad $t \to q \ell^+ \ell^-$, kjer FCNC tranzicija poteka preko istih fotonskih in $Z$ operatorjev kot v prej\v{s}njem poglavju, nadalje pa se bozon sklaplja s parom nabitih leptonov, ki jih detektiramo v kon\v{c}nem stanju. Osrednji cilj te analize je potencialna diskriminacija med razli\v{c}nimi FCNC operatorji na podlagi razli\v{c}nih kinemati\v{c}nih opazljivk, ki jih lahko definiramo na ra\v{c}un ve\v{c}jega tro-del\v{c}nega faznega prostora. Za nas bodo zlasti zanimive asimetrije definirane na podlagi smeri, pod katerimi se gibajo delci kon\v{c}nega stanja, ki so lahko senzitivne na obliko FCNC vozli\v{s}ca.

\begin{figure}[h]
\begin{center}
\includegraphics[scale=0.7]{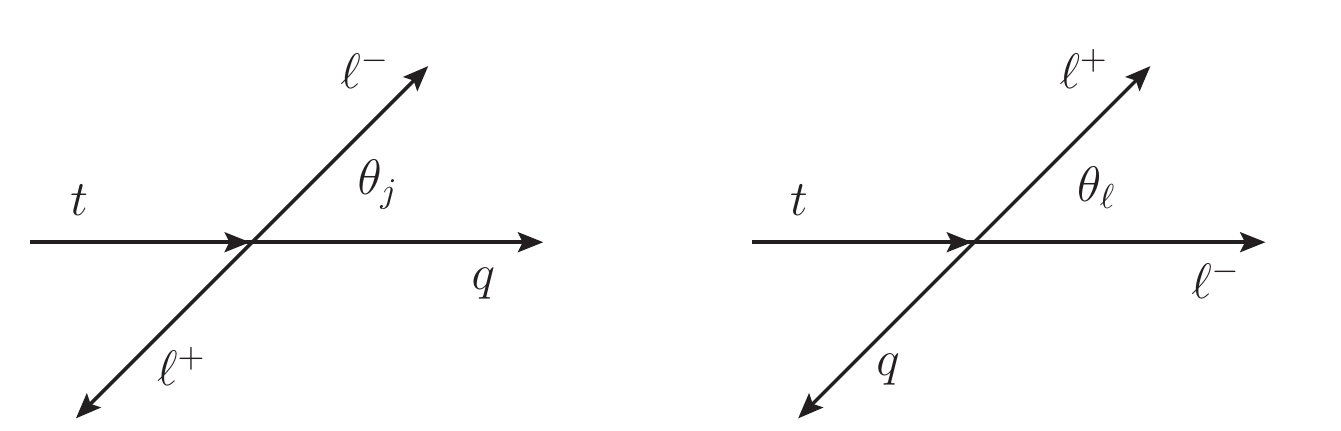}
\caption{\label{fig:angles!} Definicija dveh kotov na podlagi katerih vpeljemo dve razli\v{c}ni kinemati\v{c}ni asimetriji. Pu\v{s}\v{c}ice ozna\v{c}ujejo smeri gibalnih koli\v{c}in delcev.}
\end{center}
\end{figure}

V mirovnem sistemu leptonskega para definiramo $z_j=\cos\theta_{j}$, ki se nana\v{s}a na smeri negativno nabitega leptona in lahkega kvarka. V mirovnem sistemu pozitivno nabitega leptona in lahkega kvarka pa definiramo $z_{\ell}=\cos\theta_{\ell}$, ki se nana\v{s}a na smeri nabitih leptonov. Obe definiciji ponazarja Sl.~\ref{fig:angles!}

S pomo\v{c}jo tako definiranih kotov vpeljemo dve asimetriji
\begin{equation}
A_i = \frac{\Gamma_{z_i>0}-\Gamma_{z_i<0}}{\Gamma_{z_i>0} + \Gamma_{z_i<0}}\,.
\label{eq:As!}
\end{equation}
$A_{j}\equiv A_{\mathrm{FB}}$ imenujemo {\sl asimetrija naprej-nazaj} (FBA), $A_{\ell}\equiv A_{\mathrm{LR}}$ pa proglasimo za {\sl asimetrijo levo-desno } (LRA).

Analize asimetrij se lotimo v treh korakih. Najprej si ogledamo razpade, ki potekajo preko izmenjave fotona, nato razpade, ki potekajo preko bozona $Z$, na koncu pa analiziramo \v{s}e razpade, ki potekajo preko obeh kanalov hkrati, kar se zdi vredno raziskati, saj \v{s}tevilni modeli NF lahko generirajo znatne FCNC razpade tako v fotone kot v bozone $Z$. Skupno leptonsko kon\v{c}no stanje nam omogo\v{c}a raziskovanje interference obeh pojavov. Analiti\v{c}ne formule so podane v poglavju~\ref{chap:neutral_currents} in dodatku~\ref{app:allTB}.

\begin{figure}[h]
\begin{center}
\includegraphics[scale=0.45]{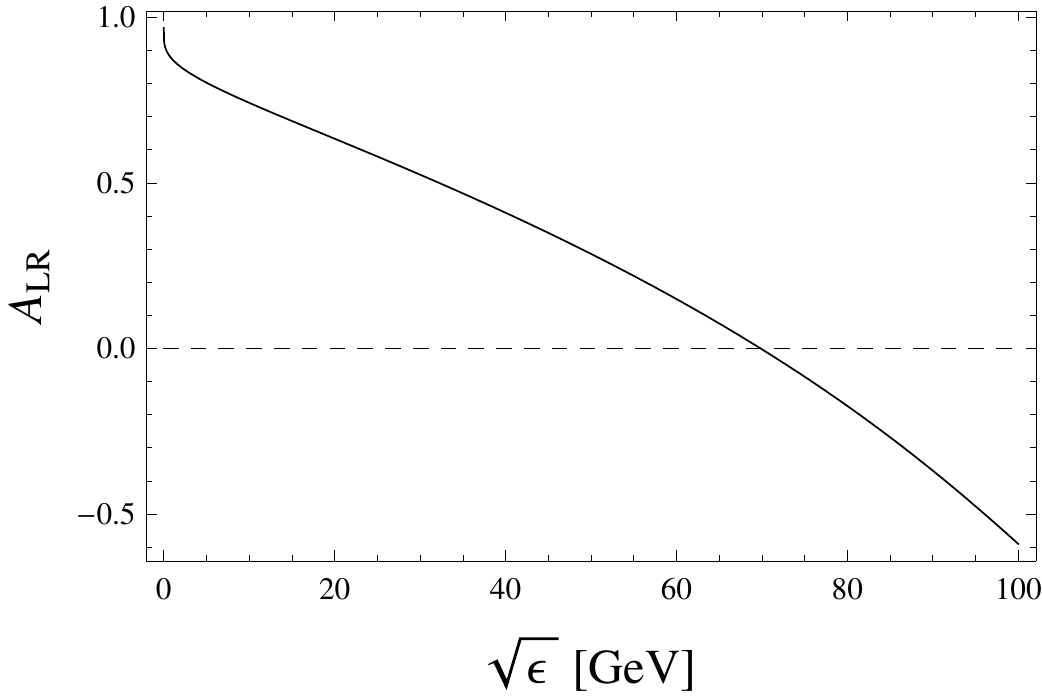}\hspace{0.5cm}
\includegraphics[scale=0.48]{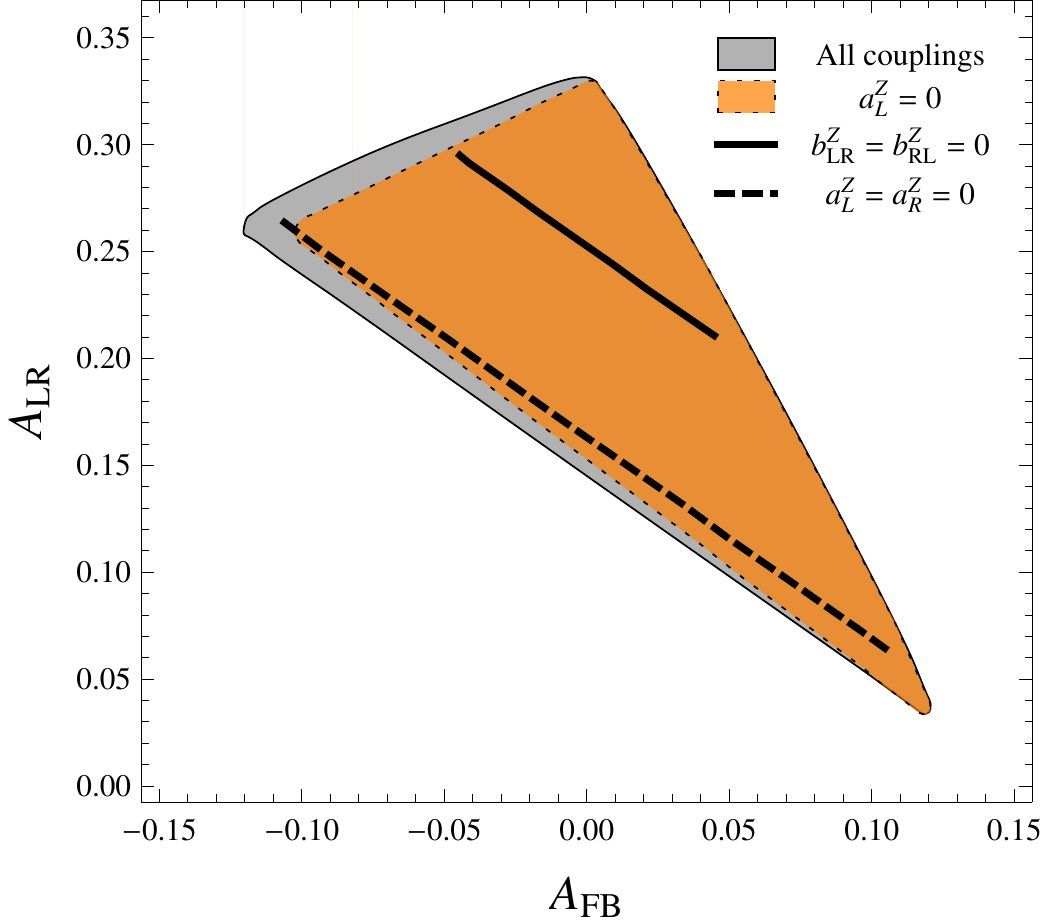}\hspace{0.5cm}
\includegraphics[scale=0.48]{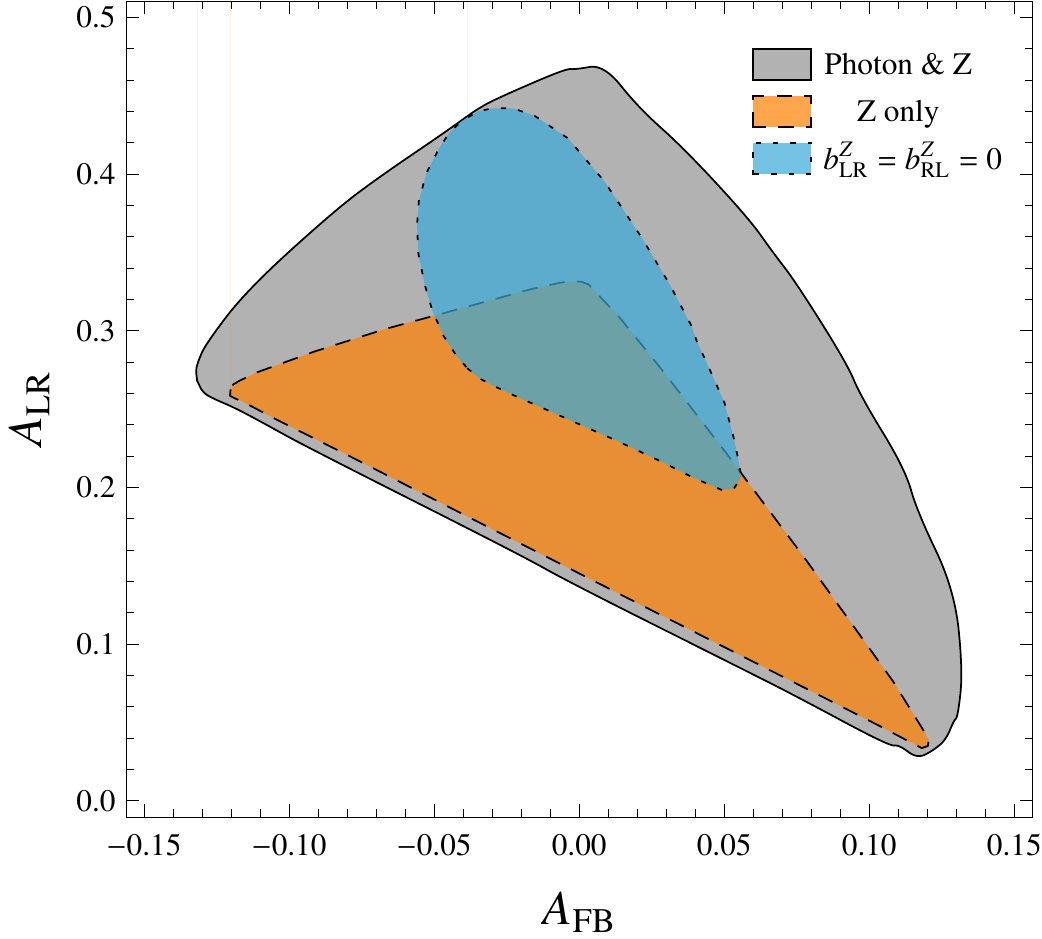}
\caption{\label{fig:assym_Z!}{\bf Levo}: Odvisnost LRA od reza na spodnjo mejo invariantne mase leptonskega para $\epsilon$ za primer razpadov preko fotona. {\bf Sredina}: Korelacija med FBA in LRA za razpade, ki potekajo preko bozona $Z$. Sivo obmo\v{c}je je dobljeno, ko so dovoljene vse sklopitve, ostala obmo\v{c}ja pa, ko so dolo\v{c}ene sklopitve postavljene na ni\v{c}. {\bf Desno:} Korelacija med FBA in LRA za razpade preko fotona in bozona $Z$, vklju\v{c}no z interferen\v{c}nimi prispevki. Sivo obmo\v{c}je je dobljeno, ko so neni\v{c}elne lahko vse FCNC sklopitve s fotonom in bozonom $Z$, ostala obmo\v{c}ja pa, ko so dolo\v{c}ene sklopitve postavljene na ni\v{c}.
}
\end{center}
\end{figure}

Zaradi izklju\v{c}no vektorske sklopitve fotona z nabitimi leptoni, v primeru fotonske mediacije razpada FBA ne more zavzeti neni\v{c}elne vrednosti. Po drugi strani se LRA izka\v{z}e za neni\v{c}elno in povsem neodvisno od paramterov nove fizike. Vsebuje pa mo\v{c}no odvisnost od reza na invariantno maso leptonskega para $\sqrt{\epsilon}$, ki jo prikazuje levi graf na Sl.~\ref{fig:assym_Z!}. Vidimo, da je glede na razli\v{c}ne kinemati\v{c}ne reze, LRA lahko celo razli\v{c}no predzna\v{c}ena.
V kolikor FCNC proces poteka preko bozona $Z$, tudi FBA zavzame netrivialne vrednosti. Obe asimetriji sedaj postaneta odvisni od parametrov NF. S pomo\v{c}jo \v{z}reba naklju\v{c}nih vrednosti parametrov NF lahko prei\v{s}\v{c}emo razpon vrednosti asimetrij in korelacijo med FBA in LRA. To prikazujeta srednji in desni graf Sl.~\ref{fig:assym_Z!}. Srednji graf se nana\v{s}a na razpade, ki potekajo izklju\v{c}no preko bozona $Z$, medtem ko desni graf vsebuje tako razpade preko bozona $Z$ kot tudi razpade preko fotona in interferen\v{c}ne prispevke obeh procesov.

Vidimo lahko, da so velike vrednosti FBA ($|A_{\mathrm{FB}}|\gg 0.1$) nedosegljive v razpadih, ki potekajo preko bozona $Z$, kjer je le ta omejena na interval $A_{\mathrm{FB}}\in[-0.12,0.12]$. Eksperimentalno izmerjena to\v{c}ka v ravnini $(A_{\mathrm{FB}},A_{\mathrm{LR}})$ bi lahko slu\v{z}ila za izklju\v{c}evanje modelov, ki generirajo le dolo\v{c}ene vrste efektivnih FCNC sklopitev. \v{C}e obravnavamo razpade preko bozona $Z$ in fotona kot nelo\v{c}ljiva se razpon dosegljivih vrednosti LRA bistveno pove\v{c}a v smeri pozitivnih vrednosti.


Predstavljeno analizo razpadov $t\to q \ell^+ \ell^-$ lahko v prihodnosti soo\v{c}imo z eksperimentalnim iskanjem $t\to q Z$, kjer identifikacija bozona $Z$ poteka preko detekcije leptonskega para, v katerega bozon $Z$ razpade. Po drugi strani je, kot smo \v{z}e omenili, eksprimentalno iskanje $t\to q \gamma$ vezano na detekcijo izoliranega fotona. V kolikor bi se \v{z}eleli oddaljiti od teh omejitev in se osredoto\v{c}iti na leptonsko kon\v{c}no stanje, bi bila potrebna nova podrobna analiza ozadij takega kon\v{c}nega stanja.


\section{NF v razpadih kvarka top: Nabiti tokovi}
\subsection{Uvod}
V tem poglavju se posvetimo nabitim tokovom, ki vsebujejo kvarke top, in analiziramo posledice odstopanja od SM v le teh. V analizi sledimo konceptu orisanem v uvodnem poglavju in najprej analiziramo implikacije v fiziki mezonov $B$. \v{S}ele nato se posvetimo glavnemu razpadnemu kanalu kvarka top in su\v{c}nostnim dele\v{z}em, ki so ob\v{c}utljivi na omenjene spremembe v nabitih tokovih.

Pri oblikovanju baze operatorjev dimenzije 6, s katerimi raz\v{s}irimo SM, najprej poi\v{s}\v{c}emo vse strukture, ki so invariantne na umeritveno grupo SM in vsebujejo nabite tokove s kvarkom top. Nato dolo\v{c}imo \v{s}e okusno strukturo teh operatorjev, pri \v{c}emer se omejimo na okvir {\sl minimalne kr\v{s}itve okusa} (MFV)~\cite{Buras:2003jf,D'Ambrosio:2002ex,Grossman:2007bd}, v katerem edina kr\v{s}itev okusa izvira iz Yukawinih sklopitev, kakor v SM. To nas privede do slede\v{c}e baze sedmih efektivnih operatorjev
\begin{subequations}
\label{eq:ops1!}
\begin{eqnarray}
 \mathcal Q_{RR}&=& V_{tb} [\bar{t}_R\gamma^{\mu}b_R] \big(\phi_u^\dagger\ii D_{\mu}\phi_d\big) \,, \\
 \mathcal Q_{LL}&=&[\bar Q^{\prime}_3\tau^a\gamma^{\mu}Q'_3] \big(\phi_d^\dagger\tau^a\ii D_{\mu}\phi_d\big)-[\bar Q'_3\gamma^{\mu}Q'_3]\big(\phi_d^\dagger\ii D_{\mu}\phi_d\big),\\
 \mathcal Q'_{LL}&=&[\bar Q_3\tau^a\gamma^{\mu}Q_3] \big(\phi_d^\dagger\tau^a\ii D_{\mu}\phi_d\big) -[\bar Q_3\gamma^{\mu}Q_3]\big(\phi_d^\dagger\ii D_{\mu}\phi_d\big),\\
 \mathcal Q^{\prime\prime}_{LL}&=&[\bar Q'_3\tau^a\gamma^{\mu}Q_3] \big(\phi_d^\dagger\tau^a\ii D_{\mu}\phi_d\big)-[\bar Q'_3\gamma^{\mu}Q_3]\big(\phi_d^\dagger\ii D_{\mu}\phi_d\big),\\
 \mathcal Q_{LRt} &=& [\bar Q'_3 \tau^a\sigma^{\mu\nu} t_R]{\phi_u}W_{\mu\nu}^a \,,\\
\mathcal Q'_{LRt} &=& [\bar Q_3 \tau^a\sigma^{\mu\nu} t_R]{\phi_u}W_{\mu\nu}^a \,,\\
\mathcal Q_{LRb} &=& [\bar Q_3 \tau^a\sigma^{\mu\nu} b_R]\phi_d W_{\mu\nu}^a \,,
\end{eqnarray}
\end{subequations}
kjer smo definirali $SU(2)_L$ dublete
\begin{eqnarray}
Q_3=(V^*_{kb} u_{Lk},b_{L})^{\mathrm{T}}\,, \hspace{0.5cm} \bar Q'_3 =\bar Q_i V^*_{ti}= (\bar{t}_L,V_{ti}^*\bar{d}_{iL})^T\,,
\end{eqnarray}
kovariantne odvode ter tenzor elektro\v{s}ibkih polj
\begin{eqnarray}
D_{\mu}&=&\partial_{\mu}+\ii \frac{g}{2}W_{\mu}^a\tau^a +\ii \frac{g'}{2}B_{\mu} Y\,,  \\
W^a_{\mu\nu}&=&\partial_{\mu}W_{\nu}^a-\partial_{\nu}W_{\mu}^a - g\epsilon_{abc}W_{\mu}^b W_{\nu}^c\,,\nn
\end{eqnarray}
in kon\v{c}no skalarna polja $\phi_{u,d}$ (v SM $\phi_u\equiv \tilde{\phi} =\ii \tau^2 \phi_d^*$).

\subsection{Indirektne posledice v fiziki mezonov $B$}
\begin{figure}[h]
\begin{center}
\includegraphics[scale=0.6]{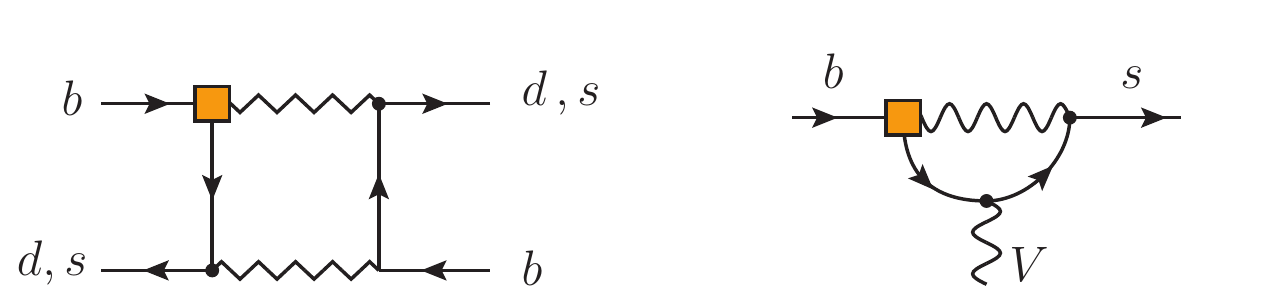}
\caption{Primer dveh diagramov, kjer prispevki efektivnih operatorjev~(\ref{eq:ops1!}), ozna\v{c}eni z ora\v{z}nim kvadratkom, vplivajo na proces me\v{c}anja mezonv $B$ preko \v{s}katlastega diagrama in razpade mezona $B$ preko pingvinskega diagrama. $V$ ozna\v{c}uje foton ali gluon, kvarki v zankah pa so $u,c,t$.}
\label{fig:dp!}
\end{center}
\end{figure}

Prisotnost operatorjev~(\ref{eq:ops1!}) spremeni izraze Wilsonovih koeficientov za $|\Delta B|=2$ in $|\Delta B|=1$ procese, saj kot prikazuje Sl.~\ref{fig:dp!} operatorji vstopijo v me\v{s}alne diagrame in diagrame za redke razpade mezonov $B$. Vselej obravnavamo vnos le enega operatorja v diagram, s \v{c}imer se konsistentno omejimo na prispevke NF ute\v{z}ene z $1/\Lambda^2$, vi\v{s}je potence pa zanemarimo.
Ko izra\v{c}unamo vse diagrame in s tem opravimo ujemanje med na\v{s}o raz\v{s}irjeno teorijo na efektivne Lagrangeove funkcije~(\ref{fig:dp!}), lahko parametriziramo efekte NF s spremembo Wilsonovih koeficientov
\begin{eqnarray}
C_i(\mu)  &=& C_i^{\mathrm{SM}}(\mu) + \delta C_i(\mu)\,, \label{eq:aaa!}\\
\delta C_i(\mu) &=&\sum_{j}\kappa_j(\mu) F_i^{(j)}(x_t,\mu) + \kappa_j^{*}(\mu) \tilde{F}^{(j)}_i (x_t,\mu)\,,
\end{eqnarray}
kjer $j=1,...,6$ in te\v{c}e po operatorjih~(\ref{eq:ops1!}), funkcije $F$, ki so odvisne od $x_t = m_t^2/m_W^2$ in skale na kateri smo opravili ujemanje $\mu$, pa so podane v dodatkih \ref{app:NP_D_B_2} za $|\Delta B| = 2$ procese (ozna\v{c}nene s $S^j$) in \ref{app:SM_D_B_1} za $|\Delta B|=1$ procese (ozna\v{c}ene s $f^j$). Definiramo tudi renormirane Wilsonove koeficiente NF, ki jih bomo uporabljali v nadaljnji analizi
\begin{eqnarray}
\kappa_{LL}^{(\prime,\prime\prime)}&=&\frac{C_{LL}^{(\prime,\prime\prime)}}{\Lambda^2\sqrt{2}G_F}\,,\hspace{0.3cm}
\kappa_{RR}=\frac{C_{RR}}{\Lambda^2 2\sqrt{2} G_F}\,,\hspace{0.3cm}
\kappa_{LRb}=\frac{C_{LRb}}{\Lambda^2 G_F}\,,\hspace{0.3cm}
\kappa_{LRt}^{(\prime)}=\frac{C_{LRt}^{(\prime)}}{\Lambda^2 G_F}\,.
\end{eqnarray}

S pomo\v{c}jo podrobnih \v{s}tudij, kako spremembe v Wilsonovih koeficientih~(\ref{eq:aaa!}) vplivajo na opazljivke v fiziki mezonov $B$, ki so bile opravljene v Ref.~\cite{Lenz:2010gu,Lenz:2012az,DescotesGenon:2011yn,Benzke:2010tq,Asner:2010qj, Huber:2007vv}, lahko izpeljemo omejitve na parametre $\kappa_j$. V kolikor se omejimo na realne $\kappa_j$, lahko izpeljemo intervale, v okviru katerih se nahajajo parametri $\kappa_j$ s 95\% {\sl stopnjo zaupnja} (C.L.), ki jih prikazuje Tab.~\ref{tab:bounds!}. 
\begin{table}[h]
\hspace{-1cm}
\begin{center}
\begin{tabular}{c|ccc|c|c}\hline\hline
&$B-\bar{B}$&$B\to X_s\gamma$&$B\to X_s \mu^{+}\mu^-$ & skupno & $C_i(2m_W)\sim 1$ \\
\hline
\LINE{$\kappa_{LL}$}{$\bs{0.08}{-0.09}$}
{$\bs{0.03}{-0.12}$}
{$\bs{0.48}{-0.49}$}
{$\bs{0.04}{-0.09}\Big(\bs{0.03}{-0.10}\Big)$}& $\Lambda> 0.82\,\, \mathrm{TeV}$\\\hline
\LINE{$\kappa_{LL}^{\pr}$}{$\bs{0.11}{-0.11}$}
{$\bs{0.17}{-0.04}$}
{$\bs{0.31}{-0.30}$}
{$\bs{0.11}{-0.06}\Big(\bs{0.10}{-0.06}\Big)$}& $\Lambda> 0.74\,\, \mathrm{TeV}$\\\hline
\LINE{$\kappa_{LL}^{\pr\pr}$}{$\bs{0.18}{-0.18}$}
{$\bs{0.06}{-0.22}$}
{$\bs{1.02}{-1.04}$}
{$\bs{0.08}{-0.17}\Big(\bs{0.05}{-0.15}\Big)$}& $\Lambda> 0.60\,\, \mathrm{TeV}$\\\hline
\LINE{$\kappa_{RR}$}{}
{$\bs{0.003}{-0.0006}$}
{$\bs{0.68}{-0.66}$}
{$\bs{0.003}{-0.0006}\Big(\bs{0.002}{-0.0006}\Big)$}& $\Lambda> 3.18\,\, \mathrm{TeV}$\\\hline
\LINE{$\kappa_{LRb}$}{}
{$\bs{0.0003}{-0.001}$}
{$\bs{0.34}{-0.35}$}
{$\bs{0.0003}{-0.001}\Big(\bs{0.003}{-0.01}\Big)$}& $\Lambda> 9.26\,\, \mathrm{TeV}$\\\hline
\LINE{$\kappa_{LRt}$}{$\bs{0.13}{-0.14}$}
{$\bs{0.51}{-0.13}$}
{$\bs{0.38}{-0.37}$}
{$\bs{0.13}{-0.07}\Big(\bs{0.12}{-0.14}\Big)$}& $\Lambda> 0.81\,\, \mathrm{TeV}$\\\hline
\LINE{$\kappa_{LRt}^{\pr}$}{$\bs{0.29}{-0.29}$}
{$\bs{0.41}{-0.11}$}
{$\bs{0.75}{-0.73}$}
{$\bs{0.27}{-0.07}\Big(\bs{0.25}{-0.06}\Big)$}& $\Lambda> 0.56\,\, \mathrm{TeV}$\\\hline\hline
\end{tabular}
\end{center}
\caption{Zgornje in spodnje $95\%$ C.L. meje za realne dele $\kappa_j$ in $\mu = 2 m_W$ ter $\mu=m_W$ (v oklepaju). Zadnji stolpec prikazuje oceno spodnje meje na skalo NF $\Lambda$, v primeru da je $C_j\sim 1$.}
\label{tab:bounds!}
\end{table}
Omeniti velja, da so meje za parametra $\kappa_{RR}$ in $\kappa_{LRb}$ za red velikosti ostrej\v{s}e kot za ostale parametre. Meji izvirata iz analize $b\to s \gamma$ razpada, kjer so prispevki operatorjev $\op_{RR}$ in $\op_{LRb}$, ki vsebujeta desno-ro\v{c}ne kvarke $b$, efektivno pove\v{c}ani za faktor $m_{W,t}/m_b$ glede na prispevke ostalih operatorjev.

\begin{figure}[h!]
\begin{center}
\includegraphics[scale=0.5]{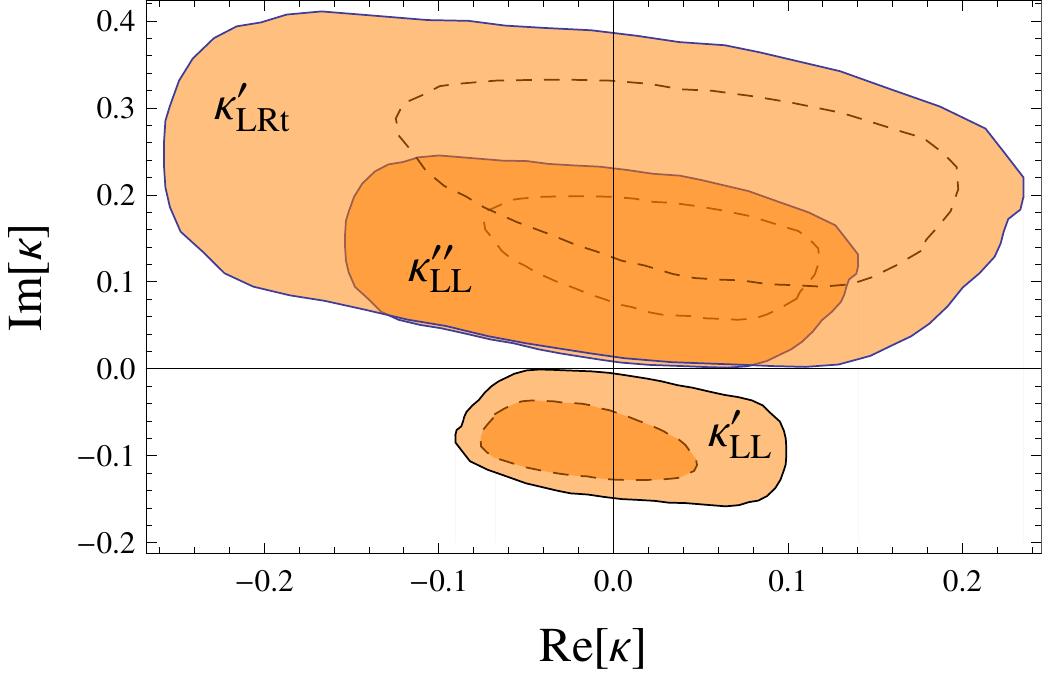}\hspace{1.5cm}
\includegraphics[scale=0.5]{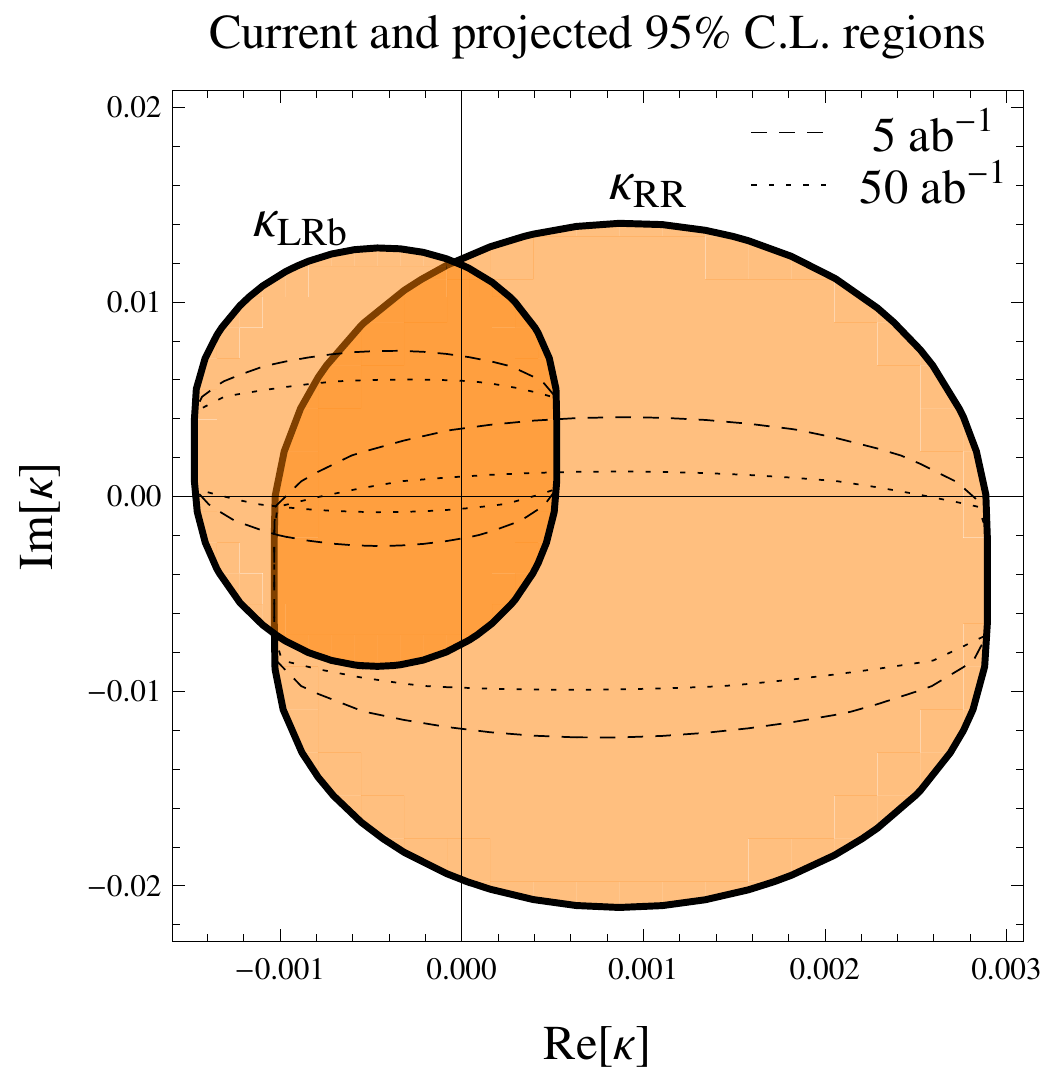}
\caption{{\bf Levo}: 95\% C.L. (polna \v{c}rta) in 68\% C.L. dovoljena obmo\v{c}ja za parametre $\kappa_{LL}^{\prime(\prime\prime)},\kappa_{LRt}^{\prime}$, dobljena analize me\v{s}anja mezonov $B$. {\bf Desno}: 95\% C.L. dovoljena obmo\v{c}ja za parametra $\kappa_{LRb}$ in $\kappa_{RR}$, dobljena iz analize CP kr\v{s}itve v $b\to s\gamma$ razapadih. Nepolne \v{c}rte ozna\v{c}ujejo potencialno zo\v{z}itev obmo\v{c}ji ob izbol\v{s}avi eksperimentalnih napak na Super-Bell.}
\label{fig:complex!}
\end{center}
\end{figure}

V primeru, da sprostimo zahtevo po realnosti $\kappa_j$ in dovolimo, da imajo tudi imaginarno komponento, lahko s pomo\v{c}jo analize efektov v me\v{s}anju in kr\v{s}itve {\sl simetrije parnosti in konjugacije naboja} (CP) v razpadih $b\to s \gamma$ izpeljemo dovoljena obmo\v{c}ja v kompleksnih ravninah parametrov $\kappa_j$. To prikazuje Sl.~\ref{fig:complex!} za vse koeficiente razen $\kappa_{LRt}$ in $\kappa_{LL}$, katerih imaginarnih komponent ne moremo omejiti preko efektov v me\v{s}anju, kjer prispevata samo z realnimi deli, niti preko CP kr\v{s}itve v razpadih $b\to s\gamma$, kjer so njuni prispevki tako majhni, da omejitve niso mogo\v{c}e.
V primeru izpeljave mej iz CP kr\v{s}itve v razpadih $b\to s\gamma$ prika\v{z}emo tudi projekcijo, koliko lahko pri\v{c}akujemo, da se bo dovoljeno obmo\v{c}je \v{s}e skr\v{c}ilo v prihodnosti, ko bo meritve izbolj\v{s}al Super-Bell~\cite{Browder:2008em,Aushev:2010bq}.

\begin{figure}[h!]
\begin{center}
\includegraphics[scale= 0.45]{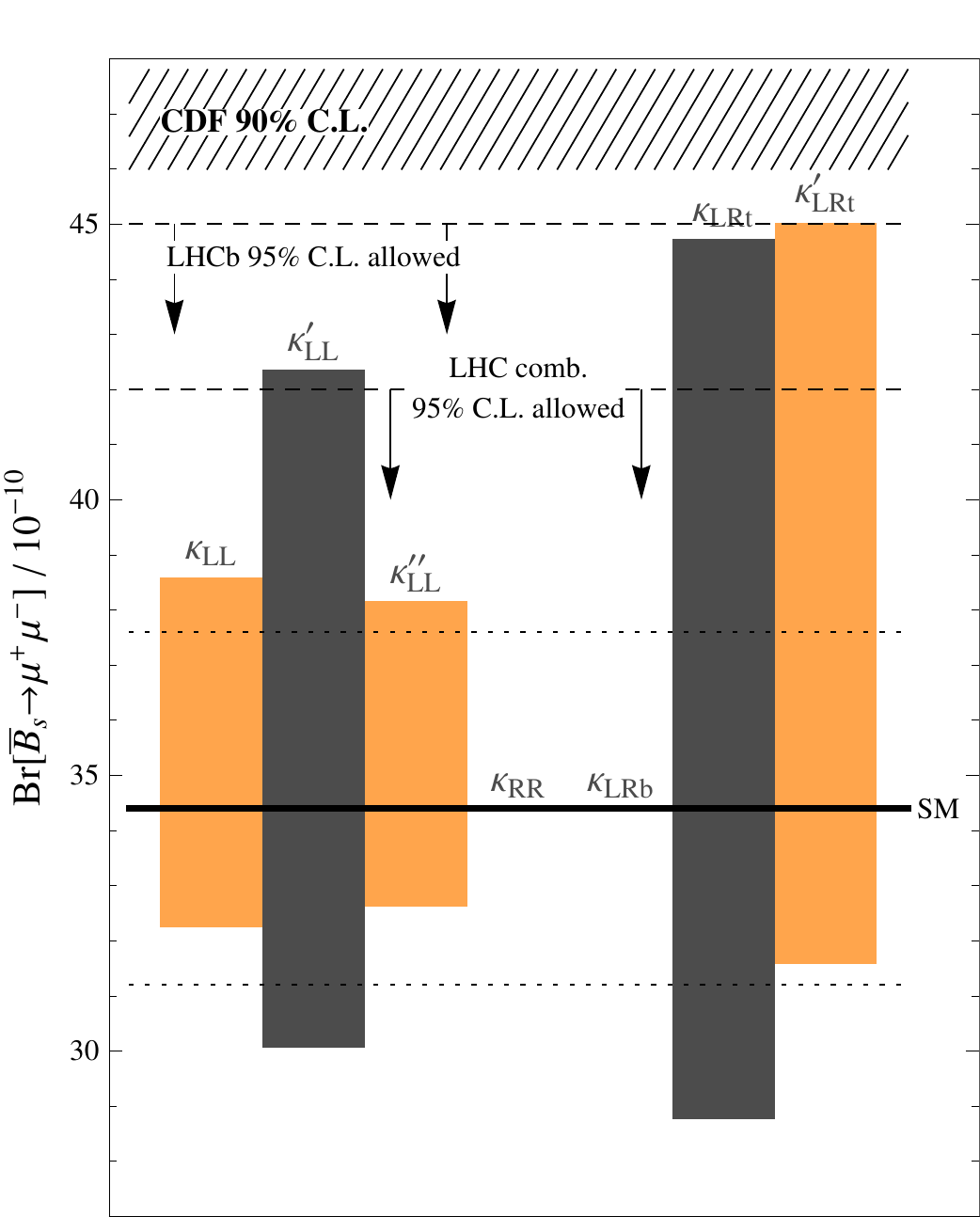}\hspace{0.8cm}
\includegraphics[scale=0.42]{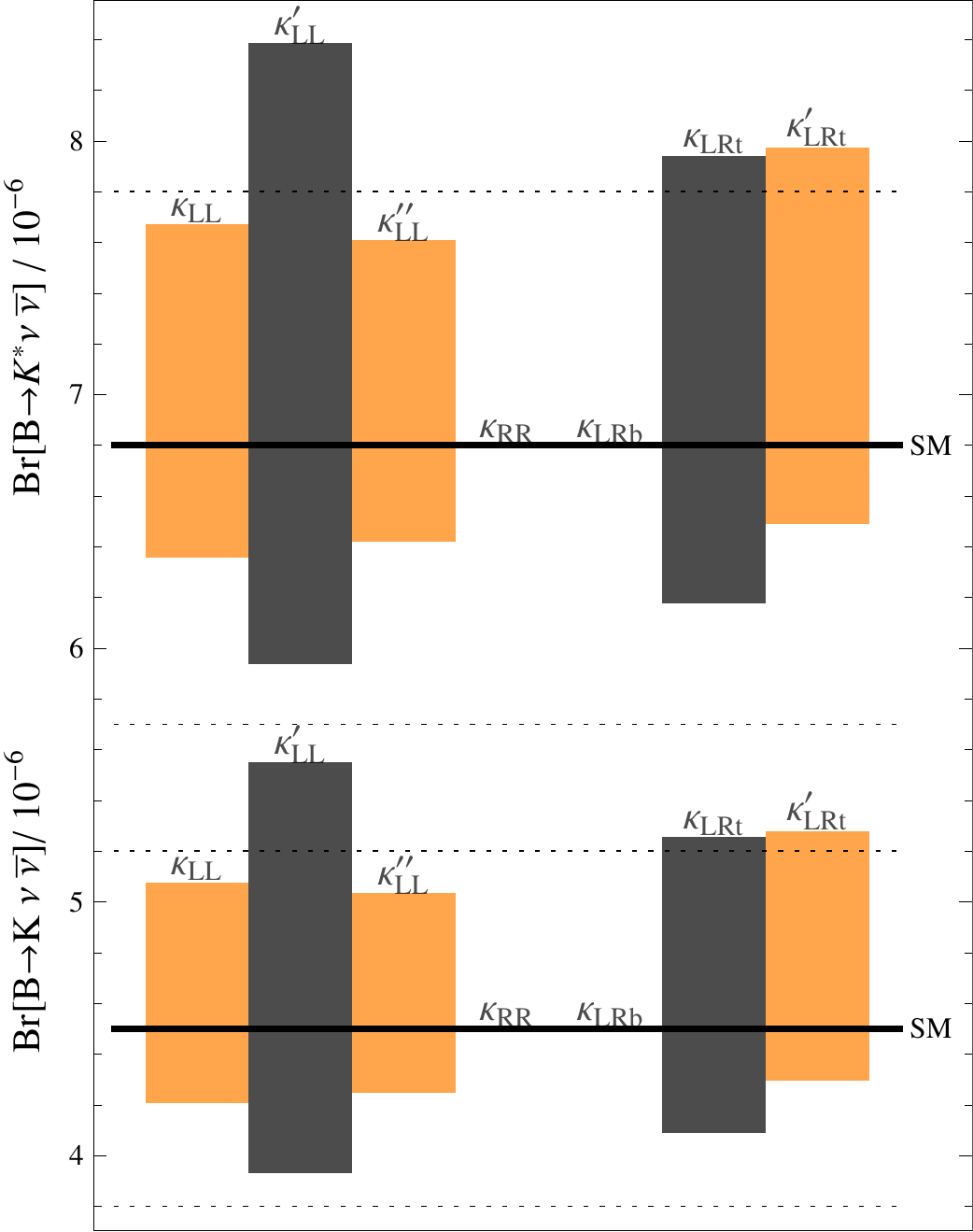}\hspace{0.5cm}
\includegraphics[scale= 0.4]{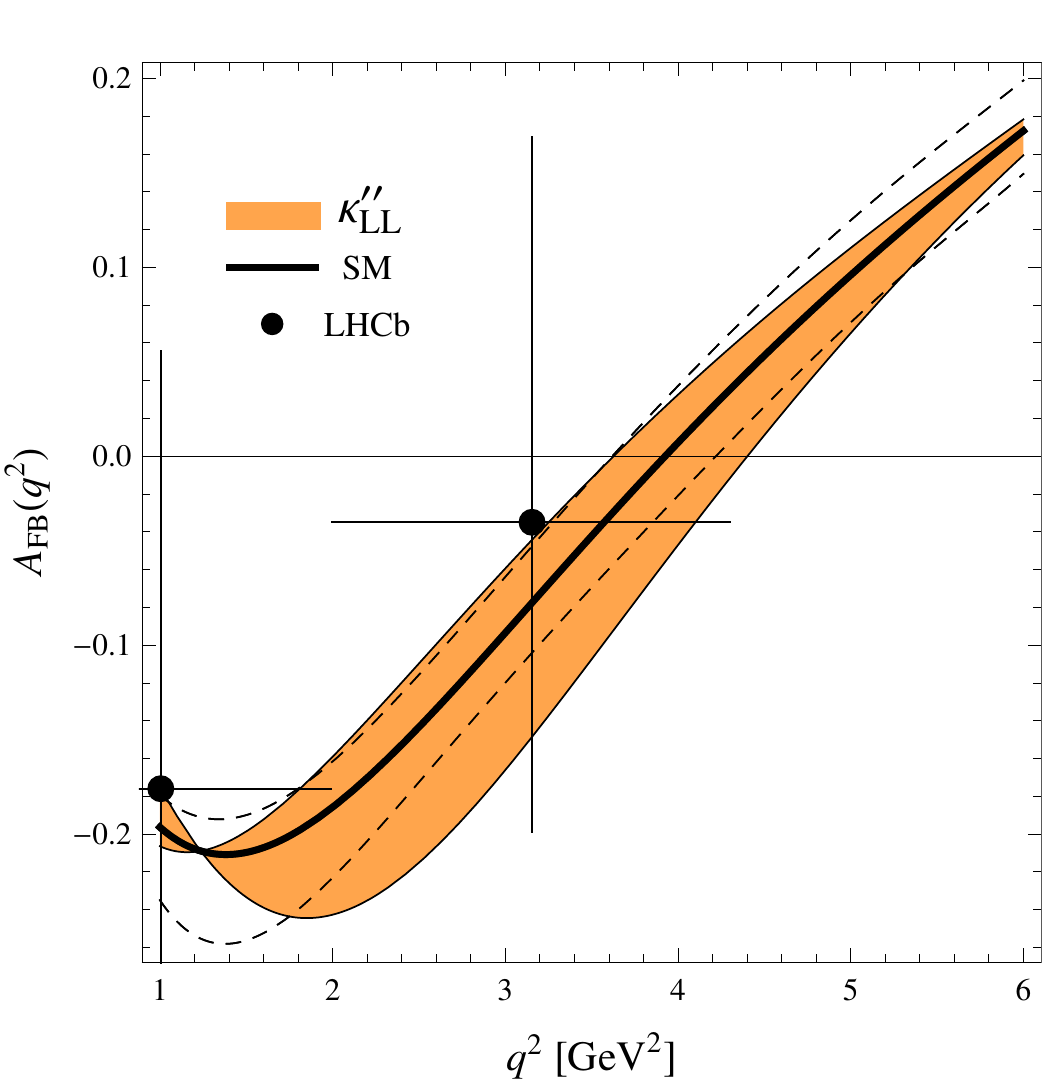}\hspace{0.8cm}
\caption{{\bf Levo, Sredina}: Razpon razvejitvenih razmerij, ko anomalne parametre $\kappa_j$ spreminjamo znotraj 95\% C.L. dovoljenih intervalov prikazanih v Tab.~\ref{tab:bounds!}. Pik\v{c}aste \v{c}rte predstavljajo $1\sigma$ teoreti\v{c}ne negotovosti SM napovedi. Za razpad v mione prikazujemo tudi spodnjo eksperimentalno mejo $90\%$ C.L. intervala~\cite{Aaltonen:2011fi}, 95\% C.L. zgornjo mejo LHCb~\cite{Aaij:2012ac} in najnovej\v{s}o zdru\v{z}eno LHC meritev \cite{CMS:Bsmumu}.
{\bf Desno}: $A_{\mathrm{FB}}(q^2)$ pas, ki ga dobimo, ko $\kappa_{LL}^{\pr\pr}$ spreminjamo znotraj 95\% dovoljenih intervalov prikazanih v Tab.~\ref{tab:bounds!}. Prikazana je tudi srednja vrednost napovedi SM (\v{c}rna) in pas $1\sigma$ teoreti\v{c}ne negotovosti (\v{c}rtkano) ter eksperimentalno izmerjeni toc\v{c}ki z pripadajo\v{c}imi napakami iz Ref.~\cite{Aaij:2011aa}.}
\label{fig:predict1!}
\end{center}
\end{figure}

Nadalje lahko rezultate za $\delta C_i$ uporabimo za analizo nekaterih opazljivk, ki \v{s}e niso merjene s tako natan\v{c}nostjo, da bi bistveno prispevale k omejevanju velikosti parametrov $\kappa_j$. Za te opazljivke lahko preverimo napovedi odstopanja od SM, ki so \v{s}e kompatibilne z izpeljanimi mejami. To prikazuje Sl.~\ref{fig:predict1!} za razvejitvena razmerja $\mathrm{Br}[\bar B_s\to\mu^+\mu^-]$,  $\mathrm{Br}[B\to K^{(*)}\nu\bar{\nu}]$ ter za asimetrijo $A_{\mathrm{FB}}(q^2)$ v razpadih $\bar{B}_d\to \bar{K}^*\ell^+\ell^-$.

V prvi vrsti lahko vidimo, da sta koeficienta $\kappa_{LRb}$ in $\kappa_{RR}$ tako mo\v{c}no omejena, da odstopanja v omenjenih razpadih ni mo\v{c} pri\v{c}akovati. Po drugi strani lahko opazimo, da najnovej\v{s}e meritve razvejitvenega razmerja za $B_s \to \mu^+ \mu^-$ postajajo uporabne za omejevanje ostalih parametrov. V primeru nadaljnjega spu\v{s}\v{c}anja zgornje eksperimentalne meje bi lahko ta opazljivka postala pomemben del omejevanja parametrov $\kappa_j$. Analiza $A_{\mathrm{FB}}(q^2)$ razkriva, da bistvenih odstopanj od SM v tej opazljivki ni pri\v{c}akovati. Predstavljen graf je za $\kappa_{LL}^{\prime\prime}$, katerega efekti so med ve\v{c}jimi. Za ostale parametre velja podobna ugotovitev.
\subsection{Su\v{c}nostni dele\v{z}i v razpadih kvarka top}
\begin{figure}[h!]
\begin{center}
\includegraphics[scale=0.6]{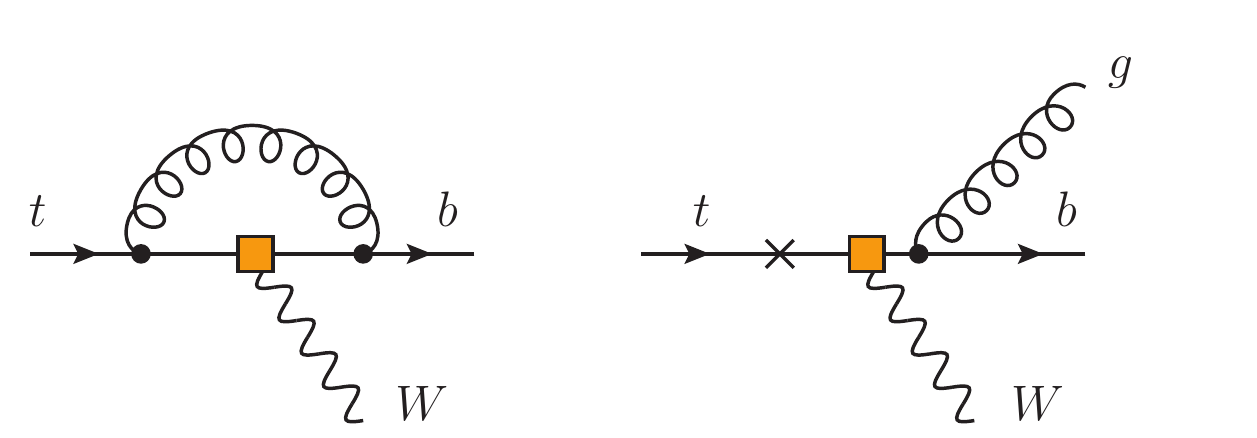}
\end{center}
\caption{Feynmanova diagrama za popravke QCD prvega red k glavnem razpadnem kanalu kvarka top $t \to W b$. Kvadratki ozna\v{c}ujejo anomalno vozli\v{s}\v{c}e, ki ga generira NF. Kri\v{z}ec ozna\v{c}uje dodatno to\v{c}ko iz katere se lahko izseva gluon.}
\label{fig:feyndiags!}
\end{figure}

\begin{figure}[h!]
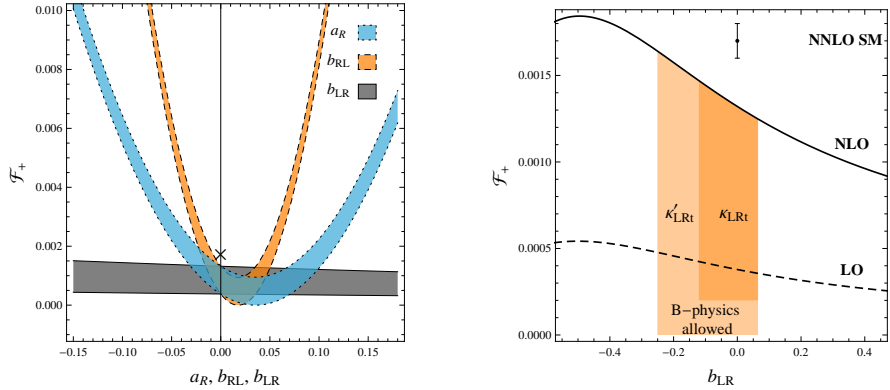

\begin{center}
\includegraphics[scale=0.5]{LOtoNLOFplusxx.pdf}\hspace{1cm}
\includegraphics[scale=0.5]{LOtoNLOFplusx2.pdf}
\end{center}
\caption{Odvisnost $\mathcal F_+$ od anomalnih sklopitev, za katere privzamemo, da so realne in da je le ena sklopitev razli\v{c}na od ni\v{c}. {\bf Levo}: Odvisnost ${\cal F}_+$ od $a_R$ (modro, pik\v{c}asto), $b_{RL}$ (oran\v{z}no, \v{c}rtkano) in $b_{LR}$ (\v{c}rno, polno). Zgornja in spodnja \v{c}rta pasu pripadata analizi brez in z prvim redom QCD popravkov. Kri\v{z}ec ozna\v{c}uje napoved SM. {\bf Desno}: Odvisnost ${\cal F}_+$ od $b_{LR}$. \v{C}rtkana \v{c}rta predstavlja rezultat brez popravkov QCD, polna \v{c}rta pa vklju\v{c}uje na\v{c}e popravke prvega reda. Prikazana je tudi SM vrednost podana v En.~(\ref{eq:e22b!}) in $95\%$ C.L. dovoljena intervala za $b_{LR}$ podana v En.~(\ref{eq:ind_translated!}).}
\label{fig:F+!}
\end{figure}

\begin{figure}[h!]
\begin{center}
\includegraphics[scale=0.5]{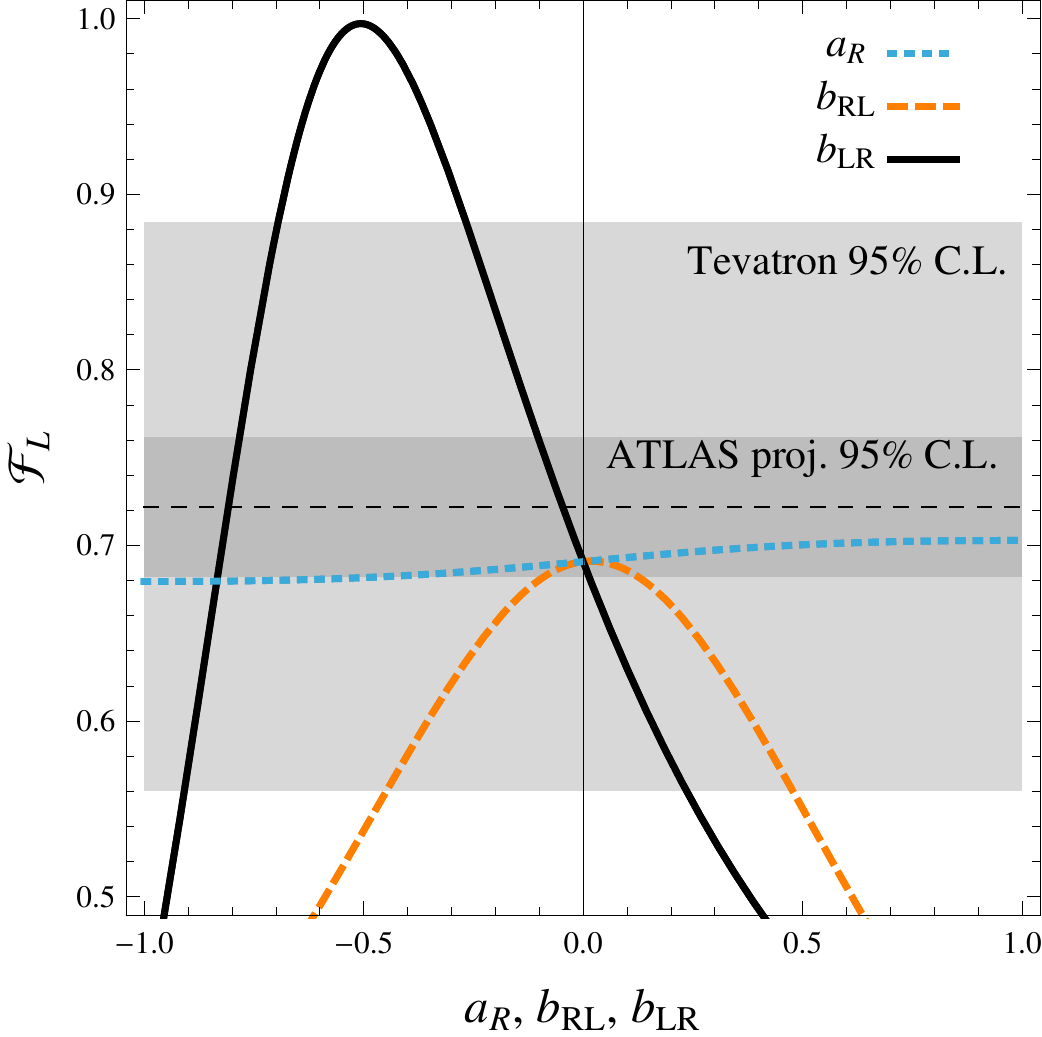}\hspace{1cm}
\includegraphics[scale=0.5]{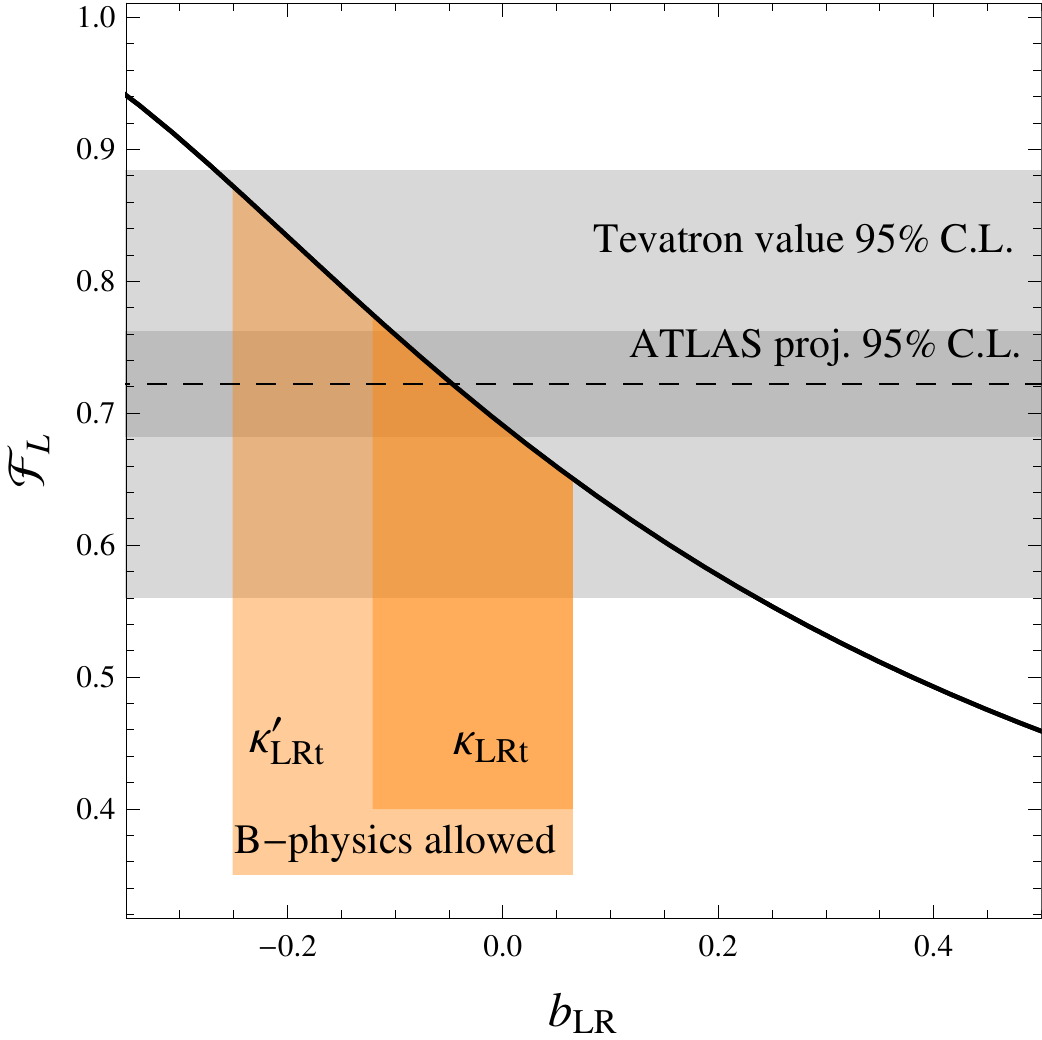}
\end{center}
\caption{Odvisnost $\mathcal F_L$ od anomalnih sklopitev, za katere privzamemo, da so realne in da je le ena sklopitev razli\v{c}na od ni\v{c}. Prikazana je tudi srednja vrednost meritve iz Tevatrona (\v{c}rtkano) in pripadajo\v{c}i 95\% C.L. interval. Poleg tega je prikazana tudi predvidena bodo\v{ca} velikost 95\% C.L. intervala iz Atlasa na podlago centralne vrednosti iz Tevatrona. {\bf Levo}: Odvisnost ${\cal F}_L$ od $a_R$ (modro, pik\v{c}asto), $b_{RL}$ (oran\v{z}no, \v{c}rtkano) and $b_{LR}$ (\v{c}rno, polno). {\bf Desno}: Odvisnost $\mathcal F_L$ od $b_{LR}$ in $95\%$ C.L. dovoljeni intervali za $b_{LR}$ podani v En.~(\ref{eq:ind_translated!}). }
\label{fig:FL!}
\end{figure}

Kot zadnje predstavimo na\v{s}o analizo vpliva operatorjev~(\ref{eq:ops1!}) na su\v{c}nostne dele\v{z}e v glavnem razpadnem kanalu kvarka top. V ta namen vpeljemo najsplo\v{s}nej\v{s}o obliko efektivnega $tWb$ vozli\v{s}\v{c}a preko slede\v{c}e Lagrangeove funkcije
\begin{eqnarray}
\mathcal L_{\mathrm{eff}} = -\frac{g}{\sqrt{2}}\bar{b}\Big[\gamma^{\mu} \big(a_L P_L +a_R P_R\big)
-(b_{RL} P_L + b_{LR} P_R)\frac{2\ii \sigma^{\mu\nu}}{m_t}q_{\nu}  \Big]t W_{\mu}\,,\label{eq:effsimple!}
\end{eqnarray}
kjer anomalne sklopitve lahko pove\v{z}emo s parametri $\kappa_j$
\begin{eqnarray}
\delta a_L =  V_{tb}^* \kappa_{LL}^{(\pr,\pr\pr)*}\,,\hspace{0.3cm} a_R = V_{tb}^{*}\kappa_{RR}^*\,,\hspace{0.3cm} b_{LR} = -\frac{m_t}{2 m_W}V_{tb}^{*}\kappa_{LRt}^{(\pr)}\,,\hspace{0.3cm} b_{RL} = -\frac{m_t}{2 m_W} V_{tb}^* \kappa_{LRb}^*\,.\label{eq:translation!}
\end{eqnarray}
V analizo vklju\v{c}imo popravke QCD prvega reda, ki jih prikazuje Sl.~\ref{fig:feyndiags!}.
Podobno kot pri analizi indirektnih efektov v fiziki mezonov $B$, bomo tudi tukaj (sprva) analizirali prispevke le enega operatorja naenkrat. Poleg tega se bomo omejili na realne vrednosti anomalnih sklopitev. Pod takimi pogoji $\delta a_L$ nima vpliva na su\v{c}nostne dele\v{z}e, saj spremlja vozli\v{s}\v{c}e z enako strukturo kot SM in se njegovi prispevki izni\v{c}ijo. Za ostale tri anomalne sklopitve analiziramo najprej spremembe v $\mathcal F_+$, kar prikazuje Sl.~\ref{fig:F+!}. Rezultati so prikazani kot pasovi, ki ponazarjajo pove\v{c}anje prispevkov k $\mathcal F_+$ ob vklju\v{c}itvi popravkov QCD prvega reda. Indirektne omejitve na parametre $\kappa_j$, ki smo jih izpeljali v prej\v{s}njem poglavju se prevedejo v 
\begin{eqnarray}
 -0.0006 \le  a_R \le 0.003\,,\hspace{0.5cm}
 -0.0004 \le b_{RL} \le 0.0016\,,\hspace{0.5cm}
 -0.14 (-0.29) \le b_{LR} \le 0.08\,.\label{eq:ind_translated!}
\end{eqnarray}
Iz Sl.~\ref{fig:F+!} lahko zaklju\v{c}imo, da tudi ob upo\v{s}tevanju QCD popravkov prvega reda, vrednosti $\mathcal F_+$ reda procent ali ve\v{c} nikakor ne morem pri\v{c}akovati. 
Ker je vrednost $\mathcal F_L$ izmerjena z bistveno bolj\v{s}no natan\v{c}nostjo, je vredno pogledati, ali morda izmerjena vrednost lahko slu\v{z}i za postavitev meje na velikost anomalnih sklopitev. Kako se $\mathcal F_L$ spreminja v odvisnosti od anomalnih sklopitev, je prikazano na Sl.~\ref{fig:FL!}. Vpliv korekcij QCD je v primeru $\mathcal F_L$ zanemarljiv. Anomalni sklopitvi  $a_R$ in $b_{RL}$ zaradi mo\v{c}nih indirektnih omejitev zopet me moreta znatno vplivati na vrednost dele\v{z}a. Po drugi strani manj omejeni $b_{LR}$ lahko znatno spremeni vrednost $\mathcal F_L$, kar je podrobneje prikazano na desnem grafu Sl.~\ref{fig:FL!}. Vidimo, da je meja, ki jo  izpeljemo iz razpadov kvarka top v tem primeru primerljiva z indirektni mejami in v prihodnosti, v kolikor bo predvidena senzitivnost Atlasa realizirana, lahko postane tudi dominantna.

\begin{figure}[h!]
\begin{center}
\includegraphics[scale= 0.6]{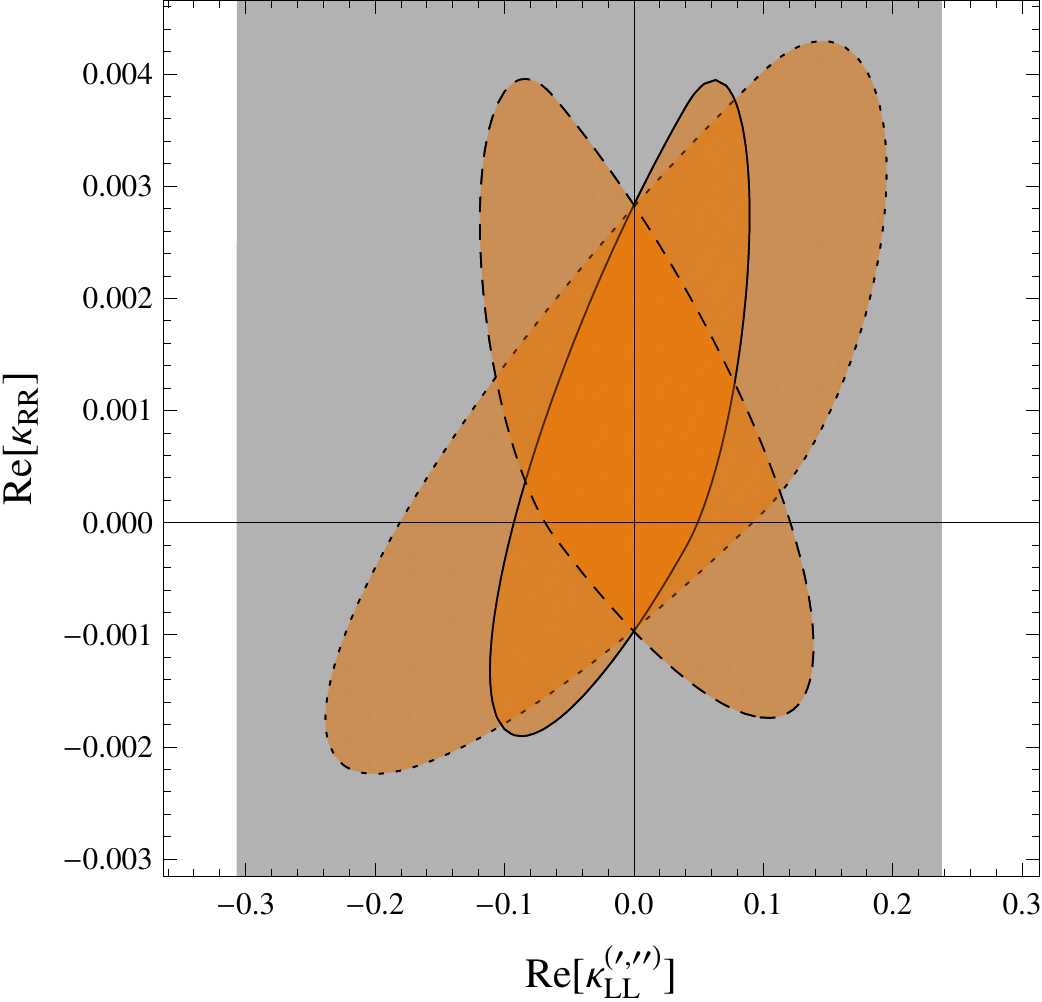}\hspace{0.8cm}
\includegraphics[scale=0.6]{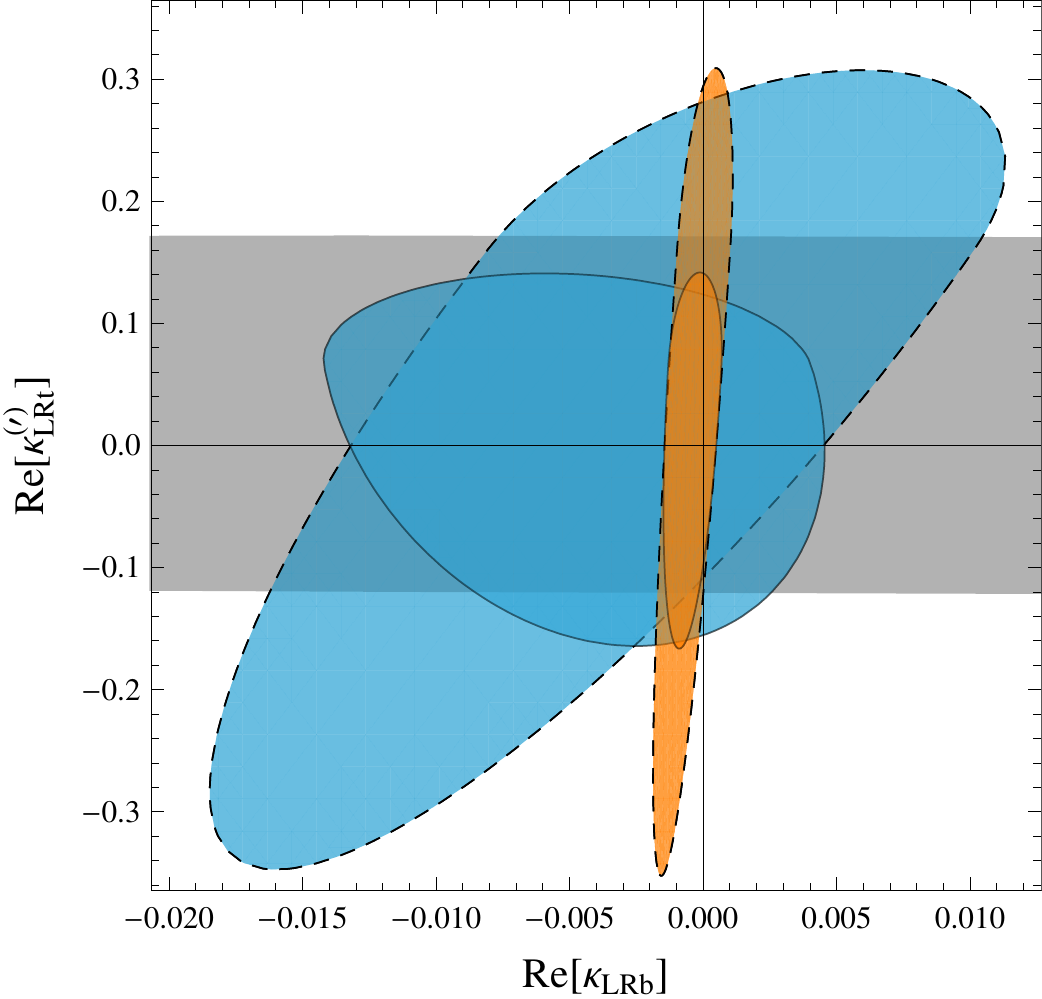}
\caption{95\% C.L. dovoljena obmo\v{c}ja v razli\v{c}nih $(\kappa_i,\kappa_j)$ ravninah. Siv pas predstavlja dovoljeno obmo\v{c}je kot dobljeno iz direktnih omejitev v Ref.~\cite{AguilarSaavedra:2011ct}. $\kappa_i$ so privzeto realni. {\bf Levo}: $\kappa_{RR}$ - $\kappa_{LL}$ (polno), $\kappa_{LL}^{\prime}$ (\v{c}rtkano), $\kappa_{LL}^{\pr\pr}$ (pik\v{c}asto) ravnina. Skala ujemanja je $\mu=2 m_W$. {\bf Desno}: $\kappa_{LRb}$ - $\kappa_{LRt}$ (polno), $\kappa_{LRt}^{\prime}$ (\v{c}rtkasto) ravnina. Skalo ujemanja spremenimo iz $\mu=2 m_W$ (o\v{z}ji obmo\v{c}ji) na $\mu=m_W$ (\v{s}ir\v{s}i obmo\v{c}ji).}
\label{fig:2d1!}
\end{center}
\end{figure}

\v{C}e bi analizirali poleg su\v{c}nostnih dele\v{z}ev \v{s}e kak\v{s}no drugo opazljivko iz fizike kvarka top, bi lahko sprostili omejitev, s katero predpostavimo le eno neni\v{c}elno anomalno sklopitev. Ravno to so naredili v Ref.~\cite{AguilarSaavedra:2011ct}, kjer so poleg su\v{c}nostnih dele\v{z}ev (brez popravkov QCD) v analizo vklju\v{c}ili tudi produkcijo enega kvarka top, ki tudi poteka preko \v{s}ibke interakcije in je ob\v{c}utljiva na anomalne $tWb$ sklopitve. Avtorji~\cite{AguilarSaavedra:2011ct} so analizirali dovoljena obmo\v{c}ja v ravninah parov anomalnih sklopitev. Zanimivo je njihove ugotovitve, ki temeljijo izklju\v{c}no na direktnih procesih, postaviti ob bok na\v{s}im ugotovitvam indirektnih omejitev. Ker smo z izpeljavo dovoljenih intervalov uporabili ve\v{c} kot eno spremenljivko, lahko prika\v{z}emo analogno analizo v ravninah parov anomalnih sklopitev, kar prikazuje Sl.~\ref{fig:2d1!}. 

Sivi obmo\v{c}ji, dobljeni iz direktnih omejitev, imata obliko pasu, ker sta v smereh $\kappa_{RR}$ in $\kappa_{LRb}$ mnogo \v{s}ir\v{s}a kot obmo\v{c}ja, ki jih dobimo iz obravnave indirektnih omejitev, kar upravi\v{c}uje na\v{s}e sklepanje, da je v primeru NF v fiziki kvarka top potrebno vedno analize direktnih opazljivk postaviti ob rob indirektnim analizam v fiziki mezonov.
\section{Zaklju\v{c}ki}
Ob koncu dobe Tevatrona smo \v{z}e globoko zakorakali v dobo LHC in lov na fiziko onkraj standardnega modela je v polnem razmahu. Iskanje novih delcev nikakor ni edini na\v{c}in, s katerim LHC teoreti\v{c}ni fiziki osnovnih delcev prina\v{s}a nova vpra\v{s}anja in odgovore. Nadejamo se razjasnitve problema fizike okusa, v kateri kvark top s svojo veliko maso igra vodilno vlogo. Ker LHC lahko smatramo kot pravo tovarno kvarka top, nam je prvi\v{c} na voljo raziskava fizike kvarka top z veliko natan\v{c}nostjo. Dolo\v{c}itve parametrov in interakcijskih struktur kvarka top nam lahko slu\v{z}i kot okno v svet nove fizike. V tem delu smo preu\v{c}ili razli\v{c}ne aspekte razpadov kvarka top in raziskovali, kako se NF, ki smo jo parametrizirali s pomo\v{c}jo efektivnih teorij, lahko v njih manifestira. 

Na eni strani smo raziskovali razpadne \v{s}irine in upo\v{s}tevali tudi popravke prvega reda v QCD, kar je smiselno, ko imamo opravka s kvarki in smo soo\v{c}eni z vedno ve\v{c}jo natan\v{c}nostjo eksperimentalnih meritev. Preu\v{c}ili smo razvejitvena razmerja razpadov $t\to q \gamma,Z$ in razli\v{c}ne kinemati\v{c}ne opazljivke v tro-del\v{c}nem razpadu $t\to q \ell^+ \ell^-$ ter glavni razpadni kanal kvarka top $t\to W b$, kjer smo pozornost usmerili v su\v{c}nostne dele\v{z}e bozona $W$, ki so ob\v{c}utljivi na strukturo $tWb$ vozli\v{s}\v{c}a.

Po drugi strani smo izpostavljali vlogo, ki jo igra top kvark v fiziki mezonov kot virtualen delec. Posledi\v{c}no dolo\v{c}ene modifikacije fizike kvarka top lahko vplivajo na teoreti\v{c}ne napovedi opazljivk v fiziki mezonov. Ker za modifikacije nabitih tokov, ki vsebujejo kvarke top celostne analize indirektnih posledic ni mo\v{c} najti v literaturi, smo podrobno preu\v{c}ili posledice v $|\Delta B|=2$ in $|\Delta B|=1$ procesih. Po pri\v{c}akovanjih smo lahko na NF postavili ne-trivialne indirektne omejitve.

Ne glede na to, ali nam LHC v razpadih kvarka top razkrije novo fiziko ali ne, bodo bodo\v{c}e meritve v fiziki kvarka top igrale pomembno vlogo v raziskovanju fizike okusa in grajenja ali omejevanja modelov fizike onkraj standardnega modela.


\end{document}